\documentclass[10pt]{article}
\usepackage{amsmath,graphicx,url,times}
\usepackage{siunitx}
\usepackage{geometry}                
\usepackage[nolist]{acronym}
\usepackage{pdflscape} 
\usepackage{afterpage}
\usepackage{capt-of} 
\usepackage{amssymb, amsmath, epsfig, rotating, multirow, url}
\usepackage{mathabx} 
\usepackage{epstopdf}
\usepackage{url}
\usepackage{hyperref}
\usepackage[super]{nth}
\usepackage{xspace} 
\usepackage{balance} 
\usepackage{authblk}
\usepackage{subfig}
\newif\ifarXiv
\arXivtrue
\title{Acoustic Characterization of Environments (ACE) Challenge Results Technical Report}
\date{}                                           
\author[1]{James Eaton\thanks{j.eaton11@imperial.ac.uk}}
\author[2]{Nikolay D. Gaubitch\thanks{n.d.gaubitch@tudelft.nl}}
\author[1]{Alastair H. Moore\thanks{alastair.h.moore@imperial.ac.uk}}
\author[1]{Patrick A. Naylor\thanks{p.naylor@imperial.ac.uk}}
\affil[1]{Dept. of Electrical and Electronic Engineering, Imperial College London, UK}
\affil[2]{SIP Lab, Delft University of Technology, Netherlands, and Pindrop Security}
\newif\ifBestPerform
\BestPerformtrue
\BestPerformfalse
\newif\ifIncludeFigsByParameter
\IncludeFigsByParametertrue
\newif\ifMakeRoomTableLandscape
\MakeRoomTableLandscapefalse
\ifarXiv
\newcommand{\expect}{E}
\newcommand{\PearsonCC}{\rho}
\newcommand{\figWidthConf}{80mm}
\newcommand{\figWidthACETR}{160mm}

\newcommand{\figWidthJournNarrow}{60mm}
\newcommand{\figMidSent}{Fig.}
\newcommand{\tabMidSent}{Table}
\newcommand{\tabStartSent}{Table}
\newcommand{\tabsStartSent}{Tables}
\newcommand{\sectMidSent}{Sec.\@\xspace}
\newcommand{\etal}{\emph{et al.}\@\xspace}
\newcommand{\etAl}[1]{{\etal}~\cite{#1}}
\newcommand{\dBel}[1]{\SI{#1}{\deci\bel}}
\newcommand{\Ex}[1]{\expect\{#1\}}
\else

\fi
\begin{document}
\begin{acronym}
\ifarXiv
\acro{ABC}{Analytical with or without Bias Compensation}
\acro{ACE}{Acoustic Characterization of Environments\acroextra{. A noisy reverberant speech corpus and IEEE challenge run by the SAP group at Imperial College}}
\acro{AIR}{Acoustic Impulse Response}
\acro{ANU}{Australian National University}
\acro{DENBE}{\ac{DRR} Estimation using a Null-Steered Beamformer}
\acro{DoA}{Direction-of-Arrival}
\acro{DRR}{Direct-to-Reverberant Ratio}
\acro{FAU}{Friedrich-Alexander-Universit{\"a}t}
\acro{FB2}[FB]{Fullband}
\acro{ISO}{Intl. Organization for Standardization}
\acro{MLMF}{Machine Learning with Multiple Features}
\acro{MSE}{Mean Square Error}
\acro{NIRA}{Non-Intrusive Room Acoustics}
\acro{NOSRMR}{Normalized Overall \ac{SRMR}}
\acro{NSRMR}{Normalised \ac{SRMR}}
\acro{NSV}{Negative-Side Variance}
\acro{OSRMR}{Overall \ac{SRMR}}
\acro{RTF}{Real-Time Factor}
\acro{SB}{Subband}
\acro{SDD}{Spectral Decay Distributions}
\acro{SDDSA}{\ac{SDD} with Mel-spaced frequency bands and selective averaging}
\acro{SDDSA-G}{\ac{SDDSA} with Gerkmann noise estimator}
\acro{SDDSA-H}{\ac{SDDSA} with Hendriks noise estimator}
\acro{SFM}{Single Feature with Mapping}
\acro{SNR}{Signal-to-Noise Ratio}
\acro{SNR2}[SNR]{Speech-to-Noise Ratio}
\acro{SRMR}{Speech-to-Reverberation Modulation Energy Ratio}
\acro{T60}[$T_\textrm{60}$]{Reverberation Time\acroextra{ to decay by $60$ dB}}
\else
\input{../SapBibTex/sapacronyms.txt}
\fi
\acro{Dev}{Development}
\acro{Eval}{Evaluation}
\end{acronym}
\maketitle
\begin{sloppy}
\begin{abstract}
This document provides supplementary information, and the results of the tests of acoustic parameter estimation algorithms on the \ac{ACE} Challenge~\cite{Eaton2015a} Evaluation dataset which were subsequently submitted and written up into papers for the Proceedings of the \ac{ACE} Challenge~\cite{Eaton2015d}.
This document is supporting material for a forthcoming journal paper on the \ac{ACE} Challenge which will provide further analysis of the results.
\end{abstract}
\clearpage
\tableofcontents
\listoffigures
\addcontentsline{toc}{section}{List of Figures}
\listoftables
\addcontentsline{toc}{section}{List of Tables}
%
\acresetall
\clearpage
\section{Introduction}
This document provides supplementary information and the results of the \ac{ACE} Challenge Phases 1 and 2 for all the tasks, \ac{T60} and \ac{DRR} estimation in both fullband and in frequency bands.
%
\subsection{Room recording procedure}
The recording procedure  in each room involved the following steps:
\begin{enumerate}
\item Install recording equipment positioning the microphones in Position 1, and document the room dimensions and the positions of all microphones, sources and seats;
\item Make empty-room \ac{AIR} measurements and noise recordings;
Empty-room measurements were for verification purposes and do not form part of the published corpus since the set is not complete, although they may be used in future experiments;
\item Participating subjects take their seating positions;
\item Make occupied \ac{AIR} measurements and noise recordings;
\item Move microphones to Position 2 and document their positions;
\item Make occupied \ac{AIR} measurements and noise recordings;
\item Participants leave the room;
\item Make unoccupied \ac{AIR} measurements and noise recordings in the second microphone position;
\item Uninstall recording equipment.
\end{enumerate}
\subsection{Room properties}
\subsubsection{Room dimension and microphone positions}
{\tabsStartSent}~\ref{tab:roomdims} and~\ref{tab:roomdimspart2} give the room dimensions and positions of the centre of each microphone array.
Also included is the position of the source and each of the fans used to create the fan noise.
Between 1 and 3 fans were used depending on the size of the room.
The microphone elements in the cruciform. mobile, linear array and Chromebook are assumed to be omnidirectional.
The look direction is provided for the source and Eigenmike since these do not have an omnidirectional directivity pattern.
This look direction also applies to the orientation of the 8-channel linear array which was always perpendicular to the look direction of the Eigenmike.
The 3-element mobile array was mounted with the longer edge with two microphones perpendicular to the look direction of the Eigenmike.
The individual elements in the Eigenmike are omnidirectional, but are mounted on a solid baffle.
The look direction is specified in degrees, where 0 degrees is in the direction of $x = \infty $, and a positive angle is towards $y=\infty$ as illustrated in {\figMidSent}~\ref{fig:coords}.
\begin{figure}[!httt]
\centerline{\epsfig{figure=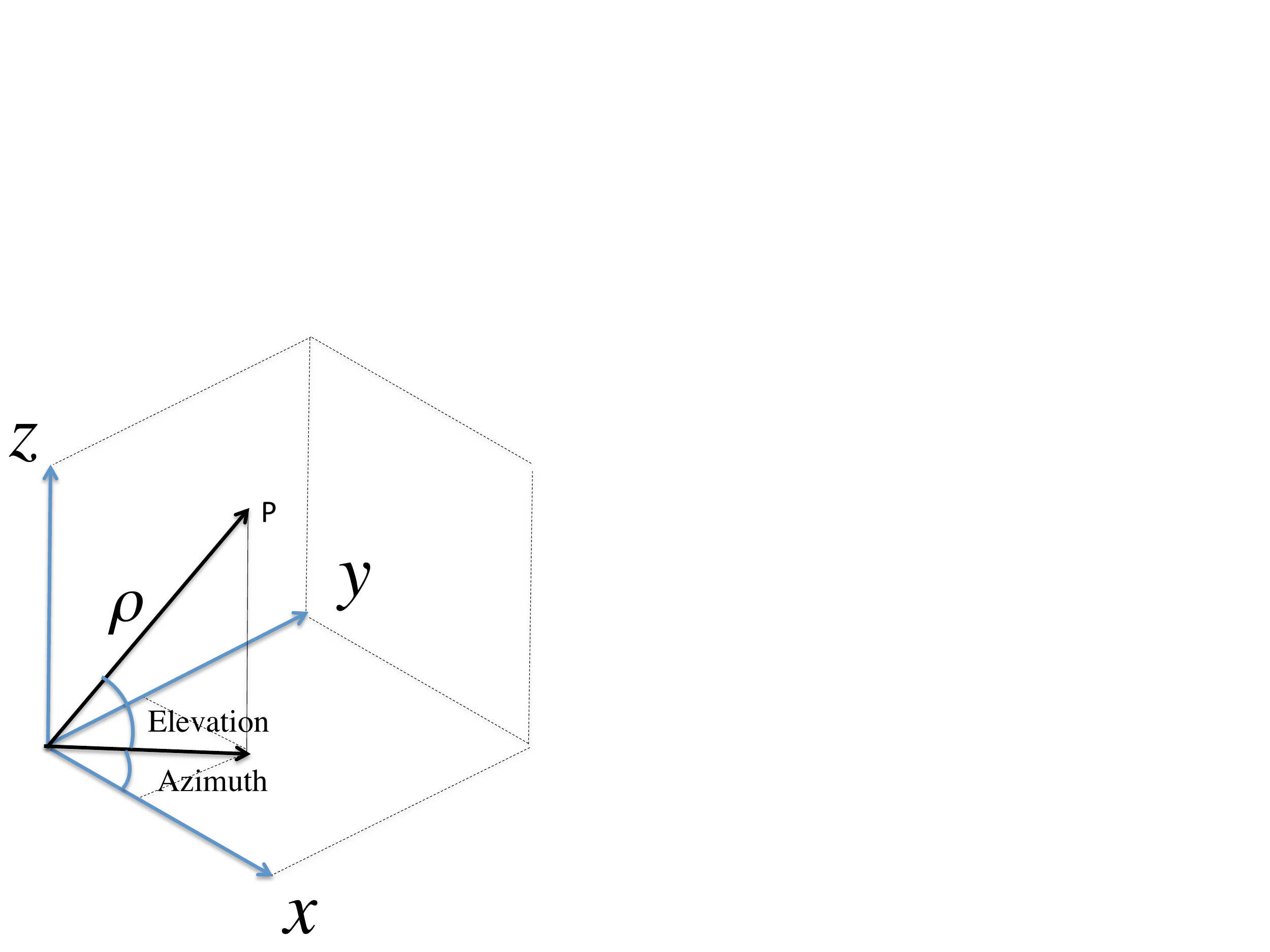,width=\figWidthConf,viewport=0 0 310 350,clip}}
\caption{{Coordinate system used in tables
}} 
\label{fig:coords}
\end{figure}
%
In the \ac{ACE} Challenge, the Dev dataset used channel 1 of the 8-channel linear array, whilst for the Eval dataset, channel 1 of the 5-channel cruciform was used.
Channel 1 of the 5-channel cruciform was the central microphone which is the same position as for the 5-channel array.
Therefore, the position of channel 1 of the 8-channel linear array is provided.
%
%
Where orientation was possible, fans faced in the same look direction as the source.
\ifMakeRoomTableLandscape
\afterpage{%
\clearpage%
\thispagestyle{empty}
\begin{landscape}
\fi
%
\begin{table*}[!h]
\small
\caption{Room dimensions, source, microphone and fan positions}
\vspace{5mm} 
\centering
\begin{tabular}{lccccccccccccccc}%
\hline%
Room
& Mic.
& Dimensions
& \multicolumn{2}{c}{Source}
& 5-channel
& 3-channel
& 8-channel
\\
Name
& Pos.
& (L, W, H)
& Position
& Look dir.
& Cruciform
& Mobile
& Linear array
\\
\hline%
Office 1 & 1 & (3.32, 4.83, 2.95) & (2.06, 1.04, 1.19) & 90 & (2.66, 2.14, 1.19) & (2.29, 2.15, 1.19) & (1.92, 2.14, 1.19) & \\ %
\hline
Office 1 & 2 & (3.32, 4.83, 2.95) & (2.06, 1.04, 1.19) & 90 & (2.49, 3.69, 1.19) & (2.15, 3.69, 1.19) & (1.79, 3.67, 1.19) & \\ %
\hline
Office 2 & 1 & (3.22, 5.1, 2.94) & (1.41, 1.73, 1.19) & 90 & (1.25, 2.81, 1.19) & (1.25, 2.62, 1.19) & (0.84, 2.78, 1.19) & \\ %
\hline
Office 2 & 2 & (3.22, 5.1, 2.94) & (1.41, 1.73, 1.19) & 90 & (2.25, 4.35, 1.19) & (2.05, 4.16, 1.19) & (1.58, 4.16, 1.19) & \\ %
\hline
Meeting Room 1 & 1 & (6.61, 5.11, 2.95) & (1.39, 1.26, 1.19) & 0 & (2.74, 0.48, 1.19) & (2.74, 0.82, 1.19) & (2.74, 1.14, 1.19) & \\ %
\hline
Meeting Room 1 & 2 & (6.61, 5.11, 2.95) & (1.39, 1.26, 1.19) & 0 & (3.96, 0.52, 1.19) & (3.96, 0.85, 1.19) & (3.96, 1.14, 1.19) & \\ %
\hline
Meeting Room 2 & 1 & (10.3, 9.07, 2.63) & (4.65, 4.07, 1.19) & 180 & (3, 4.39, 1.19) & (3, 3.99, 1.19) & (3, 3.59, 1.19) & \\ %
\hline
Meeting Room 2 & 2 & (10.3, 9.07, 2.63) & (4.65, 4.07, 1.19) & 180 & (2, 4.39, 1.19) & (2, 3.99, 1.19) & (2, 3.59, 1.19) & \\ %
\hline
Lecture Room 1 & 1 & (6.93, 9.73, 3) & (3.65, 3.73, 1.19) & 180 & (2.81, 3.84, 1.19) & (2.8, 3.44, 1.19) & (2.89, 3.04, 1.19) & \\ %
\hline
Lecture Room 1 & 2 & (6.93, 9.73, 3) & (3.65, 3.73, 1.19) & 180 & (1.07, 3.92, 1.19) & (1.07, 3.52, 1.19) & (1.07, 3.12, 1.19) & \\ %
\hline
Lecture Room 2 & 1 & (13.6, 9.29, 2.94) & (6.03, 3.14, 1.19) & 180 & (5.09, 5.87, 1.19) & (5.09, 5.47, 1.19) & (5.09, 5.07, 1.19) & \\ %
\hline
Lecture Room 2 & 2 & (13.6, 9.29, 2.94) & (6.03, 3.14, 1.19) & 180 & (3.93, 5.87, 1.19) & (3.93, 5.47, 1.19) & (3.93, 5.07, 1.19) & \\ %
\hline
Building Lobby & 1 & (4.47, 5.13, 3.18) & (1.98, 0.61, 1.19) & 90 & (2.69, 2.02, 1.19) & (2.33, 2, 1.19) & (1.95, 2.03, 1.19) & \\ %
\hline
Building Lobby & 2 & (4.47, 5.13, 3.18) & (1.98, 0.61, 1.19) & 90 & (2.62, 3.51, 1.19) & (2.25, 3.49, 1.19) & (1.86, 3.54, 1.19) & \\ %
\hline

\hline
\end{tabular}%
\label{tab:roomdims}
\end{table*}%
%
\begin{table*}[!h]
\small
\caption{Room dimensions, source, microphone and fan positions continued}
\vspace{5mm} 
\centering
\begin{tabular}{lccccccccccccccc}%
\hline%
Room
& Mic.
& \multicolumn{2}{c}{32-ch. Eigenmike}
& 2-channel
& Ch 1 of
& Fan
& Fan
& Fan\\
Name
& Pos.
& Position
& Look dir.
& Chromebook
& 8-ch. linear
& 1
& 2
& 3\\
\hline%
Office 1 & 1 & (2.94, 2.15, 1.19) & -90 & (3.02, 2.15, 0.68) & (1.68, 2.14, 1.19) & (0.56, 1.43, 1.19) \\ %
\hline
Office 1 & 2 & (2.92, 3.69, 1.19) & -90 & (2.97, 3.49, 0.68) & (1.55, 3.67, 1.19) & (0.56, 1.43, 1.19) \\ %
\hline
Office 2 & 1 & (2.69, 2.84, 1.19) & -90 & (2.04, 2.84, 0.68) & (0.6, 2.78, 1.19) & (2.75, 1.25, 1.19) \\ %
\hline
Office 2 & 2 & (1.25, 4.25, 1.19) & -90 & (0.83, 4.13, 0.68) & (1.34, 4.16, 1.19) & (2.75, 1.25, 1.19) \\ %
\hline
Meeting Room 1 & 1 & (2.74, 0.17, 1.19) & 180 & (2.74, 1.65, 0.68) & (2.74, 1.38, 1.19) & (1.62, 2.2, 1.19) \\ %
\hline
Meeting Room 1 & 2 & (3.96, 0.21, 1.19) & 180 & (3.96, 1.55, 0.68) & (3.96, 1.38, 1.19) & (1.62, 2.2, 1.19) \\ %
\hline
Meeting Room 2 & 1 & (3, 4.79, 1.19) & 0 & (3, 5.19, 0.68) & (3, 3.35, 1.19) & (3.6, 3.15, 0.35) & (3.9, 3.25, 0.35) \\ %
\hline
Meeting Room 2 & 2 & (2, 4.79, 1.19) & 0 & (2, 5.19, 0.68) & (2, 3.35, 1.19) & (3.6, 3.15, 0.35) & (3.9, 3.25, 0.35) \\ %
\hline
Lecture Room 1 & 1 & (2.79, 4.24, 1.19) & 0 & (2.83, 4.64, 0.68) & (2.89, 2.8, 1.19) & (3.65, 3.98, 0.35) & (3.65, 3.48, 0.35) & (3.65, 3.25, 0.1) \\ %
\hline
Lecture Room 1 & 2 & (1.16, 4.32, 1.19) & 0 & (1.16, 4.72, 0.68) & (1.07, 2.88, 1.19) & (3.65, 3.98, 0.35) & (3.65, 3.48, 0.35) & (3.65, 3.25, 0.1) \\ %
\hline
Lecture Room 2 & 1 & (5.09, 6.27, 1.19) & 0 & (5.09, 6.67, 0.68) & (5.09, 4.83, 1.19) & (6.1, 2.82, 0.35) & (6.1, 3.43, 0.35) \\ %
\hline
Lecture Room 2 & 2 & (3.93, 6.27, 1.19) & 0 & (3.93, 6.67, 0.68) & (3.93, 4.83, 1.19) & (6.1, 2.82, 0.35) & (6.1, 3.43, 0.35) \\ %
\hline
Building Lobby & 1 & (3.1, 2.04, 1.19) & -90 & (2.1, 3.49, 0.72) & (1.71, 2.03, 1.19) & (1.74, 0.68, 0.35) \\ %
\hline
Building Lobby & 2 & (2.95, 3.49, 1.19) & -90 & (3.35, 3.41, 0.72) & (1.62, 3.54, 1.19) & (1.74, 0.68, 0.35) \\ %
\hline

\hline
\end{tabular}%
\label{tab:roomdimspart2}
\end{table*}%
\ifMakeRoomTableLandscape
\end{landscape}
\clearpage
}
\fi
\subsubsection{Talker positions for babble noise}
{\tabStartSent}~\ref{tab:talkpos} provides each of the talker positions used to produce the babble noise.
The $z$ coordinates are not provided since these were not captured.
However, the talkers were seated and their mouths were situated at approximately the same height as the microphone arrays which were all at \SI{1.19}{\metre} above the floor.
\begin{table*}[!h]
\small
\caption{Talker positions used to produce babble noise}
\vspace{5mm} 
\centering
\begin{tabular}{lllllllll}%
\hline%
Room
& \multicolumn{7}{c}{Talker ID and associated $x$-$y$ coordinates}\\
name
& 1
& 2
& 3
& 4
& 5
& 6
& 7\\
\hline%
Office 1 & F6:(2.95, 0.85) & M10:(2.37, 0.65) & M11:(1.64, 0.68) & M17:(1.25, 1.18) \\ %
\hline
Office 2 & F7:(0.84, 0.55) & M16:(0.6, 1.39) & M10:(0.55, 2.15) & M12:(2.12, 0.4) & M20:(2.07, 1.25) & M15:(2.48, 1.9) \\ %
\hline
Meeting Room 1 & F7:(0.4, 0.95) & F8:(1.15, 0.4) & M10:(0.65, 3.25) & M11:(0.55, 0.3) & M12:(0.37, 1.78) & M23:(0.37, 2.7) \\ %
\hline
Meeting Room 2 & F8:(5.8, 4.53) & M10:(4.39, 2.89) & M11:(5.45, 5.32) & M12:(5.45, 2.95) & M13:(4.98, 5.68) & M14:(4.37, 5.96) & M23:(5.65, 3.73) \\ %
\hline
Lecture Room 1 & F6:(4.75, 3.55) & M11:(4.65, 3.25) & M13:(3.65, 5.08) & M14:(4.45, 2.75) & M18:(4.65, 3.9) & M19:(4.55, 4.38) \\ %
\hline
Lecture Room 2 & F6:(7.2, 2.75) & M10:(6.27, 4.36) & M11:(6.65, 2.12) & M12:(7.07, 3.49) & M13:(6.11, 1.77) & M14:(5.68, 4.52) & M23:(6.82, 4.01) \\ %
\hline
Building Lobby & M10:(1.23, 0.53) & M13:(2.72, 0.53) & M14:(2.23, 0.53) & M21:(0.93, 0.53) & M22:(3.21, 0.53) \\ %
\hline

\hline
\end{tabular}%
\label{tab:talkpos}
\end{table*}%
\subsubsection{Distances and look directions}
{\tabStartSent}~\ref{tab:roomdists} provides the source-microphone distances and \acp{DoA} in spherical coordinates,
whilst {\tabMidSent}~\ref{tab:fandists} provides the fan-microphone distances and \acp{DoA} in spherical coordinates.
%
\begin{table*}[!h]
\tiny
\caption{Source--microphone distances and \acp{DoA} in spherical coordinates}
\vspace{5mm} 
\centering
\begin{tabular}{lcccccccccccccccccccc}%
\hline%
&
&\multicolumn{3}{c}{5-ch. cruciform}
&\multicolumn{3}{c}{3-ch. mobile}
&\multicolumn{3}{c}{8-ch. linear}
&\multicolumn{3}{c}{32-ch. Eigenmike}
&\multicolumn{3}{c}{2-ch. Chromebook}
&\multicolumn{3}{c}{Ch-1. of 8-ch. lin.}
\\

& 
& $\rho$
& Azi
& Elev
& $\rho$
& Azi
& Elev
& $\rho$
& Azi
& Elev
& $\rho$
& Azi
& Elev
& $\rho$
& Azi
& Elev
& $\rho$
& Azi
& Elev
\\
Name
& Pos.
& (m)
& ($^{\circ}$)
& ($^{\circ}$)
& (m)
& ($^{\circ}$)
& ($^{\circ}$)
& (m)
& ($^{\circ}$)
& ($^{\circ}$)
& (m)
& ($^{\circ}$)
& ($^{\circ}$)
& (m)
& ($^{\circ}$)
& ($^{\circ}$)
& (m)
& ($^{\circ}$)
& ($^{\circ}$)
\\
\hline%
Office 1 & 1 & 1.25 & -28.6 & 0 & 1.13 & -11.7 & 0 & 1.11 & 7.25 & 0 & 1.42 & -38.4 & 0 & 1.55 & -40.9 & 19.2 & 1.16 & 19.1 & 0 & \\ %
\hline
Office 1 & 2 & 2.68 & -9.22 & 0 & 2.65 & -1.95 & 0 & 2.64 & 5.86 & 0 & 2.79 & -18 & 0 & 2.66 & -20.4 & 11 & 2.68 & 11 & 0 & \\ %
\hline
Office 2 & 1 & 1.09 & 8.43 & 0 & 0.904 & 10.2 & 0 & 1.19 & 28.5 & 0 & 1.69 & -49.1 & 0 & 1.37 & -29.6 & 21.8 & 1.33 & 37.6 & 0 & \\ %
\hline
Office 2 & 2 & 2.75 & -17.8 & 0 & 2.51 & -14.8 & 0 & 2.44 & -4 & 0 & 2.53 & 3.63 & 0 & 2.52 & 13.6 & 11.7 & 2.43 & 1.65 & 0 & \\ %
\hline
Meeting Room 1 & 1 & 1.55 & -30.1 & 0 & 1.41 & -18.1 & 0 & 1.35 & -5.07 & 0 & 1.73 & -39 & 0 & 1.49 & -344 & 20.1 & 1.35 & -355 & 0 & \\ %
\hline
Meeting Room 1 & 2 & 2.67 & -16.1 & 0 & 2.59 & -9.07 & 0 & 2.56 & -2.66 & 0 & 2.77 & -22.2 & 0 & 2.63 & -354 & 11.2 & 2.56 & -357 & 0 & \\ %
\hline
Meeting Room 2 & 1 & 1.68 & -11 & 0 & 1.65 & 2.78 & 0 & 1.72 & 16.2 & 0 & 1.8 & -23.6 & 0 & 2.06 & -34.2 & 14.3 & 1.8 & 23.6 & 0 & \\ %
\hline
Meeting Room 2 & 2 & 2.67 & -6.89 & 0 & 2.65 & 1.73 & 0 & 2.69 & 10.3 & 0 & 2.75 & -15.2 & 0 & 2.92 & -22.9 & 10.1 & 2.75 & 15.2 & 0 & \\ %
\hline
Lecture Room 1 & 1 & 0.847 & -7.46 & 0 & 0.898 & 18.8 & 0 & 1.03 & 42.2 & 0 & 1 & -30.7 & 0 & 1.33 & -48 & 22.6 & 1.2 & 50.7 & 0 & \\ %
\hline
Lecture Room 1 & 2 & 2.59 & -4.21 & 0 & 2.59 & 4.65 & 0 & 2.65 & 13.3 & 0 & 2.56 & -13.3 & 0 & 2.73 & -21.7 & 10.8 & 2.72 & 18.2 & 0 & \\ %
\hline
Lecture Room 2 & 1 & 2.89 & -71 & 0 & 2.51 & -68 & 0 & 2.15 & -64 & 0 & 3.27 & -73.3 & 0 & 3.69 & -75.1 & 7.95 & 1.93 & -60.9 & 0 & \\ %
\hline
Lecture Room 2 & 2 & 3.44 & -52.4 & 0 & 3.14 & -48 & 0 & 2.85 & -42.6 & 0 & 3.77 & -56.1 & 0 & 4.14 & -59.3 & 7.08 & 2.7 & -38.8 & 0 & \\ %
\hline
Building Lobby & 1 & 1.58 & -26.7 & 0 & 1.43 & -14.1 & 0 & 1.42 & 1.21 & 0 & 1.82 & -38.1 & 0 & 2.92 & -2.39 & 9.26 & 1.45 & 10.8 & 0 & \\ %
\hline
Building Lobby & 2 & 2.97 & -12.4 & 0 & 2.89 & -5.36 & 0 & 2.93 & 2.35 & 0 & 3.04 & -18.6 & 0 & 3.15 & -26.1 & 8.57 & 2.95 & 7 & 0 & \\ %
\hline

\hline
\end{tabular}%
\label{tab:roomdists}
\end{table*}%
%
\begin{table*}[!h]
\tiny
\caption{Fan--microphone distances and \acp{DoA}}
\vspace{5mm} 
\centering
\begin{tabular}{lccccccccccccccccccccc}%
\hline%
&
&
&\multicolumn{3}{c}{Crucif}
&\multicolumn{3}{c}{Mobile}
&\multicolumn{3}{c}{Lin8Ch}
&\multicolumn{3}{c}{Eigenmike}
&\multicolumn{3}{c}{Chromebook}
&\multicolumn{3}{c}{Ch-1. of 8-ch. lin.}
\\

& 
& 
& $\rho$
& Azi
& Elev
& $\rho$
& Azi
& Elev
& $\rho$
& Azi
& Elev
& $\rho$
& Azi
& Elev
& $\rho$
& Azi
& Elev
& $\rho$
& Azi
& Elev
\\
Name
& Pos.
& Fan
& (m)
& ($^{\circ}$)
& ($^{\circ}$)
& (m)
& ($^{\circ}$)
& ($^{\circ}$)
& (m)
& ($^{\circ}$)
& ($^{\circ}$)
& (m)
& ($^{\circ}$)
& ($^{\circ}$)
& (m)
& ($^{\circ}$)
& ($^{\circ}$)
& (m)
& ($^{\circ}$)
& ($^{\circ}$)
\\
\hline%
Office 1 & 1 & 1 & 2.22 & -71.3 & 0 & 1.87 & -67.4 & 0 & 1.53 & -62.4 & 0 & 2.49 & -73.2 & 0 & 2.61 & -73.7 & 11.3 & 1.33 & -57.6 & 0 & \\ %
\hline
Office 1 & 2 & 1 & 2.97 & -40.5 & 0 & 2.76 & -35.1 & 0 & 2.56 & -28.8 & 0 & 3.27 & -46.2 & 0 & 3.21 & -49.5 & 9.14 & 2.45 & -23.8 & 0 & \\ %
\hline
Office 2 & 1 & 1 & 2.16 & 43.9 & 0 & 2.03 & 47.6 & 0 & 2.45 & 51.3 & 0 & 1.59 & 2.16 & 0 & 1.81 & 24.1 & 16.3 & 2.64 & 54.6 & 0 & \\ %
\hline
Office 2 & 2 & 1 & 3.14 & 9.16 & 0 & 2.99 & 13.5 & 0 & 3.14 & 21.9 & 0 & 3.35 & 26.6 & 0 & 3.5 & 33.7 & 8.38 & 3.23 & 25.9 & 0 & \\ %
\hline
Meeting Room 1 & 1 & 1 & 2.05 & -56.9 & 0 & 1.77 & -50.9 & 0 & 1.54 & -43.4 & 0 & 2.32 & -61 & 0 & 1.35 & -26.2 & 22.3 & 1.38 & -36.2 & 0 & \\ %
\hline
Meeting Room 1 & 2 & 1 & 2.88 & -35.7 & 0 & 2.7 & -29.9 & 0 & 2.57 & -24.3 & 0 & 3.07 & -40.3 & 0 & 2.48 & -15.5 & 11.9 & 2.48 & -19.3 & 0 & \\ %
\hline
Meeting Room 2 & 1 & 1 & 1.61 & -64.2 & -31.4 & 1.33 & -54.5 & -39.1 & 1.12 & -36.3 & -48.5 & 1.94 & -69.9 & -25.7 & 2.15 & -73.6 & -8.82 & 1.05 & -18.4 & -53 & \\ %
\hline
Meeting Room 2 & 1 & 2 & 1.68 & -51.7 & -30 & 1.44 & -39.4 & -35.8 & 1.28 & -20.7 & -41.1 & 1.97 & -59.7 & -25.2 & 2.16 & -65.1 & -8.77 & 1.24 & -6.34 & -42.8 & \\ %
\hline
Meeting Room 2 & 2 & 1 & 2.19 & -37.8 & -22.5 & 1.99 & -27.7 & -24.9 & 1.86 & -15.4 & -26.8 & 2.44 & -45.7 & -20.1 & 2.61 & -51.9 & -7.25 & 1.82 & -7.13 & -27.5 & \\ %
\hline
Meeting Room 2 & 2 & 2 & 2.37 & -31 & -20.8 & 2.21 & -21.3 & -22.4 & 2.11 & -10.1 & -23.5 & 2.59 & -39 & -19 & 2.74 & -45.6 & -6.93 & 2.08 & -3.01 & -23.8 & \\ %
\hline
Lecture Room 1 & 1 & 1 & 1.2 & 9.46 & -44.6 & 1.31 & 32.4 & -39.8 & 1.47 & 51 & -34.8 & 1.23 & -16.8 & -43.1 & 1.1 & -38.8 & -17.4 & 1.64 & 57.2 & -30.9 & \\ %
\hline
Lecture Room 1 & 1 & 2 & 1.24 & -23.2 & -42.6 & 1.2 & 2.69 & -44.6 & 1.22 & 30.1 & -43.7 & 1.42 & -41.5 & -36.2 & 1.46 & -54.7 & -13.1 & 1.32 & 41.8 & -39.5 & \\ %
\hline
Lecture Room 1 & 1 & 3 & 1.5 & -35.1 & -46.7 & 1.4 & -12.6 & -51.4 & 1.35 & 15.4 & -54.1 & 1.71 & -49 & -39.7 & 1.71 & -59.5 & -19.8 & 1.4 & 30.6 & -51 & \\ %
\hline
Lecture Room 1 & 2 & 1 & 2.71 & 1.33 & -18 & 2.75 & 10.1 & -17.8 & 2.85 & 18.4 & -17.2 & 2.65 & -7.78 & -18.5 & 2.62 & -16.6 & -7.24 & 2.93 & 23.1 & -16.7 & \\ %
\hline
Lecture Room 1 & 2 & 2 & 2.75 & -9.68 & -17.8 & 2.71 & -0.888 & -18 & 2.74 & 7.94 & -17.9 & 2.76 & -18.6 & -17.7 & 2.8 & -26.5 & -6.77 & 2.78 & 13.1 & -17.6 & \\ %
\hline
Lecture Room 1 & 2 & 3 & 2.88 & -14.6 & -22.2 & 2.81 & -5.97 & -22.8 & 2.8 & 2.88 & -22.9 & 2.92 & -23.3 & -21.9 & 2.95 & -30.6 & -11.3 & 2.83 & 8.16 & -22.7 & \\ %
\hline
Lecture Room 2 & 1 & 1 & 3.32 & -71.7 & -14.7 & 2.96 & -69.1 & -16.5 & 2.61 & -65.8 & -18.8 & 3.69 & -73.7 & -13.2 & 3.99 & -75.3 & -4.74 & 2.4 & -63.3 & -20.5 & \\ %
\hline
Lecture Room 2 & 1 & 2 & 2.77 & -67.5 & -17.6 & 2.43 & -63.7 & -20.3 & 2.1 & -58.4 & -23.6 & 3.13 & -70.4 & -15.6 & 3.41 & -72.7 & -5.55 & 1.92 & -54.2 & -25.9 & \\ %
\hline
Lecture Room 2 & 2 & 1 & 3.84 & -54.6 & -12.6 & 3.53 & -50.7 & -13.8 & 3.24 & -46 & -15 & 4.16 & -57.8 & -11.6 & 4.43 & -60.6 & -4.27 & 3.07 & -42.8 & -15.9 & \\ %
\hline
Lecture Room 2 & 2 & 2 & 3.37 & -48.4 & -14.4 & 3.09 & -43.2 & -15.8 & 2.85 & -37.1 & -17.2 & 3.67 & -52.6 & -13.2 & 3.91 & -56.2 & -4.84 & 2.72 & -32.8 & -18 & \\ %
\hline
Building Lobby & 1 & 1 & 1.84 & -35.3 & -27.1 & 1.67 & -24.1 & -30.2 & 1.6 & -8.84 & -31.6 & 2.1 & -45 & -23.6 & 2.86 & -7.3 & -7.44 & 1.59 & 1.27 & -31.9 & \\ %
\hline
Building Lobby & 2 & 1 & 3.08 & -17.3 & -15.8 & 2.98 & -10.3 & -16.4 & 2.98 & -2.4 & -16.4 & 3.17 & -23.3 & -15.4 & 3.19 & -30.5 & -6.66 & 2.98 & 2.4 & -16.4 & \\ %
\hline

\hline
\end{tabular}%
\label{tab:fandists}
\end{table*}%
%
%
\subsection{Taxonomy of algorithms submitted}
\label{sec:taxonomy}
There were three main classes of algorithms submitted to the \ac{ACE} Challenge:
\begin{enumerate}
\item {\ac{ABC}};
\item {\ac{SFM}};
\item {\ac{MLMF}}.
\end{enumerate}
The \ac{ABC} approaches derive the estimate for the acoustic parameter directly from the signal without requiring any prior information.
Bias compensation may be performed in order to account for noise or specific aspects of the source material.
An example of this is the maximum likelihood method~\cite{Lollmann2010} which directly produces the \ac{T60} estimate.

The \ac{SFM} approaches estimate a parameter from a signal that is correlated with the acoustic parameter to be estimated, and then apply a mapping function to give the acoustic parameter estimate.
An example of this is the \ac{SDD} method which determines \ac{NSV} from STFT bins and then applies a mapping to obtain the \ac{T60}.

The \ac{MLMF} approaches typically use many features of the source material to train a neural network which then estimates the acoustic parameter from the features of a test signal.
An example of this is the \ac{NIRA}~\cite{Parada2015} algorithm.

There were no hybrid approaches submitted to the \ac{ACE} Challenge although several participants applied noise reduction to the source signals before performing parameter estimation.

Algorithms are further classified as being either providing an estimate in \ac{FB2}, in frequency bands, or \acp{SB}.
\subsection{Results}
The participating institutions in the \ac{ACE} challenge along with their respective algorithms are listed in Table~\ref{tab:ACE_Participants} in order of appearance of their algorithms in the results tables.
%
%
\begin{table*}[]
\caption{\ac{ACE} Challenge participants}
\vspace{5mm} 
\centering
\begin{tabular}{lll}%
\hline%
\multirow{2}{*}{Participant} &
\multicolumn{2}{l}{Algorithms submitted (see results tables)} \\
& \ac{T60} & \ac{DRR} \\
\hline%
\hline%
\multirow{1}{*}{Federal University of Rio de Janeiro} & 
 \multirow{1}{*}{A} &{$\alpha$}  \\
\hline
\multirow{1}{*}{\ac{FAU}} &
\multirow{1}{*}{B, C, D, E} &\\
\hline%
\multirow{1}{*}{Imperial College London}  & 
\multirow{1}{*}{F, G} & \multirow{1}{*}{v, w, x, y, z}\\
\hline
\multirow{1}{*}{Fraunhofer IDMT} & 
\multirow{1}{*}{H, I, J, K, L, M, N, O}  & \multirow{1}{*}{p, q, r, s, t, u} \\
\hline
\multirow{1}{*}{MuSAELab} & 
\multirow{1}{*}{O, P, Q, R, S, T, U, V, W} &  \multirow{1}{*}{0, 1, 2, 3, 4, 5, 6, 7, 8, 9} \\
\hline
\multirow{1}{*}{Nuance Communications Inc.} & 
\multirow{1}{*}{X, Y, Z} &  \multirow{1}{*}{k, l, m}\\
\hline
\multirow{1}{*}{Microsoft Research} & 
 \multirow{1}{*}{a, b} & \\
\hline
University of Auckland/NTT & 
& {f, g, h, i, j} \\
\hline
\multirow{1}{*}{\ac{ANU}} & 
& \multirow{1}{*}{n} \\
\hline
\end{tabular}%
\label{tab:ACE_Participants}
\end{table*}%
For the fullband tasks the results are presented as box plots where there is a box shown for each algorithm.
Both single and multi-channel algorithms are shown in the same figures and tables.
On each box in the box plot, the central notch is the median, the edges of the box are the \nth{25} and \nth{75} percentiles, the whiskers extend to the most extreme data points not considered outliers.
Boxes are colour-coded according to algorithm class: 
 \ac{ABC}: yellow; 
 \ac{MLMF}: cyan;
 \ac{SFM}: green.

Outliers are plotted individually.
The algorithms are identified on the box plot by a single character which corresponds to the character in the table after the figure.
The results are sorted by the research group which achieved the highest correlation coefficient in the results across all noises and \acp{SNR} in fullband.
For the \ac{T60} fullband task, the last three algorithms are those compared in Gaubitch \etAl{Gaubitch2012} and are included as baselines to enable the progress made in blind \ac{T60} estimation since 2012 to be assessed.
Similarly, for the \ac{DRR} fullband task, Jeub \etAl{Jeub2011} is included as the last algorithm since this was a freely available estimator prior to the \ac{ACE} Challenge.
The correlation coefficient for each algorithm is plotted as a black cross in the same column as the algorithm.
The value is provided on the right hand $y$-axis.

A table of numerical results is also provided following each figure which also provides the legend for the algorithm identifiers, A, B, C, etc.
%
The columns in the table are as follows:
\begin{enumerate}
\item Ref., the identifier for each algorithm used on the $x$-axis of the preceding figure;
\item Algorithm, the name used by the respective \ac{ACE} Challenge participant to refer to their algorithm;
\item Class, the class of algorithm according to {\sectMidSent}~\ref{sec:taxonomy};
\item Mic. Config, the microphone configuration of the Evaluation dataset used to test the algorithm. Valid values are Single (1-channel), Chromebook (2-channel), Mobile (3-channel), Crucif (5-channel), Lin8Ch (8-channel), and EM32 (32-channel);
Further details of the microphone configurations can be found in~\cite{Eaton2015a};
\item Bias, the mean error in the results. ;
Let $X = [x_0, x_1, \dots x_{N-1}]$ equal the set of $N$ ground truth \ac{T60} and \ac{DRR} measurements, and let $\hat{X}$ equal the set of estimated results defined similarly.
Then
\begin{equation}
\text{Bias} = \frac{1}{N}\sum\limits^{N-1}_{n=0} \hat{x_n} - x_n;
\end{equation}
\item \acs{MSE}, the mean squared error in the estimation results defined as
\begin{equation}
\text{MSE} = \frac{1}{N}\sum\limits^{N-1}_{n=0} (\hat{x_n} - x_n)^2.
\end{equation}
\item $\PearsonCC$, the Pearson correlation coefficient between the estimated and the ground truth results defined as
\begin{equation}
\PearsonCC = \frac{\Ex{\hat{X}X} - \Ex{\hat{X}}\Ex{X}}
{\sqrt{(\Ex{\hat{X^2}}-\Ex{\hat{X}}^2)(\Ex{X^2}-\Ex{X}^2 )}},
\end{equation}
where $\Ex{\cdot}$ is the mathematical expectation;
\item \acsu{RTF}, the real-time factor, the total computation time divided by the total duration of all processed speech files.
All implementations were in Matlab except for those marked with a $^\dagger$ which used Matlab for feature extraction and C++ for the machine learning-based mapping, and those marked with a $^\ddagger$ which were implemented entirely in C++.
\end{enumerate}
By considering the bias, \acs{MSE}, and $\PearsonCC$, it is possible to determine how well the estimator works.
For example, an estimator with a low bias and \acs{MSE} might simply be giving an estimate close to the median for every speech file.
However, by examining the $\PearsonCC$, it will be possible to distinguish between such an algorithm, which will have a low correlation, and a better algorithm which is more accurately estimating the parameter concerned which will have a higher correlation.
The \ac{RTF} is useful for determining whether the algorithm has practical applications requiring low computational complexity such as hearing aids and mobile devices.
%

For the frequency-dependent tasks, a box plot is provided per algorithm with each box representing the performance in a particular frequency band.
Frequency dependent algorithms have also been included in the fullband plots.
Where those algorithms themselves produce a fullband estimate, this has been used directly as in the case of the \ac{DENBE}~\cite{Eaton2015c} and Particle Velocity~\cite{Chen2015} algorithms.
Where no fullband estimate is produced, a fullband estimate was obtained by taking the mean of the results over the \num{400} to \SI{1250}{\hertz} frequency bands as in the case of the Model-based subband RTE~\cite{Lollmann2015} algorithm as recommended in \acs*{ISO} 3382~\cite{ISO_3382}.
\clearpage
\section{Overall results summary}
%
\subsection{Fullband \ac{T60} estimation overall results}
The overall results for fullband \ac{T60} estimation are shown in {\figMidSent}~\ref{fig:ACE_T60_A2_All} and {\tabMidSent}~\ref{tab:ACE_T60_A2_All}.
%
\begin{figure}[!ht]
	\ifarXiv
\centerline{\epsfig{figure=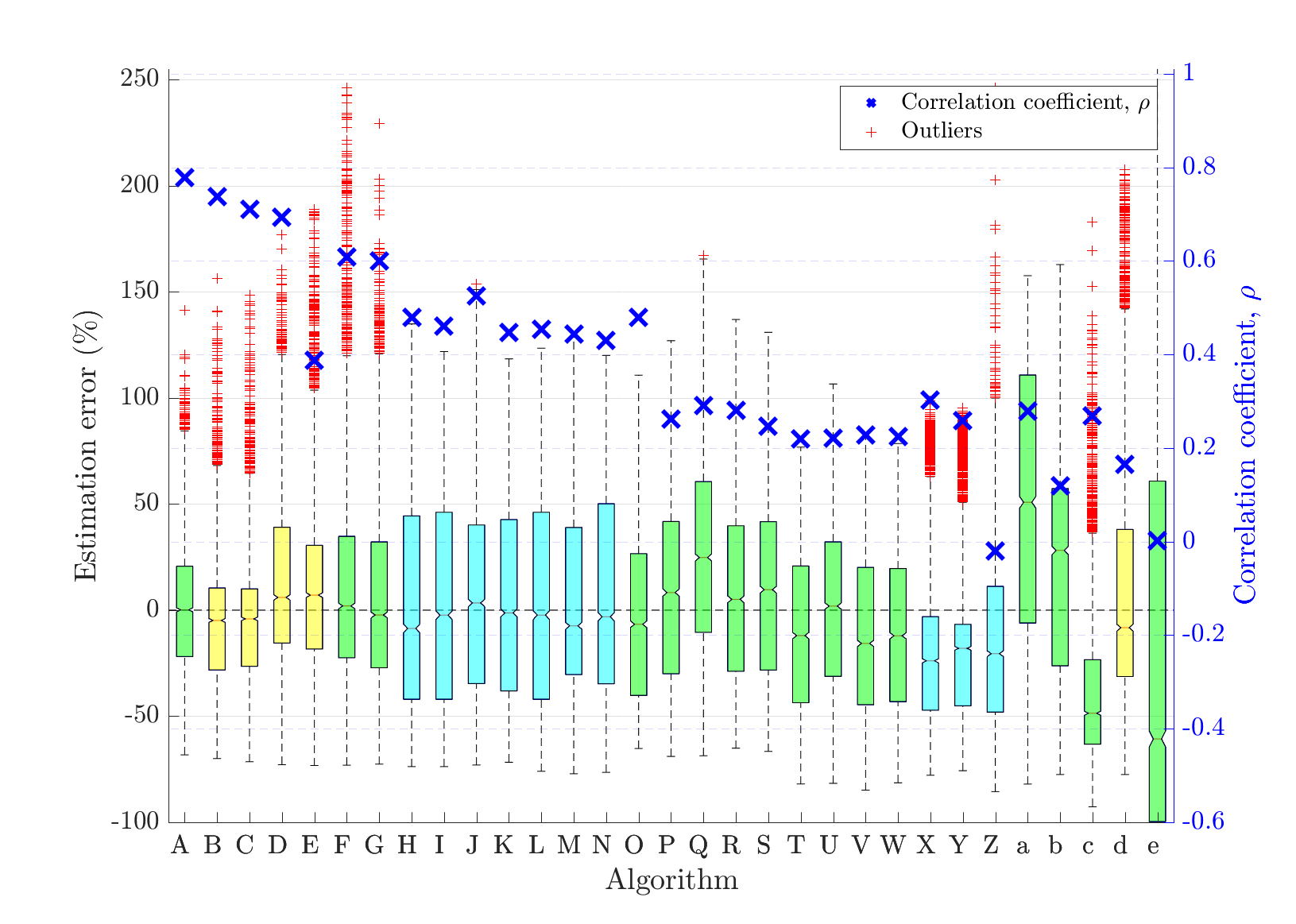,
	width=\figWidthACETR,viewport=45 10 765 530,clip}}%
	\else
	\centerline{\epsfig{figure=FigsACE/ana_eval_gt_partic_results_combined_Phase3_All_WASPAA_P3_T60_Perc_A2_All_Noises.png,
	width=\figWidthACETR,viewport=45 10 765 530,clip}}%
	\fi
	\caption{Fullband {\ac{T60} estimation error in all noises for all \acp{SNR}}}%
\label{fig:ACE_T60_A2_All}%
\end{figure}%
\acused{RTF}%
\acused{NSRMR}%
\acused{SRMR}%
\acused{SDDSA-G}%
\begin{table*}[!htb]\small
\caption{\ac{T60} estimation algorithm performance in all noises for all \acp{SNR}}
\vspace{5mm} 
\centering
\begin{tabular}{cllllllll}%
\hline%
Ref.
& Algorithm
& Class
& Mic. Config.
& Bias
& \acs{MSE}
& $\PearsonCC$
& \ac{RTF}
\\
\hline%
\hline%
\ifarXiv
A & QA Reverb~\cite{Prego2015} & \ac{SFM} & Single & -0.068 & 0.0648 & 0.778 & 0.4\\ 
\hline
B & Octave \ac{SB}-based \ac{FB2} RTE~\cite{Lollmann2015} & \ac{ABC} & Single & -0.104 & 0.0731 & 0.738 & 0.939\\ 
\hline
C & DCT-based \ac{FB2} RTE~\cite{Lollmann2015} & \ac{ABC} & Single & -0.104 & 0.0766 & 0.71 & 1\\ 
\hline
D & Model-based \ac{SB} RTE~\cite{Lollmann2015} & \ac{ABC} & Single & -0.0363 & 0.102 & 0.693 & 0.451\\ 
\hline
E & Baseline algorithm for \ac{FB2} RTE~\cite{Lollmann2015} & \ac{ABC} & Single & -0.0432 & 0.11 & 0.387 & 0.0424\\ 
\hline
F & \ac{SDDSA-G} retrained~\cite{Eaton2015b} & \ac{SFM} & Single & 0.0167 & 0.0937 & 0.608 & 0.0152\\ 
\hline
G & \ac{SDDSA-G}~\cite{Eaton2013} & \ac{SFM} & Single & -0.0423 & 0.0803 & 0.6 & 0.0164\\ 
\hline
H & Multi-layer perceptron~\cite{Xiong2015} & \ac{MLMF} & Single & -0.0967 & 0.104 & 0.48 & 0.0578$^\ddagger$\\ 
\hline
I & Multi-layer perceptron P2~\cite{Xiong2015} & \ac{MLMF} & Single & -0.0497 & 0.0992 & 0.46 & 0.0578$^\ddagger$\\ 
\hline
J & Multi-layer perceptron P2~\cite{Xiong2015} & \ac{MLMF} & Chromebook & -0.054 & 0.0933 & 0.525 & 0.0589$^\ddagger$\\ 
\hline
K & Multi-layer perceptron P2~\cite{Xiong2015} & \ac{MLMF} & Mobile & -0.0299 & 0.082 & 0.447 & 0.0556$^\ddagger$\\ 
\hline
L & Multi-layer perceptron P2~\cite{Xiong2015} & \ac{MLMF} & Crucif & -0.0503 & 0.1 & 0.454 & 0.0569$^\ddagger$\\ 
\hline
M & Multi-layer perceptron P2~\cite{Xiong2015} & \ac{MLMF} & Lin8Ch & -0.0468 & 0.0868 & 0.443 & 0.0618$^\ddagger$\\ 
\hline
N & Multi-layer perceptron P2~\cite{Xiong2015} & \ac{MLMF} & EM32 & -0.0602 & 0.0879 & 0.43 & 0.0576$^\ddagger$\\ 
\hline
O & Per acoust. band \ac{SRMR} {\sectMidSent} 2.5.~\cite{Senoussaoui2015} & \ac{SFM} & Single & -0.114 & 0.109 & 0.48 & 0.578\\ 
\hline
P & \ac{NSRMR} {\sectMidSent} 2.4.~\cite{Santos2014,Senoussaoui2015} & \ac{SFM} & Single & -0.0646 & 0.119 & 0.261 & 0.571\\ 
\hline
Q & \ac{NSRMR} {\sectMidSent} 2.4.~\cite{Santos2014,Senoussaoui2015} & \ac{SFM} & Chromebook & 0.012 & 0.116 & 0.291 & 1.04\\ 
\hline
R & \ac{NSRMR} {\sectMidSent} 2.4.~\cite{Santos2014,Senoussaoui2015} & \ac{SFM} & Mobile & -0.0504 & 0.0958 & 0.281 & 1.58\\ 
\hline
S & \ac{NSRMR} {\sectMidSent} 2.4.~\cite{Santos2014,Senoussaoui2015} & \ac{SFM} & Crucif & -0.0516 & 0.107 & 0.246 & 2.62\\ 
\hline
T & \ac{SRMR} {\sectMidSent} 2.3.~\cite{Senoussaoui2015} & \ac{SFM} & Single & -0.16 & 0.144 & 0.22 & 0.457\\ 
\hline
U & \ac{SRMR} {\sectMidSent} 2.3.~\cite{Senoussaoui2015} & \ac{SFM} & Chromebook & -0.105 & 0.132 & 0.221 & 0.829\\ 
\hline
V & \ac{SRMR} {\sectMidSent} 2.3.~\cite{Senoussaoui2015} & \ac{SFM} & Mobile & -0.153 & 0.12 & 0.228 & 1.26\\ 
\hline
W & \ac{SRMR} {\sectMidSent} 2.3.~\cite{Senoussaoui2015} & \ac{SFM} & Crucif & -0.153 & 0.128 & 0.225 & 2.09\\ 
\hline
X & NIRAv3~\cite{Parada2015} & \ac{MLMF} & Single & -0.192 & 0.151 & 0.302 & 0.899$^\dagger$\\ 
\hline
Y & NIRAv1~\cite{Parada2015} & \ac{MLMF} & Single & -0.184 & 0.151 & 0.258 & 0.899$^\dagger$\\ 
\hline
Z & NIRAv2~\cite{Parada2015} & \ac{MLMF} & Single & -0.179 & 0.198 & -0.0199 & 0.907$^\dagger$\\ 
\hline
a & Blur kernel~\cite{Lim2015} & \ac{SFM} & Single & 0.173 & 0.15 & 0.279 & 8.46\\ 
\hline
b & Blur kernel with sliding window~\cite{Lim2015a} & \ac{SFM} & Single & -0.00555 & 0.139 & 0.12 & 0.421\\ 
\hline
c & Temporal dynamics~\cite{Falk2010a} & \ac{SFM} & Single & -0.304 & 0.211 & 0.269 & 0.362\\ 
\hline
d & Improved blind RTE~\cite{Lollmann2010} & \ac{ABC} & Single & -0.0635 & 0.165 & 0.166 & 0.0259\\ 
\hline
e & \ac{SDD}~\cite{Wen2008} & \ac{SFM} & Single & 0.463 & 305 & 0.00158 & 0.0221\\ 
\hline
\else
\fi
\end{tabular}%
\label{tab:ACE_T60_A2_All}
\end{table*}%
\clearpage
\subsection{Fullband \ac{DRR} estimation overall results}
The overall results for fullband \ac{DRR} estimation are shown in {\figMidSent}~\ref{fig:ACE_DRR_A2_All}  and {\tabMidSent}~\ref{tab:ACE_DRR_A2_All}.
\begin{figure}[!ht]
	\ifarXiv
\centerline{\epsfig{figure=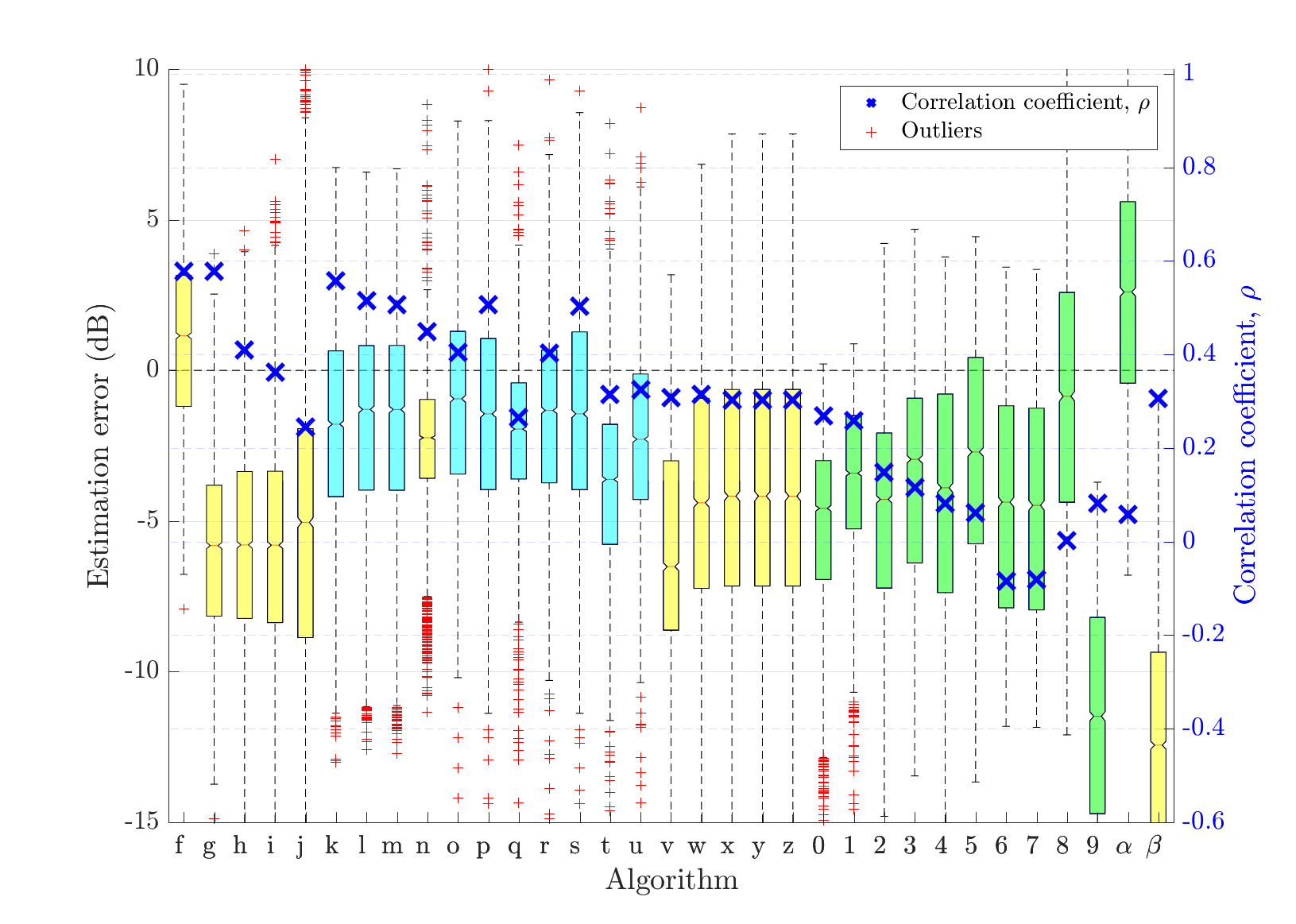,
	width=\figWidthACETR,viewport=45 10 765 530,clip}}%
	\else
	\centerline{\epsfig{figure=FigsACE/ana_eval_gt_partic_results_combined_Phase3_All_WASPAA_P3_DRR_dB_A2_All_Noises.png,
	width=\figWidthACETR,viewport=45 10 765 530,clip}}%
	\fi
	\caption{Fullband {\ac{DRR} estimation error in all noises for all \acp{SNR}}}%
\label{fig:ACE_DRR_A2_All}%
\end{figure}%
\acused{PSD}%
\begin{table*}[!ht]
\small
\caption{\ac{DRR} estimation algorithm performance in all noises for all \acp{SNR}}
\vspace{5mm} 
\centering
\begin{tabular}{cllllllll}%
\hline%
Ref.
& Algorithm
& Class
& Mic. Config.
& Bias
& \acs{MSE}
& $\PearsonCC$
& \ac{RTF}
\\
\hline
\hline
\ifarXiv
f & PSD est. in beamspace, bias comp.~\cite{Hioka2015} & \ac{ABC} & Mobile & 1.07 & 8.14 & 0.577 & 0.757\\ 
\hline
g & PSD est. in beamspace (Raw)~\cite{Hioka2015} & \ac{ABC} & Mobile & -5.9 & 41.8 & 0.577 & 3.17\\ 
\hline
h & PSD est. in beamspace v2~\cite{Hioka2015} & \ac{ABC} & Mobile & -5.7 & 43 & 0.41 & 0.844\\ 
\hline
i & PSD est. by twin BF~\cite{Hioka2012} & \ac{ABC} & Mobile & -5.71 & 44.9 & 0.362 & 0.614\\ 
\hline
j & Spatial Covariance in matrix mode~\cite{Hioka2011} & \ac{ABC} & Mobile & -5.37 & 61.2 & 0.244 & 0.627\\ 
\hline
k & NIRAv2~\cite{Parada2015} & \ac{MLMF} & Single & -1.85 & 14.8 & 0.558 & 0.899$^\dagger$\\ 
\hline
l & NIRAv3~\cite{Parada2015} & \ac{MLMF} & Single & -1.62 & 14.7 & 0.515 & 0.899$^\dagger$\\ 
\hline
m & NIRAv1~\cite{Parada2015} & \ac{MLMF} & Single & -1.64 & 15 & 0.507 & 0.899$^\dagger$\\ 
\hline
n & Particle velocity~\cite{Chen2015} & \ac{ABC} & EM32 & -2.38 & 10.4 & 0.449 & 0.134\\ 
\hline
o & Multi-layer perceptron~\cite{Xiong2015} & \ac{MLMF} & Single & -1.14 & 15.9 & 0.405 & 0.0578$^\ddagger$\\ 
\hline
p & Multi-layer perceptron P2~\cite{Xiong2015} & \ac{MLMF} & Single & -1.52 & 16.1 & 0.507 & 0.0578$^\ddagger$\\ 
\hline
q & Multi-layer perceptron P2~\cite{Xiong2015} & \ac{MLMF} & Chromebook & -2.43 & 13.6 & 0.265 & 0.0589$^\ddagger$\\ 
\hline
r & Multi-layer perceptron P2~\cite{Xiong2015} & \ac{MLMF} & Mobile & -1.67 & 15 & 0.403 & 0.0556$^\ddagger$\\ 
\hline
s & Multi-layer perceptron P2~\cite{Xiong2015} & \ac{MLMF} & Crucif & -1.5 & 16 & 0.503 & 0.0569$^\ddagger$\\ 
\hline
t & Multi-layer perceptron P2~\cite{Xiong2015} & \ac{MLMF} & Lin8Ch & -3.64 & 25.7 & 0.314 & 0.0618$^\ddagger$\\ 
\hline
u & Multi-layer perceptron P2~\cite{Xiong2015} & \ac{MLMF} & EM32 & -2.22 & 14.6 & 0.325 & 0.0576$^\ddagger$\\ 
\hline
v & \ac{DENBE} no noise reduction~\cite{Eaton2015} & \ac{ABC} & Chromebook & -6.04 & 51.2 & 0.308 & 0.0323\\ 
\hline
w & \ac{DENBE} spectral subtraction~\cite{Eaton2015c} & \ac{ABC} & Chromebook & -4.25 & 34.1 & 0.314 & 0.0589\\ 
\hline
x & \ac{DENBE} spec. sub. Gerkmann~\cite{Eaton2015} & \ac{ABC} & Chromebook & -4.01 & 32.8 & 0.303 & 0.0477\\ 
\hline
y & \ac{DENBE} filtered subbands~\cite{Eaton2015c} & \ac{ABC} & Chromebook & -4.01 & 32.8 & 0.303 & 0.775\\ 
\hline
z & \ac{DENBE} FFT derived subbands~\cite{Eaton2015c} & \ac{ABC} & Chromebook & -4.01 & 32.8 & 0.303 & 0.0449\\ 
\hline
0 & \ac{NOSRMR} {\sectMidSent} 2.2.~\cite{Senoussaoui2015} & \ac{SFM} & Chromebook & -5.1 & 34.3 & 0.269 & 1.04\\ 
\hline
1 & \ac{OSRMR} {\sectMidSent} 2.2.~\cite{Senoussaoui2015} & \ac{SFM} & Chromebook & -3.71 & 20.6 & 0.259 & 0.829\\ 
\hline
2 & \ac{NOSRMR} {\sectMidSent} 2.2.~\cite{Senoussaoui2015} & \ac{SFM} & Mobile & -4.47 & 32 & 0.148 & 1.58\\ 
\hline
3 & \ac{OSRMR} {\sectMidSent} 2.2.~\cite{Senoussaoui2015} & \ac{SFM} & Mobile & -3.28 & 22.2 & 0.116 & 1.26\\ 
\hline
4 & \ac{NOSRMR} {\sectMidSent} 2.2.~\cite{Senoussaoui2015} & \ac{SFM} & Crucif & -4.05 & 31.1 & 0.0814 & 2.62\\ 
\hline
5 & \ac{OSRMR} {\sectMidSent} 2.2.~\cite{Senoussaoui2015} & \ac{SFM} & Crucif & -2.88 & 22.3 & 0.0616 & 2.09\\ 
\hline
6 & \ac{NOSRMR} {\sectMidSent} 2.2.~\cite{Senoussaoui2015} & \ac{SFM} & Single & -4.16 & 33.9 & -0.0841 & 0.54\\ 
\hline
7 & \ac{OSRMR} {\sectMidSent} 2.2.~\cite{Senoussaoui2015} & \ac{SFM} & Single & -4.24 & 34.6 & -0.0815 & 0.446\\ 
\hline
8 & Per acoust. band SRMR {\sectMidSent} 2.5.~\cite{Senoussaoui2015} & \ac{SFM} & Single & -0.9 & 22.8 & 0.00192 & 0.578\\ 
\hline
9 & Temporal dynamics~\cite{Falk2009} & \ac{SFM} & Single & -11.4 & 147 & 0.0815 & 0.082\\ 
\hline
$\alpha$ & QA Reverb~\cite{Prego2015} & \ac{SFM} & Single & 2.51 & 23.6 & 0.0576 & 0.391\\ 
\hline
$\beta$ & Blind est. of coherent-to-diffuse energy ratio~\cite{Jeub2011} & \ac{ABC} & Chromebook & -12.1 & 162 & 0.305 & 0.019\\ 
\hline

\else

\fi
\end{tabular}
\label{tab:ACE_DRR_A2_All}
\end{table*}
\clearpage
%
%
%
\ifIncludeFigsByParameter
\subsubsection{Fullband \ac{T60} estimation results by parameter}
\begin{figure}[!ht]
\centering
	\ifarXiv
\subfloat[]{\epsfig{figure=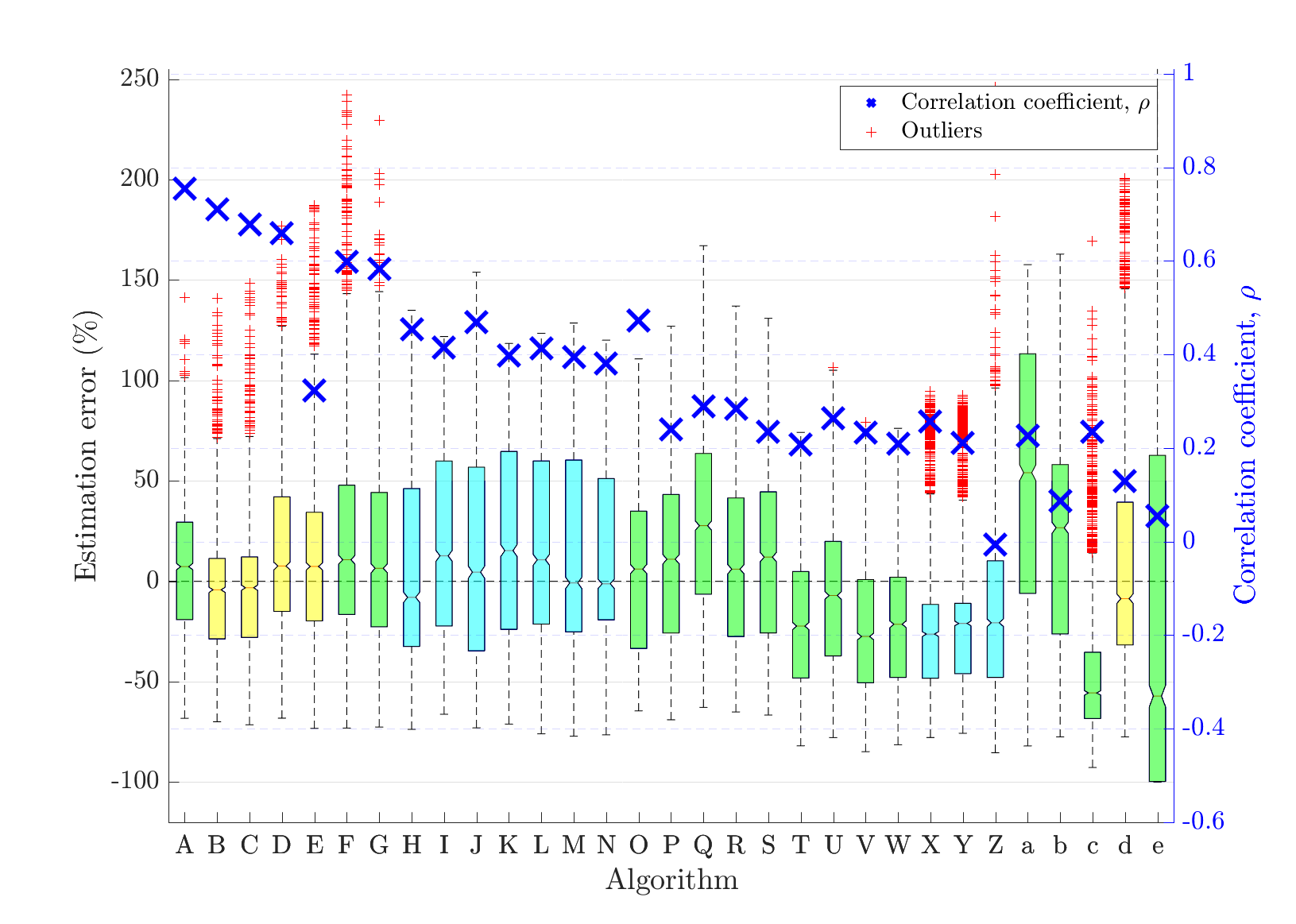,
	width=\figWidthConf,viewport=45 10 765 530,clip}}%
	\else
\subfloat[]{\epsfig{figure=FigsACE/ana_eval_gt_partic_results_combined_Phase3_All_TR3_T60_Perc_MOF_FEM_All_Noises.png,
	width=\figWidthConf,viewport=45 10 765 530,clip}}%
	\fi
\hfil
	\ifarXiv
\subfloat[]{\epsfig{figure=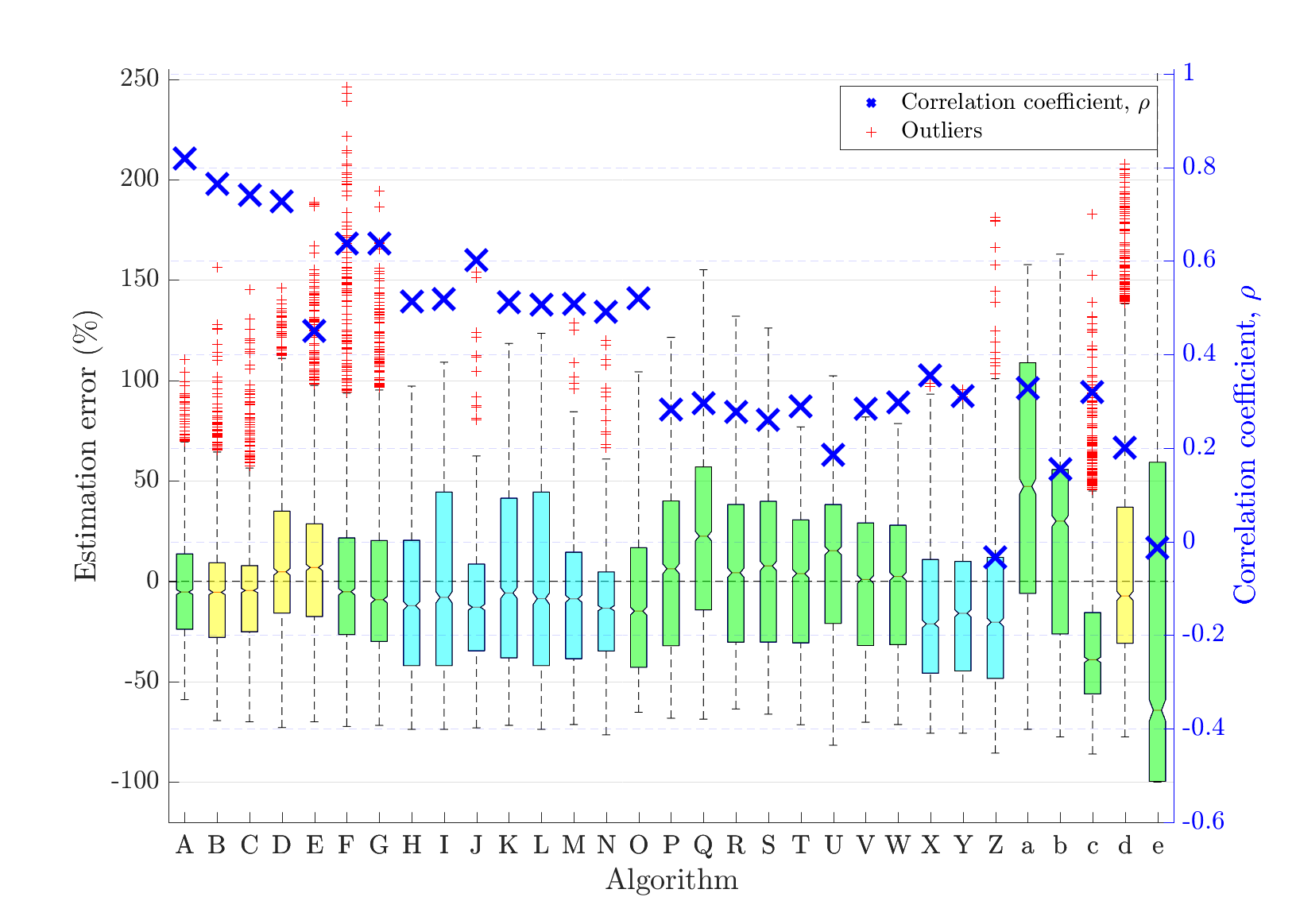,
	width=\figWidthConf,viewport=45 10 765 530,clip}}%
	\else
\subfloat[]{\epsfig{figure=FigsACE/ana_eval_gt_partic_results_combined_Phase3_All_TR3_T60_Perc_MOF_MAL_All_Noises.png,
	width=\figWidthConf,viewport=45 10 765 530,clip}}%
	\fi
\caption{\ac{FB2} \ac{T60} estimation error in all noises and all \acp{SNR} for a) female talkers and b) male talkers}%
\label{fig:ACE_T60_MOF_All}%
\end{figure}%
%
\begin{figure}[!ht]
\centering
	\ifarXiv
\subfloat[]{\epsfig{figure=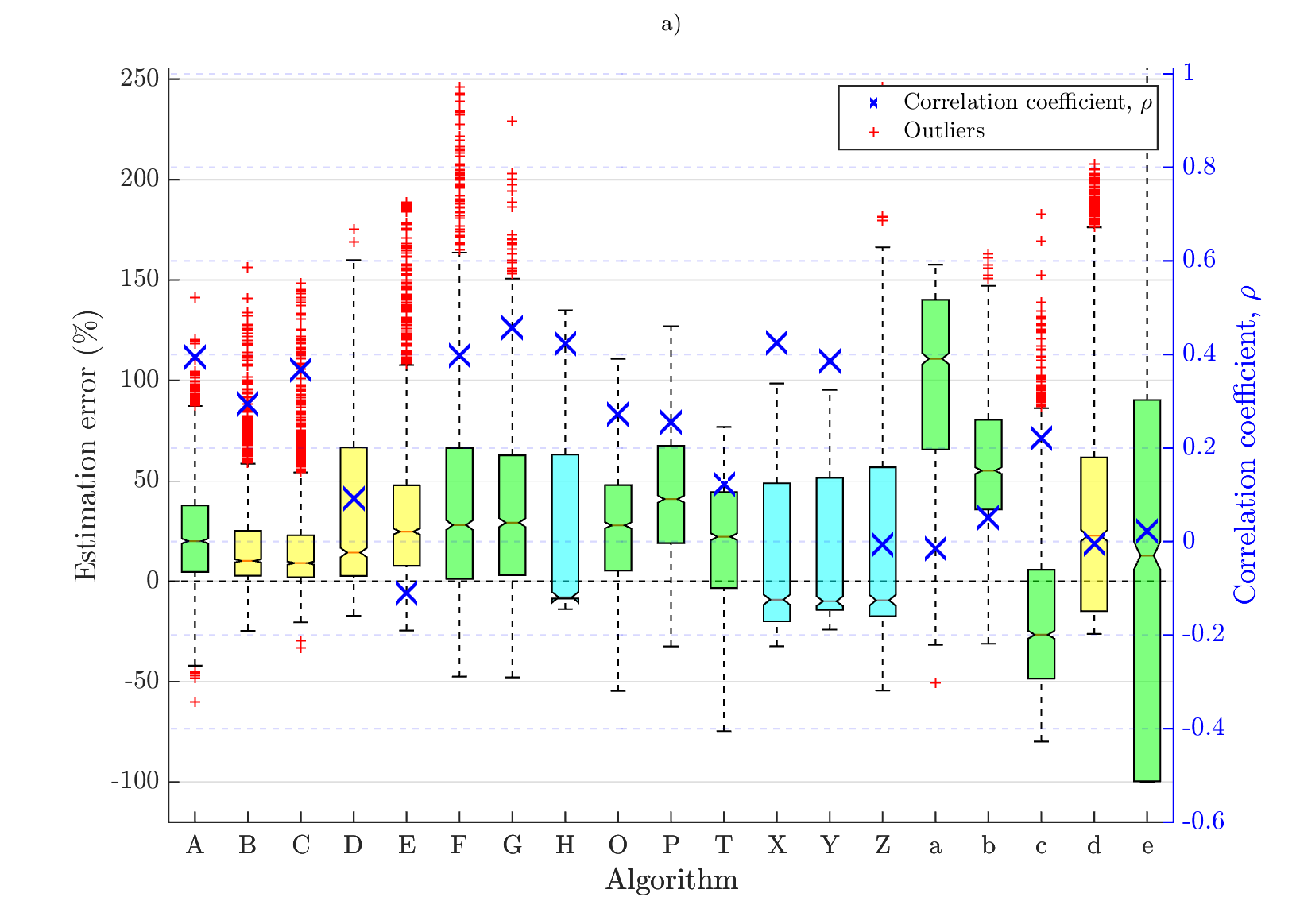, width=\figWidthConf,viewport=20 10 765 530,clip}}%
	\else
\subfloat[]{\epsfig{figure=FigsACE/ana_eval_gt_partic_results_combined_Phase3_All_TR3_Single_T60_Perc_GT_L_All_Noises.png, width=\figWidthConf,viewport=20 10 765 530,clip}}%
	\fi
\hfil
	\ifarXiv
\subfloat[]{\epsfig{figure=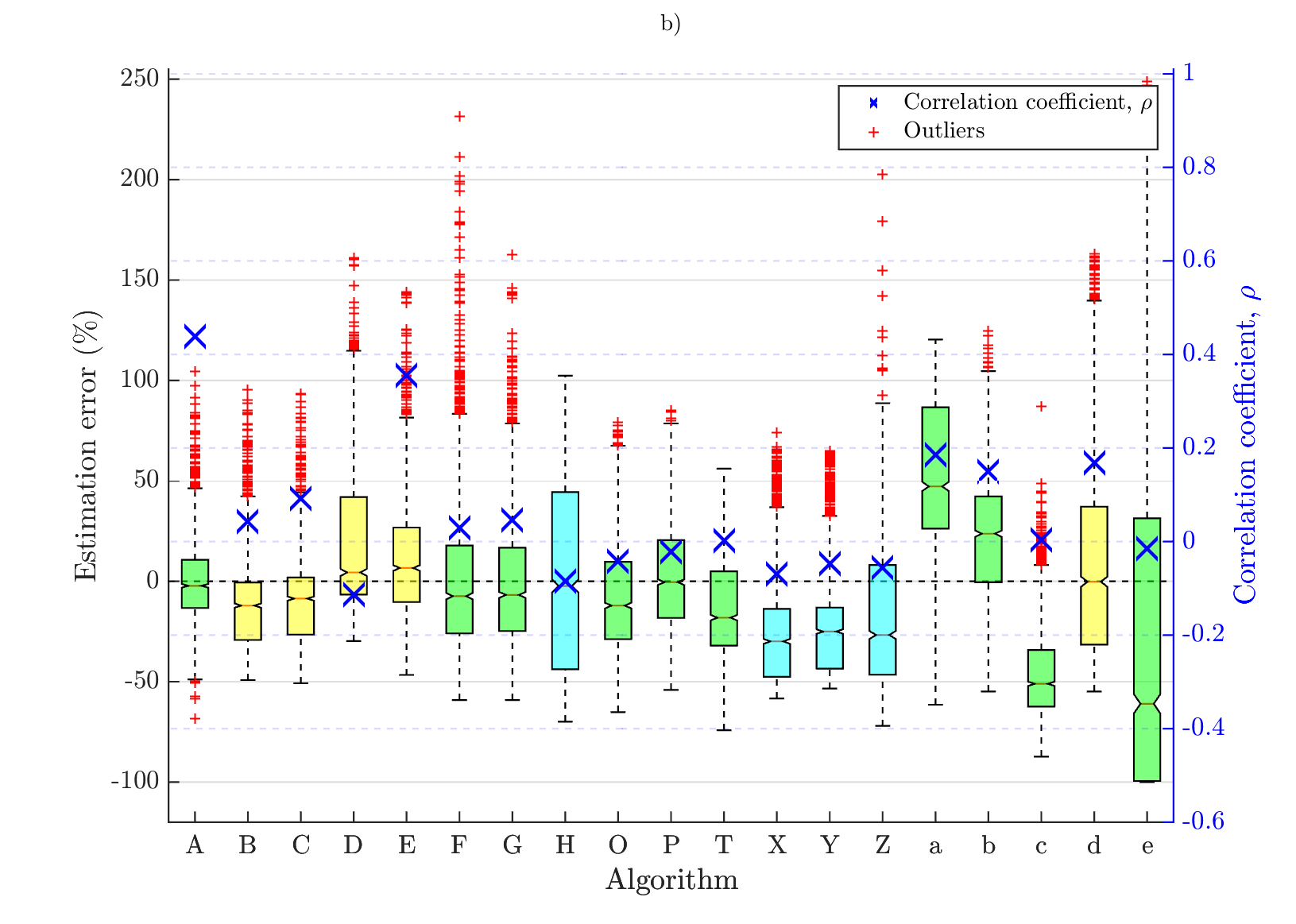,width=\figWidthConf,viewport=20 10 765 530,clip}}%
	\else
\subfloat[]{\epsfig{figure=FigsACE/ana_eval_gt_partic_results_combined_Phase3_All_TR3_Single_T60_Perc_GT_M_All_Noises.png,width=\figWidthConf,viewport=20 10 765 530,clip}}%
	\fi
\hfil
	\ifarXiv
\subfloat[]{\epsfig{figure=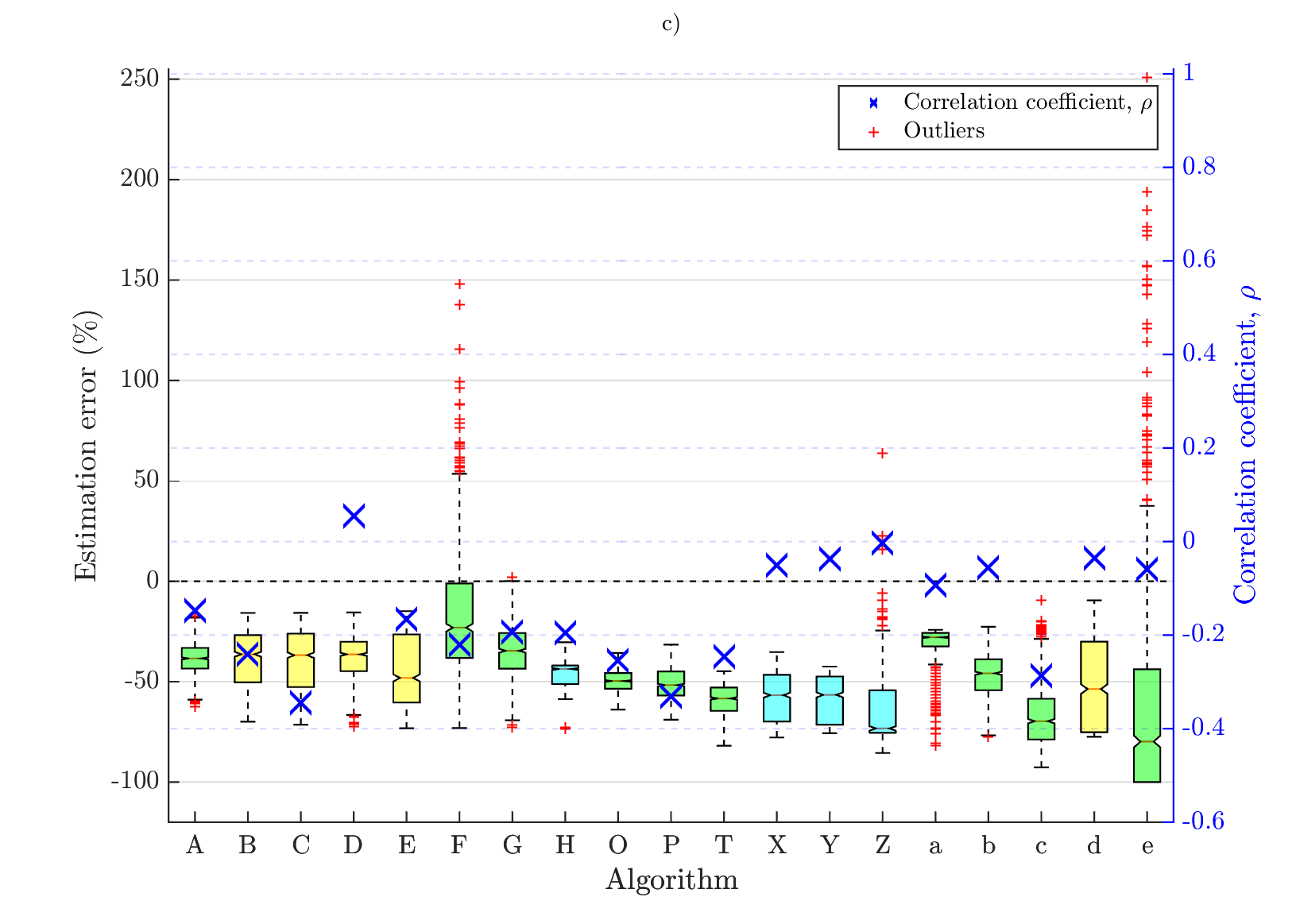, width=\figWidthConf,viewport=20 10 765 530,clip}}%
	\else
\subfloat[]{\epsfig{figure=FigsACE/ana_eval_gt_partic_results_combined_Phase3_All_TR3_Single_T60_Perc_GT_H_All_Noises.png, width=\figWidthConf,viewport=20 10 765 530,clip}}%
	\fi
\caption{Single channel \ac{FB2} \ac{T60} estimation error in all noises and all \acp{SNR} for a) \ac{T60} $<$\SI{0.43}{\second} b) $0.43\leq$\ac{T60} $<$~\SI{0.75}{\second}  and c) \ac{T60} $\geq$~\SI{0.75}{\second}. Observe that $\PearsonCC<0$ for all except algorithm D}%
\label{fig:ACE_T60_GT_All}%
\end{figure}%
%
%
\begin{figure}[!ht]
\centering
	\ifarXiv
\subfloat[]{\epsfig{figure=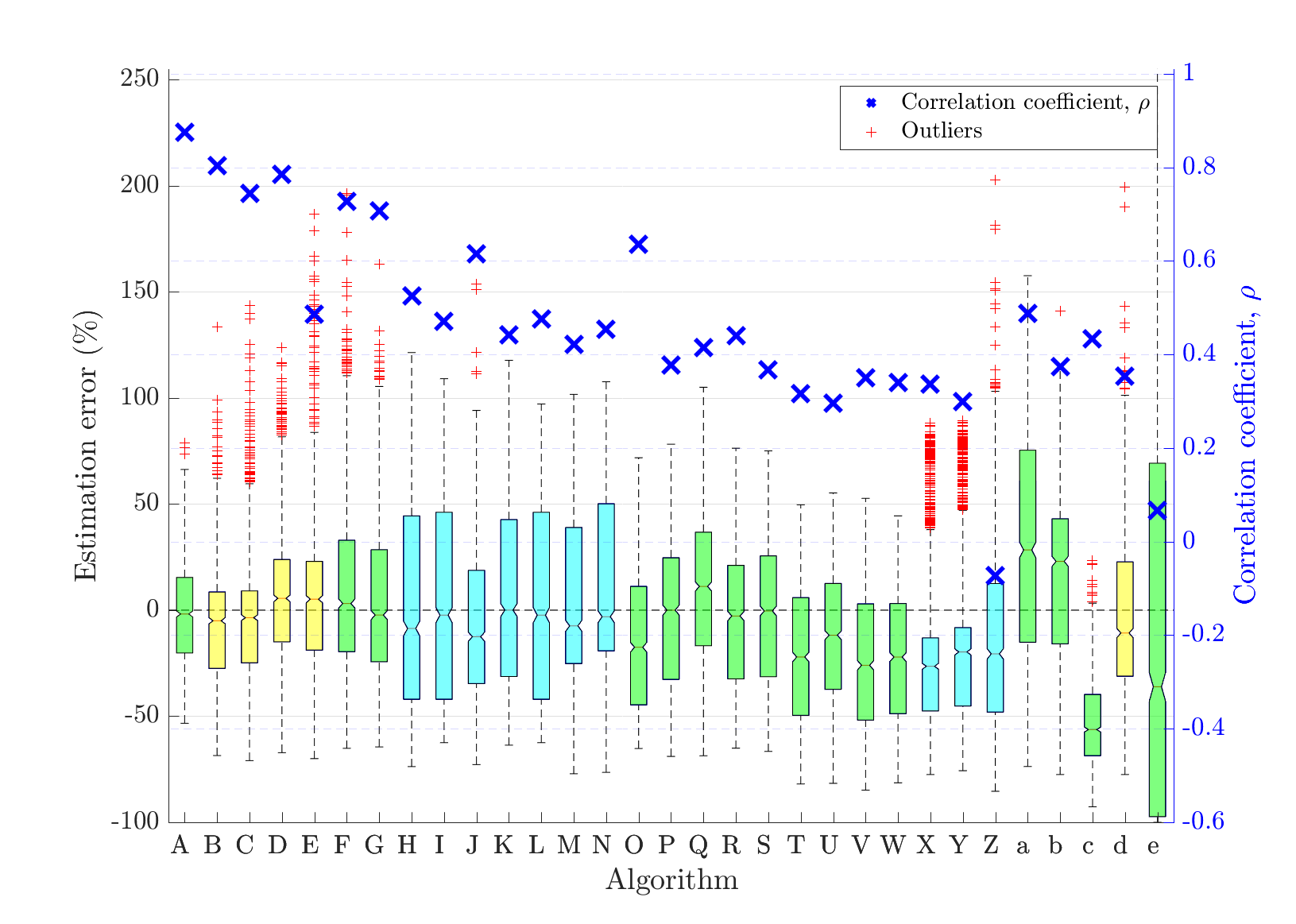, 	width=\figWidthConf,viewport=45 10 765 530,clip}}%
	\else
\subfloat[]{\epsfig{figure=FigsACE/ana_eval_gt_partic_results_combined_Phase3_All_WASPAA_P3_T60_Perc_H_All_Noises.png, 	width=\figWidthConf,viewport=45 10 765 530,clip}}%
	\fi
\hfil
	\ifarXiv
\subfloat[]{\epsfig{figure=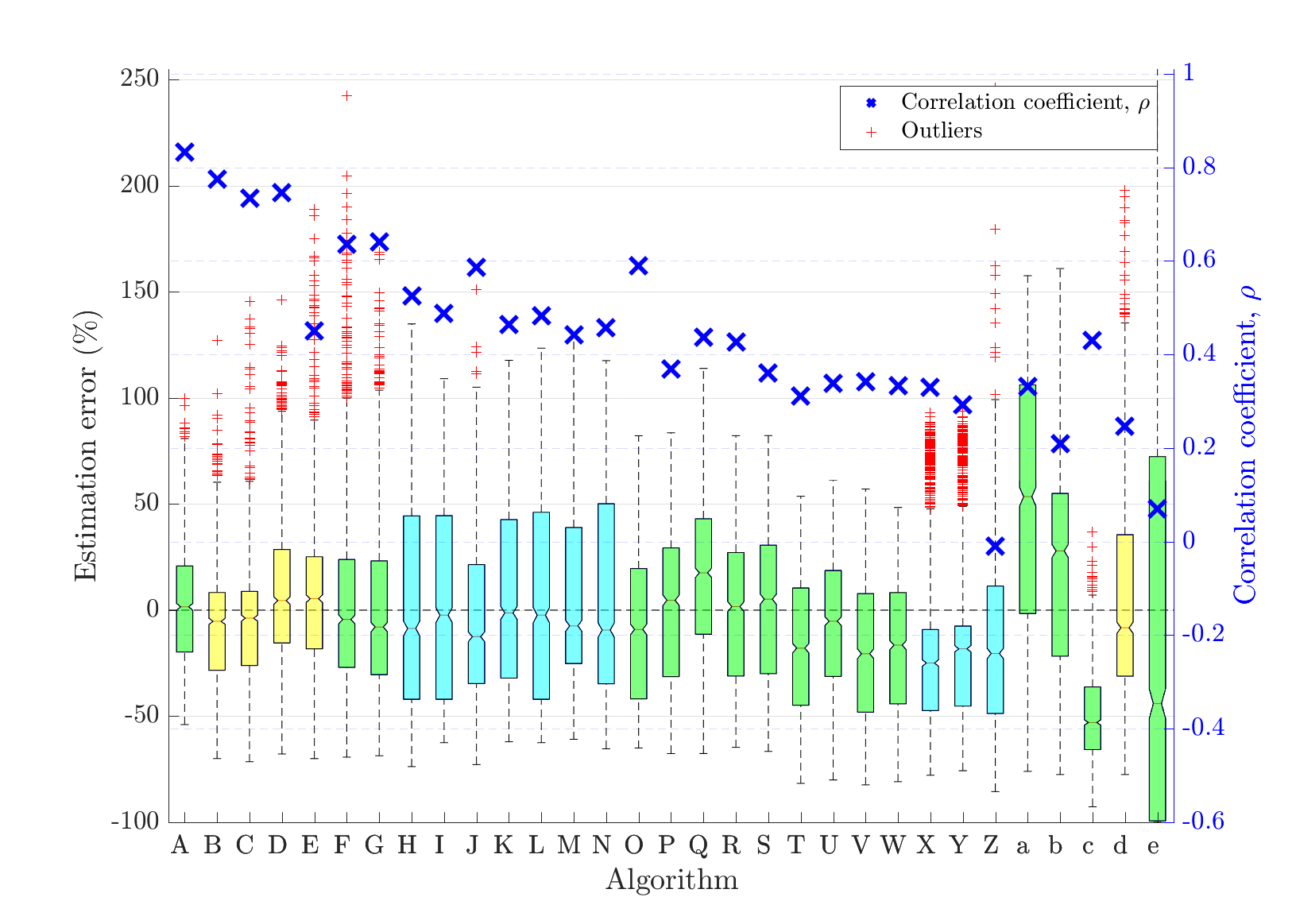, 	width=\figWidthConf,viewport=45 10 765 530,clip}}%
	\else
\subfloat[]{\epsfig{figure=FigsACE/ana_eval_gt_partic_results_combined_Phase3_All_WASPAA_P3_T60_Perc_M_All_Noises.png, 	width=\figWidthConf,viewport=45 10 765 530,clip}}%
	\fi
\hfil
	\ifarXiv
\subfloat[]{\epsfig{figure=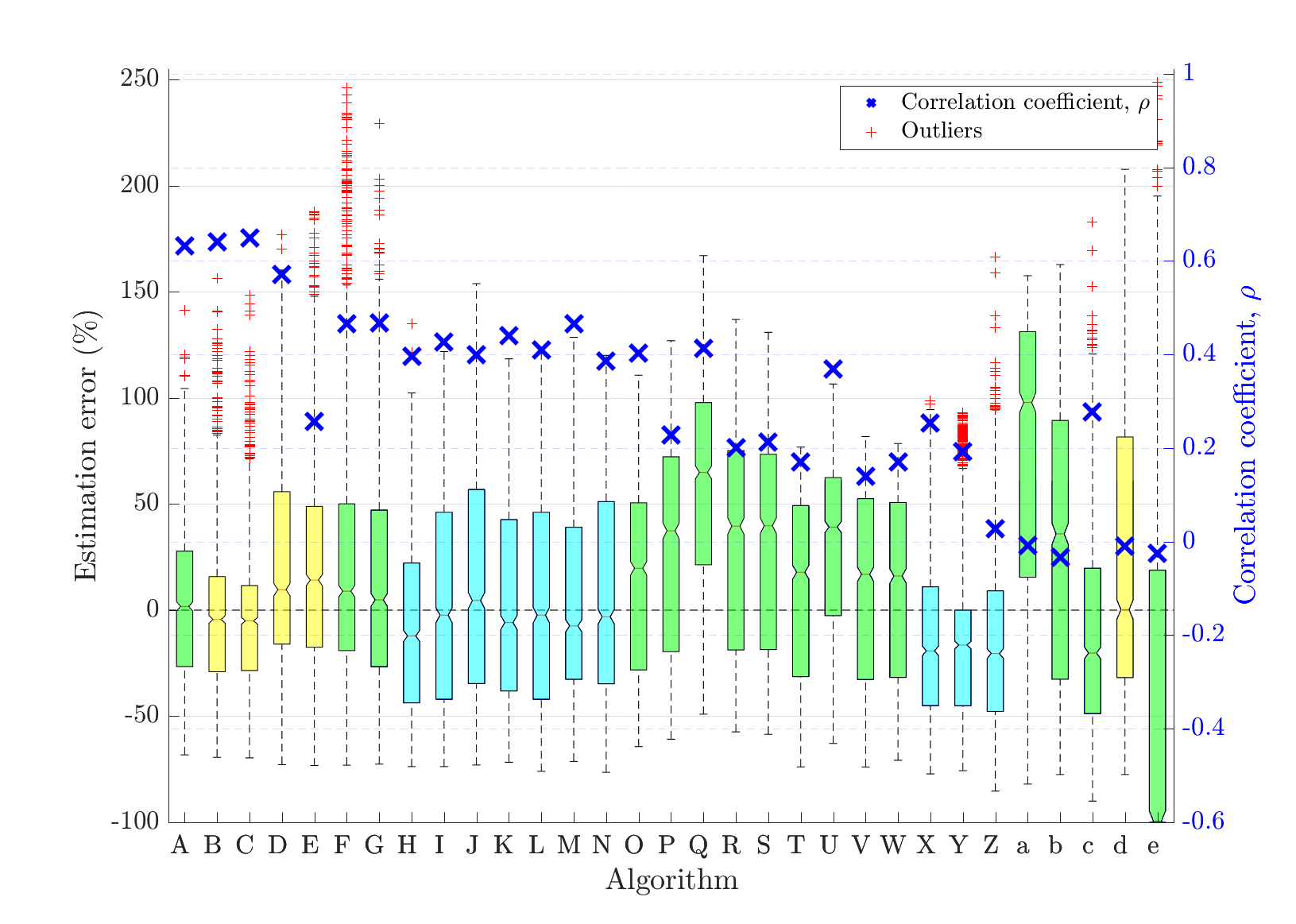, 	width=\figWidthConf,viewport=45 10 765 530,clip}}%
	\else
\subfloat[]{\epsfig{figure=FigsACE/ana_eval_gt_partic_results_combined_Phase3_All_WASPAA_P3_T60_Perc_L_All_Noises.png, 	width=\figWidthConf,viewport=45 10 765 530,clip}}%
	\fi
\caption{\ac{FB2} \ac{T60} estimation error in all noises at a), \dBel{18} \ac{SNR}, b), \dBel{12} \ac{SNR}, and c) \dBel{-1} \ac{SNR}}%
\label{fig:ACE_T60_SNR_All}%
\end{figure}%
%
%
\begin{figure}[!ht]
\centering
	\ifarXiv
\subfloat[]{\epsfig{figure=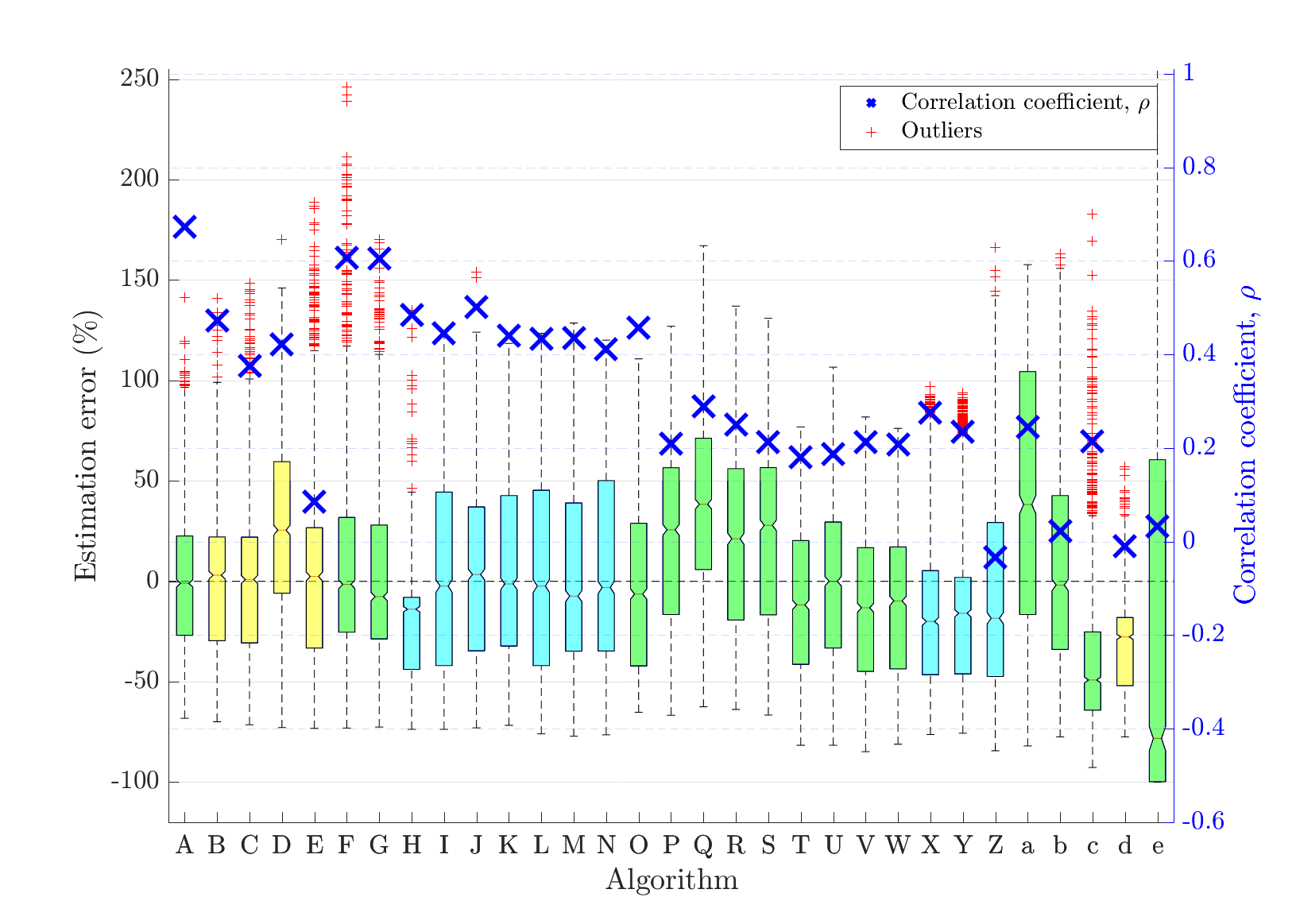,
	width=\figWidthConf,viewport=20 10 765 530,clip}}%
	\else
\subfloat[]{\epsfig{figure=FigsACE/ana_eval_gt_partic_results_combined_Phase3_All_TR3_T60_Perc_LEN_L_All_Noises.png,
	width=\figWidthConf,viewport=20 10 765 530,clip}}%
	\fi
\hfil
	\ifarXiv
\subfloat[]{\epsfig{figure=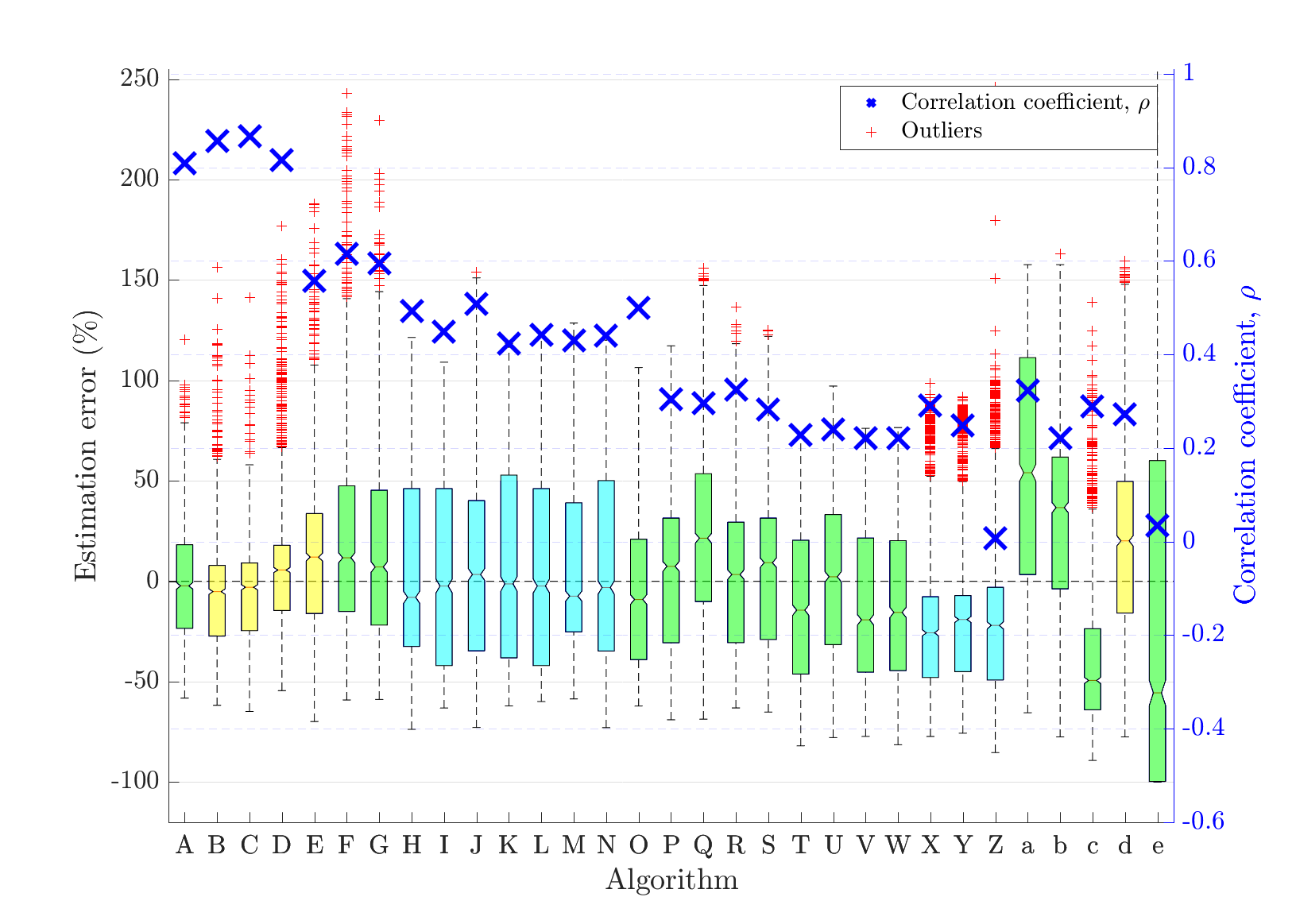,
	width=\figWidthConf,viewport=20 10 765 530,clip}}%
	\else
\subfloat[]{\epsfig{figure=FigsACE/ana_eval_gt_partic_results_combined_Phase3_All_TR3_T60_Perc_LEN_M_All_Noises.png,
	width=\figWidthConf,viewport=20 10 765 530,clip}}%
	\fi
\hfil
	\ifarXiv
\subfloat[]{\epsfig{figure=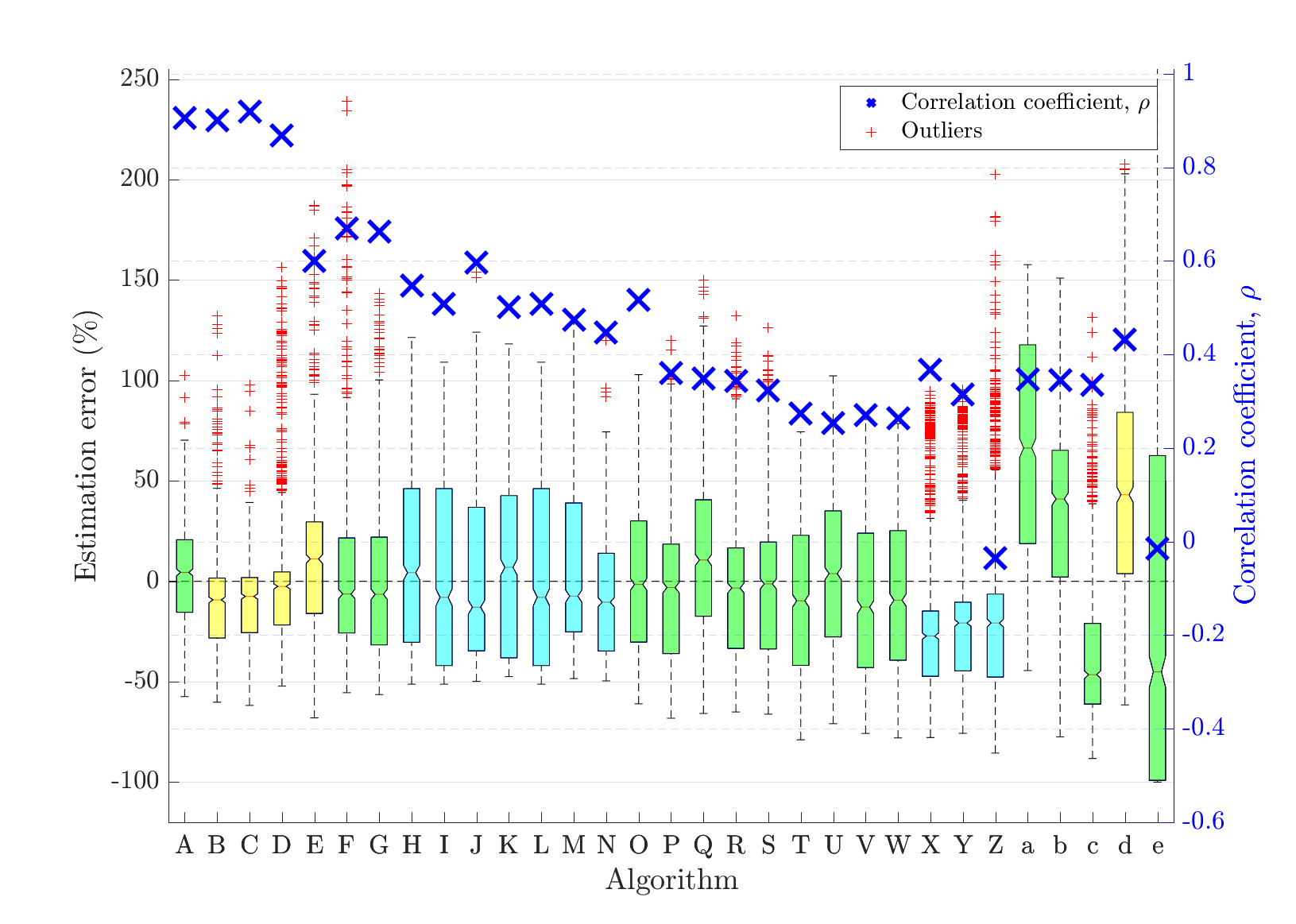,
	width=\figWidthConf,viewport=20 10 765 530,clip}}%
	\else
\subfloat[]{\epsfig{figure=FigsACE/ana_eval_gt_partic_results_combined_Phase3_All_TR3_T60_Perc_LEN_H_All_Noises.png,
	width=\figWidthConf,viewport=20 10 765 530,clip}}%
	\fi
\caption{\ac{FB2} \ac{T60} estimation error in all noises and all \acp{SNR} for a) utterance length $<$\SI{5}{\second} b) utterance length $<$~\SI{15}{\second}  and c) utterance length $\geq$~\SI{15}{\second}}%
\label{fig:ACE_T60_LEN_All}%
\end{figure}%
%
%
\clearpage
\subsubsection{Fullband \ac{DRR} estimation results by parameter}
\begin{figure}[!ht]
\centering
	\ifarXiv
\subfloat[]{\epsfig{figure=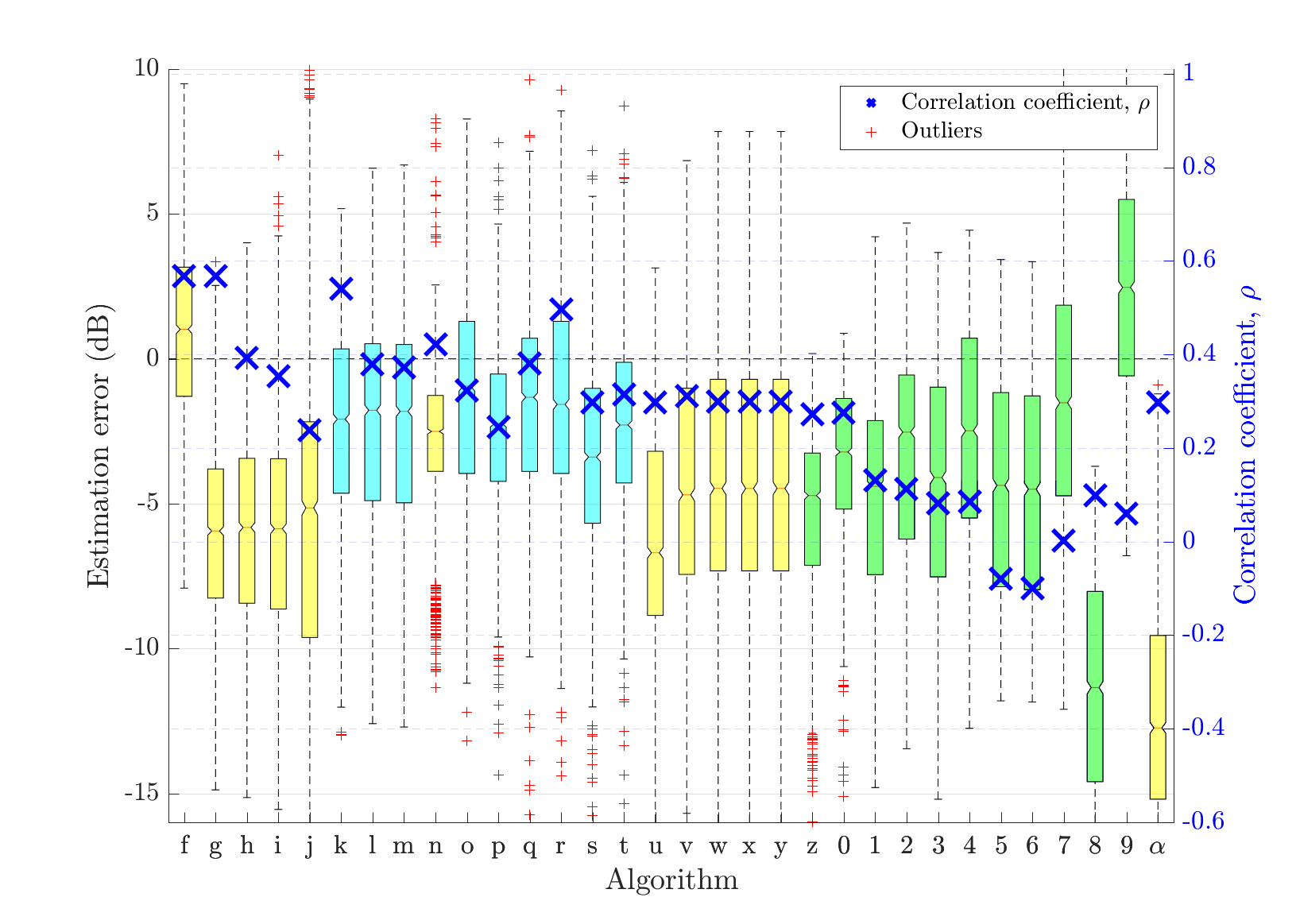,
	width=\figWidthConf,viewport=45 10 765 530,clip}}%
	\else
\subfloat[]{\epsfig{figure=FigsACE/ana_eval_gt_partic_results_combined_Phase3_All_TR3_DRR_dB_MOF_FEM_All_Noises.png,
	width=\figWidthConf,viewport=45 10 765 530,clip}}%
	\fi
\hfil
	\ifarXiv
\subfloat[]{\epsfig{figure=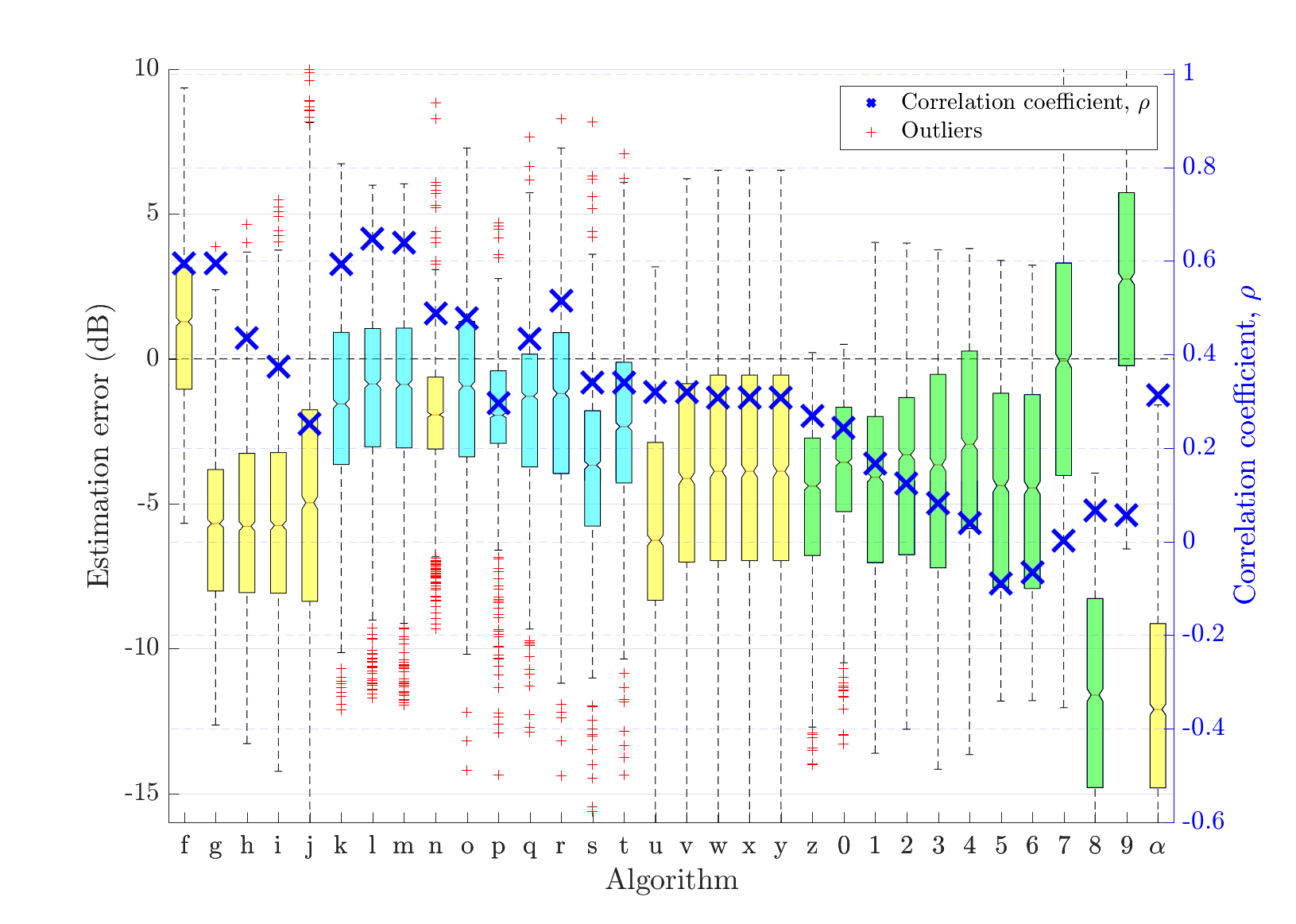,
	width=\figWidthConf,viewport=45 10 765 530,clip}}%
	\else
\subfloat[]{\epsfig{figure=FigsACE/ana_eval_gt_partic_results_combined_Phase3_All_TR3_DRR_dB_MOF_MAL_All_Noises.png,
	width=\figWidthConf,viewport=45 10 765 530,clip}}%
	\fi
\caption{\ac{FB2} \ac{DRR} estimation error in all noises and all \acp{SNR} for a) female talkers and b) male talkers}%
\label{fig:ACE_DRR_MOF_All}%
\end{figure}%
%
\begin{figure}[!ht]
\centering
	\ifarXiv
\subfloat[]{\epsfig{figure=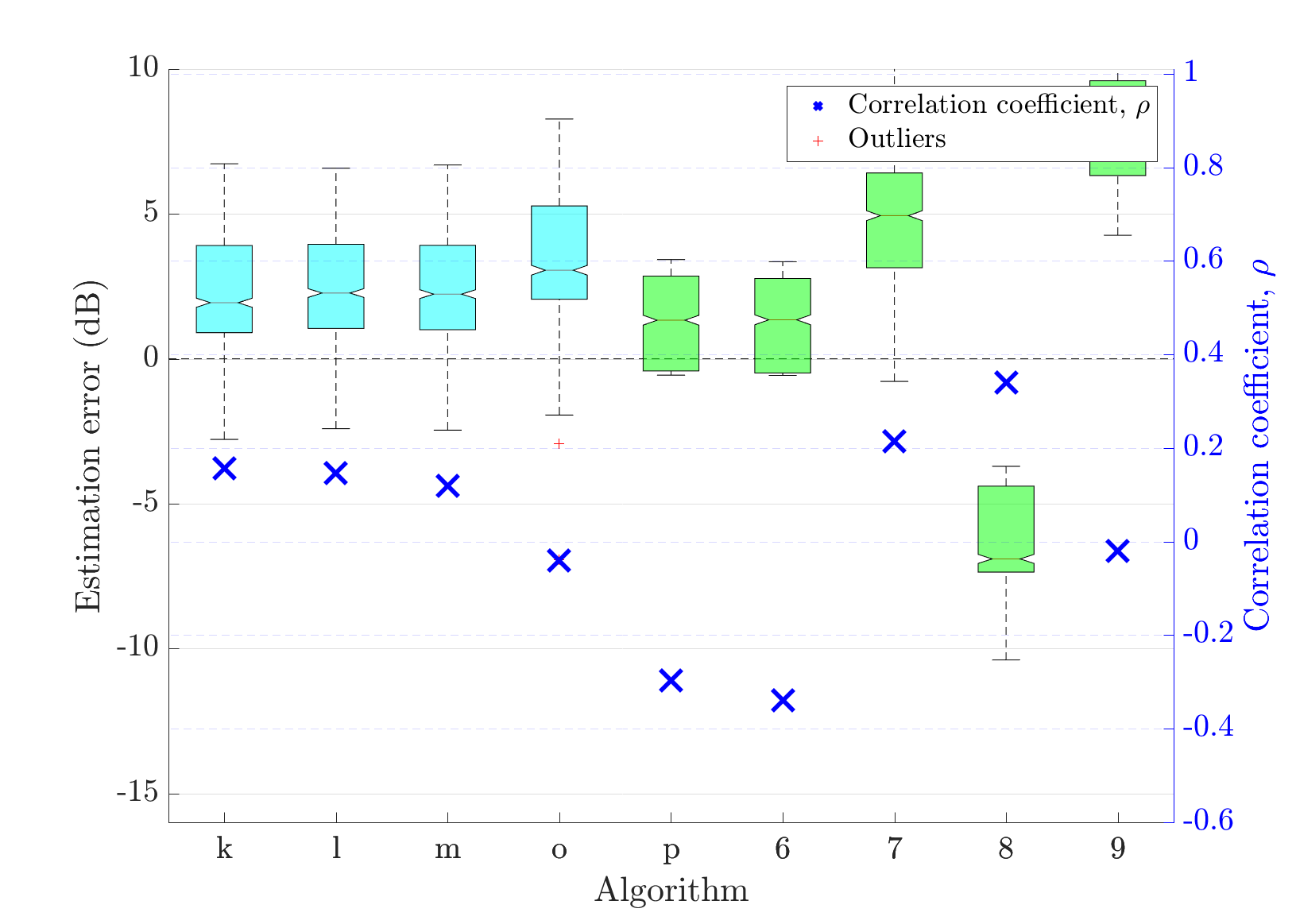,
	width=\figWidthJournNarrow,viewport=45 10 772 530,clip}}%
	\else
\subfloat[]{\epsfig{figure=FigsACE/ana_eval_gt_partic_results_combined_Phase3_All_TR3_Single_DRR_dB_GT_L_All_Noises.png,
	width=\figWidthJournNarrow,viewport=45 10 772 530,clip}}%
	\fi
\hfil
	\ifarXiv
\subfloat[]{\epsfig{figure=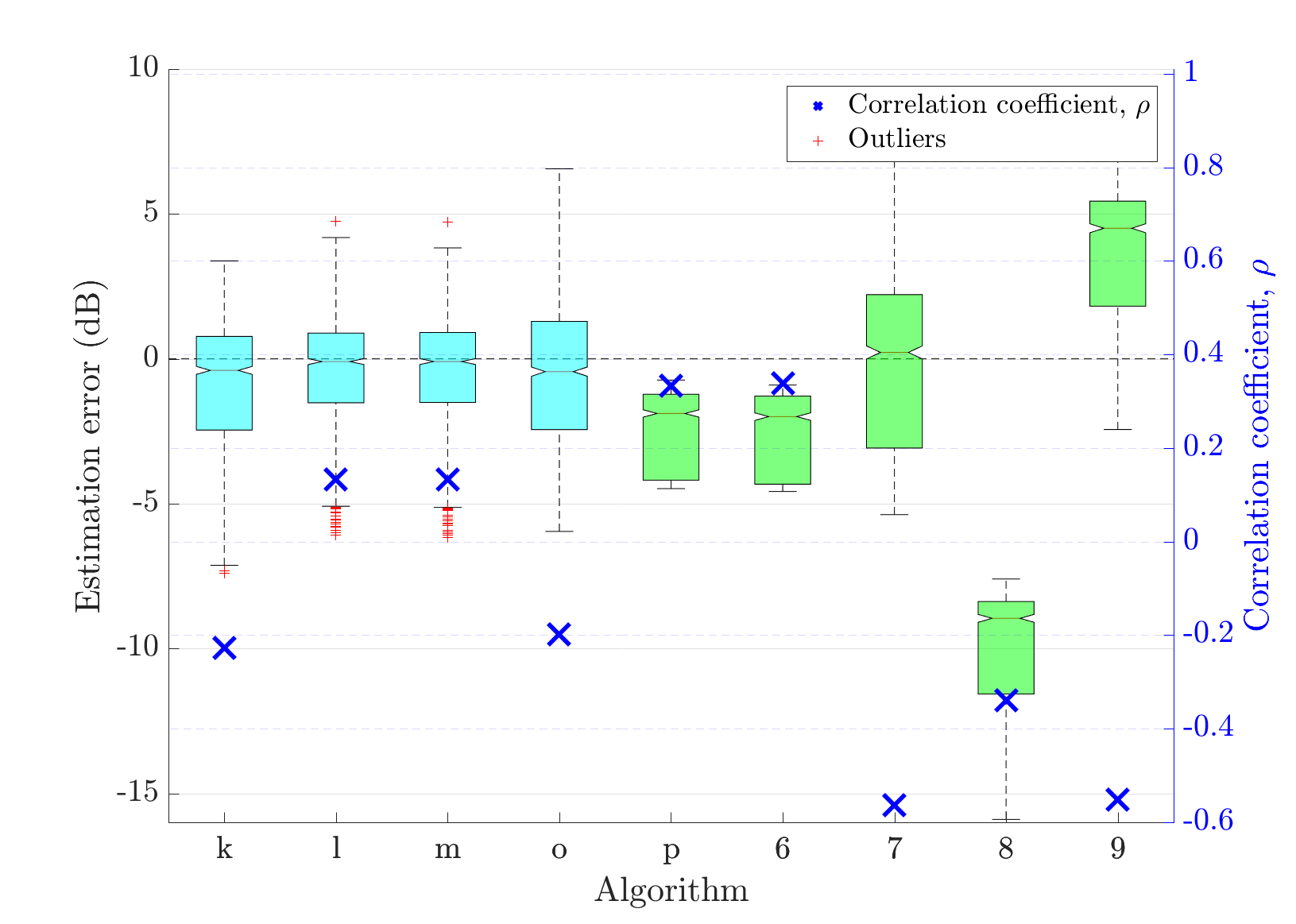,
	width=\figWidthJournNarrow,viewport=45 10 772 530,clip}}%
	\else
\subfloat[]{\epsfig{figure=FigsACE/ana_eval_gt_partic_results_combined_Phase3_All_TR3_Single_DRR_dB_GT_M_All_Noises.png,
	width=\figWidthJournNarrow,viewport=45 10 772 530,clip}}%
	\fi
\hfil
	\ifarXiv
\subfloat[]{\epsfig{figure=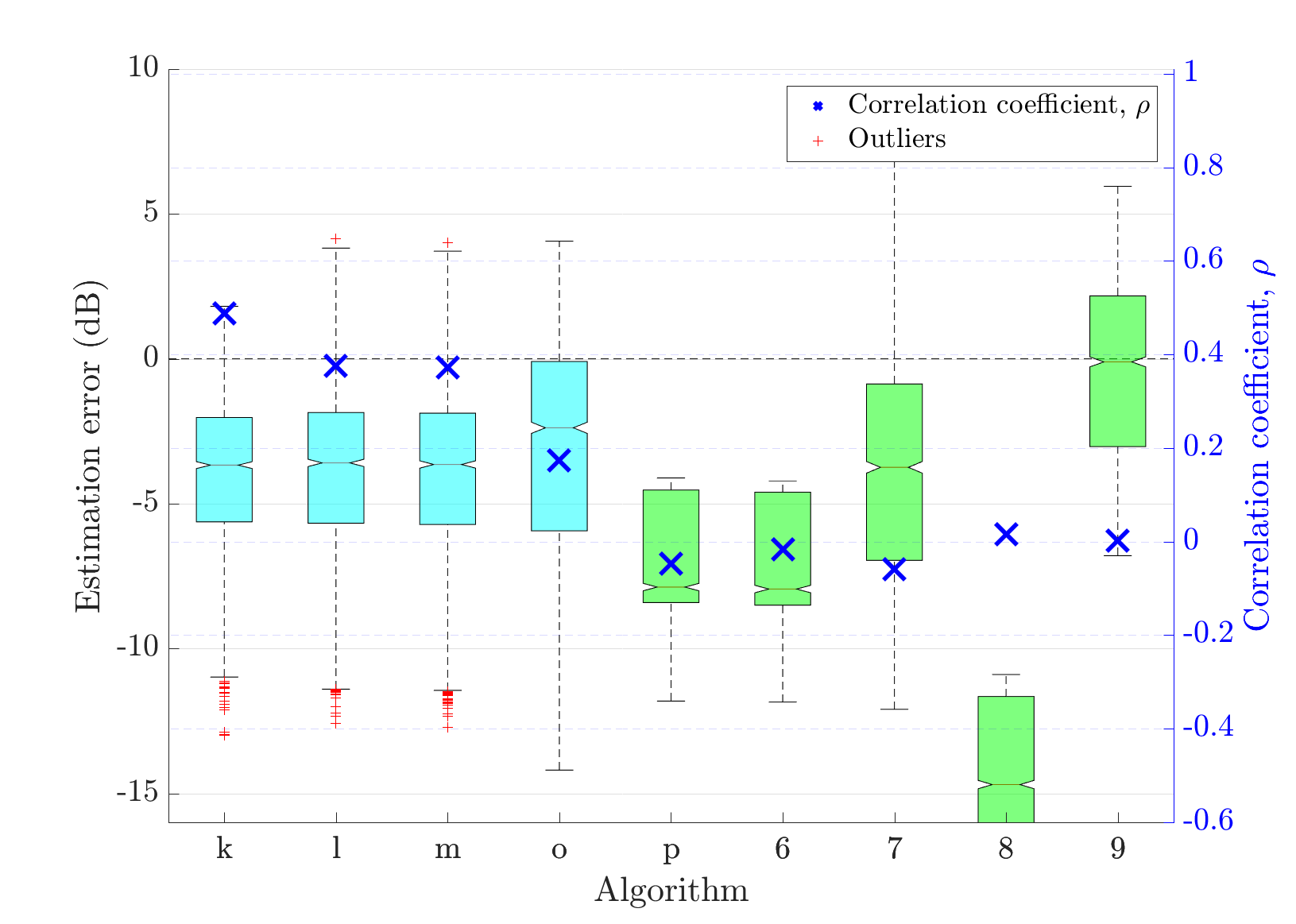,
	width=\figWidthJournNarrow,viewport=45 10 772 530,clip}}%
	\else
\subfloat[]{\epsfig{figure=FigsACE/ana_eval_gt_partic_results_combined_Phase3_All_TR3_Single_DRR_dB_GT_H_All_Noises.png,
	width=\figWidthJournNarrow,viewport=45 10 772 530,clip}}%
	\fi
\caption{Single channel \ac{FB2} \ac{DRR} estimation error in all noises and all \acp{SNR} for a) \ac{DRR} $<$\dBel{2} b) $2\leq$\ac{DRR} $<$~\dBel{5} and c) \ac{DRR} $\geq$~\dBel{5}}%
\label{fig:ACE_DRR_GT_Single_All}%
\end{figure}%
%
\begin{figure}[!ht]
\centering	
	\ifarXiv
\subfloat[]{\epsfig{figure=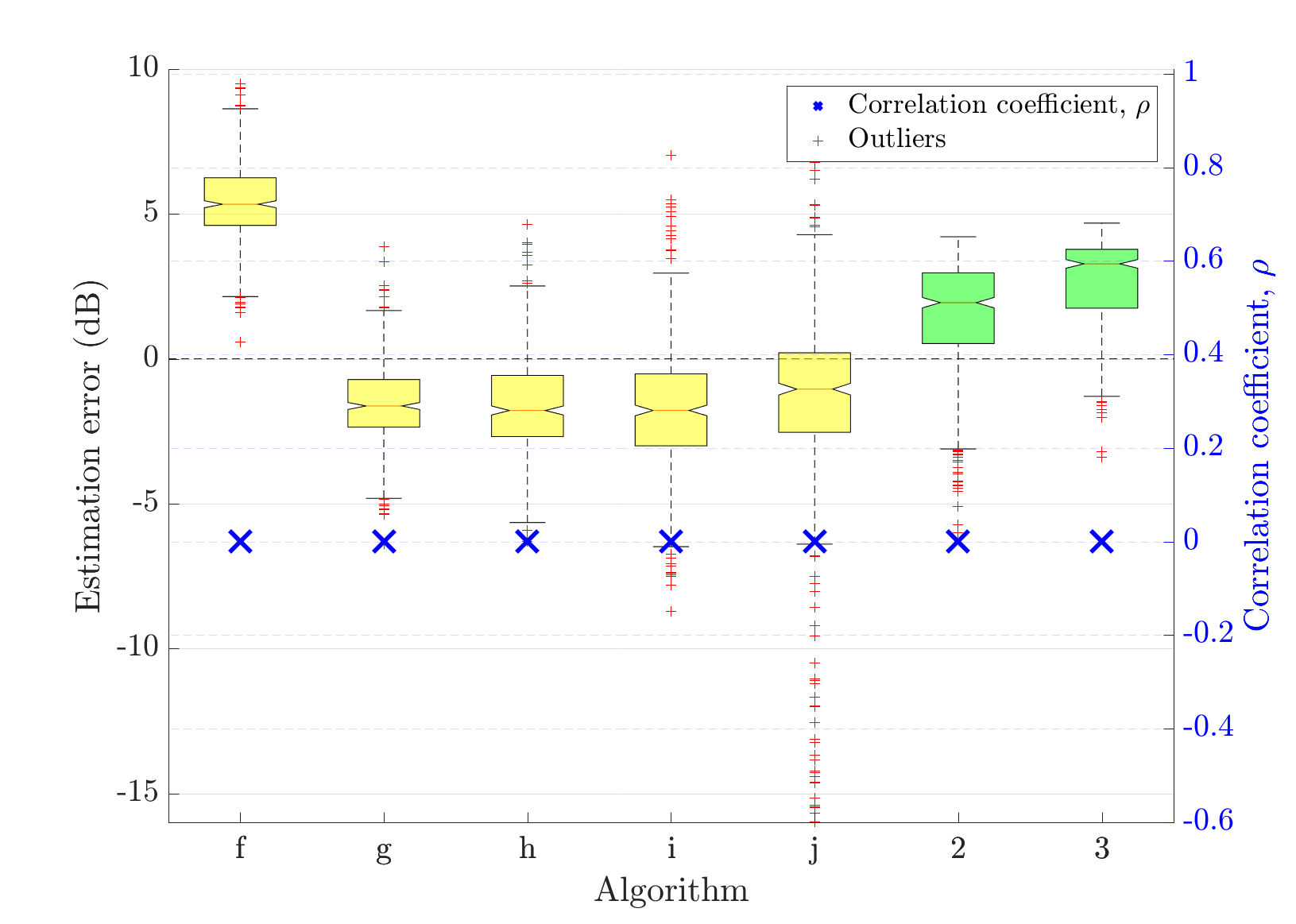,
	width=\figWidthJournNarrow,viewport=45 10 772 530,clip}}%
	\else
\subfloat[]{\epsfig{figure=FigsACE/ana_eval_gt_partic_results_combined_Phase3_All_TR3_Mobile_DRR_dB_GT_L_All_Noises.png,
	width=\figWidthJournNarrow,viewport=45 10 772 530,clip}}%
	\fi
\hfil
	\ifarXiv
\subfloat[]{\epsfig{figure=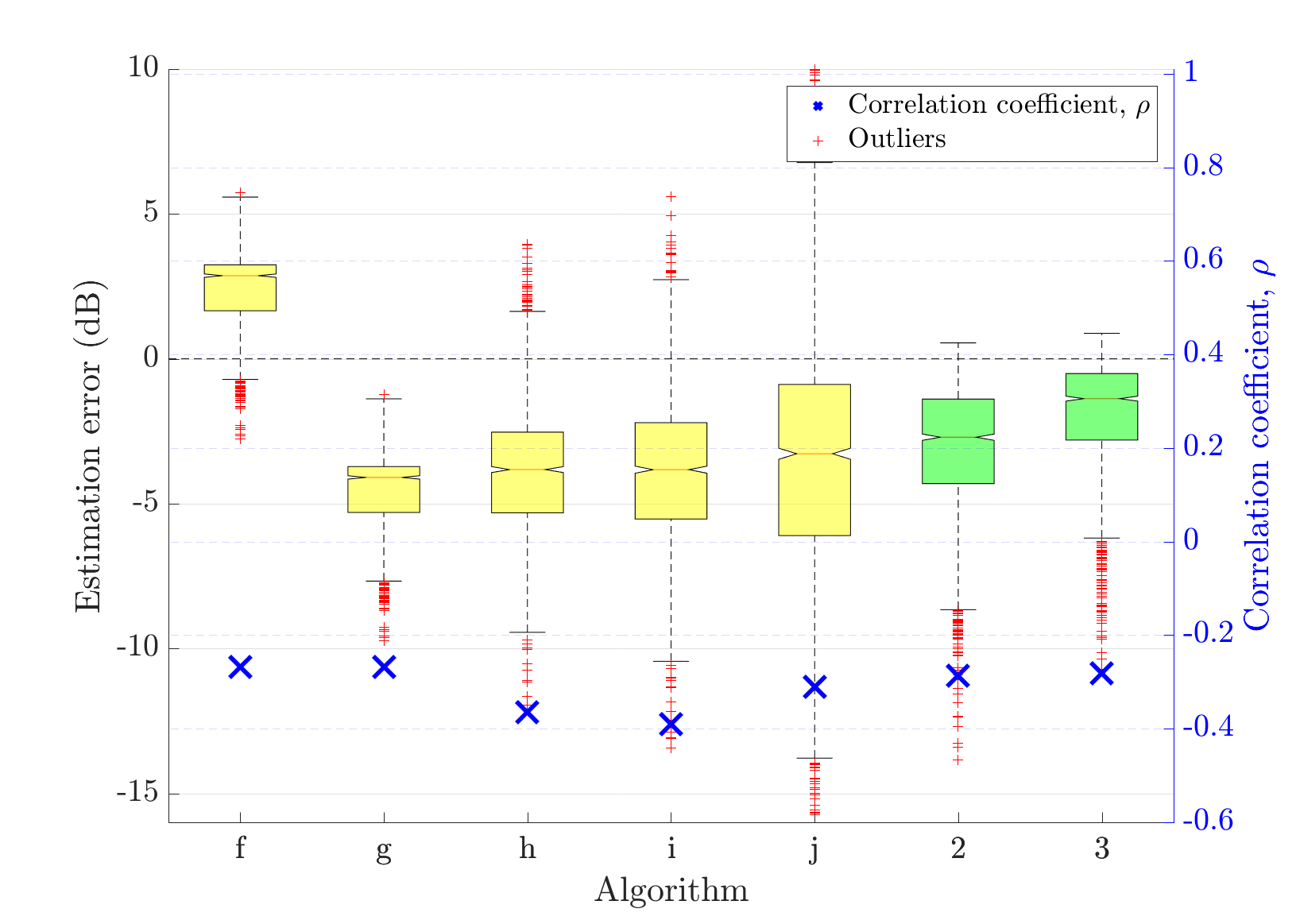,width=\figWidthJournNarrow,viewport=45 10 772 530,clip}}%
	\else
\subfloat[]{\epsfig{figure=FigsACE/ana_eval_gt_partic_results_combined_Phase3_All_TR3_Mobile_DRR_dB_GT_M_All_Noises.png,width=\figWidthJournNarrow,viewport=45 10 772 530,clip}}%
\fi
\hfil
	\ifarXiv
\subfloat[]{\epsfig{figure=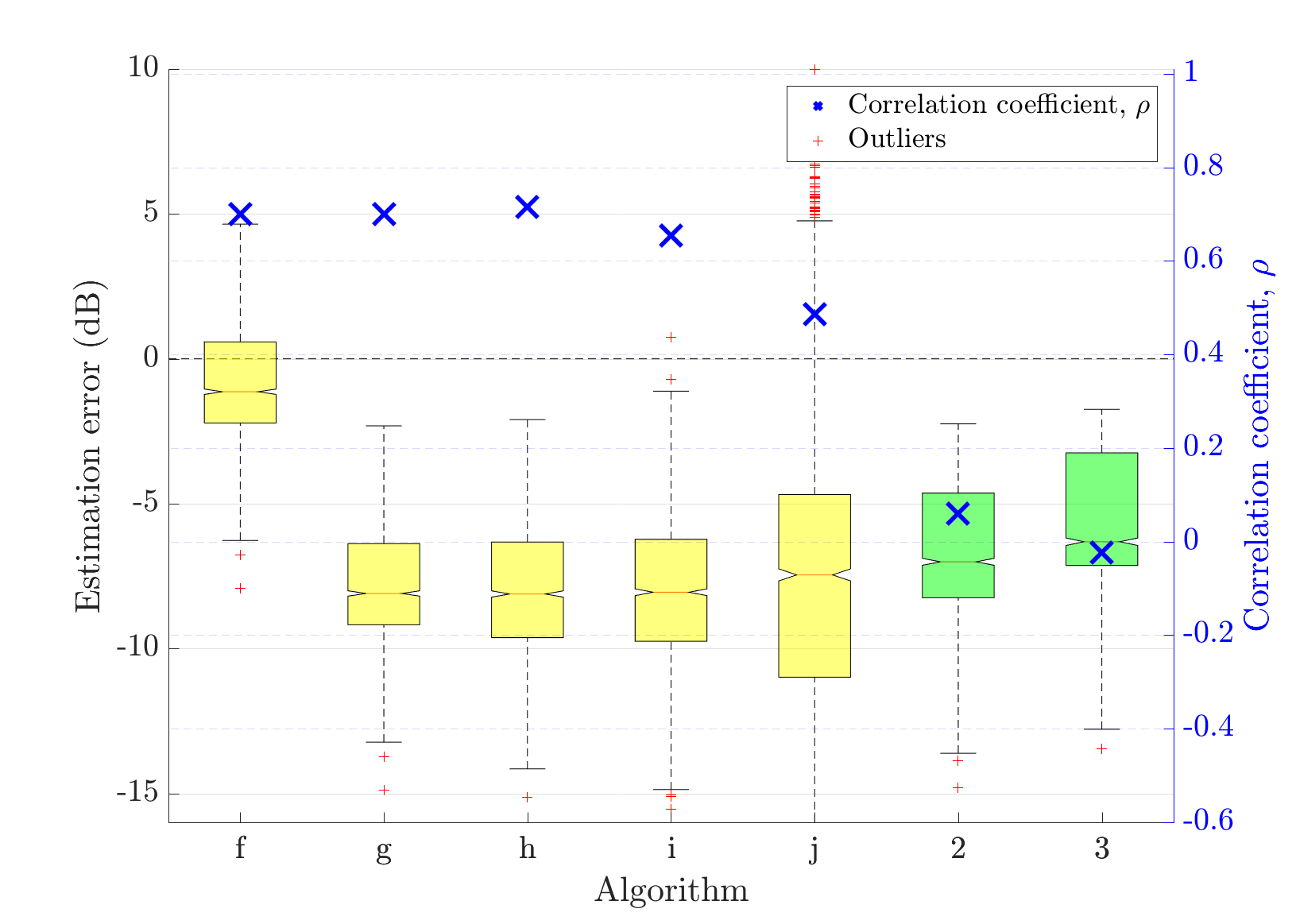,
	width=\figWidthJournNarrow,viewport=45 10 772 530,clip}}%
	\else
\subfloat[]{\epsfig{figure=FigsACE/ana_eval_gt_partic_results_combined_Phase3_All_TR3_Mobile_DRR_dB_GT_H_All_Noises.png,
	width=\figWidthJournNarrow,viewport=45 10 772 530,clip}}%
	\fi
\caption{Mobile (3-channel) \ac{FB2} \ac{DRR} estimation error in all noises and all \acp{SNR} for a) \ac{DRR} $<$\dBel{2} b) $2\leq$\ac{DRR} $<$~\dBel{5} and c) \ac{DRR} $\geq$~\dBel{5}. Note that for b) there are strong negative correlations for all algorithms}%
\label{fig:ACE_DRR_GT_Mobile_All}%
\end{figure}%
%
\begin{figure}[!ht]
\centering
	\ifarXiv
\subfloat[]{\epsfig{figure=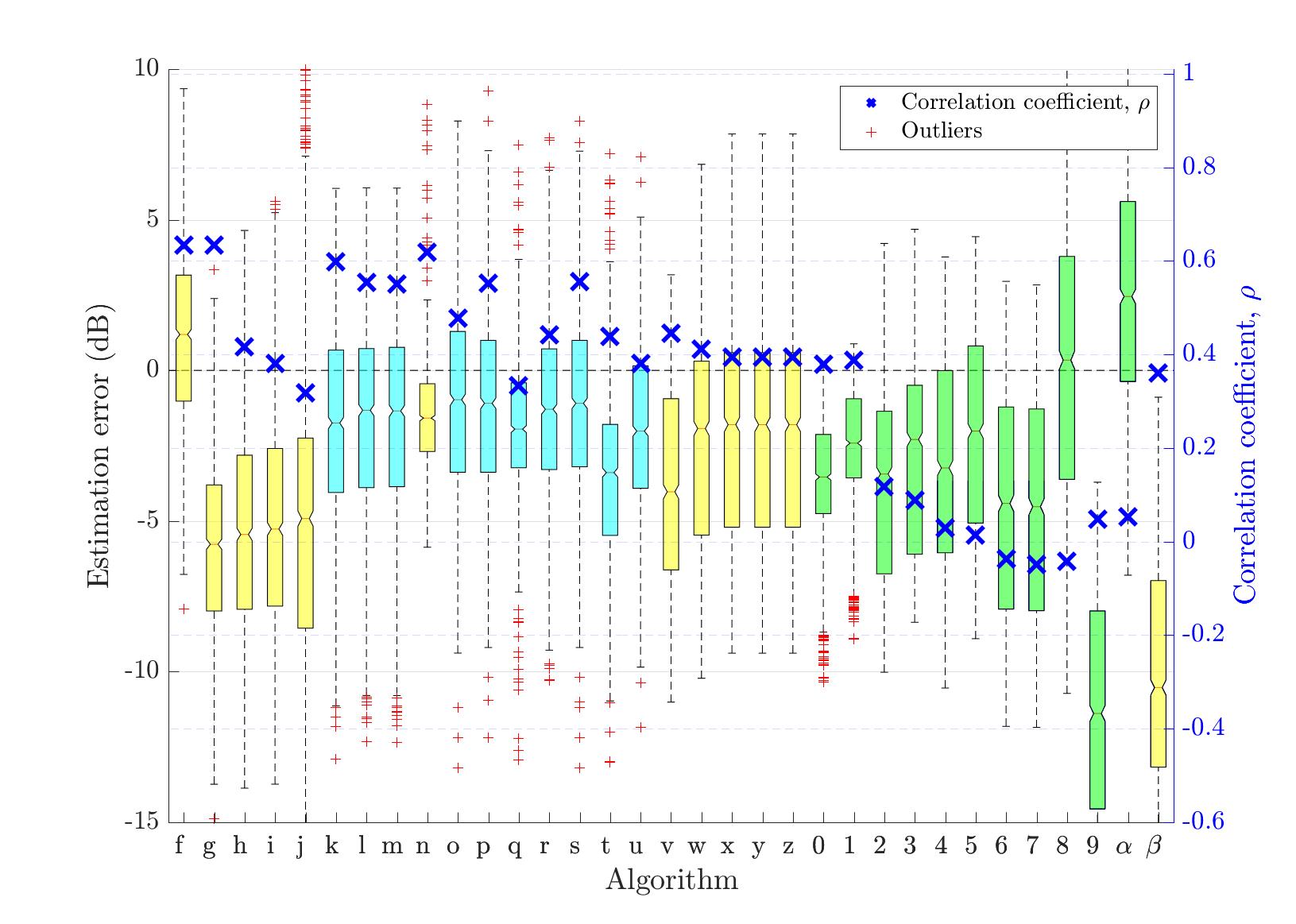,
	width=\figWidthConf,viewport=45 10 765 530,clip}}%
	\else
\subfloat[]{\epsfig{figure=FigsACE/ana_eval_gt_partic_results_combined_Phase3_All_WASPAA_P3_DRR_dB_H_All_Noises.png,
	width=\figWidthConf,viewport=45 10 765 530,clip}}%
	\fi
\hfil
	\ifarXiv
\subfloat[]{\epsfig{figure=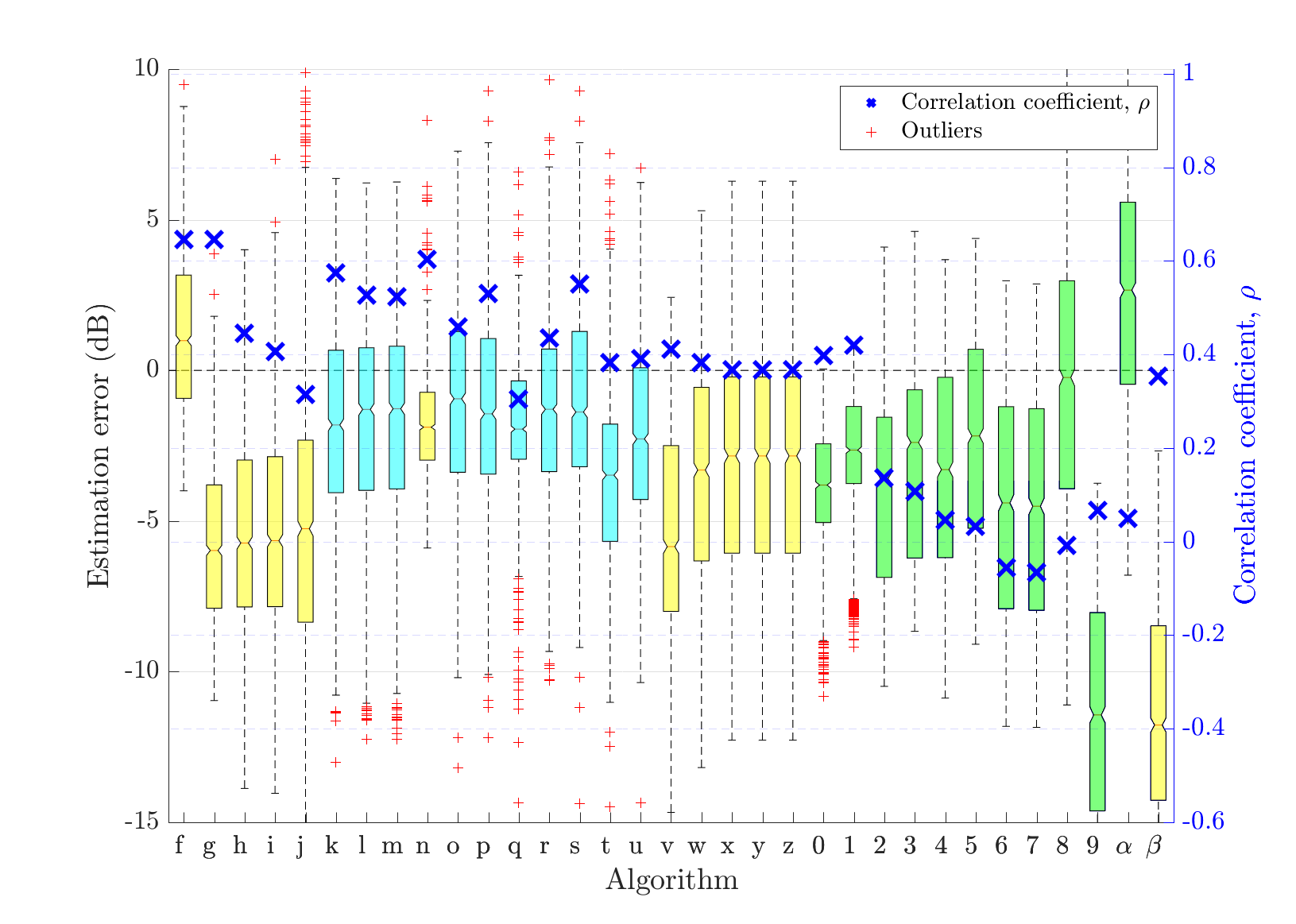,
	width=\figWidthConf,viewport=45 10 765 530,clip}}%
	\else
\subfloat[]{\epsfig{figure=FigsACE/ana_eval_gt_partic_results_combined_Phase3_All_WASPAA_P3_DRR_dB_M_All_Noises.png,
	width=\figWidthConf,viewport=45 10 765 530,clip}}%
	\fi
\hfil
	\ifarXiv
\subfloat[]{\epsfig{figure=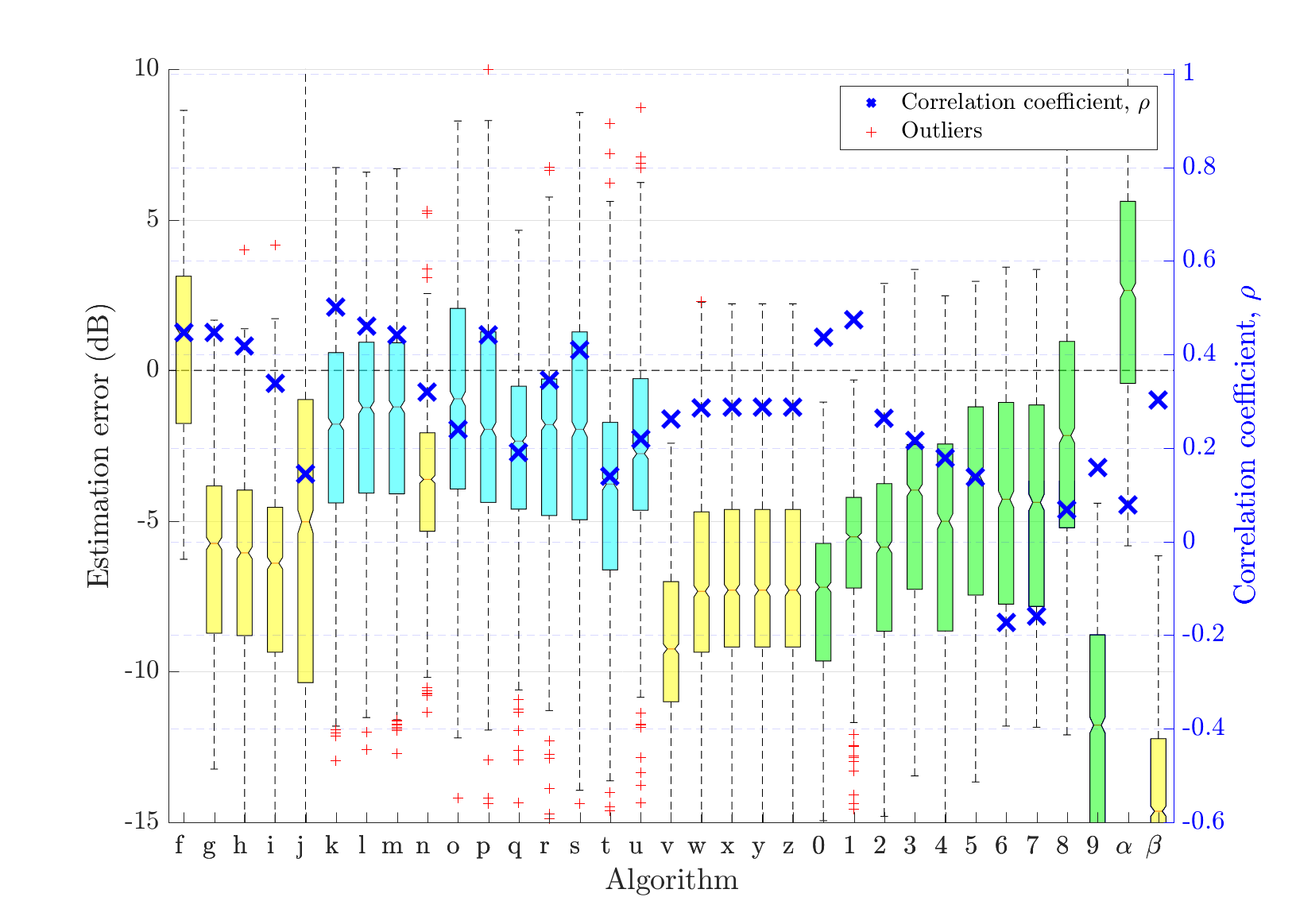,
	width=\figWidthConf,viewport=45 10 765 530,clip}}%
	\else
\subfloat[]{\epsfig{figure=FigsACE/ana_eval_gt_partic_results_combined_Phase3_All_WASPAA_P3_DRR_dB_L_All_Noises.png,
	width=\figWidthConf,viewport=45 10 765 530,clip}}%
	\fi
\caption{\ac{FB2} \ac{DRR} estimation error in all noises at a), \dBel{18} \ac{SNR}, b), \dBel{12} \ac{SNR}, and c) \dBel{-1} \ac{SNR}}%
\label{fig:ACE_DRR_SNR_All}%
\end{figure}%
%
\begin{figure}[!ht]
\centering	
	\ifarXiv
\subfloat[]{\epsfig{figure=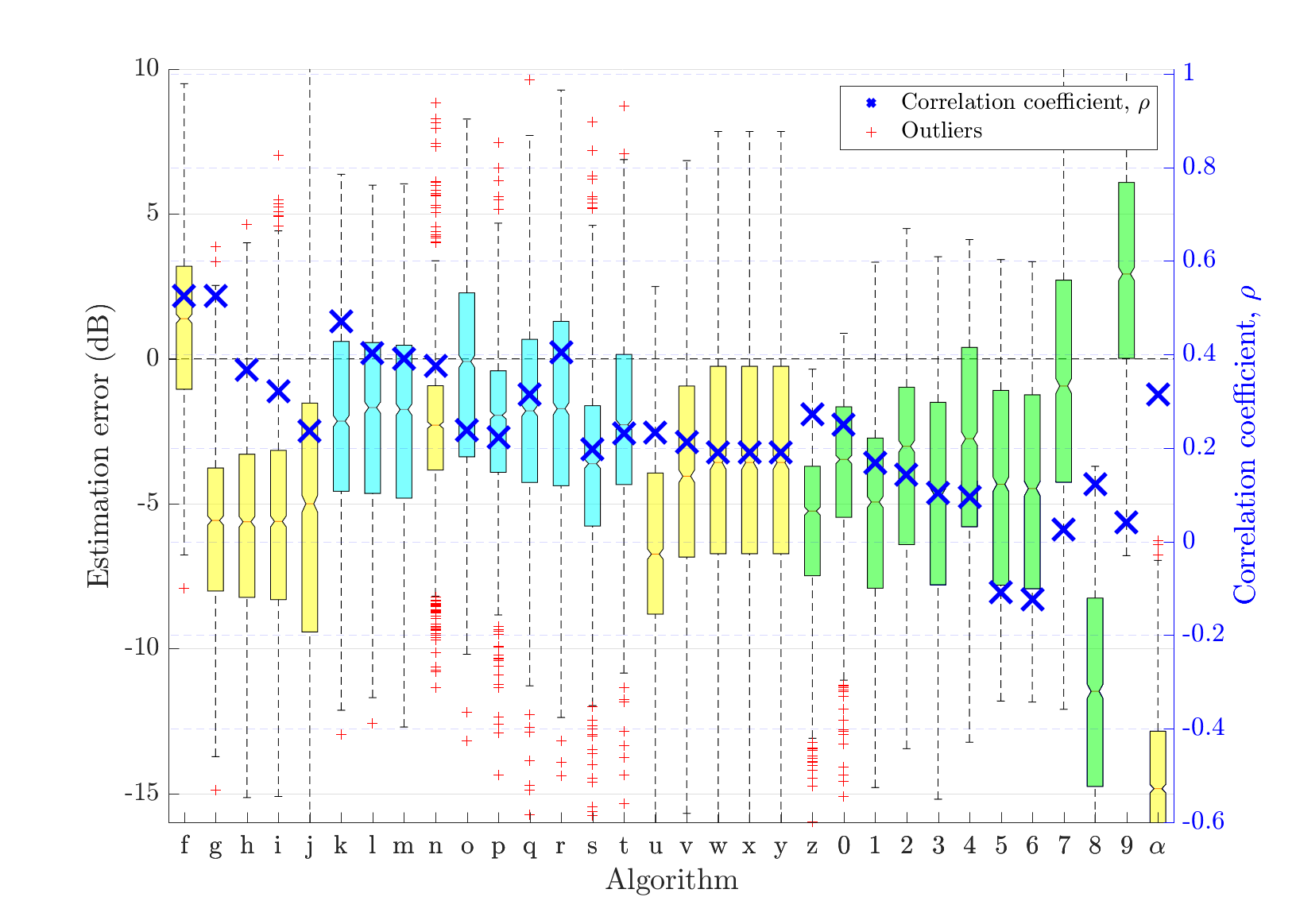,
	width=\figWidthConf,viewport=45 10 765 530,clip}}%
	\else
\subfloat[]{\epsfig{figure=FigsACE/ana_eval_gt_partic_results_combined_Phase3_All_TR3_DRR_dB_LEN_L_All_Noises.png,
	width=\figWidthConf,viewport=45 10 765 530,clip}}%
	\fi
\hfil
	\ifarXiv
\subfloat[]{\epsfig{figure=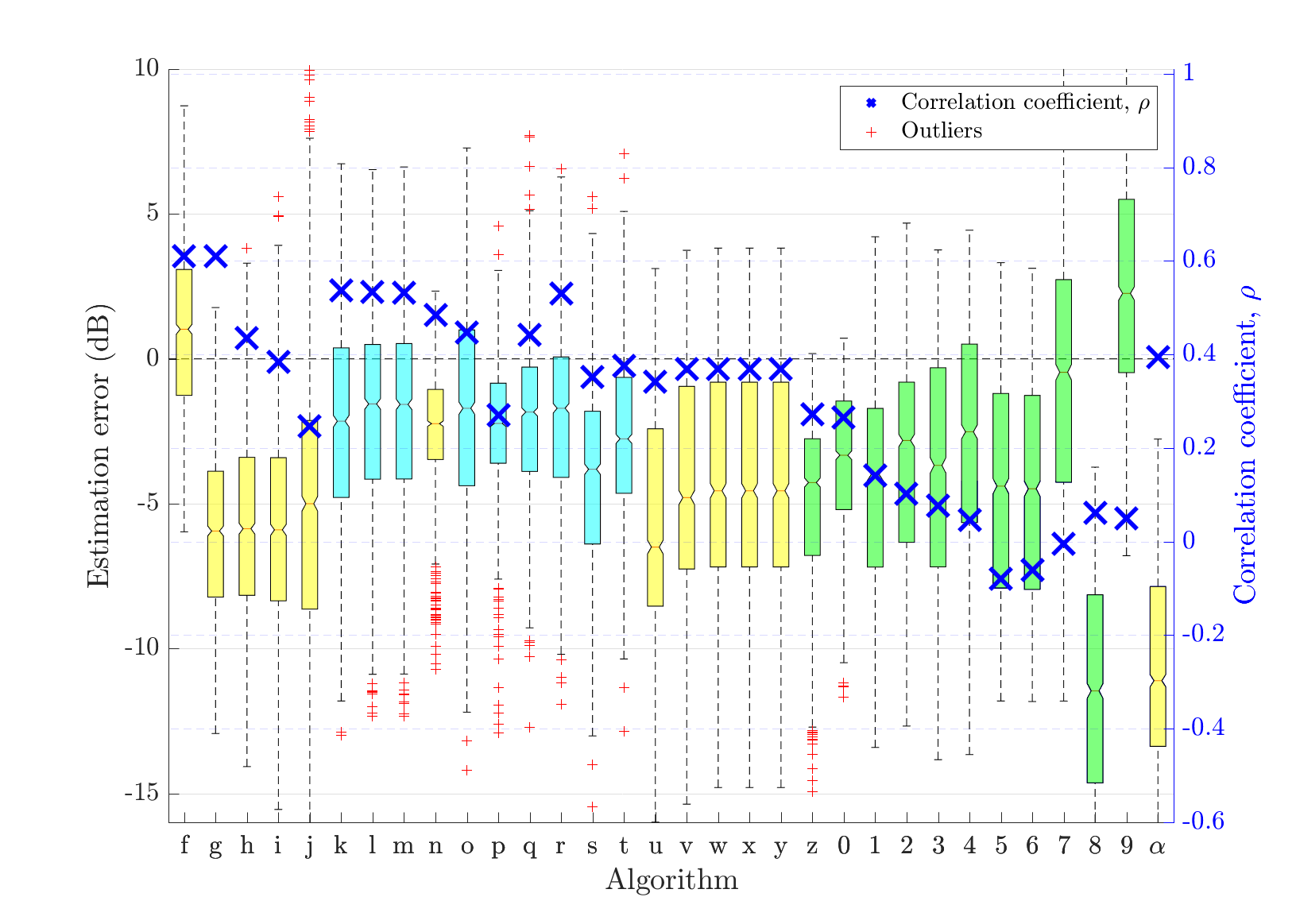,
	width=\figWidthConf,viewport=45 10 765 530,clip}}%
	\else
\subfloat[]{\epsfig{figure=FigsACE/ana_eval_gt_partic_results_combined_Phase3_All_TR3_DRR_dB_LEN_M_All_Noises.png,
	width=\figWidthConf,viewport=45 10 765 530,clip}}%
	\fi
\hfil
	\ifarXiv
\subfloat[]{\epsfig{figure=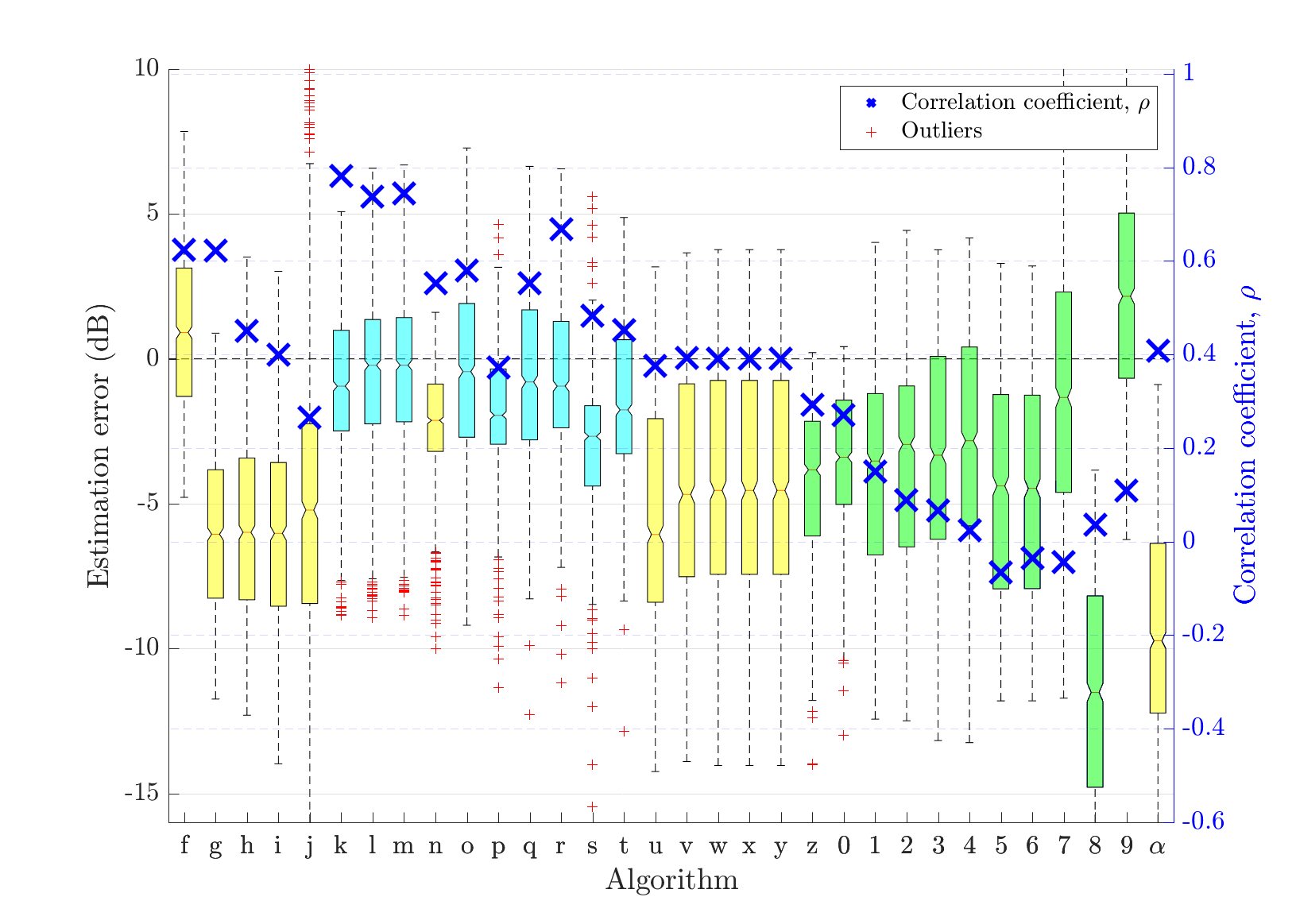,
	width=\figWidthConf,viewport=45 10 765 530,clip}}%
	\else
\subfloat[]{\epsfig{figure=FigsACE/ana_eval_gt_partic_results_combined_Phase3_All_TR3_DRR_dB_LEN_H_All_Noises.png,
	width=\figWidthConf,viewport=45 10 765 530,clip}}%
	\fi
\caption{\ac{FB2} \ac{DRR} estimation error in all noises and all \acp{SNR} for a) utterance length $<$\SI{5}{\second} b) utterance length $<$~\SI{15}{\second}  and c) utterance length $\geq$~\SI{15}{\second}}%
\label{fig:ACE_DRR_LEN_All}%
\end{figure}%
%
%
\fi
\clearpage
\section{\ac{T60} estimation results}
\subsection{Fullband \ac{T60} estimation results by noise type}
\subsubsection{Ambient noise}
%
\begin{figure}[!ht]
	\ifarXiv
\centerline{\epsfig{figure=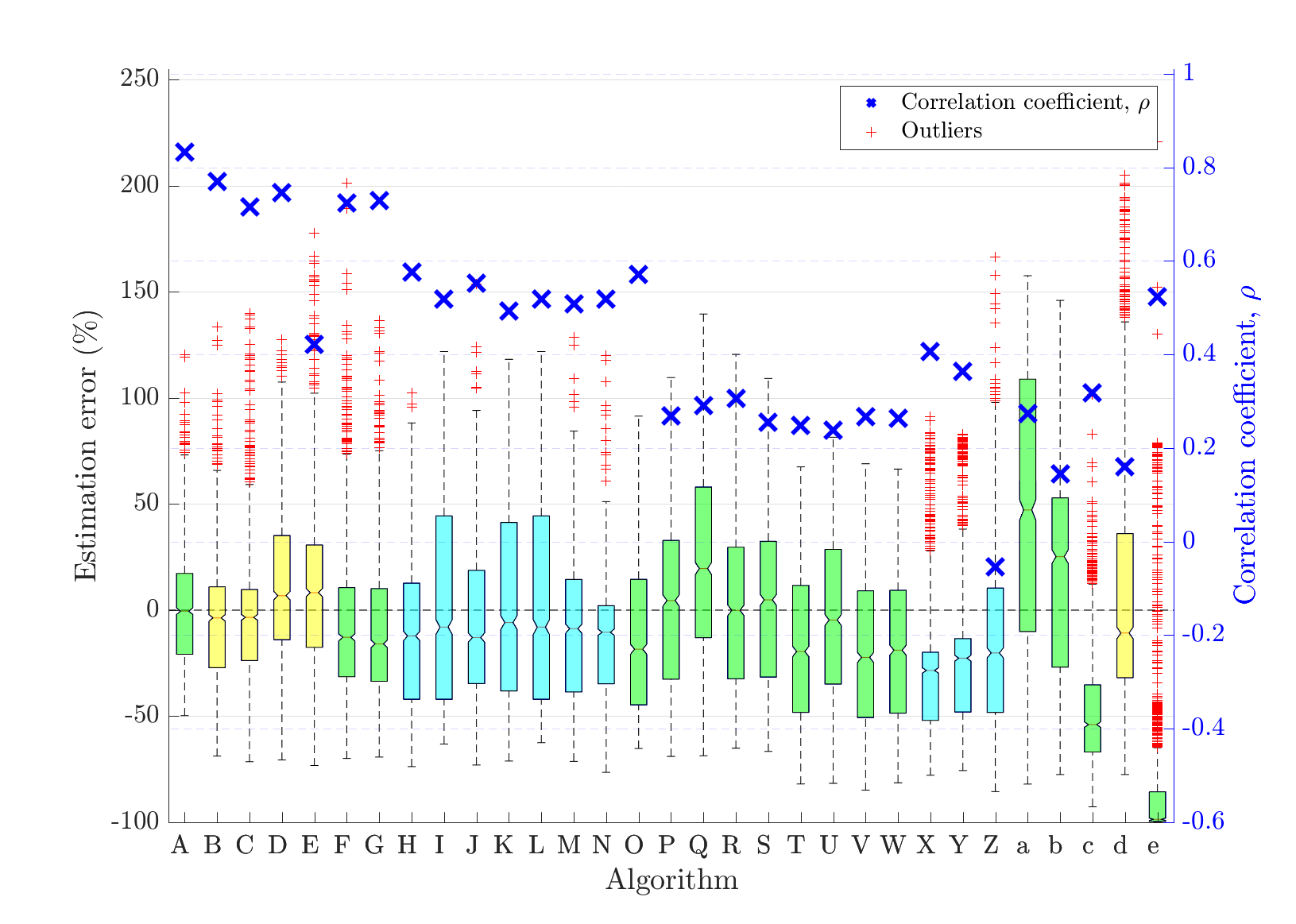,
	width=\figWidthACETR,viewport=45 10 765 530,clip}}%
	\else
	\centerline{\epsfig{figure=FigsACE/ana_eval_gt_partic_results_combined_Phase3_All_WASPAA_P3_T60_Perc_A2_Ambient.png,
	width=\figWidthACETR,viewport=45 10 765 530,clip}}%
	\fi
	\caption{Fullband {\ac{T60} estimation error in ambient noise for all \acp{SNR}}}%
\label{fig:ACE_T60_Ambient}%
\end{figure}%
\begin{table*}[!htb]\small
\caption{\ac{T60}  estimation algorithm performance in ambient noise for all \acp{SNR}}
\vspace{5mm} 
\centering
\begin{tabular}{cllllllll}%
\hline%
Ref.
& Algorithm
& Class
& Mic. Config.
& Bias
& \acs{MSE}
& $\PearsonCC$
& \ac{RTF}
\\
\hline%
\hline%
\ifarXiv
A & QA Reverb~\cite{Prego2015} & \ac{SFM} & Single & -0.0682 & 0.0565 & 0.833 & 0.401\\ 
\hline
B & Octave \ac{SB}-based \ac{FB2} RTE~\cite{Lollmann2015} & \ac{ABC} & Single & -0.0993 & 0.068 & 0.769 & 1\\ 
\hline
C & DCT-based \ac{FB2} RTE~\cite{Lollmann2015} & \ac{ABC} & Single & -0.0978 & 0.0738 & 0.715 & 1.04\\ 
\hline
D & Model-based \ac{SB} RTE~\cite{Lollmann2015} & \ac{ABC} & Single & -0.0378 & 0.0936 & 0.746 & 0.478\\ 
\hline
E & Baseline algorithm for \ac{FB2} RTE~\cite{Lollmann2015} & \ac{ABC} & Single & -0.0411 & 0.105 & 0.422 & 0.0421\\ 
\hline
F & \ac{SDDSA-G} retrained~\cite{Eaton2015b} & \ac{SFM} & Single & -0.0817 & 0.0676 & 0.723 & 0.0153\\ 
\hline
G & \ac{SDDSA-G}~\cite{Eaton2013} & \ac{SFM} & Single & -0.117 & 0.0738 & 0.729 & 0.0166\\ 
\hline
H & Multi-layer perceptron~\cite{Xiong2015} & \ac{MLMF} & Single & -0.125 & 0.0977 & 0.576 & 0.0578$^\ddagger$\\ 
\hline
I & Multi-layer perceptron P2~\cite{Xiong2015} & \ac{MLMF} & Single & -0.0844 & 0.0969 & 0.518 & 0.0578$^\ddagger$\\ 
\hline
J & Multi-layer perceptron P2~\cite{Xiong2015} & \ac{MLMF} & Chromebook & -0.0704 & 0.0917 & 0.553 & 0.0589$^\ddagger$\\ 
\hline
K & Multi-layer perceptron P2~\cite{Xiong2015} & \ac{MLMF} & Mobile & -0.0581 & 0.0798 & 0.492 & 0.0557$^\ddagger$\\ 
\hline
L & Multi-layer perceptron P2~\cite{Xiong2015} & \ac{MLMF} & Crucif & -0.0852 & 0.0971 & 0.518 & 0.0569$^\ddagger$\\ 
\hline
M & Multi-layer perceptron P2~\cite{Xiong2015} & \ac{MLMF} & Lin8Ch & -0.0818 & 0.084 & 0.508 & 0.062$^\ddagger$\\ 
\hline
N & Multi-layer perceptron P2~\cite{Xiong2015} & \ac{MLMF} & EM32 & -0.0968 & 0.084 & 0.519 & 0.0578$^\ddagger$\\ 
\hline
O & Per acoust. band \ac{SRMR} {\sectMidSent} 2.5.~\cite{Senoussaoui2015} & \ac{SFM} & Single & -0.16 & 0.113 & 0.572 & 0.58\\ 
\hline
P & \ac{NSRMR} {\sectMidSent} 2.4.~\cite{Santos2014,Senoussaoui2015} & \ac{SFM} & Single & -0.0964 & 0.123 & 0.27 & 0.571\\ 
\hline
Q & \ac{NSRMR} {\sectMidSent} 2.4.~\cite{Santos2014,Senoussaoui2015} & \ac{SFM} & Chromebook & -0.00429 & 0.116 & 0.291 & 1.04\\ 
\hline
R & \ac{NSRMR} {\sectMidSent} 2.4.~\cite{Santos2014,Senoussaoui2015} & \ac{SFM} & Mobile & -0.0837 & 0.0976 & 0.306 & 1.59\\ 
\hline
S & \ac{NSRMR} {\sectMidSent} 2.4.~\cite{Santos2014,Senoussaoui2015} & \ac{SFM} & Crucif & -0.0838 & 0.11 & 0.256 & 2.63\\ 
\hline
T & \ac{SRMR} {\sectMidSent} 2.3.~\cite{Senoussaoui2015} & \ac{SFM} & Single & -0.195 & 0.153 & 0.249 & 0.457\\ 
\hline
U & \ac{SRMR} {\sectMidSent} 2.3.~\cite{Senoussaoui2015} & \ac{SFM} & Chromebook & -0.13 & 0.136 & 0.239 & 0.831\\ 
\hline
V & \ac{SRMR} {\sectMidSent} 2.3.~\cite{Senoussaoui2015} & \ac{SFM} & Mobile & -0.189 & 0.129 & 0.268 & 1.26\\ 
\hline
W & \ac{SRMR} {\sectMidSent} 2.3.~\cite{Senoussaoui2015} & \ac{SFM} & Crucif & -0.188 & 0.137 & 0.263 & 2.09\\ 
\hline
X & NIRAv3~\cite{Parada2015} & \ac{MLMF} & Single & -0.263 & 0.172 & 0.406 & 0.897$^\dagger$\\ 
\hline
Y & NIRAv1~\cite{Parada2015} & \ac{MLMF} & Single & -0.243 & 0.166 & 0.363 & 0.897$^\dagger$\\ 
\hline
Z & NIRAv2~\cite{Parada2015} & \ac{MLMF} & Single & -0.183 & 0.198 & -0.0532 & 0.912$^\dagger$\\ 
\hline
a & Blur kernel~\cite{Lim2015} & \ac{SFM} & Single & 0.164 & 0.15 & 0.274 & 8.16\\ 
\hline
b & Blur kernel with sliding window~\cite{Lim2015a} & \ac{SFM} & Single & -0.0155 & 0.137 & 0.144 & 0.413\\ 
\hline
c & Temporal dynamics~\cite{Falk2010a} & \ac{SFM} & Single & -0.359 & 0.239 & 0.319 & 0.362\\ 
\hline
d & Improved blind RTE~\cite{Lollmann2010} & \ac{ABC} & Single & -0.0752 & 0.168 & 0.159 & 0.0255\\ 
\hline
e & \ac{SDD}~\cite{Wen2008} & \ac{SFM} & Single & -0.515 & 0.355 & 0.524 & 0.0219\\ 
\hline
\else
\fi
\end{tabular}%
\end{table*}%
\clearpage
\subsubsection{Babble noise}
%
\begin{figure}[!ht]
	\ifarXiv
\centerline{\epsfig{figure=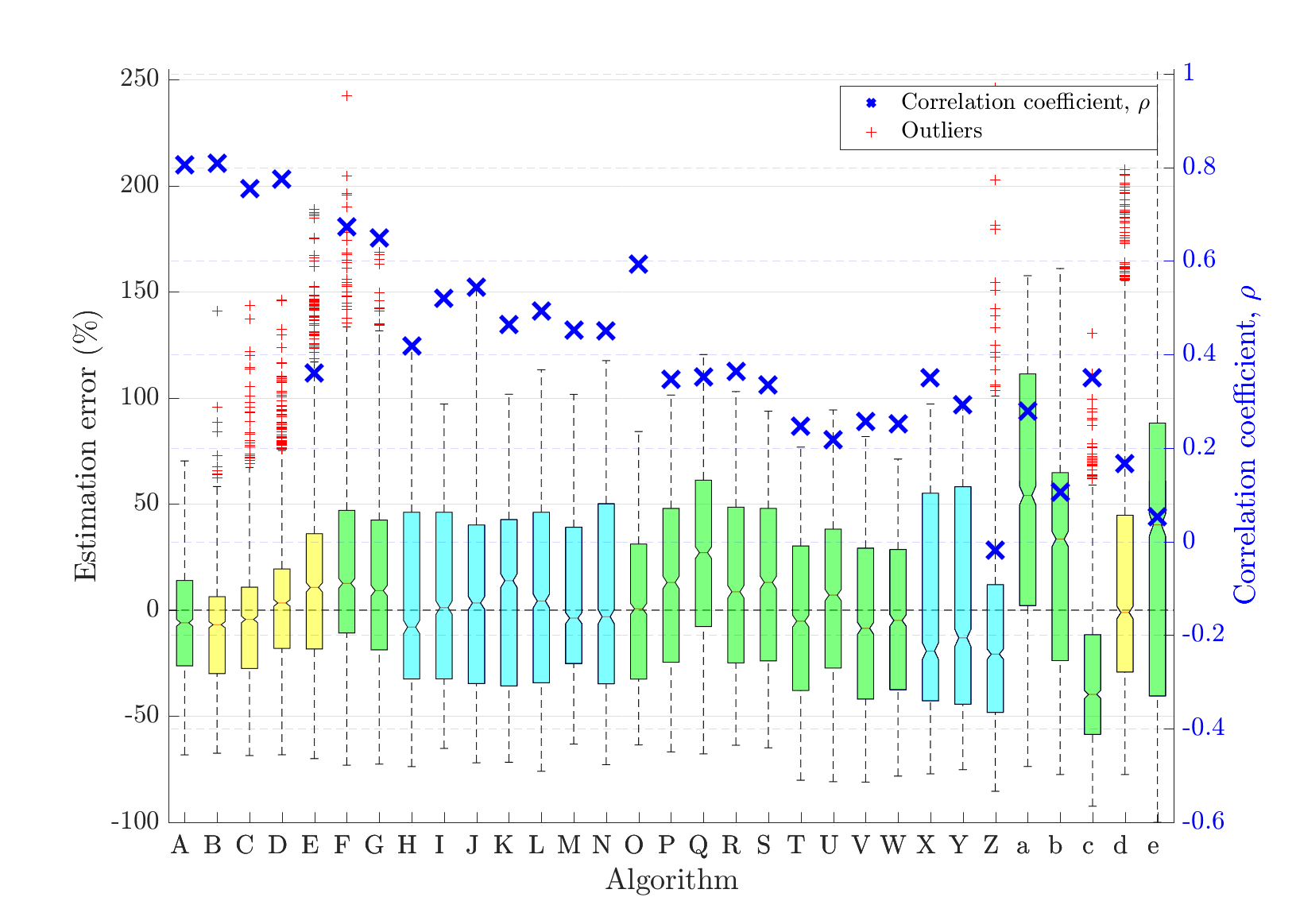,
	width=\figWidthACETR,viewport=45 10 765 530,clip}}%
	\else
	\centerline{\epsfig{figure=FigsACE/ana_eval_gt_partic_results_combined_Phase3_All_WASPAA_P3_T60_Perc_A2_Babble.png,
	width=\figWidthACETR,viewport=45 10 765 530,clip}}%
	\fi
	\caption{Fullband {\ac{T60} estimation error in babble noise for all \acp{SNR}}}%
\label{fig:ACE_T60_Babble}%
\end{figure}%
\begin{table*}[!htb]\small
\caption{\ac{T60} estimation algorithm performance in babble noise for all \acp{SNR}}
\vspace{5mm} 
\centering
\begin{tabular}{cllllllll}%
\hline%
Ref.
& Algorithm
& Class
& Mic. Config.
& Bias
& \acs{MSE}
& $\PearsonCC$
& \ac{RTF}
\\
\hline%
\hline%
\ifarXiv
A & QA Reverb~\cite{Prego2015} & \ac{SFM} & Single & -0.109 & 0.0707 & 0.805 & 0.398\\ 
\hline
B & Octave \ac{SB}-based \ac{FB2} RTE~\cite{Lollmann2015} & \ac{ABC} & Single & -0.124 & 0.0701 & 0.809 & 0.911\\ 
\hline
C & DCT-based \ac{FB2} RTE~\cite{Lollmann2015} & \ac{ABC} & Single & -0.106 & 0.0718 & 0.755 & 0.99\\ 
\hline
D & Model-based \ac{SB} RTE~\cite{Lollmann2015} & \ac{ABC} & Single & -0.0699 & 0.0933 & 0.774 & 0.443\\ 
\hline
E & Baseline algorithm for \ac{FB2} RTE~\cite{Lollmann2015} & \ac{ABC} & Single & -0.0236 & 0.112 & 0.36 & 0.0428\\ 
\hline
F & \ac{SDDSA-G} retrained~\cite{Eaton2015b} & \ac{SFM} & Single & 0.0688 & 0.0836 & 0.673 & 0.0155\\ 
\hline
G & \ac{SDDSA-G}~\cite{Eaton2013} & \ac{SFM} & Single & -0.000784 & 0.0718 & 0.649 & 0.0162\\ 
\hline
H & Multi-layer perceptron~\cite{Xiong2015} & \ac{MLMF} & Single & -0.0684 & 0.106 & 0.419 & 0.0579$^\ddagger$\\ 
\hline
I & Multi-layer perceptron P2~\cite{Xiong2015} & \ac{MLMF} & Single & -0.045 & 0.092 & 0.52 & 0.0579$^\ddagger$\\ 
\hline
J & Multi-layer perceptron P2~\cite{Xiong2015} & \ac{MLMF} & Chromebook & -0.0534 & 0.0912 & 0.543 & 0.0588$^\ddagger$\\ 
\hline
K & Multi-layer perceptron P2~\cite{Xiong2015} & \ac{MLMF} & Mobile & -0.0244 & 0.0796 & 0.465 & 0.0555$^\ddagger$\\ 
\hline
L & Multi-layer perceptron P2~\cite{Xiong2015} & \ac{MLMF} & Crucif & -0.0432 & 0.0948 & 0.494 & 0.057$^\ddagger$\\ 
\hline
M & Multi-layer perceptron P2~\cite{Xiong2015} & \ac{MLMF} & Lin8Ch & -0.0277 & 0.084 & 0.452 & 0.0618$^\ddagger$\\ 
\hline
N & Multi-layer perceptron P2~\cite{Xiong2015} & \ac{MLMF} & EM32 & -0.0569 & 0.0853 & 0.451 & 0.0576$^\ddagger$\\ 
\hline
O & Per acoust. band \ac{SRMR} {\sectMidSent} 2.5.~\cite{Senoussaoui2015} & \ac{SFM} & Single & -0.0967 & 0.0992 & 0.593 & 0.579\\ 
\hline
P & \ac{NSRMR} {\sectMidSent} 2.4.~\cite{Santos2014,Senoussaoui2015} & \ac{SFM} & Single & -0.0435 & 0.11 & 0.347 & 0.572\\ 
\hline
Q & \ac{NSRMR} {\sectMidSent} 2.4.~\cite{Santos2014,Senoussaoui2015} & \ac{SFM} & Chromebook & 0.00512 & 0.11 & 0.353 & 1.04\\ 
\hline
R & \ac{NSRMR} {\sectMidSent} 2.4.~\cite{Santos2014,Senoussaoui2015} & \ac{SFM} & Mobile & -0.0287 & 0.0873 & 0.364 & 1.58\\ 
\hline
S & \ac{NSRMR} {\sectMidSent} 2.4.~\cite{Santos2014,Senoussaoui2015} & \ac{SFM} & Crucif & -0.03 & 0.098 & 0.335 & 2.63\\ 
\hline
T & \ac{SRMR} {\sectMidSent} 2.3.~\cite{Senoussaoui2015} & \ac{SFM} & Single & -0.129 & 0.133 & 0.246 & 0.457\\ 
\hline
U & \ac{SRMR} {\sectMidSent} 2.3.~\cite{Senoussaoui2015} & \ac{SFM} & Chromebook & -0.0928 & 0.13 & 0.217 & 0.833\\ 
\hline
V & \ac{SRMR} {\sectMidSent} 2.3.~\cite{Senoussaoui2015} & \ac{SFM} & Mobile & -0.12 & 0.109 & 0.257 & 1.26\\ 
\hline
W & \ac{SRMR} {\sectMidSent} 2.3.~\cite{Senoussaoui2015} & \ac{SFM} & Crucif & -0.121 & 0.118 & 0.252 & 2.1\\ 
\hline
X & NIRAv3~\cite{Parada2015} & \ac{MLMF} & Single & -0.0965 & 0.121 & 0.35 & 0.906$^\dagger$\\ 
\hline
Y & NIRAv1~\cite{Parada2015} & \ac{MLMF} & Single & -0.0899 & 0.124 & 0.292 & 0.906$^\dagger$\\ 
\hline
Z & NIRAv2~\cite{Parada2015} & \ac{MLMF} & Single & -0.176 & 0.203 & -0.0191 & 0.901$^\dagger$\\ 
\hline
a & Blur kernel~\cite{Lim2015} & \ac{SFM} & Single & 0.184 & 0.152 & 0.279 & 8.88\\ 
\hline
b & Blur kernel with sliding window~\cite{Lim2015a} & \ac{SFM} & Single & 0.0187 & 0.138 & 0.106 & 0.438\\ 
\hline
c & Temporal dynamics~\cite{Falk2010a} & \ac{SFM} & Single & -0.257 & 0.178 & 0.35 & 0.365\\ 
\hline
d & Improved blind RTE~\cite{Lollmann2010} & \ac{ABC} & Single & -0.0357 & 0.164 & 0.167 & 0.0269\\ 
\hline
e & \ac{SDD}~\cite{Wen2008} & \ac{SFM} & Single & 0.593 & 52.8 & 0.0524 & 0.0224\\ 
\hline
\else
\fi
\end{tabular}%
\end{table*}%
\clearpage
\subsubsection{Fan noise}
%
%
\begin{figure}[!ht]
	\ifarXiv
\centerline{\epsfig{figure=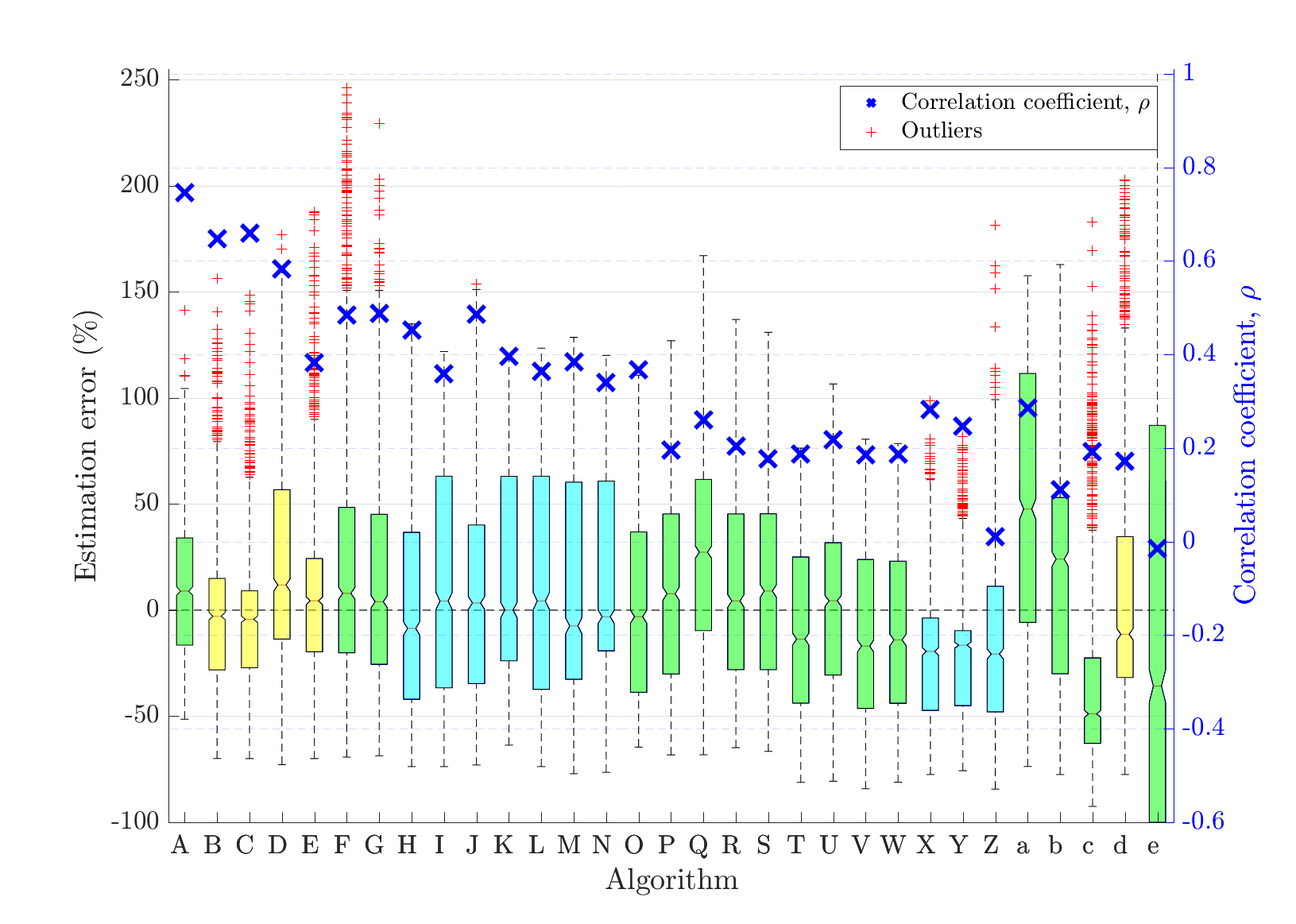,
	width=\figWidthACETR,viewport=45 10 765 530,clip}}%
	\else
	\centerline{\epsfig{figure=FigsACE/ana_eval_gt_partic_results_combined_Phase3_All_WASPAA_P3_T60_Perc_A2_Fan.png,
	width=\figWidthACETR,viewport=45 10 765 530,clip}}%
	\fi
	\caption{Fullband {\ac{T60} estimation error in fan noise for all \acp{SNR}}}%
\label{fig:ACE_T60_All}%
\end{figure}%
\begin{table*}[!htb]\small
\caption{\ac{T60} estimation algorithm performance in fan noise for all \acp{SNR}}
\vspace{5mm} 
\centering
\begin{tabular}{cllllllll}%
\hline%
Ref.
& Algorithm
& Class
& Mic. Config.
& Bias
& \acs{MSE}
& $\PearsonCC$
& \ac{RTF}
\\
\hline%
\hline%
\ifarXiv
A & QA Reverb~\cite{Prego2015} & \ac{SFM} & Single & -0.0267 & 0.0672 & 0.746 & 0.4\\ 
\hline
B & Octave \ac{SB}-based \ac{FB2} RTE~\cite{Lollmann2015} & \ac{ABC} & Single & -0.0881 & 0.0811 & 0.647 & 0.903\\ 
\hline
C & DCT-based \ac{FB2} RTE~\cite{Lollmann2015} & \ac{ABC} & Single & -0.109 & 0.0843 & 0.659 & 0.984\\ 
\hline
D & Model-based \ac{SB} RTE~\cite{Lollmann2015} & \ac{ABC} & Single & -0.00134 & 0.119 & 0.583 & 0.433\\ 
\hline
E & Baseline algorithm for \ac{FB2} RTE~\cite{Lollmann2015} & \ac{ABC} & Single & -0.065 & 0.112 & 0.383 & 0.0421\\ 
\hline
F & \ac{SDDSA-G} retrained~\cite{Eaton2015b} & \ac{SFM} & Single & 0.0629 & 0.13 & 0.484 & 0.0148\\ 
\hline
G & \ac{SDDSA-G}~\cite{Eaton2013} & \ac{SFM} & Single & -0.00884 & 0.0952 & 0.488 & 0.0164\\ 
\hline
H & Multi-layer perceptron~\cite{Xiong2015} & \ac{MLMF} & Single & -0.097 & 0.108 & 0.451 & 0.0578$^\ddagger$\\ 
\hline
I & Multi-layer perceptron P2~\cite{Xiong2015} & \ac{MLMF} & Single & -0.0197 & 0.109 & 0.359 & 0.0578$^\ddagger$\\ 
\hline
J & Multi-layer perceptron P2~\cite{Xiong2015} & \ac{MLMF} & Chromebook & -0.0382 & 0.0971 & 0.486 & 0.059$^\ddagger$\\ 
\hline
K & Multi-layer perceptron P2~\cite{Xiong2015} & \ac{MLMF} & Mobile & -0.00714 & 0.0864 & 0.396 & 0.0555$^\ddagger$\\ 
\hline
L & Multi-layer perceptron P2~\cite{Xiong2015} & \ac{MLMF} & Crucif & -0.0224 & 0.108 & 0.364 & 0.0569$^\ddagger$\\ 
\hline
M & Multi-layer perceptron P2~\cite{Xiong2015} & \ac{MLMF} & Lin8Ch & -0.031 & 0.0925 & 0.384 & 0.0617$^\ddagger$\\ 
\hline
N & Multi-layer perceptron P2~\cite{Xiong2015} & \ac{MLMF} & EM32 & -0.0268 & 0.0945 & 0.339 & 0.0574$^\ddagger$\\ 
\hline
O & Per acoust. band \ac{SRMR} {\sectMidSent} 2.5.~\cite{Senoussaoui2015} & \ac{SFM} & Single & -0.0853 & 0.114 & 0.367 & 0.576\\ 
\hline
P & \ac{NSRMR} {\sectMidSent} 2.4.~\cite{Santos2014,Senoussaoui2015} & \ac{SFM} & Single & -0.054 & 0.125 & 0.195 & 0.569\\ 
\hline
Q & \ac{NSRMR} {\sectMidSent} 2.4.~\cite{Santos2014,Senoussaoui2015} & \ac{SFM} & Chromebook & 0.0352 & 0.122 & 0.26 & 1.03\\ 
\hline
R & \ac{NSRMR} {\sectMidSent} 2.4.~\cite{Santos2014,Senoussaoui2015} & \ac{SFM} & Mobile & -0.0389 & 0.102 & 0.204 & 1.58\\ 
\hline
S & \ac{NSRMR} {\sectMidSent} 2.4.~\cite{Santos2014,Senoussaoui2015} & \ac{SFM} & Crucif & -0.0411 & 0.113 & 0.177 & 2.61\\ 
\hline
T & \ac{SRMR} {\sectMidSent} 2.3.~\cite{Senoussaoui2015} & \ac{SFM} & Single & -0.156 & 0.145 & 0.188 & 0.455\\ 
\hline
U & \ac{SRMR} {\sectMidSent} 2.3.~\cite{Senoussaoui2015} & \ac{SFM} & Chromebook & -0.0922 & 0.131 & 0.218 & 0.824\\ 
\hline
V & \ac{SRMR} {\sectMidSent} 2.3.~\cite{Senoussaoui2015} & \ac{SFM} & Mobile & -0.149 & 0.123 & 0.185 & 1.26\\ 
\hline
W & \ac{SRMR} {\sectMidSent} 2.3.~\cite{Senoussaoui2015} & \ac{SFM} & Crucif & -0.149 & 0.13 & 0.188 & 2.08\\ 
\hline
X & NIRAv3~\cite{Parada2015} & \ac{MLMF} & Single & -0.215 & 0.159 & 0.283 & 0.895$^\dagger$\\ 
\hline
Y & NIRAv1~\cite{Parada2015} & \ac{MLMF} & Single & -0.22 & 0.164 & 0.247 & 0.895$^\dagger$\\ 
\hline
Z & NIRAv2~\cite{Parada2015} & \ac{MLMF} & Single & -0.179 & 0.192 & 0.0105 & 0.906$^\dagger$\\ 
\hline
a & Blur kernel~\cite{Lim2015} & \ac{SFM} & Single & 0.172 & 0.149 & 0.285 & 8.36\\ 
\hline
b & Blur kernel with sliding window~\cite{Lim2015a} & \ac{SFM} & Single & -0.0198 & 0.142 & 0.111 & 0.412\\ 
\hline
c & Temporal dynamics~\cite{Falk2010a} & \ac{SFM} & Single & -0.295 & 0.217 & 0.191 & 0.358\\ 
\hline
d & Improved blind RTE~\cite{Lollmann2010} & \ac{ABC} & Single & -0.0795 & 0.165 & 0.172 & 0.0254\\ 
\hline
e & \ac{SDD}~\cite{Wen2008} & \ac{SFM} & Single & 1.31 & 861 & -0.0141 & 0.0221\\ 
\hline
\else
\fi
\end{tabular}%
\end{table*}%
\clearpage
\subsection{Fullband \ac{T60} estimation results by noise type and \ac{SNR}}
\subsubsection{Ambient noise at \dBel{18} \ac{SNR}}
\begin{figure}[!ht]
	\ifarXiv
\centerline{\epsfig{figure=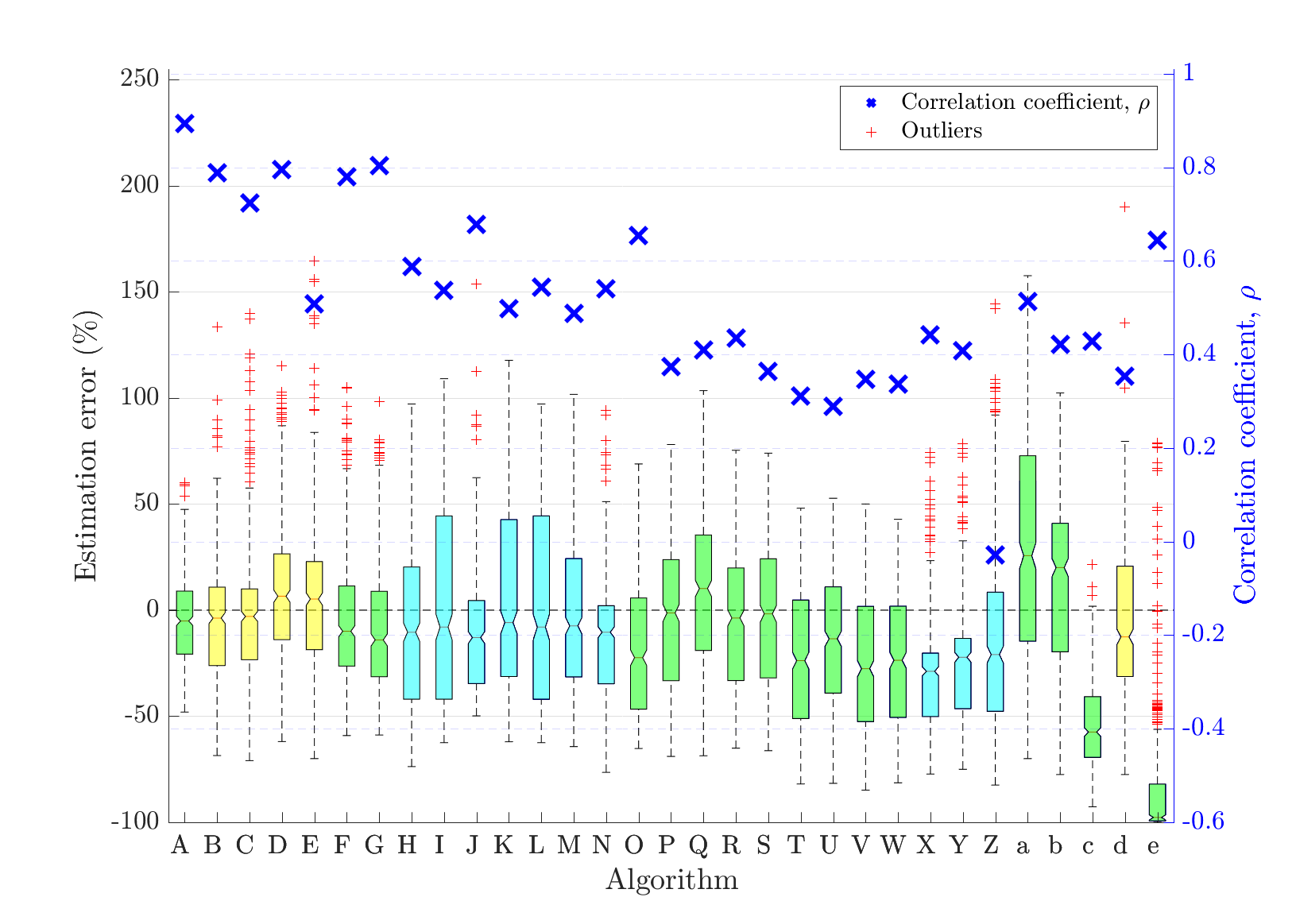,
	width=\figWidthACETR,viewport=45 10 765 530,clip}}%
	\else
	\centerline{\epsfig{figure=FigsACE/ana_eval_gt_partic_results_combined_Phase3_All_WASPAA_P3_T60_Perc_H_Ambient.png,
	width=\figWidthACETR,viewport=45 10 765 530,clip}}%
	\fi
	\caption{Fullband {\ac{T60} estimation error in ambient noise at \dBel{18} \ac{SNR}}}%
\label{fig:ACE_T60_Ambient_H}%
\end{figure}%
\begin{table*}[!ht]\small
\caption{\ac{T60} estimation algorithm performance in ambient noise at \dBel{18} \ac{SNR}}
\vspace{5mm} 
\centering
\begin{tabular}{clllllll}%
\hline%
Ref.
& Algorithm
& Class
& Mic. Config.
& Bias
& MSE
&  $\PearsonCC$
& \ac{RTF}
\\
\hline
\hline
\ifarXiv
A & QA Reverb~\cite{Prego2015} & \ac{SFM} & Single & -0.0913 & 0.0519 & 0.893 & 0.401\\ 
\hline
B & Octave \ac{SB}-based \ac{FB2} RTE~\cite{Lollmann2015} & \ac{ABC} & Single & -0.0979 & 0.0647 & 0.788 & 1\\ 
\hline
C & DCT-based \ac{FB2} RTE~\cite{Lollmann2015} & \ac{ABC} & Single & -0.0934 & 0.0712 & 0.724 & 1.04\\ 
\hline
D & Model-based \ac{SB} RTE~\cite{Lollmann2015} & \ac{ABC} & Single & -0.044 & 0.0857 & 0.795 & 0.478\\ 
\hline
E & Baseline algorithm for \ac{FB2} RTE~\cite{Lollmann2015} & \ac{ABC} & Single & -0.0705 & 0.0959 & 0.509 & 0.0421\\ 
\hline
F & \ac{SDDSA-G} retrained~\cite{Eaton2015b} & \ac{SFM} & Single & -0.0554 & 0.0593 & 0.78 & 0.0153\\ 
\hline
G & \ac{SDDSA-G}~\cite{Eaton2013} & \ac{SFM} & Single & -0.107 & 0.0591 & 0.804 & 0.0166\\ 
\hline
H & Multi-layer perceptron~\cite{Xiong2015} & \ac{MLMF} & Single & -0.11 & 0.0927 & 0.588 & 0.0578$^\ddagger$\\ 
\hline
I & Multi-layer perceptron P2~\cite{Xiong2015} & \ac{MLMF} & Single & -0.0815 & 0.0947 & 0.537 & 0.0578$^\ddagger$\\ 
\hline
J & Multi-layer perceptron P2~\cite{Xiong2015} & \ac{MLMF} & Chromebook & -0.1 & 0.0816 & 0.678 & 0.0589$^\ddagger$\\ 
\hline
K & Multi-layer perceptron P2~\cite{Xiong2015} & \ac{MLMF} & Mobile & -0.0493 & 0.0781 & 0.499 & 0.0557$^\ddagger$\\ 
\hline
L & Multi-layer perceptron P2~\cite{Xiong2015} & \ac{MLMF} & Crucif & -0.0853 & 0.0946 & 0.543 & 0.0569$^\ddagger$\\ 
\hline
M & Multi-layer perceptron P2~\cite{Xiong2015} & \ac{MLMF} & Lin8Ch & -0.0793 & 0.0855 & 0.488 & 0.062$^\ddagger$\\ 
\hline
N & Multi-layer perceptron P2~\cite{Xiong2015} & \ac{MLMF} & EM32 & -0.0906 & 0.0806 & 0.54 & 0.0578$^\ddagger$\\ 
\hline
O & Per acoust. band \ac{SRMR} {\sectMidSent} 2.5.~\cite{Senoussaoui2015} & \ac{SFM} & Single & -0.191 & 0.118 & 0.655 & 0.58\\ 
\hline
P & \ac{NSRMR} {\sectMidSent} 2.4.~\cite{Santos2014,Senoussaoui2015} & \ac{SFM} & Single & -0.128 & 0.123 & 0.374 & 0.571\\ 
\hline
Q & \ac{NSRMR} {\sectMidSent} 2.4.~\cite{Santos2014,Senoussaoui2015} & \ac{SFM} & Chromebook & -0.0736 & 0.113 & 0.41 & 1.04\\ 
\hline
R & \ac{NSRMR} {\sectMidSent} 2.4.~\cite{Santos2014,Senoussaoui2015} & \ac{SFM} & Mobile & -0.118 & 0.0961 & 0.436 & 1.59\\ 
\hline
S & \ac{NSRMR} {\sectMidSent} 2.4.~\cite{Santos2014,Senoussaoui2015} & \ac{SFM} & Crucif & -0.115 & 0.109 & 0.363 & 2.63\\ 
\hline
T & \ac{SRMR} {\sectMidSent} 2.3.~\cite{Senoussaoui2015} & \ac{SFM} & Single & -0.221 & 0.16 & 0.312 & 0.457\\ 
\hline
U & \ac{SRMR} {\sectMidSent} 2.3.~\cite{Senoussaoui2015} & \ac{SFM} & Chromebook & -0.186 & 0.15 & 0.29 & 0.831\\ 
\hline
V & \ac{SRMR} {\sectMidSent} 2.3.~\cite{Senoussaoui2015} & \ac{SFM} & Mobile & -0.219 & 0.136 & 0.346 & 1.26\\ 
\hline
W & \ac{SRMR} {\sectMidSent} 2.3.~\cite{Senoussaoui2015} & \ac{SFM} & Crucif & -0.215 & 0.144 & 0.336 & 2.09\\ 
\hline
X & NIRAv3~\cite{Parada2015} & \ac{MLMF} & Single & -0.268 & 0.172 & 0.442 & 0.897$^\dagger$\\ 
\hline
Y & NIRAv1~\cite{Parada2015} & \ac{MLMF} & Single & -0.245 & 0.164 & 0.408 & 0.897$^\dagger$\\ 
\hline
Z & NIRAv2~\cite{Parada2015} & \ac{MLMF} & Single & -0.183 & 0.199 & -0.0283 & 0.912$^\dagger$\\ 
\hline
a & Blur kernel~\cite{Lim2015} & \ac{SFM} & Single & 0.0888 & 0.0989 & 0.513 & 8.16\\ 
\hline
b & Blur kernel with sliding window~\cite{Lim2015a} & \ac{SFM} & Single & -0.045 & 0.104 & 0.421 & 0.413\\ 
\hline
c & Temporal dynamics~\cite{Falk2010a} & \ac{SFM} & Single & -0.387 & 0.253 & 0.429 & 0.362\\ 
\hline
d & Improved blind RTE~\cite{Lollmann2010} & \ac{ABC} & Single & -0.128 & 0.132 & 0.354 & 0.0255\\ 
\hline
e & \ac{SDD}~\cite{Wen2008} & \ac{SFM} & Single & -0.508 & 0.329 & 0.644 & 0.0219\\ 
\hline

\else

\fi
\end{tabular}
\end{table*}
\clearpage
\subsubsection{Ambient noise at \dBel{12} \ac{SNR}}
\begin{figure}[!ht]
	\ifarXiv
\centerline{\epsfig{figure=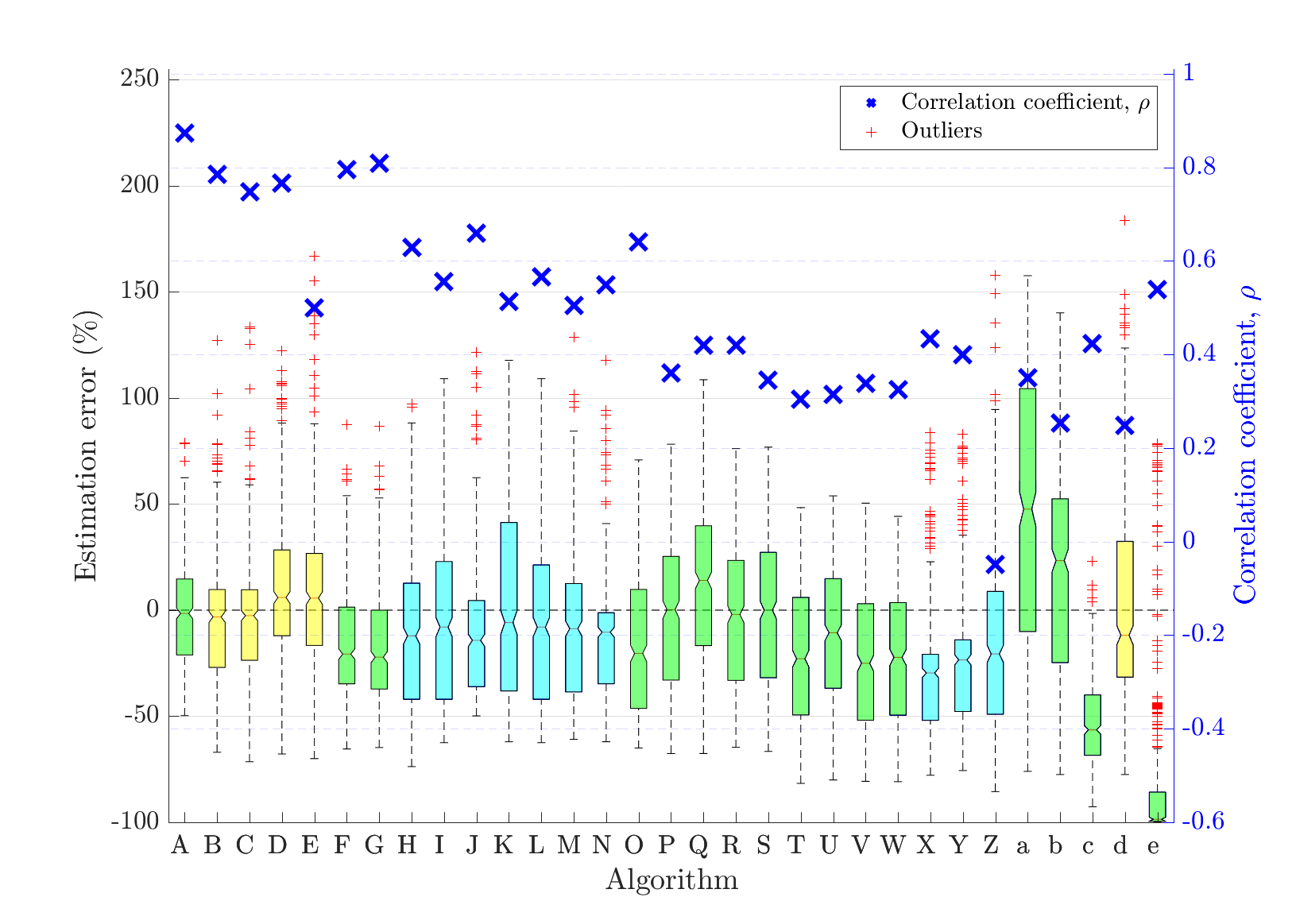,
	width=\figWidthACETR,viewport=45 10 765 530,clip}}%
	\else
	\centerline{\epsfig{figure=FigsACE/ana_eval_gt_partic_results_combined_Phase3_All_WASPAA_P3_T60_Perc_M_Ambient.png,
	width=\figWidthACETR,viewport=45 10 765 530,clip}}%
	\fi
	\caption{Fullband {\ac{T60} estimation error in ambient noise at \dBel{12} \ac{SNR}}}%
\label{fig:ACE_T60_Ambient_M}%
\end{figure}%
\begin{table*}[!ht]\small
\caption{\ac{T60} estimation algorithm performance in ambient noise at \dBel{12} \ac{SNR}}
\vspace{5mm} 
\centering
\begin{tabular}{clllllll}%
\hline%
Ref.
& Algorithm
& Class
& Mic. Config.
& Bias
& MSE
&  $\PearsonCC$
& \ac{RTF}
\\
\hline
\hline
\ifarXiv
A & QA Reverb~\cite{Prego2015} & \ac{SFM} & Single & -0.0795 & 0.0543 & 0.873 & 0.401\\ 
\hline
B & Octave \ac{SB}-based \ac{FB2} RTE~\cite{Lollmann2015} & \ac{ABC} & Single & -0.1 & 0.0657 & 0.786 & 1\\ 
\hline
C & DCT-based \ac{FB2} RTE~\cite{Lollmann2015} & \ac{ABC} & Single & -0.0967 & 0.069 & 0.748 & 1.04\\ 
\hline
D & Model-based \ac{SB} RTE~\cite{Lollmann2015} & \ac{ABC} & Single & -0.0433 & 0.0905 & 0.767 & 0.478\\ 
\hline
E & Baseline algorithm for \ac{FB2} RTE~\cite{Lollmann2015} & \ac{ABC} & Single & -0.0527 & 0.0955 & 0.499 & 0.0421\\ 
\hline
F & \ac{SDDSA-G} retrained~\cite{Eaton2015b} & \ac{SFM} & Single & -0.133 & 0.0629 & 0.796 & 0.0153\\ 
\hline
G & \ac{SDDSA-G}~\cite{Eaton2013} & \ac{SFM} & Single & -0.157 & 0.075 & 0.808 & 0.0166\\ 
\hline
H & Multi-layer perceptron~\cite{Xiong2015} & \ac{MLMF} & Single & -0.117 & 0.0888 & 0.629 & 0.0578$^\ddagger$\\ 
\hline
I & Multi-layer perceptron P2~\cite{Xiong2015} & \ac{MLMF} & Single & -0.0899 & 0.0935 & 0.556 & 0.0578$^\ddagger$\\ 
\hline
J & Multi-layer perceptron P2~\cite{Xiong2015} & \ac{MLMF} & Chromebook & -0.097 & 0.0827 & 0.659 & 0.0589$^\ddagger$\\ 
\hline
K & Multi-layer perceptron P2~\cite{Xiong2015} & \ac{MLMF} & Mobile & -0.0631 & 0.0781 & 0.514 & 0.0557$^\ddagger$\\ 
\hline
L & Multi-layer perceptron P2~\cite{Xiong2015} & \ac{MLMF} & Crucif & -0.0917 & 0.0927 & 0.565 & 0.0569$^\ddagger$\\ 
\hline
M & Multi-layer perceptron P2~\cite{Xiong2015} & \ac{MLMF} & Lin8Ch & -0.0873 & 0.0852 & 0.505 & 0.062$^\ddagger$\\ 
\hline
N & Multi-layer perceptron P2~\cite{Xiong2015} & \ac{MLMF} & EM32 & -0.103 & 0.0821 & 0.549 & 0.0578$^\ddagger$\\ 
\hline
O & Per acoust. band \ac{SRMR} {\sectMidSent} 2.5.~\cite{Senoussaoui2015} & \ac{SFM} & Single & -0.18 & 0.115 & 0.64 & 0.58\\ 
\hline
P & \ac{NSRMR} {\sectMidSent} 2.4.~\cite{Santos2014,Senoussaoui2015} & \ac{SFM} & Single & -0.121 & 0.122 & 0.36 & 0.571\\ 
\hline
Q & \ac{NSRMR} {\sectMidSent} 2.4.~\cite{Santos2014,Senoussaoui2015} & \ac{SFM} & Chromebook & -0.0575 & 0.11 & 0.42 & 1.04\\ 
\hline
R & \ac{NSRMR} {\sectMidSent} 2.4.~\cite{Santos2014,Senoussaoui2015} & \ac{SFM} & Mobile & -0.111 & 0.0956 & 0.42 & 1.59\\ 
\hline
S & \ac{NSRMR} {\sectMidSent} 2.4.~\cite{Santos2014,Senoussaoui2015} & \ac{SFM} & Crucif & -0.109 & 0.109 & 0.345 & 2.63\\ 
\hline
T & \ac{SRMR} {\sectMidSent} 2.3.~\cite{Senoussaoui2015} & \ac{SFM} & Single & -0.215 & 0.158 & 0.305 & 0.457\\ 
\hline
U & \ac{SRMR} {\sectMidSent} 2.3.~\cite{Senoussaoui2015} & \ac{SFM} & Chromebook & -0.171 & 0.143 & 0.314 & 0.831\\ 
\hline
V & \ac{SRMR} {\sectMidSent} 2.3.~\cite{Senoussaoui2015} & \ac{SFM} & Mobile & -0.212 & 0.133 & 0.338 & 1.26\\ 
\hline
W & \ac{SRMR} {\sectMidSent} 2.3.~\cite{Senoussaoui2015} & \ac{SFM} & Crucif & -0.209 & 0.142 & 0.324 & 2.09\\ 
\hline
X & NIRAv3~\cite{Parada2015} & \ac{MLMF} & Single & -0.273 & 0.176 & 0.433 & 0.897$^\dagger$\\ 
\hline
Y & NIRAv1~\cite{Parada2015} & \ac{MLMF} & Single & -0.25 & 0.167 & 0.399 & 0.897$^\dagger$\\ 
\hline
Z & NIRAv2~\cite{Parada2015} & \ac{MLMF} & Single & -0.189 & 0.196 & -0.0487 & 0.912$^\dagger$\\ 
\hline
a & Blur kernel~\cite{Lim2015} & \ac{SFM} & Single & 0.161 & 0.138 & 0.35 & 8.16\\ 
\hline
b & Blur kernel with sliding window~\cite{Lim2015a} & \ac{SFM} & Single & -0.0199 & 0.117 & 0.254 & 0.413\\ 
\hline
c & Temporal dynamics~\cite{Falk2010a} & \ac{SFM} & Single & -0.382 & 0.25 & 0.423 & 0.362\\ 
\hline
d & Improved blind RTE~\cite{Lollmann2010} & \ac{ABC} & Single & -0.0994 & 0.147 & 0.249 & 0.0255\\ 
\hline
e & \ac{SDD}~\cite{Wen2008} & \ac{SFM} & Single & -0.518 & 0.356 & 0.539 & 0.0219\\ 
\hline

\else

\fi
\end{tabular}
\end{table*}
\clearpage
\subsubsection{Ambient noise at \dBel{-1} \ac{SNR}}
\begin{figure}[!ht]
	\ifarXiv
\centerline{\epsfig{figure=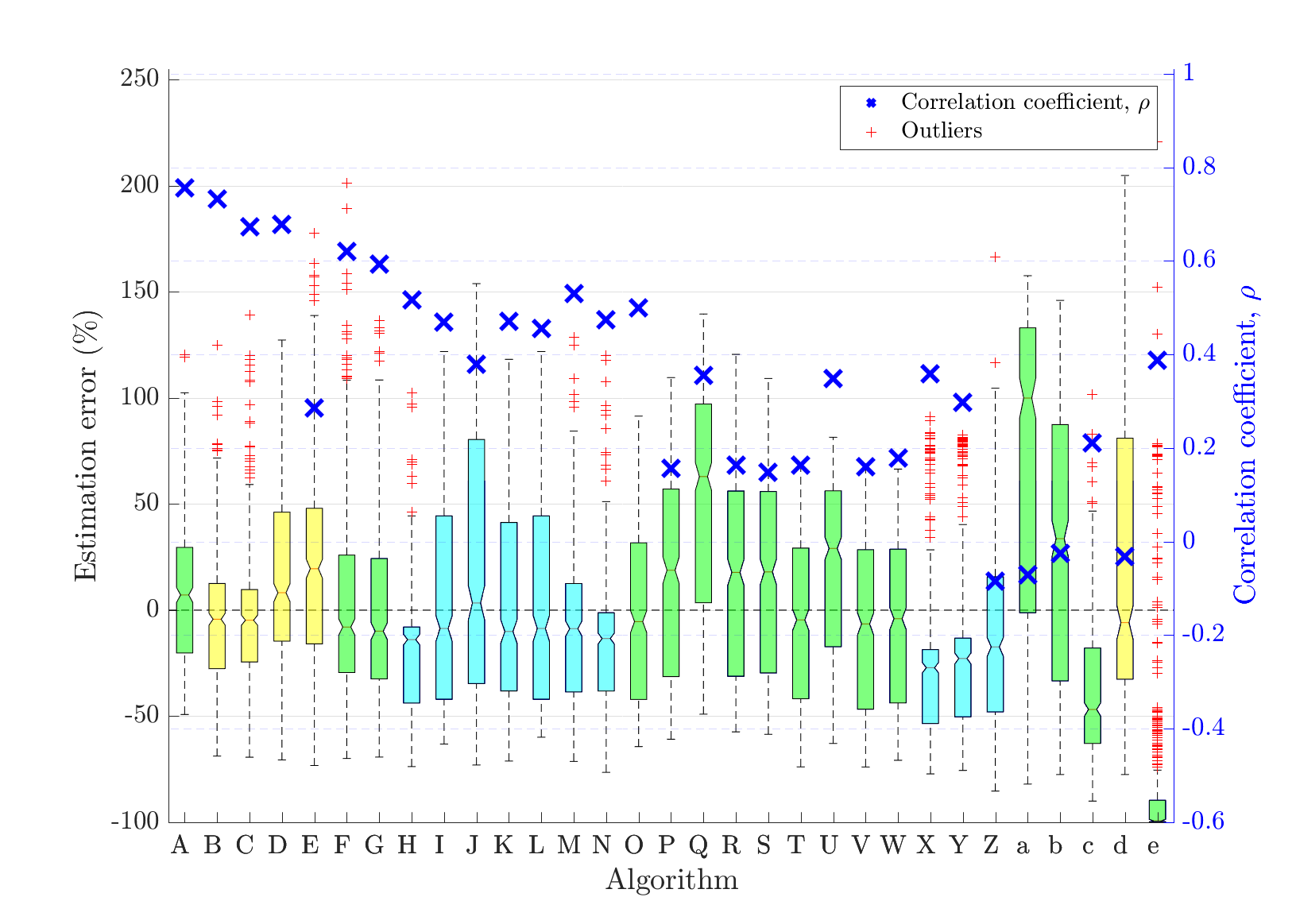,
	width=\figWidthACETR,viewport=45 10 765 530,clip}}%
	\else
	\centerline{\epsfig{figure=FigsACE/ana_eval_gt_partic_results_combined_Phase3_All_WASPAA_P3_T60_Perc_L_Ambient.png,
	width=\figWidthACETR,viewport=45 10 765 530,clip}}%
	\fi
	\caption{Fullband {\ac{T60} estimation error in ambient noise at \dBel{-1} \ac{SNR}}}%
\label{fig:ACE_T60_Ambient_L}%
\end{figure}%
\begin{table*}[!ht]\small
\caption{\ac{T60} estimation algorithm performance in ambient noise at \dBel{-1} \ac{SNR}}
\vspace{5mm} 
\centering
\begin{tabular}{clllllll}%
\hline%
Ref.
& Algorithm
& Class
& Mic. Config.
& Bias
& MSE
&  $\PearsonCC$
& \ac{RTF}
\\
\hline
\hline
\ifarXiv
A & QA Reverb~\cite{Prego2015} & \ac{SFM} & Single & -0.0339 & 0.0634 & 0.757 & 0.401\\ 
\hline
B & Octave \ac{SB}-based \ac{FB2} RTE~\cite{Lollmann2015} & \ac{ABC} & Single & -0.0997 & 0.0735 & 0.733 & 1\\ 
\hline
C & DCT-based \ac{FB2} RTE~\cite{Lollmann2015} & \ac{ABC} & Single & -0.103 & 0.0812 & 0.673 & 1.04\\ 
\hline
D & Model-based \ac{SB} RTE~\cite{Lollmann2015} & \ac{ABC} & Single & -0.0259 & 0.105 & 0.678 & 0.478\\ 
\hline
E & Baseline algorithm for \ac{FB2} RTE~\cite{Lollmann2015} & \ac{ABC} & Single & -0.0001 & 0.124 & 0.285 & 0.0421\\ 
\hline
F & \ac{SDDSA-G} retrained~\cite{Eaton2015b} & \ac{SFM} & Single & -0.0564 & 0.0806 & 0.62 & 0.0153\\ 
\hline
G & \ac{SDDSA-G}~\cite{Eaton2013} & \ac{SFM} & Single & -0.0867 & 0.0874 & 0.592 & 0.0166\\ 
\hline
H & Multi-layer perceptron~\cite{Xiong2015} & \ac{MLMF} & Single & -0.147 & 0.112 & 0.516 & 0.0578$^\ddagger$\\ 
\hline
I & Multi-layer perceptron P2~\cite{Xiong2015} & \ac{MLMF} & Single & -0.0819 & 0.103 & 0.47 & 0.0578$^\ddagger$\\ 
\hline
J & Multi-layer perceptron P2~\cite{Xiong2015} & \ac{MLMF} & Chromebook & -0.0138 & 0.111 & 0.379 & 0.0589$^\ddagger$\\ 
\hline
K & Multi-layer perceptron P2~\cite{Xiong2015} & \ac{MLMF} & Mobile & -0.0619 & 0.0834 & 0.471 & 0.0557$^\ddagger$\\ 
\hline
L & Multi-layer perceptron P2~\cite{Xiong2015} & \ac{MLMF} & Crucif & -0.0787 & 0.104 & 0.455 & 0.0569$^\ddagger$\\ 
\hline
M & Multi-layer perceptron P2~\cite{Xiong2015} & \ac{MLMF} & Lin8Ch & -0.0787 & 0.0812 & 0.53 & 0.062$^\ddagger$\\ 
\hline
N & Multi-layer perceptron P2~\cite{Xiong2015} & \ac{MLMF} & EM32 & -0.0972 & 0.0894 & 0.474 & 0.0578$^\ddagger$\\ 
\hline
O & Per acoust. band \ac{SRMR} {\sectMidSent} 2.5.~\cite{Senoussaoui2015} & \ac{SFM} & Single & -0.108 & 0.106 & 0.499 & 0.58\\ 
\hline
P & \ac{NSRMR} {\sectMidSent} 2.4.~\cite{Santos2014,Senoussaoui2015} & \ac{SFM} & Single & -0.0403 & 0.124 & 0.157 & 0.571\\ 
\hline
Q & \ac{NSRMR} {\sectMidSent} 2.4.~\cite{Santos2014,Senoussaoui2015} & \ac{SFM} & Chromebook & 0.118 & 0.125 & 0.356 & 1.04\\ 
\hline
R & \ac{NSRMR} {\sectMidSent} 2.4.~\cite{Santos2014,Senoussaoui2015} & \ac{SFM} & Mobile & -0.022 & 0.101 & 0.163 & 1.59\\ 
\hline
S & \ac{NSRMR} {\sectMidSent} 2.4.~\cite{Santos2014,Senoussaoui2015} & \ac{SFM} & Crucif & -0.0271 & 0.111 & 0.148 & 2.63\\ 
\hline
T & \ac{SRMR} {\sectMidSent} 2.3.~\cite{Senoussaoui2015} & \ac{SFM} & Single & -0.148 & 0.142 & 0.164 & 0.457\\ 
\hline
U & \ac{SRMR} {\sectMidSent} 2.3.~\cite{Senoussaoui2015} & \ac{SFM} & Chromebook & -0.0332 & 0.114 & 0.348 & 0.831\\ 
\hline
V & \ac{SRMR} {\sectMidSent} 2.3.~\cite{Senoussaoui2015} & \ac{SFM} & Mobile & -0.135 & 0.117 & 0.159 & 1.26\\ 
\hline
W & \ac{SRMR} {\sectMidSent} 2.3.~\cite{Senoussaoui2015} & \ac{SFM} & Crucif & -0.14 & 0.126 & 0.178 & 2.09\\ 
\hline
X & NIRAv3~\cite{Parada2015} & \ac{MLMF} & Single & -0.248 & 0.168 & 0.359 & 0.897$^\dagger$\\ 
\hline
Y & NIRAv1~\cite{Parada2015} & \ac{MLMF} & Single & -0.235 & 0.167 & 0.298 & 0.897$^\dagger$\\ 
\hline
Z & NIRAv2~\cite{Parada2015} & \ac{MLMF} & Single & -0.177 & 0.198 & -0.085 & 0.912$^\dagger$\\ 
\hline
a & Blur kernel~\cite{Lim2015} & \ac{SFM} & Single & 0.242 & 0.214 & -0.0718 & 8.16\\ 
\hline
b & Blur kernel with sliding window~\cite{Lim2015a} & \ac{SFM} & Single & 0.0183 & 0.188 & -0.0244 & 0.413\\ 
\hline
c & Temporal dynamics~\cite{Falk2010a} & \ac{SFM} & Single & -0.307 & 0.214 & 0.211 & 0.362\\ 
\hline
d & Improved blind RTE~\cite{Lollmann2010} & \ac{ABC} & Single & 0.00207 & 0.224 & -0.0319 & 0.0255\\ 
\hline
e & \ac{SDD}~\cite{Wen2008} & \ac{SFM} & Single & -0.518 & 0.381 & 0.387 & 0.0219\\ 
\hline

\else

\fi
\end{tabular}
\end{table*}
\clearpage
\subsubsection{Babble noise at \dBel{18} \ac{SNR}}
\begin{figure}[!ht]
	\ifarXiv
\centerline{\epsfig{figure=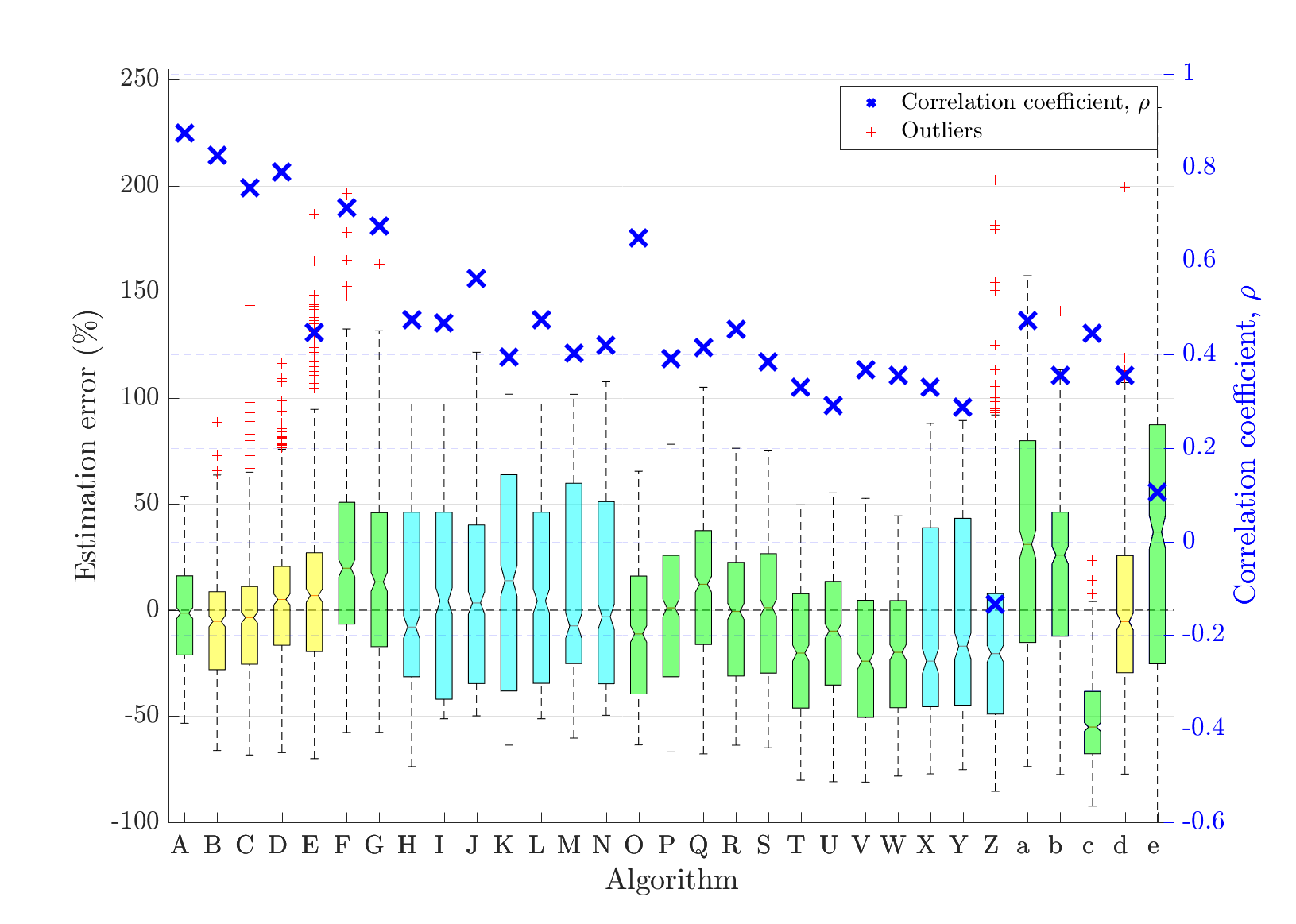,
	width=\figWidthACETR,viewport=45 10 765 530,clip}}%
	\else
	\centerline{\epsfig{figure=FigsACE/ana_eval_gt_partic_results_combined_Phase3_All_WASPAA_P3_T60_Perc_H_Babble.png,
	width=\figWidthACETR,viewport=45 10 765 530,clip}}%
	\fi
	\caption{Fullband {\ac{T60} estimation error in babble noise at \dBel{18} \ac{SNR}}}%
\label{fig:ACE_T60_Babble_H}%
\end{figure}%
\begin{table*}[!ht]\small
\caption{\ac{T60} estimation algorithm performance in babble noise at \dBel{18} \ac{SNR}}
\vspace{5mm} 
\centering
\begin{tabular}{clllllll}%
\hline%
Ref.
& Algorithm
& Class
& Mic. Config.
& Bias
& MSE
&  $\PearsonCC$
& \ac{RTF}
\\
\hline
\hline
\ifarXiv
A & QA Reverb~\cite{Prego2015} & \ac{SFM} & Single & -0.0854 & 0.058 & 0.873 & 0.398\\ 
\hline
B & Octave \ac{SB}-based \ac{FB2} RTE~\cite{Lollmann2015} & \ac{ABC} & Single & -0.112 & 0.064 & 0.826 & 0.911\\ 
\hline
C & DCT-based \ac{FB2} RTE~\cite{Lollmann2015} & \ac{ABC} & Single & -0.1 & 0.0698 & 0.756 & 0.99\\ 
\hline
D & Model-based \ac{SB} RTE~\cite{Lollmann2015} & \ac{ABC} & Single & -0.0665 & 0.0905 & 0.79 & 0.443\\ 
\hline
E & Baseline algorithm for \ac{FB2} RTE~\cite{Lollmann2015} & \ac{ABC} & Single & -0.0546 & 0.102 & 0.448 & 0.0428\\ 
\hline
F & \ac{SDDSA-G} retrained~\cite{Eaton2015b} & \ac{SFM} & Single & 0.0931 & 0.0836 & 0.713 & 0.0155\\ 
\hline
G & \ac{SDDSA-G}~\cite{Eaton2013} & \ac{SFM} & Single & 0.0104 & 0.0681 & 0.674 & 0.0162\\ 
\hline
H & Multi-layer perceptron~\cite{Xiong2015} & \ac{MLMF} & Single & -0.054 & 0.098 & 0.474 & 0.0579$^\ddagger$\\ 
\hline
I & Multi-layer perceptron P2~\cite{Xiong2015} & \ac{MLMF} & Single & -0.0373 & 0.0974 & 0.467 & 0.0579$^\ddagger$\\ 
\hline
J & Multi-layer perceptron P2~\cite{Xiong2015} & \ac{MLMF} & Chromebook & -0.056 & 0.0891 & 0.562 & 0.0588$^\ddagger$\\ 
\hline
K & Multi-layer perceptron P2~\cite{Xiong2015} & \ac{MLMF} & Mobile & -0.0197 & 0.086 & 0.394 & 0.0555$^\ddagger$\\ 
\hline
L & Multi-layer perceptron P2~\cite{Xiong2015} & \ac{MLMF} & Crucif & -0.0431 & 0.097 & 0.474 & 0.057$^\ddagger$\\ 
\hline
M & Multi-layer perceptron P2~\cite{Xiong2015} & \ac{MLMF} & Lin8Ch & -0.0359 & 0.0897 & 0.403 & 0.0618$^\ddagger$\\ 
\hline
N & Multi-layer perceptron P2~\cite{Xiong2015} & \ac{MLMF} & EM32 & -0.0518 & 0.0876 & 0.42 & 0.0576$^\ddagger$\\ 
\hline
O & Per acoust. band \ac{SRMR} {\sectMidSent} 2.5.~\cite{Senoussaoui2015} & \ac{SFM} & Single & -0.152 & 0.108 & 0.648 & 0.579\\ 
\hline
P & \ac{NSRMR} {\sectMidSent} 2.4.~\cite{Santos2014,Senoussaoui2015} & \ac{SFM} & Single & -0.116 & 0.119 & 0.391 & 0.572\\ 
\hline
Q & \ac{NSRMR} {\sectMidSent} 2.4.~\cite{Santos2014,Senoussaoui2015} & \ac{SFM} & Chromebook & -0.0642 & 0.111 & 0.414 & 1.04\\ 
\hline
R & \ac{NSRMR} {\sectMidSent} 2.4.~\cite{Santos2014,Senoussaoui2015} & \ac{SFM} & Mobile & -0.106 & 0.0924 & 0.454 & 1.58\\ 
\hline
S & \ac{NSRMR} {\sectMidSent} 2.4.~\cite{Santos2014,Senoussaoui2015} & \ac{SFM} & Crucif & -0.104 & 0.105 & 0.385 & 2.63\\ 
\hline
T & \ac{SRMR} {\sectMidSent} 2.3.~\cite{Senoussaoui2015} & \ac{SFM} & Single & -0.207 & 0.152 & 0.329 & 0.457\\ 
\hline
U & \ac{SRMR} {\sectMidSent} 2.3.~\cite{Senoussaoui2015} & \ac{SFM} & Chromebook & -0.173 & 0.145 & 0.29 & 0.833\\ 
\hline
V & \ac{SRMR} {\sectMidSent} 2.3.~\cite{Senoussaoui2015} & \ac{SFM} & Mobile & -0.204 & 0.128 & 0.367 & 1.26\\ 
\hline
W & \ac{SRMR} {\sectMidSent} 2.3.~\cite{Senoussaoui2015} & \ac{SFM} & Crucif & -0.2 & 0.137 & 0.356 & 2.1\\ 
\hline
X & NIRAv3~\cite{Parada2015} & \ac{MLMF} & Single & -0.14 & 0.133 & 0.33 & 0.906$^\dagger$\\ 
\hline
Y & NIRAv1~\cite{Parada2015} & \ac{MLMF} & Single & -0.126 & 0.132 & 0.288 & 0.906$^\dagger$\\ 
\hline
Z & NIRAv2~\cite{Parada2015} & \ac{MLMF} & Single & -0.176 & 0.228 & -0.134 & 0.901$^\dagger$\\ 
\hline
a & Blur kernel~\cite{Lim2015} & \ac{SFM} & Single & 0.102 & 0.107 & 0.472 & 8.88\\ 
\hline
b & Blur kernel with sliding window~\cite{Lim2015a} & \ac{SFM} & Single & -0.026 & 0.108 & 0.356 & 0.438\\ 
\hline
c & Temporal dynamics~\cite{Falk2010a} & \ac{SFM} & Single & -0.375 & 0.243 & 0.445 & 0.365\\ 
\hline
d & Improved blind RTE~\cite{Lollmann2010} & \ac{ABC} & Single & -0.104 & 0.126 & 0.355 & 0.0269\\ 
\hline
e & \ac{SDD}~\cite{Wen2008} & \ac{SFM} & Single & 0.793 & 141 & 0.105 & 0.0224\\ 
\hline

\else

\fi
\end{tabular}
\end{table*}
\clearpage
\subsubsection{Babble noise at \dBel{12} \ac{SNR}}
\begin{figure}[!ht]
	\ifarXiv
\centerline{\epsfig{figure=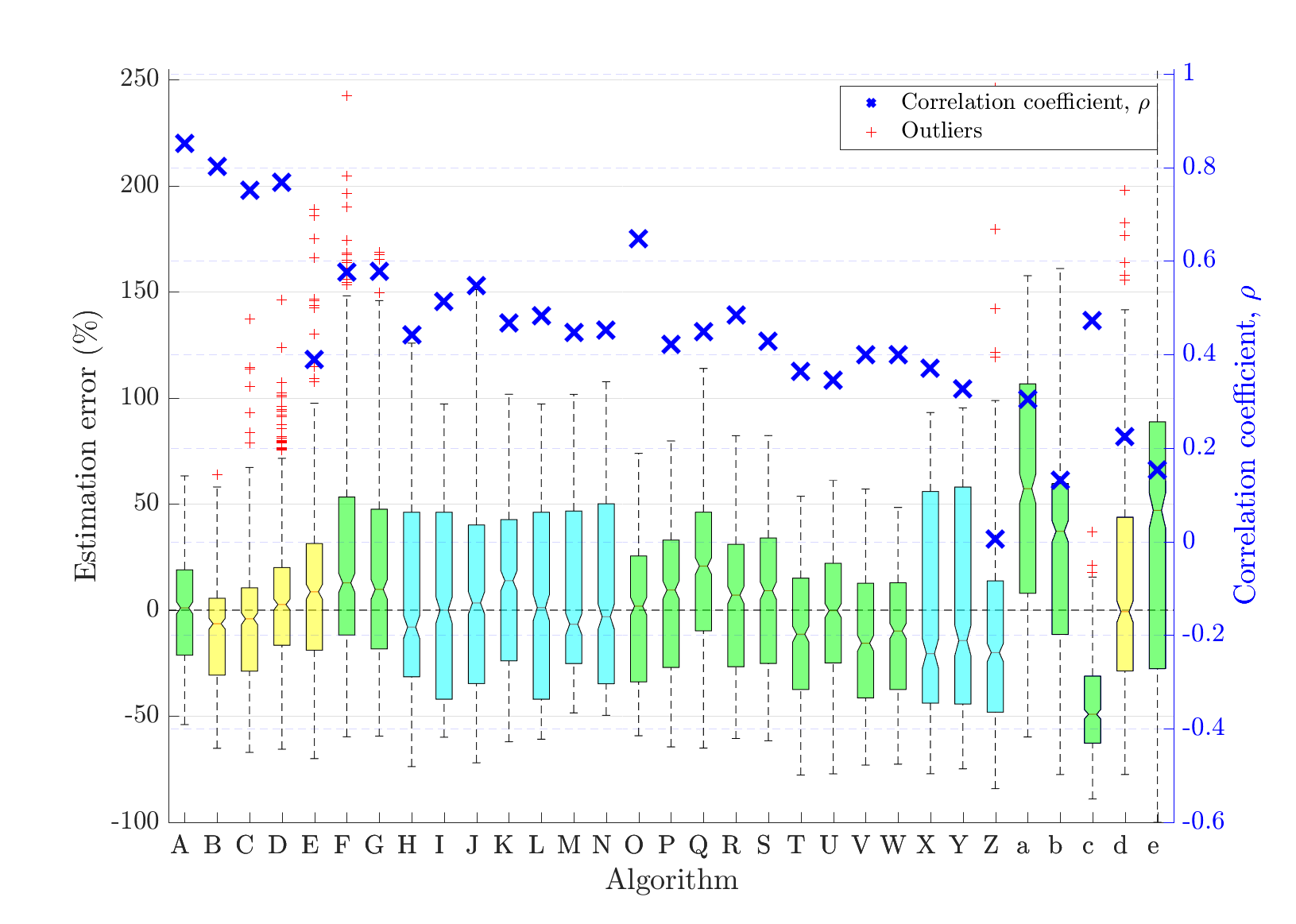,
	width=\figWidthACETR,viewport=45 10 765 530,clip}}%
	\else
	\centerline{\epsfig{figure=FigsACE/ana_eval_gt_partic_results_combined_Phase3_All_WASPAA_P3_T60_Perc_M_Babble.png,
	width=\figWidthACETR,viewport=45 10 765 530,clip}}%
	\fi
	\caption{Fullband {\ac{T60} estimation error in babble noise at \dBel{12} \ac{SNR}}}%
\label{fig:ACE_T60_Babble_M}%
\end{figure}%
\begin{table*}[!ht]\small
\caption{\ac{T60} estimation algorithm performance in babble noise at \dBel{12} \ac{SNR}}
\vspace{5mm} 
\centering
\begin{tabular}{clllllll}%
\hline%
Ref.
& Algorithm
& Class
& Mic. Config.
& Bias
& MSE
&  $\PearsonCC$
& \ac{RTF}
\\
\hline
\hline
\ifarXiv
A & QA Reverb~\cite{Prego2015} & \ac{SFM} & Single & -0.0796 & 0.0609 & 0.851 & 0.398\\ 
\hline
B & Octave \ac{SB}-based \ac{FB2} RTE~\cite{Lollmann2015} & \ac{ABC} & Single & -0.128 & 0.0735 & 0.802 & 0.911\\ 
\hline
C & DCT-based \ac{FB2} RTE~\cite{Lollmann2015} & \ac{ABC} & Single & -0.11 & 0.0737 & 0.751 & 0.99\\ 
\hline
D & Model-based \ac{SB} RTE~\cite{Lollmann2015} & \ac{ABC} & Single & -0.0701 & 0.095 & 0.768 & 0.443\\ 
\hline
E & Baseline algorithm for \ac{FB2} RTE~\cite{Lollmann2015} & \ac{ABC} & Single & -0.0458 & 0.108 & 0.389 & 0.0428\\ 
\hline
F & \ac{SDDSA-G} retrained~\cite{Eaton2015b} & \ac{SFM} & Single & 0.0875 & 0.113 & 0.577 & 0.0155\\ 
\hline
G & \ac{SDDSA-G}~\cite{Eaton2013} & \ac{SFM} & Single & 0.0122 & 0.0819 & 0.578 & 0.0162\\ 
\hline
H & Multi-layer perceptron~\cite{Xiong2015} & \ac{MLMF} & Single & -0.0609 & 0.103 & 0.443 & 0.0579$^\ddagger$\\ 
\hline
I & Multi-layer perceptron P2~\cite{Xiong2015} & \ac{MLMF} & Single & -0.0513 & 0.0934 & 0.514 & 0.0579$^\ddagger$\\ 
\hline
J & Multi-layer perceptron P2~\cite{Xiong2015} & \ac{MLMF} & Chromebook & -0.0474 & 0.0898 & 0.548 & 0.0588$^\ddagger$\\ 
\hline
K & Multi-layer perceptron P2~\cite{Xiong2015} & \ac{MLMF} & Mobile & -0.0195 & 0.079 & 0.468 & 0.0555$^\ddagger$\\ 
\hline
L & Multi-layer perceptron P2~\cite{Xiong2015} & \ac{MLMF} & Crucif & -0.0445 & 0.0961 & 0.483 & 0.057$^\ddagger$\\ 
\hline
M & Multi-layer perceptron P2~\cite{Xiong2015} & \ac{MLMF} & Lin8Ch & -0.0231 & 0.0843 & 0.447 & 0.0618$^\ddagger$\\ 
\hline
N & Multi-layer perceptron P2~\cite{Xiong2015} & \ac{MLMF} & EM32 & -0.0616 & 0.0856 & 0.453 & 0.0576$^\ddagger$\\ 
\hline
O & Per acoust. band \ac{SRMR} {\sectMidSent} 2.5.~\cite{Senoussaoui2015} & \ac{SFM} & Single & -0.11 & 0.101 & 0.647 & 0.579\\ 
\hline
P & \ac{NSRMR} {\sectMidSent} 2.4.~\cite{Santos2014,Senoussaoui2015} & \ac{SFM} & Single & -0.0862 & 0.112 & 0.422 & 0.572\\ 
\hline
Q & \ac{NSRMR} {\sectMidSent} 2.4.~\cite{Santos2014,Senoussaoui2015} & \ac{SFM} & Chromebook & -0.0337 & 0.107 & 0.449 & 1.04\\ 
\hline
R & \ac{NSRMR} {\sectMidSent} 2.4.~\cite{Santos2014,Senoussaoui2015} & \ac{SFM} & Mobile & -0.0739 & 0.0856 & 0.484 & 1.58\\ 
\hline
S & \ac{NSRMR} {\sectMidSent} 2.4.~\cite{Santos2014,Senoussaoui2015} & \ac{SFM} & Crucif & -0.0736 & 0.098 & 0.429 & 2.63\\ 
\hline
T & \ac{SRMR} {\sectMidSent} 2.3.~\cite{Senoussaoui2015} & \ac{SFM} & Single & -0.173 & 0.138 & 0.363 & 0.457\\ 
\hline
U & \ac{SRMR} {\sectMidSent} 2.3.~\cite{Senoussaoui2015} & \ac{SFM} & Chromebook & -0.135 & 0.132 & 0.344 & 0.833\\ 
\hline
V & \ac{SRMR} {\sectMidSent} 2.3.~\cite{Senoussaoui2015} & \ac{SFM} & Mobile & -0.166 & 0.113 & 0.4 & 1.26\\ 
\hline
W & \ac{SRMR} {\sectMidSent} 2.3.~\cite{Senoussaoui2015} & \ac{SFM} & Crucif & -0.164 & 0.122 & 0.4 & 2.1\\ 
\hline
X & NIRAv3~\cite{Parada2015} & \ac{MLMF} & Single & -0.108 & 0.12 & 0.371 & 0.906$^\dagger$\\ 
\hline
Y & NIRAv1~\cite{Parada2015} & \ac{MLMF} & Single & -0.0969 & 0.122 & 0.327 & 0.906$^\dagger$\\ 
\hline
Z & NIRAv2~\cite{Parada2015} & \ac{MLMF} & Single & -0.167 & 0.196 & 0.00486 & 0.901$^\dagger$\\ 
\hline
a & Blur kernel~\cite{Lim2015} & \ac{SFM} & Single & 0.187 & 0.148 & 0.305 & 8.88\\ 
\hline
b & Blur kernel with sliding window~\cite{Lim2015a} & \ac{SFM} & Single & 0.00866 & 0.127 & 0.131 & 0.438\\ 
\hline
c & Temporal dynamics~\cite{Falk2010a} & \ac{SFM} & Single & -0.344 & 0.218 & 0.472 & 0.365\\ 
\hline
d & Improved blind RTE~\cite{Lollmann2010} & \ac{ABC} & Single & -0.0563 & 0.145 & 0.224 & 0.0269\\ 
\hline
e & \ac{SDD}~\cite{Wen2008} & \ac{SFM} & Single & 0.458 & 5.11 & 0.153 & 0.0224\\ 
\hline

\else

\fi
\end{tabular}
\end{table*}
\clearpage
\subsubsection{Babble noise at \dBel{-1} \ac{SNR}}
\begin{figure}[!ht]
	\ifarXiv
\centerline{\epsfig{figure=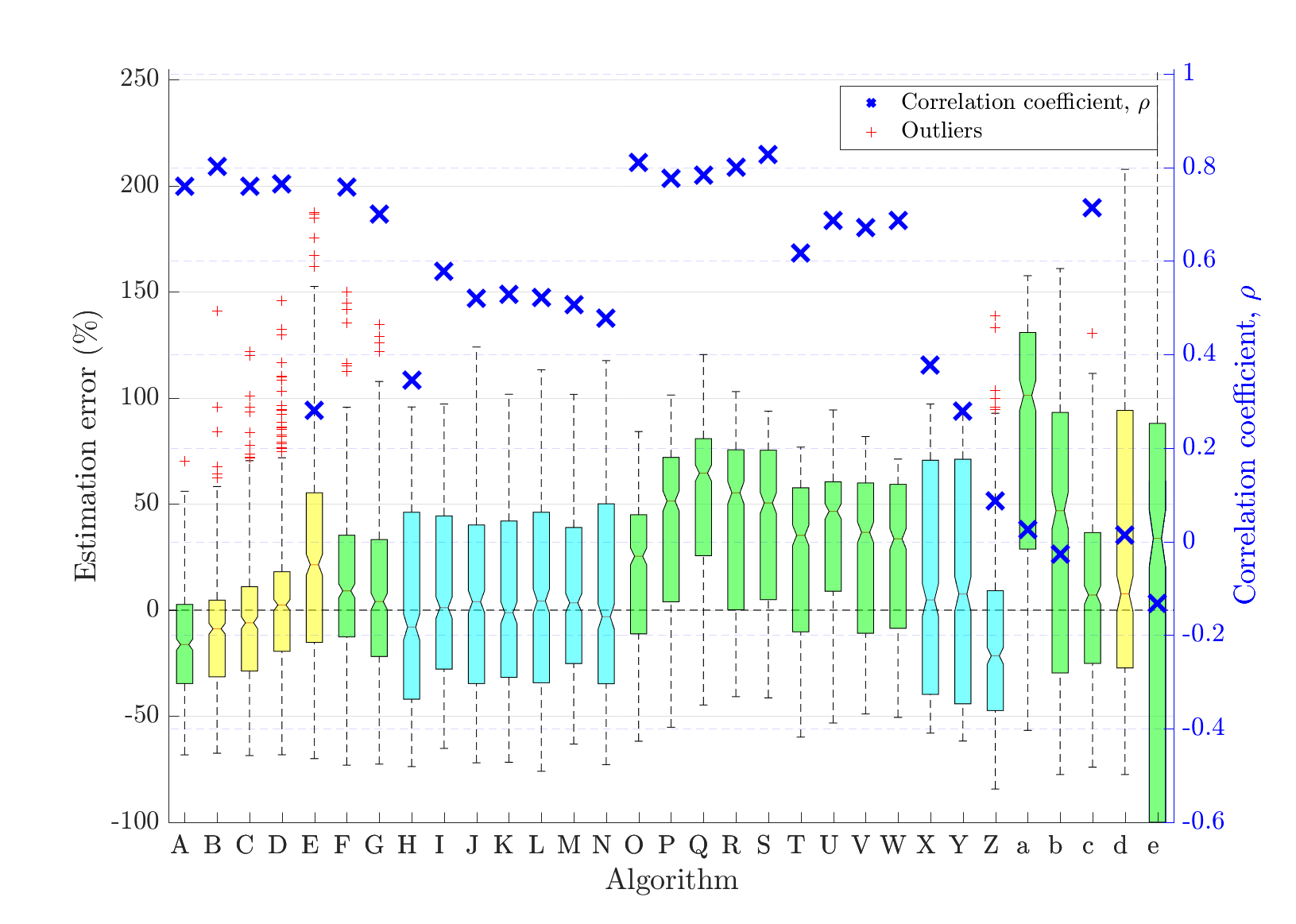,
	width=\figWidthACETR,viewport=45 10 765 530,clip}}%
	\else
	\centerline{\epsfig{figure=FigsACE/ana_eval_gt_partic_results_combined_Phase3_All_WASPAA_P3_T60_Perc_L_Babble.png,
	width=\figWidthACETR,viewport=45 10 765 530,clip}}%
	\fi
	\caption{Fullband {\ac{T60} estimation error in babble noise at \dBel{-1} \ac{SNR}}}%
\label{fig:ACE_T60_Babble_L}%
\end{figure}%
\begin{table*}[!ht]\small
\caption{\ac{T60} estimation algorithm performance in babble noise at \dBel{-1} \ac{SNR}}
\vspace{5mm} 
\centering
\begin{tabular}{clllllll}%
\hline%
Ref.
& Algorithm
& Class
& Mic. Config.
& Bias
& MSE
&  $\PearsonCC$
& \ac{RTF}
\\
\hline
\hline
\ifarXiv
A & QA Reverb~\cite{Prego2015} & \ac{SFM} & Single & -0.162 & 0.0934 & 0.759 & 0.398\\ 
\hline
B & Octave \ac{SB}-based \ac{FB2} RTE~\cite{Lollmann2015} & \ac{ABC} & Single & -0.13 & 0.0727 & 0.802 & 0.911\\ 
\hline
C & DCT-based \ac{FB2} RTE~\cite{Lollmann2015} & \ac{ABC} & Single & -0.108 & 0.072 & 0.759 & 0.99\\ 
\hline
D & Model-based \ac{SB} RTE~\cite{Lollmann2015} & \ac{ABC} & Single & -0.073 & 0.0943 & 0.765 & 0.443\\ 
\hline
E & Baseline algorithm for \ac{FB2} RTE~\cite{Lollmann2015} & \ac{ABC} & Single & 0.0297 & 0.127 & 0.281 & 0.0428\\ 
\hline
F & \ac{SDDSA-G} retrained~\cite{Eaton2015b} & \ac{SFM} & Single & 0.0259 & 0.0541 & 0.757 & 0.0155\\ 
\hline
G & \ac{SDDSA-G}~\cite{Eaton2013} & \ac{SFM} & Single & -0.0249 & 0.0655 & 0.7 & 0.0162\\ 
\hline
H & Multi-layer perceptron~\cite{Xiong2015} & \ac{MLMF} & Single & -0.0903 & 0.119 & 0.345 & 0.0579$^\ddagger$\\ 
\hline
I & Multi-layer perceptron P2~\cite{Xiong2015} & \ac{MLMF} & Single & -0.0465 & 0.0853 & 0.577 & 0.0579$^\ddagger$\\ 
\hline
J & Multi-layer perceptron P2~\cite{Xiong2015} & \ac{MLMF} & Chromebook & -0.0568 & 0.0945 & 0.52 & 0.0588$^\ddagger$\\ 
\hline
K & Multi-layer perceptron P2~\cite{Xiong2015} & \ac{MLMF} & Mobile & -0.0339 & 0.0738 & 0.528 & 0.0555$^\ddagger$\\ 
\hline
L & Multi-layer perceptron P2~\cite{Xiong2015} & \ac{MLMF} & Crucif & -0.0419 & 0.0912 & 0.523 & 0.057$^\ddagger$\\ 
\hline
M & Multi-layer perceptron P2~\cite{Xiong2015} & \ac{MLMF} & Lin8Ch & -0.0241 & 0.078 & 0.506 & 0.0618$^\ddagger$\\ 
\hline
N & Multi-layer perceptron P2~\cite{Xiong2015} & \ac{MLMF} & EM32 & -0.0572 & 0.0826 & 0.478 & 0.0576$^\ddagger$\\ 
\hline
O & Per acoust. band \ac{SRMR} {\sectMidSent} 2.5.~\cite{Senoussaoui2015} & \ac{SFM} & Single & -0.0282 & 0.0888 & 0.81 & 0.579\\ 
\hline
P & \ac{NSRMR} {\sectMidSent} 2.4.~\cite{Santos2014,Senoussaoui2015} & \ac{SFM} & Single & 0.0712 & 0.0987 & 0.777 & 0.572\\ 
\hline
Q & \ac{NSRMR} {\sectMidSent} 2.4.~\cite{Santos2014,Senoussaoui2015} & \ac{SFM} & Chromebook & 0.113 & 0.113 & 0.783 & 1.04\\ 
\hline
R & \ac{NSRMR} {\sectMidSent} 2.4.~\cite{Santos2014,Senoussaoui2015} & \ac{SFM} & Mobile & 0.0935 & 0.0841 & 0.801 & 1.58\\ 
\hline
S & \ac{NSRMR} {\sectMidSent} 2.4.~\cite{Santos2014,Senoussaoui2015} & \ac{SFM} & Crucif & 0.0871 & 0.0905 & 0.828 & 2.63\\ 
\hline
T & \ac{SRMR} {\sectMidSent} 2.3.~\cite{Senoussaoui2015} & \ac{SFM} & Single & -0.00916 & 0.108 & 0.617 & 0.457\\ 
\hline
U & \ac{SRMR} {\sectMidSent} 2.3.~\cite{Senoussaoui2015} & \ac{SFM} & Chromebook & 0.029 & 0.111 & 0.687 & 0.833\\ 
\hline
V & \ac{SRMR} {\sectMidSent} 2.3.~\cite{Senoussaoui2015} & \ac{SFM} & Mobile & 0.00906 & 0.0867 & 0.671 & 1.26\\ 
\hline
W & \ac{SRMR} {\sectMidSent} 2.3.~\cite{Senoussaoui2015} & \ac{SFM} & Crucif & 0.00276 & 0.0948 & 0.687 & 2.1\\ 
\hline
X & NIRAv3~\cite{Parada2015} & \ac{MLMF} & Single & -0.0413 & 0.109 & 0.377 & 0.906$^\dagger$\\ 
\hline
Y & NIRAv1~\cite{Parada2015} & \ac{MLMF} & Single & -0.0465 & 0.119 & 0.279 & 0.906$^\dagger$\\ 
\hline
Z & NIRAv2~\cite{Parada2015} & \ac{MLMF} & Single & -0.184 & 0.185 & 0.0865 & 0.901$^\dagger$\\ 
\hline
a & Blur kernel~\cite{Lim2015} & \ac{SFM} & Single & 0.263 & 0.201 & 0.0261 & 8.88\\ 
\hline
b & Blur kernel with sliding window~\cite{Lim2015a} & \ac{SFM} & Single & 0.0733 & 0.178 & -0.0263 & 0.438\\ 
\hline
c & Temporal dynamics~\cite{Falk2010a} & \ac{SFM} & Single & -0.053 & 0.0728 & 0.713 & 0.365\\ 
\hline
d & Improved blind RTE~\cite{Lollmann2010} & \ac{ABC} & Single & 0.0536 & 0.219 & 0.0134 & 0.0269\\ 
\hline
e & \ac{SDD}~\cite{Wen2008} & \ac{SFM} & Single & 0.529 & 12.5 & -0.131 & 0.0224\\ 
\hline

\else

\fi
\end{tabular}
\end{table*}
\clearpage
\subsubsection{Fan noise at \dBel{18} \ac{SNR}}
\begin{figure}[!ht]
	\ifarXiv
\centerline{\epsfig{figure=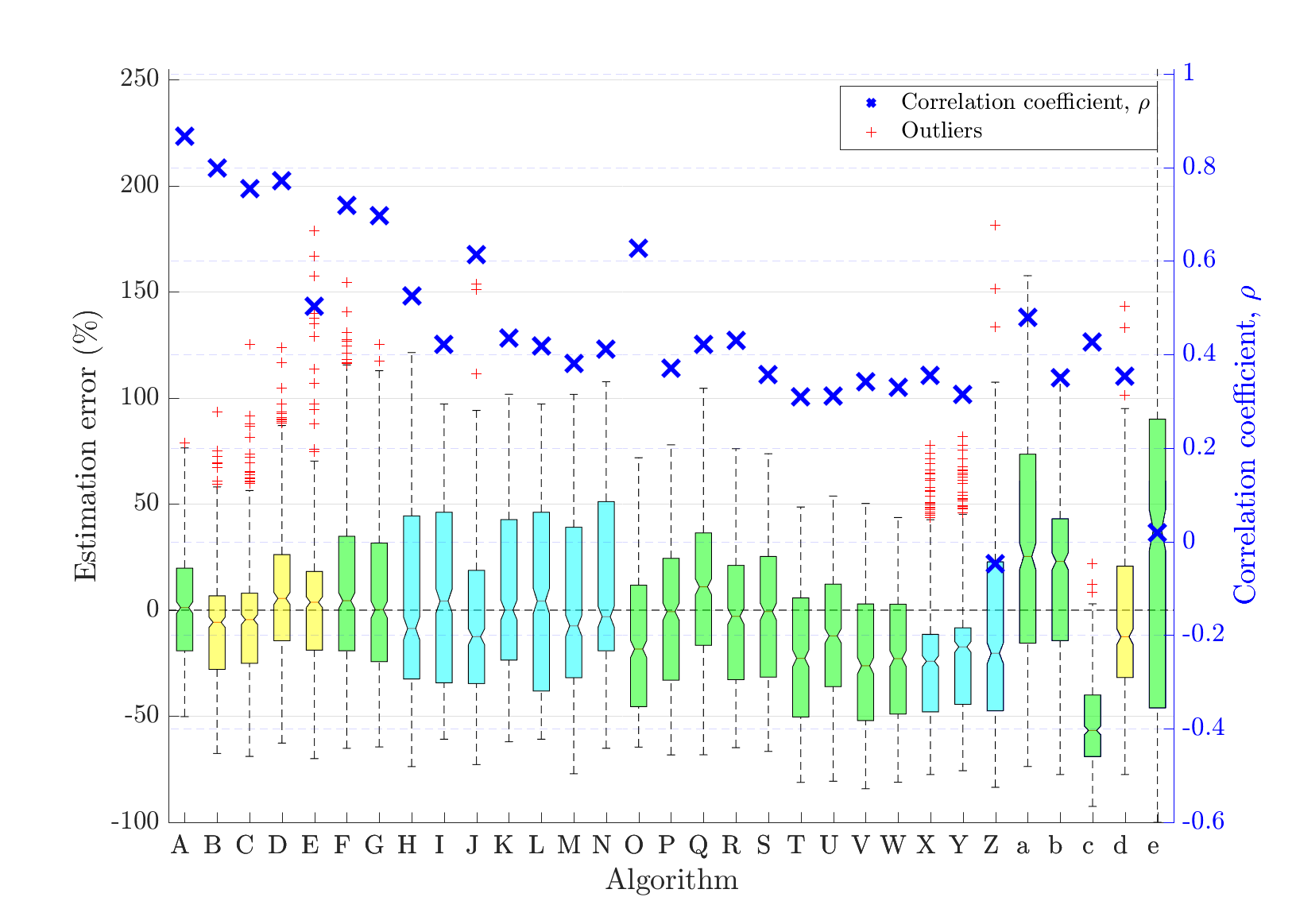,
	width=\figWidthACETR,viewport=45 10 765 530,clip}}%
	\else
	\centerline{\epsfig{figure=FigsACE/ana_eval_gt_partic_results_combined_Phase3_All_WASPAA_P3_T60_Perc_H_Fan.png,
	width=\figWidthACETR,viewport=45 10 765 530,clip}}%
	\fi
	\caption{Fullband {\ac{T60} estimation error in fan noise at \dBel{18} \ac{SNR}}}%
\label{fig:ACE_T60_Fan_H}%
\end{figure}%
\begin{table*}[!ht]\small
\caption{\ac{T60} estimation algorithm performance in fan noise at \dBel{18} \ac{SNR}}
\vspace{5mm} 
\centering
\begin{tabular}{clllllll}%
\hline%
Ref.
& Algorithm
& Class
& Mic. Config.
& Bias
& MSE
&  $\PearsonCC$
& \ac{RTF}
\\
\hline
\hline
\ifarXiv
A & QA Reverb~\cite{Prego2015} & \ac{SFM} & Single & -0.0649 & 0.055 & 0.867 & 0.4\\ 
\hline
B & Octave \ac{SB}-based \ac{FB2} RTE~\cite{Lollmann2015} & \ac{ABC} & Single & -0.111 & 0.0666 & 0.798 & 0.903\\ 
\hline
C & DCT-based \ac{FB2} RTE~\cite{Lollmann2015} & \ac{ABC} & Single & -0.106 & 0.0705 & 0.755 & 0.984\\ 
\hline
D & Model-based \ac{SB} RTE~\cite{Lollmann2015} & \ac{ABC} & Single & -0.0527 & 0.0917 & 0.772 & 0.433\\ 
\hline
E & Baseline algorithm for \ac{FB2} RTE~\cite{Lollmann2015} & \ac{ABC} & Single & -0.079 & 0.098 & 0.503 & 0.0421\\ 
\hline
F & \ac{SDDSA-G} retrained~\cite{Eaton2015b} & \ac{SFM} & Single & 0.0258 & 0.0717 & 0.719 & 0.0148\\ 
\hline
G & \ac{SDDSA-G}~\cite{Eaton2013} & \ac{SFM} & Single & -0.0387 & 0.066 & 0.696 & 0.0164\\ 
\hline
H & Multi-layer perceptron~\cite{Xiong2015} & \ac{MLMF} & Single & -0.0699 & 0.0938 & 0.525 & 0.0578$^\ddagger$\\ 
\hline
I & Multi-layer perceptron P2~\cite{Xiong2015} & \ac{MLMF} & Single & -0.0325 & 0.102 & 0.421 & 0.0578$^\ddagger$\\ 
\hline
J & Multi-layer perceptron P2~\cite{Xiong2015} & \ac{MLMF} & Chromebook & -0.0706 & 0.0843 & 0.614 & 0.059$^\ddagger$\\ 
\hline
K & Multi-layer perceptron P2~\cite{Xiong2015} & \ac{MLMF} & Mobile & -0.0177 & 0.0821 & 0.435 & 0.0555$^\ddagger$\\ 
\hline
L & Multi-layer perceptron P2~\cite{Xiong2015} & \ac{MLMF} & Crucif & -0.0371 & 0.103 & 0.418 & 0.0569$^\ddagger$\\ 
\hline
M & Multi-layer perceptron P2~\cite{Xiong2015} & \ac{MLMF} & Lin8Ch & -0.0477 & 0.0931 & 0.381 & 0.0617$^\ddagger$\\ 
\hline
N & Multi-layer perceptron P2~\cite{Xiong2015} & \ac{MLMF} & EM32 & -0.0406 & 0.0872 & 0.411 & 0.0574$^\ddagger$\\ 
\hline
O & Per acoust. band \ac{SRMR} {\sectMidSent} 2.5.~\cite{Senoussaoui2015} & \ac{SFM} & Single & -0.174 & 0.115 & 0.627 & 0.576\\ 
\hline
P & \ac{NSRMR} {\sectMidSent} 2.4.~\cite{Santos2014,Senoussaoui2015} & \ac{SFM} & Single & -0.124 & 0.122 & 0.37 & 0.569\\ 
\hline
Q & \ac{NSRMR} {\sectMidSent} 2.4.~\cite{Santos2014,Senoussaoui2015} & \ac{SFM} & Chromebook & -0.0681 & 0.111 & 0.422 & 1.03\\ 
\hline
R & \ac{NSRMR} {\sectMidSent} 2.4.~\cite{Santos2014,Senoussaoui2015} & \ac{SFM} & Mobile & -0.114 & 0.0956 & 0.431 & 1.58\\ 
\hline
S & \ac{NSRMR} {\sectMidSent} 2.4.~\cite{Santos2014,Senoussaoui2015} & \ac{SFM} & Crucif & -0.112 & 0.109 & 0.358 & 2.61\\ 
\hline
T & \ac{SRMR} {\sectMidSent} 2.3.~\cite{Senoussaoui2015} & \ac{SFM} & Single & -0.217 & 0.158 & 0.309 & 0.455\\ 
\hline
U & \ac{SRMR} {\sectMidSent} 2.3.~\cite{Senoussaoui2015} & \ac{SFM} & Chromebook & -0.179 & 0.146 & 0.312 & 0.824\\ 
\hline
V & \ac{SRMR} {\sectMidSent} 2.3.~\cite{Senoussaoui2015} & \ac{SFM} & Mobile & -0.215 & 0.134 & 0.342 & 1.26\\ 
\hline
W & \ac{SRMR} {\sectMidSent} 2.3.~\cite{Senoussaoui2015} & \ac{SFM} & Crucif & -0.211 & 0.142 & 0.33 & 2.08\\ 
\hline
X & NIRAv3~\cite{Parada2015} & \ac{MLMF} & Single & -0.223 & 0.157 & 0.355 & 0.895$^\dagger$\\ 
\hline
Y & NIRAv1~\cite{Parada2015} & \ac{MLMF} & Single & -0.207 & 0.153 & 0.315 & 0.895$^\dagger$\\ 
\hline
Z & NIRAv2~\cite{Parada2015} & \ac{MLMF} & Single & -0.164 & 0.201 & -0.0474 & 0.906$^\dagger$\\ 
\hline
a & Blur kernel~\cite{Lim2015} & \ac{SFM} & Single & 0.0893 & 0.103 & 0.48 & 8.36\\ 
\hline
b & Blur kernel with sliding window~\cite{Lim2015a} & \ac{SFM} & Single & -0.0394 & 0.109 & 0.35 & 0.412\\ 
\hline
c & Temporal dynamics~\cite{Falk2010a} & \ac{SFM} & Single & -0.384 & 0.251 & 0.427 & 0.358\\ 
\hline
d & Improved blind RTE~\cite{Lollmann2010} & \ac{ABC} & Single & -0.134 & 0.133 & 0.354 & 0.0254\\ 
\hline
e & \ac{SDD}~\cite{Wen2008} & \ac{SFM} & Single & 0.666 & 28.5 & 0.0185 & 0.0221\\ 
\hline

\else

\fi
\end{tabular}
\end{table*}
\clearpage
\subsubsection{Fan noise at \dBel{12} \ac{SNR}}
\begin{figure}[!ht]
	\ifarXiv
\centerline{\epsfig{figure=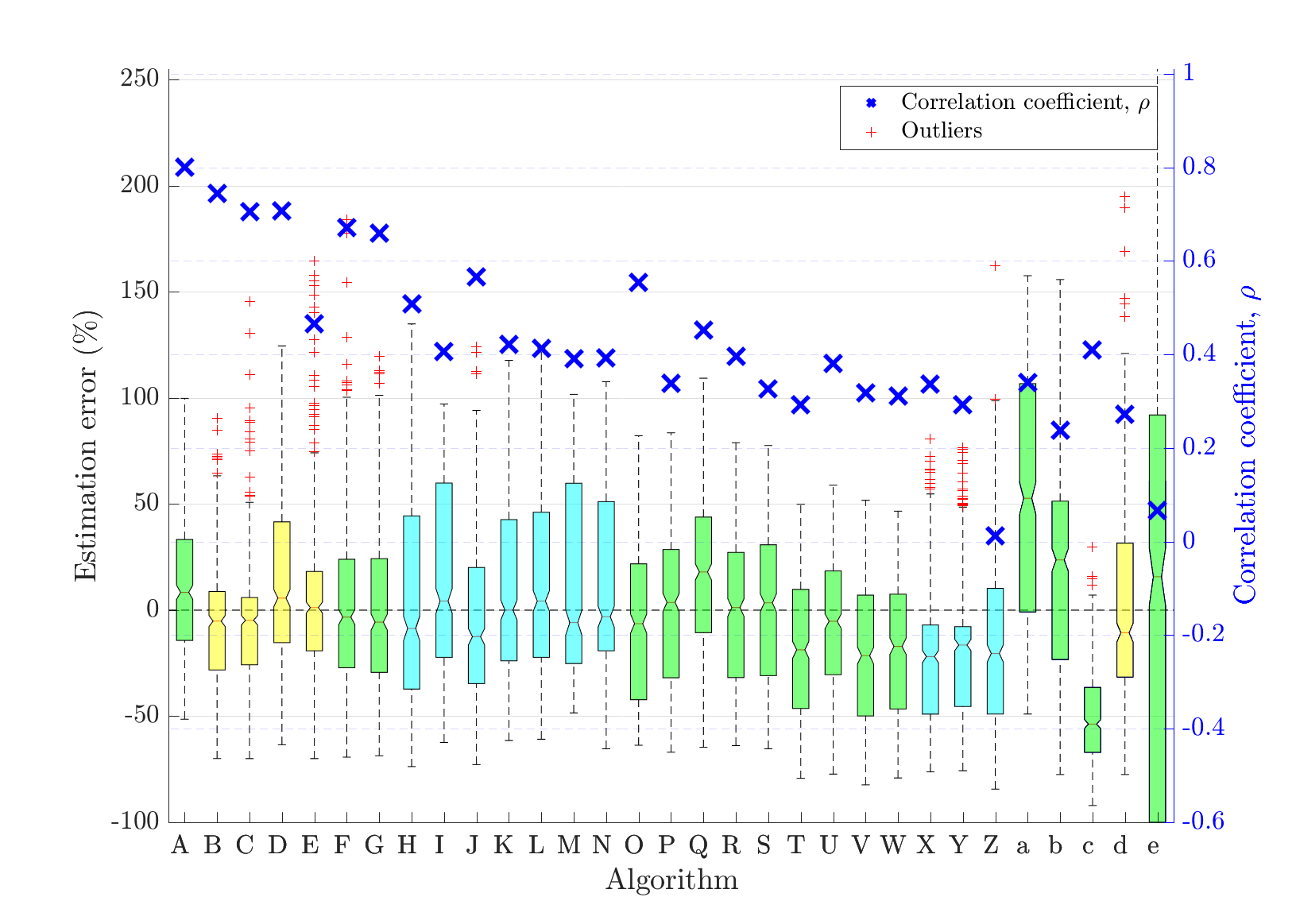,
	width=\figWidthACETR,viewport=45 10 765 530,clip}}%
	\else
	\centerline{\epsfig{figure=FigsACE/ana_eval_gt_partic_results_combined_Phase3_All_WASPAA_P3_T60_Perc_M_Fan.png,
	width=\figWidthACETR,viewport=45 10 765 530,clip}}%
	\fi
	\caption{Fullband {\ac{T60} estimation error in fan noise at \dBel{12} \ac{SNR}}}%
\label{fig:ACE_T60_Fan_M}%
\end{figure}%
\begin{table*}[!ht]\small
\caption{\ac{T60} estimation algorithm performance in fan noise at \dBel{12} \ac{SNR}}
\vspace{5mm} 
\centering
\begin{tabular}{clllllll}%
\hline%
Ref.
& Algorithm
& Class
& Mic. Config.
& Bias
& MSE
& $\PearsonCC$
& \ac{RTF}
\\
\hline
\hline
\ifarXiv
A & QA Reverb~\cite{Prego2015} & \ac{SFM} & Single & -0.0315 & 0.0622 & 0.8 & 0.4\\ 
\hline
B & Octave \ac{SB}-based \ac{FB2} RTE~\cite{Lollmann2015} & \ac{ABC} & Single & -0.114 & 0.0755 & 0.744 & 0.903\\ 
\hline
C & DCT-based \ac{FB2} RTE~\cite{Lollmann2015} & \ac{ABC} & Single & -0.117 & 0.0803 & 0.705 & 0.984\\ 
\hline
D & Model-based \ac{SB} RTE~\cite{Lollmann2015} & \ac{ABC} & Single & -0.0419 & 0.102 & 0.706 & 0.433\\ 
\hline
E & Baseline algorithm for \ac{FB2} RTE~\cite{Lollmann2015} & \ac{ABC} & Single & -0.0834 & 0.103 & 0.465 & 0.0421\\ 
\hline
F & \ac{SDDSA-G} retrained~\cite{Eaton2015b} & \ac{SFM} & Single & -0.05 & 0.0707 & 0.671 & 0.0148\\ 
\hline
G & \ac{SDDSA-G}~\cite{Eaton2013} & \ac{SFM} & Single & -0.0808 & 0.0782 & 0.66 & 0.0164\\ 
\hline
H & Multi-layer perceptron~\cite{Xiong2015} & \ac{MLMF} & Single & -0.08 & 0.0975 & 0.508 & 0.0578$^\ddagger$\\ 
\hline
I & Multi-layer perceptron P2~\cite{Xiong2015} & \ac{MLMF} & Single & -0.0275 & 0.103 & 0.407 & 0.0578$^\ddagger$\\ 
\hline
J & Multi-layer perceptron P2~\cite{Xiong2015} & \ac{MLMF} & Chromebook & -0.0644 & 0.0894 & 0.565 & 0.059$^\ddagger$\\ 
\hline
K & Multi-layer perceptron P2~\cite{Xiong2015} & \ac{MLMF} & Mobile & -0.00734 & 0.0834 & 0.422 & 0.0555$^\ddagger$\\ 
\hline
L & Multi-layer perceptron P2~\cite{Xiong2015} & \ac{MLMF} & Crucif & -0.0295 & 0.103 & 0.414 & 0.0569$^\ddagger$\\ 
\hline
M & Multi-layer perceptron P2~\cite{Xiong2015} & \ac{MLMF} & Lin8Ch & -0.0329 & 0.0913 & 0.391 & 0.0617$^\ddagger$\\ 
\hline
N & Multi-layer perceptron P2~\cite{Xiong2015} & \ac{MLMF} & EM32 & -0.0338 & 0.0891 & 0.392 & 0.0574$^\ddagger$\\ 
\hline
O & Per acoust. band \ac{SRMR} {\sectMidSent} 2.5.~\cite{Senoussaoui2015} & \ac{SFM} & Single & -0.132 & 0.109 & 0.554 & 0.576\\ 
\hline
P & \ac{NSRMR} {\sectMidSent} 2.4.~\cite{Santos2014,Senoussaoui2015} & \ac{SFM} & Single & -0.109 & 0.121 & 0.339 & 0.569\\ 
\hline
Q & \ac{NSRMR} {\sectMidSent} 2.4.~\cite{Santos2014,Senoussaoui2015} & \ac{SFM} & Chromebook & -0.0388 & 0.106 & 0.452 & 1.03\\ 
\hline
R & \ac{NSRMR} {\sectMidSent} 2.4.~\cite{Santos2014,Senoussaoui2015} & \ac{SFM} & Mobile & -0.0978 & 0.0944 & 0.396 & 1.58\\ 
\hline
S & \ac{NSRMR} {\sectMidSent} 2.4.~\cite{Santos2014,Senoussaoui2015} & \ac{SFM} & Crucif & -0.0963 & 0.107 & 0.326 & 2.61\\ 
\hline
T & \ac{SRMR} {\sectMidSent} 2.3.~\cite{Senoussaoui2015} & \ac{SFM} & Single & -0.201 & 0.153 & 0.293 & 0.455\\ 
\hline
U & \ac{SRMR} {\sectMidSent} 2.3.~\cite{Senoussaoui2015} & \ac{SFM} & Chromebook & -0.149 & 0.133 & 0.38 & 0.824\\ 
\hline
V & \ac{SRMR} {\sectMidSent} 2.3.~\cite{Senoussaoui2015} & \ac{SFM} & Mobile & -0.197 & 0.129 & 0.318 & 1.26\\ 
\hline
W & \ac{SRMR} {\sectMidSent} 2.3.~\cite{Senoussaoui2015} & \ac{SFM} & Crucif & -0.195 & 0.137 & 0.312 & 2.08\\ 
\hline
X & NIRAv3~\cite{Parada2015} & \ac{MLMF} & Single & -0.216 & 0.155 & 0.337 & 0.895$^\dagger$\\ 
\hline
Y & NIRAv1~\cite{Parada2015} & \ac{MLMF} & Single & -0.206 & 0.155 & 0.293 & 0.895$^\dagger$\\ 
\hline
Z & NIRAv2~\cite{Parada2015} & \ac{MLMF} & Single & -0.188 & 0.188 & 0.0117 & 0.906$^\dagger$\\ 
\hline
a & Blur kernel~\cite{Lim2015} & \ac{SFM} & Single & 0.18 & 0.143 & 0.34 & 8.36\\ 
\hline
b & Blur kernel with sliding window~\cite{Lim2015a} & \ac{SFM} & Single & -0.0215 & 0.119 & 0.238 & 0.412\\ 
\hline
c & Temporal dynamics~\cite{Falk2010a} & \ac{SFM} & Single & -0.371 & 0.242 & 0.409 & 0.358\\ 
\hline
d & Improved blind RTE~\cite{Lollmann2010} & \ac{ABC} & Single & -0.106 & 0.143 & 0.271 & 0.0254\\ 
\hline
e & \ac{SDD}~\cite{Wen2008} & \ac{SFM} & Single & 0.918 & 57.2 & 0.0662 & 0.0221\\ 
\hline

\else

\fi
\end{tabular}
\end{table*}
\clearpage
\subsubsection{Fan noise at \dBel{-1} \ac{SNR}}
\begin{figure}[!ht]
	\ifarXiv
\centerline{\epsfig{figure=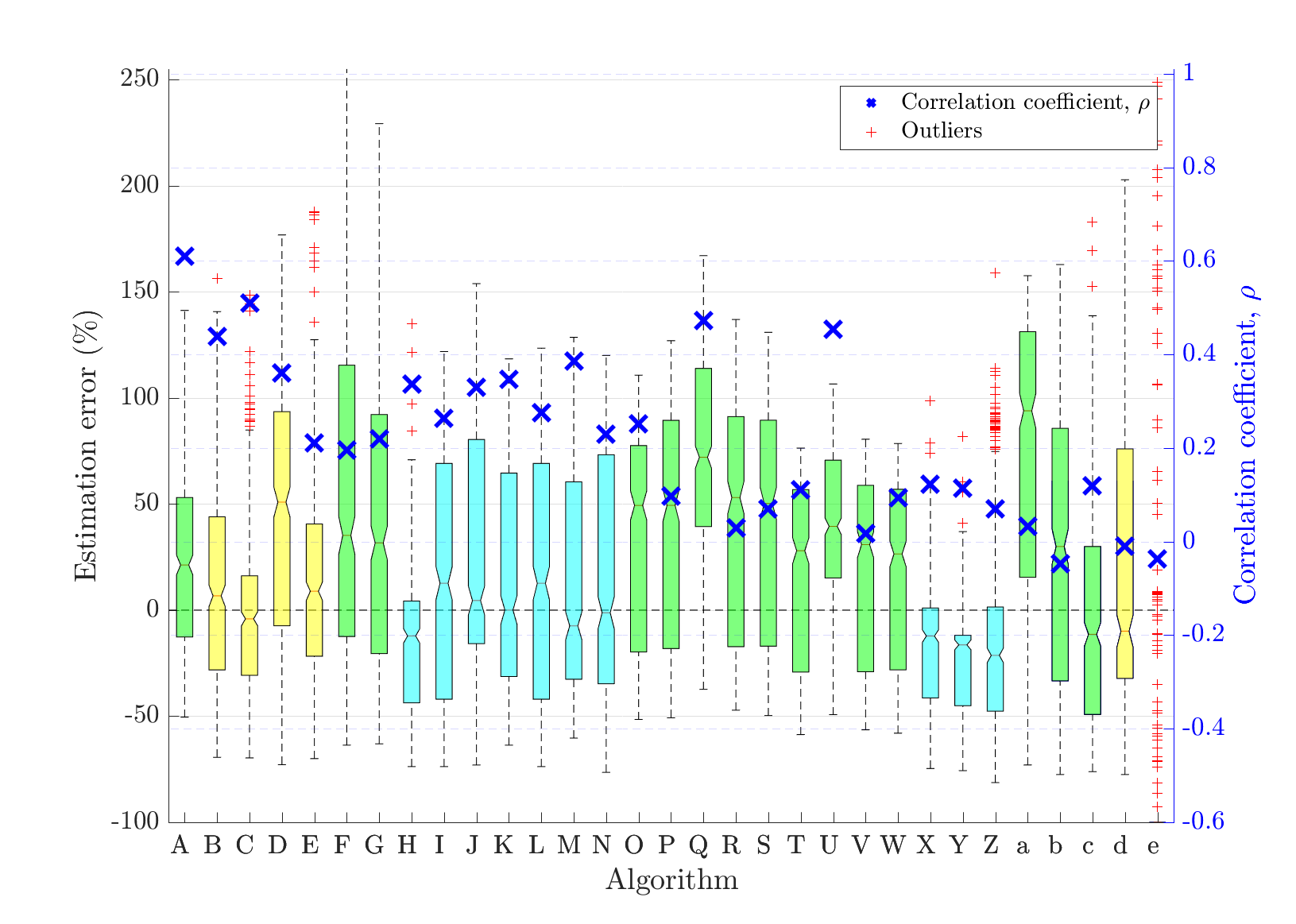,
	width=\figWidthACETR,viewport=45 10 765 530,clip}}%
	\else
	\centerline{\epsfig{figure=FigsACE/ana_eval_gt_partic_results_combined_Phase3_All_WASPAA_P3_T60_Perc_L_Fan.png,
	width=\figWidthACETR,viewport=45 10 765 530,clip}}%
	\fi
	\caption{Fullband {\ac{T60} estimation error in fan noise at \dBel{-1} \ac{SNR}}}%
\label{fig:ACE_T60_Fan_L}%
\end{figure}%
\begin{table*}[!ht]\small
\caption{\ac{T60} estimation algorithm performance in fan noise at \dBel{-1} \ac{SNR}}
\vspace{5mm} 
\centering
\begin{tabular}{clllllll}%
\hline%
Ref.
& Algorithm
& Class
& Mic. Config.
& Bias
& MSE
&  $\PearsonCC$
& \ac{RTF}
\\
\hline
\hline
\ifarXiv
A & QA Reverb~\cite{Prego2015} & \ac{SFM} & Single & 0.0162 & 0.0844 & 0.609 & 0.4\\ 
\hline
B & Octave \ac{SB}-based \ac{FB2} RTE~\cite{Lollmann2015} & \ac{ABC} & Single & -0.0396 & 0.101 & 0.438 & 0.903\\ 
\hline
C & DCT-based \ac{FB2} RTE~\cite{Lollmann2015} & \ac{ABC} & Single & -0.105 & 0.102 & 0.509 & 0.984\\ 
\hline
D & Model-based \ac{SB} RTE~\cite{Lollmann2015} & \ac{ABC} & Single & 0.0906 & 0.162 & 0.36 & 0.433\\ 
\hline
E & Baseline algorithm for \ac{FB2} RTE~\cite{Lollmann2015} & \ac{ABC} & Single & -0.0326 & 0.135 & 0.21 & 0.0421\\ 
\hline
F & \ac{SDDSA-G} retrained~\cite{Eaton2015b} & \ac{SFM} & Single & 0.213 & 0.247 & 0.196 & 0.0148\\ 
\hline
G & \ac{SDDSA-G}~\cite{Eaton2013} & \ac{SFM} & Single & 0.093 & 0.141 & 0.22 & 0.0164\\ 
\hline
H & Multi-layer perceptron~\cite{Xiong2015} & \ac{MLMF} & Single & -0.141 & 0.133 & 0.337 & 0.0578$^\ddagger$\\ 
\hline
I & Multi-layer perceptron P2~\cite{Xiong2015} & \ac{MLMF} & Single & 0.000904 & 0.12 & 0.264 & 0.0578$^\ddagger$\\ 
\hline
J & Multi-layer perceptron P2~\cite{Xiong2015} & \ac{MLMF} & Chromebook & 0.0206 & 0.118 & 0.329 & 0.059$^\ddagger$\\ 
\hline
K & Multi-layer perceptron P2~\cite{Xiong2015} & \ac{MLMF} & Mobile & 0.00366 & 0.0937 & 0.347 & 0.0555$^\ddagger$\\ 
\hline
L & Multi-layer perceptron P2~\cite{Xiong2015} & \ac{MLMF} & Crucif & -0.000496 & 0.119 & 0.276 & 0.0569$^\ddagger$\\ 
\hline
M & Multi-layer perceptron P2~\cite{Xiong2015} & \ac{MLMF} & Lin8Ch & -0.0123 & 0.093 & 0.386 & 0.0617$^\ddagger$\\ 
\hline
N & Multi-layer perceptron P2~\cite{Xiong2015} & \ac{MLMF} & EM32 & -0.00585 & 0.107 & 0.229 & 0.0574$^\ddagger$\\ 
\hline
O & Per acoust. band \ac{SRMR} {\sectMidSent} 2.5.~\cite{Senoussaoui2015} & \ac{SFM} & Single & 0.0497 & 0.117 & 0.251 & 0.576\\ 
\hline
P & \ac{NSRMR} {\sectMidSent} 2.4.~\cite{Santos2014,Senoussaoui2015} & \ac{SFM} & Single & 0.0712 & 0.132 & 0.0973 & 0.569\\ 
\hline
Q & \ac{NSRMR} {\sectMidSent} 2.4.~\cite{Santos2014,Senoussaoui2015} & \ac{SFM} & Chromebook & 0.212 & 0.151 & 0.472 & 1.03\\ 
\hline
R & \ac{NSRMR} {\sectMidSent} 2.4.~\cite{Santos2014,Senoussaoui2015} & \ac{SFM} & Mobile & 0.0952 & 0.117 & 0.0296 & 1.58\\ 
\hline
S & \ac{NSRMR} {\sectMidSent} 2.4.~\cite{Santos2014,Senoussaoui2015} & \ac{SFM} & Crucif & 0.0847 & 0.122 & 0.0702 & 2.61\\ 
\hline
T & \ac{SRMR} {\sectMidSent} 2.3.~\cite{Senoussaoui2015} & \ac{SFM} & Single & -0.0493 & 0.124 & 0.111 & 0.455\\ 
\hline
U & \ac{SRMR} {\sectMidSent} 2.3.~\cite{Senoussaoui2015} & \ac{SFM} & Chromebook & 0.0513 & 0.113 & 0.454 & 0.824\\ 
\hline
V & \ac{SRMR} {\sectMidSent} 2.3.~\cite{Senoussaoui2015} & \ac{SFM} & Mobile & -0.0333 & 0.105 & 0.0169 & 1.26\\ 
\hline
W & \ac{SRMR} {\sectMidSent} 2.3.~\cite{Senoussaoui2015} & \ac{SFM} & Crucif & -0.0408 & 0.111 & 0.0942 & 2.08\\ 
\hline
X & NIRAv3~\cite{Parada2015} & \ac{MLMF} & Single & -0.207 & 0.166 & 0.122 & 0.895$^\dagger$\\ 
\hline
Y & NIRAv1~\cite{Parada2015} & \ac{MLMF} & Single & -0.246 & 0.184 & 0.115 & 0.895$^\dagger$\\ 
\hline
Z & NIRAv2~\cite{Parada2015} & \ac{MLMF} & Single & -0.185 & 0.188 & 0.0698 & 0.906$^\dagger$\\ 
\hline
a & Blur kernel~\cite{Lim2015} & \ac{SFM} & Single & 0.248 & 0.201 & 0.0335 & 8.36\\ 
\hline
b & Blur kernel with sliding window~\cite{Lim2015a} & \ac{SFM} & Single & 0.00154 & 0.198 & -0.0472 & 0.412\\ 
\hline
c & Temporal dynamics~\cite{Falk2010a} & \ac{SFM} & Single & -0.131 & 0.158 & 0.119 & 0.358\\ 
\hline
d & Improved blind RTE~\cite{Lollmann2010} & \ac{ABC} & Single & 0.000833 & 0.219 & -0.00975 & 0.0254\\ 
\hline
e & \ac{SDD}~\cite{Wen2008} & \ac{SFM} & Single & 2.35 & 2.5e+03 & -0.0369 & 0.0221\\ 
\hline

\else

\fi
\end{tabular}
\end{table*}
\clearpage
\clearpage
\subsection{Frequency-dependent \ac{T60} estimation results}
%
\begin{figure}[!ht]
	\ifarXiv
\centerline{\epsfig{figure=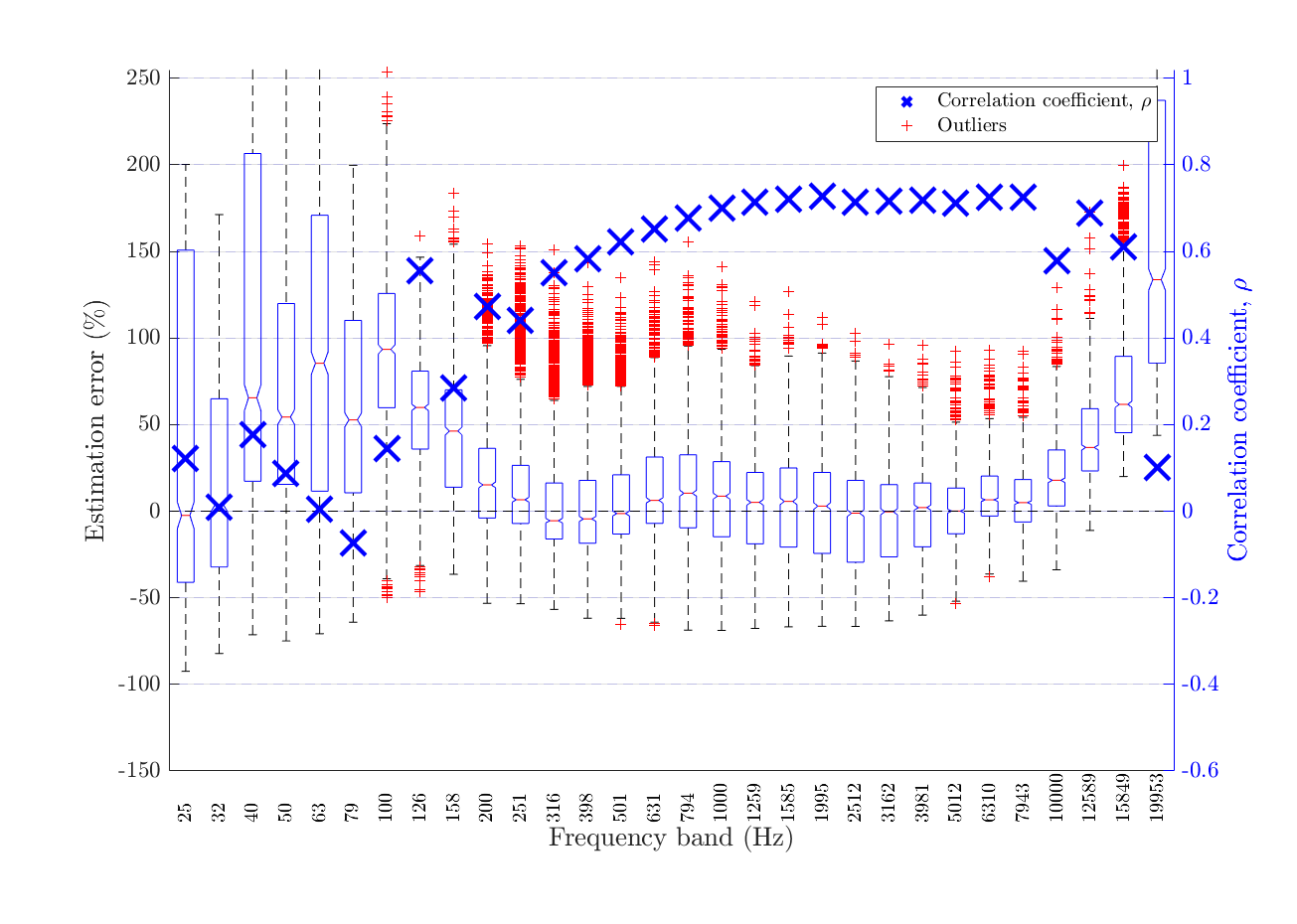,
	width=\figWidthACETR,viewport=45 10 765 530,clip}}%
	\else
	\centerline{\epsfig{figure=FigsACE/ana_eval_gt_partic_results_combined_Phase3_TR_P3S_T60_Perc_All_SNR_All_Noises_sub_Frequency-dependent-RTE.png,
	width=\figWidthACETR,viewport=45 10 765 530,clip}}%
	\fi
	\caption{{Frequency-dependent \ac{T60} estimation error in all noises for all \acp{SNR} for algorithm Model-based \ac{SB} RTE~\cite{Lollmann2015}}}%
\label{fig:ACE_T60_Sub_All_Lollman}%
\end{figure}%
\begin{table*}[!htb]\small
\caption{\ac{T60} estimation algorithm performance for all noises for all \acp{SNR}}
\vspace{5mm} 
\centering
\begin{tabular}{crrrl}%
\hline%
Freq. band
& Centre Freq. (Hz)
& Bias
& \acs{MSE}
& $\PearsonCC$

\\
\hline%
\hline%
\ifarXiv
 1 & 25.12 & -0.5197 & 2.901 & 0.1224 \\ 
\hline
 2 & 31.62 & -0.02769 & 0.5496 & 0.009101 \\ 
\hline
 3 & 39.81 & 0.4369 & 0.6085 & 0.177 \\ 
\hline
 4 & 50.12 & 0.4183 & 0.5136 & 0.08707 \\ 
\hline
 5 & 63.10 & 0.471 & 0.5336 & 0.004263 \\ 
\hline
 6 & 79.43 & 0.3703 & 0.3121 & -0.07279 \\ 
\hline
 7 & 100.00 & 0.4955 & 0.3206 & 0.145 \\ 
\hline
 8 & 125.89 & 0.3541 & 0.1641 & 0.5561 \\ 
\hline
 9 & 158.49 & 0.2743 & 0.136 & 0.2838 \\ 
\hline
10 & 199.53 & 0.139 & 0.08303 & 0.4733 \\ 
\hline
11 & 251.19 & 0.07525 & 0.08501 & 0.4412 \\ 
\hline
12 & 316.23 & 0.03405 & 0.06716 & 0.5516 \\ 
\hline
13 & 398.11 & 0.01442 & 0.07244 & 0.5838 \\ 
\hline
14 & 501.19 & 0.002211 & 0.08705 & 0.6224 \\ 
\hline
15 & 630.96 & -0.005208 & 0.1117 & 0.651 \\ 
\hline
16 & 794.33 & -0.02265 & 0.1192 & 0.6773 \\ 
\hline
17 & 1000.00 & -0.05042 & 0.128 & 0.6991 \\ 
\hline
18 & 1258.93 & -0.05597 & 0.1051 & 0.7128 \\ 
\hline
19 & 1584.89 & -0.05515 & 0.09516 & 0.7203 \\ 
\hline
20 & 1995.26 & -0.06708 & 0.09227 & 0.7271 \\ 
\hline
21 & 2511.89 & -0.09857 & 0.09601 & 0.7138 \\ 
\hline
22 & 3162.28 & -0.07575 & 0.0704 & 0.7165 \\ 
\hline
23 & 3981.07 & -0.04222 & 0.04788 & 0.7184 \\ 
\hline
24 & 5011.87 & -0.00961 & 0.0256 & 0.7104 \\ 
\hline
25 & 6309.57 & 0.0536 & 0.01633 & 0.7254 \\ 
\hline
26 & 7943.28 & 0.04453 & 0.01525 & 0.7262 \\ 
\hline
27 & 10000.00 & 0.0958 & 0.02727 & 0.5792 \\ 
\hline
28 & 12589.25 & 0.1761 & 0.04667 & 0.6879 \\ 
\hline
29 & 15848.93 & 0.2424 & 0.07895 & 0.611 \\ 
\hline
30 & 19952.62 & 0.3372 & 0.1412 & 0.1018 \\ 
\hline
\else
\fi
\end{tabular}%
\end{table*}%
\clearpage
\subsection{Frequency-dependent \ac{T60} estimation results by noise type}
\subsubsection{Ambient noise}
\begin{figure}[!ht]
	\ifarXiv
\centerline{\epsfig{figure=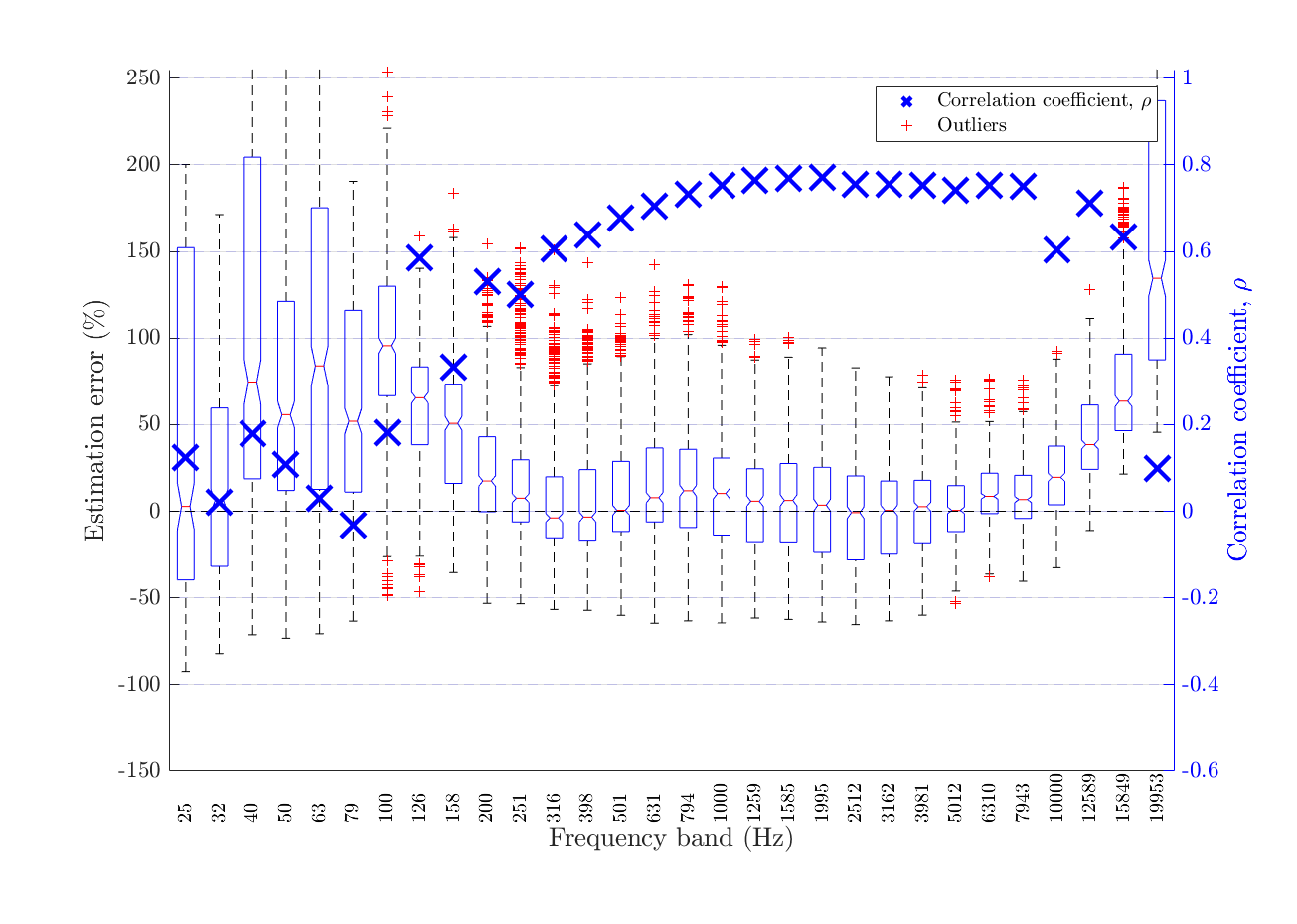,
	width=\figWidthACETR,viewport=45 10 765 530,clip}}%
	\else
	\centerline{\epsfig{figure=FigsACE/ana_eval_gt_partic_results_combined_Phase3_TR_P3S_T60_Perc_All_SNR_Ambient_sub_Frequency-dependent-RTE.png,
	width=\figWidthACETR,viewport=45 10 765 530,clip}}%
	\fi
	\caption{{Frequency-dependent \ac{T60} estimation error in ambient noise for all \acp{SNR} for algorithm Model-based \ac{SB} RTE~\cite{Lollmann2015}}}%
\label{fig:ACE_T60_Sub_Ambient_Lollman}%
\end{figure}%
\begin{table*}[!htb]\small
\caption{Frequency-dependent \ac{T60} estimation error in ambient noise for all \acp{SNR} for algorithm Model-based \ac{SB} RTE~\cite{Lollmann2015}}
\vspace{5mm} 
\centering
\begin{tabular}{crrrl}%
\hline%
Freq. band
& Centre Freq. (Hz)
& Bias
& \acs{MSE}
& $\PearsonCC$

\\
\hline%
\hline%
\ifarXiv
 1 & 25.12 & -0.5118 & 2.892 & 0.1237 \\ 
\hline
 2 & 31.62 & -0.0155 & 0.5479 & 0.02024 \\ 
\hline
 3 & 39.81 & 0.451 & 0.6237 & 0.1794 \\ 
\hline
 4 & 50.12 & 0.4318 & 0.5217 & 0.1085 \\ 
\hline
 5 & 63.10 & 0.482 & 0.5402 & 0.03066 \\ 
\hline
 6 & 79.43 & 0.3776 & 0.3146 & -0.03175 \\ 
\hline
 7 & 100.00 & 0.4989 & 0.3227 & 0.1825 \\ 
\hline
 8 & 125.89 & 0.3543 & 0.1621 & 0.5862 \\ 
\hline
 9 & 158.49 & 0.2722 & 0.1309 & 0.3326 \\ 
\hline
10 & 199.53 & 0.1355 & 0.0744 & 0.5299 \\ 
\hline
11 & 251.19 & 0.07118 & 0.07479 & 0.5004 \\ 
\hline
12 & 316.23 & 0.02995 & 0.05818 & 0.605 \\ 
\hline
13 & 398.11 & 0.01065 & 0.06323 & 0.6376 \\ 
\hline
14 & 501.19 & -0.001016 & 0.07766 & 0.676 \\ 
\hline
15 & 630.96 & -0.007819 & 0.1019 & 0.705 \\ 
\hline
16 & 794.33 & -0.02463 & 0.1098 & 0.7319 \\ 
\hline
17 & 1000.00 & -0.0518 & 0.1192 & 0.7521 \\ 
\hline
18 & 1258.93 & -0.05678 & 0.09748 & 0.7635 \\ 
\hline
19 & 1584.89 & -0.05546 & 0.08837 & 0.7677 \\ 
\hline
20 & 1995.26 & -0.06695 & 0.08603 & 0.7711 \\ 
\hline
21 & 2511.89 & -0.09804 & 0.09027 & 0.7541 \\ 
\hline
22 & 3162.28 & -0.07489 & 0.06565 & 0.7539 \\ 
\hline
23 & 3981.07 & -0.04107 & 0.04412 & 0.7529 \\ 
\hline
24 & 5011.87 & -0.008213 & 0.02323 & 0.7417 \\ 
\hline
25 & 6309.57 & 0.05521 & 0.01547 & 0.7528 \\ 
\hline
26 & 7943.28 & 0.04631 & 0.01457 & 0.7503 \\ 
\hline
27 & 10000.00 & 0.09773 & 0.02723 & 0.6039 \\ 
\hline
28 & 12589.25 & 0.1781 & 0.04744 & 0.7118 \\ 
\hline
29 & 15848.93 & 0.2446 & 0.08045 & 0.6337 \\ 
\hline
30 & 19952.62 & 0.3395 & 0.1436 & 0.09808 \\ 
\hline
\else
\fi
\end{tabular}%
\end{table*}%
\clearpage
\subsubsection{Babble noise}
\begin{figure}[!ht]
	\ifarXiv
\centerline{\epsfig{figure=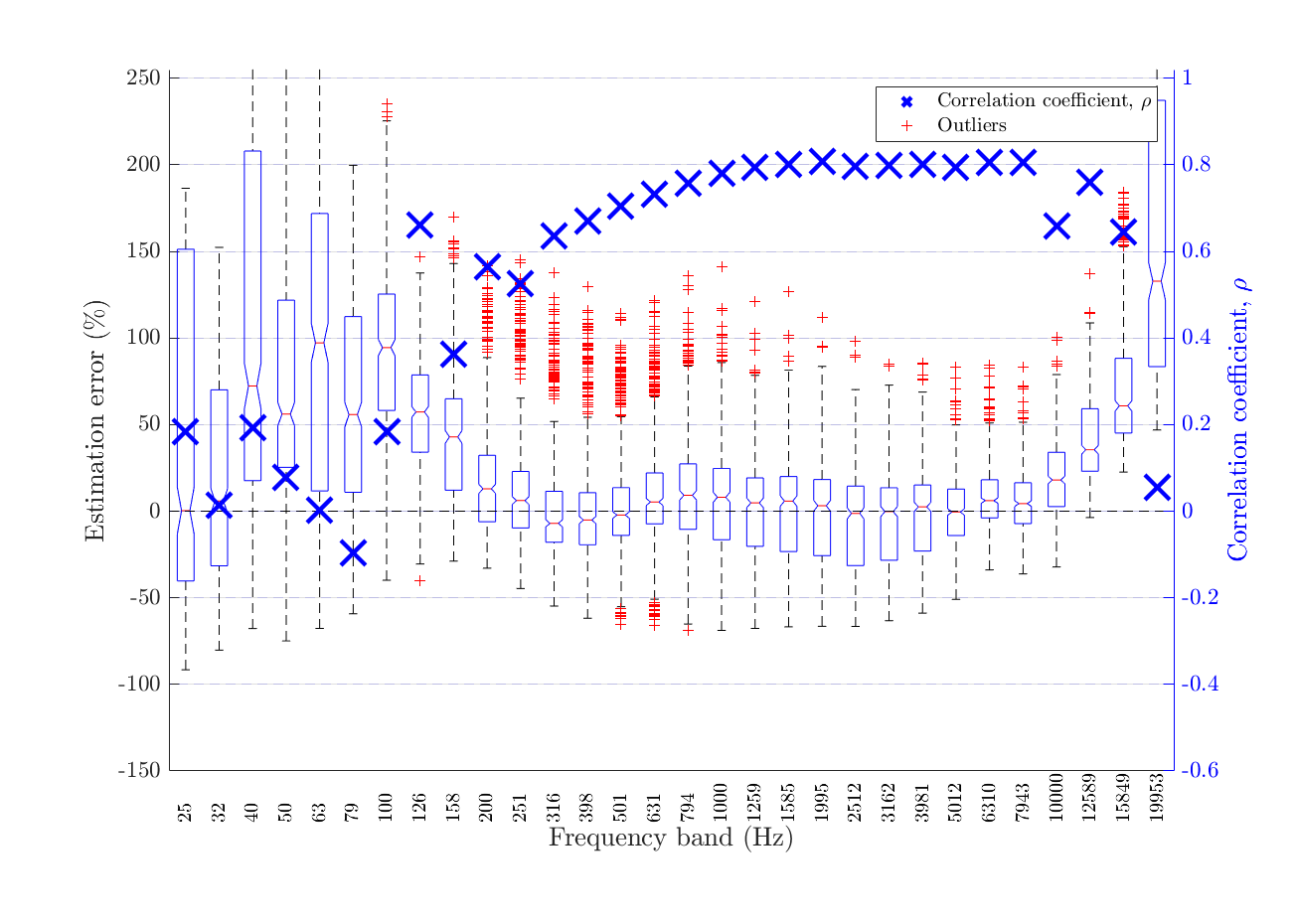,
	width=\figWidthACETR,viewport=45 10 765 530,clip}}%
	\else
	\centerline{\epsfig{figure=FigsACE/ana_eval_gt_partic_results_combined_Phase3_TR_P3S_T60_Perc_All_SNR_Babble_sub_Frequency-dependent-RTE.png,
	width=\figWidthACETR,viewport=45 10 765 530,clip}}%
	\fi
	\caption{{Frequency-dependent \ac{T60} estimation error in babble noise for all \acp{SNR} for algorithm Model-based \ac{SB} RTE~\cite{Lollmann2015}}}%
\label{fig:ACE_T60_Sub_Babble_Lollman}%
\end{figure}%
\begin{table*}[!htb]\small
\caption{Frequency-dependent \ac{T60} estimation error in babble noise for all \acp{SNR} for algorithm Model-based \ac{SB} RTE~\cite{Lollmann2015}}
\vspace{5mm} 
\centering
\begin{tabular}{crrrl}%
\hline%
Freq. band
& Centre Freq. (Hz)
& Bias
& \acs{MSE}
& $\PearsonCC$

\\
\hline%
\hline%
\ifarXiv
 1 & 25.12 & -0.5021 & 2.837 & 0.1826 \\ 
\hline
 2 & 31.62 & 0.01671 & 0.5267 & 0.01443 \\ 
\hline
 3 & 39.81 & 0.4939 & 0.6333 & 0.1917 \\ 
\hline
 4 & 50.12 & 0.4729 & 0.5405 & 0.07811 \\ 
\hline
 5 & 63.10 & 0.5109 & 0.5531 & 0.001786 \\ 
\hline
 6 & 79.43 & 0.3889 & 0.3159 & -0.09623 \\ 
\hline
 7 & 100.00 & 0.4919 & 0.3071 & 0.1825 \\ 
\hline
 8 & 125.89 & 0.3316 & 0.1374 & 0.6609 \\ 
\hline
 9 & 158.49 & 0.2383 & 0.1079 & 0.3631 \\ 
\hline
10 & 199.53 & 0.09501 & 0.05918 & 0.5637 \\ 
\hline
11 & 251.19 & 0.0279 & 0.06538 & 0.5263 \\ 
\hline
12 & 316.23 & -0.01336 & 0.05215 & 0.6363 \\ 
\hline
13 & 398.11 & -0.03105 & 0.05875 & 0.6695 \\ 
\hline
14 & 501.19 & -0.04029 & 0.07434 & 0.7055 \\ 
\hline
15 & 630.96 & -0.04439 & 0.09955 & 0.7309 \\ 
\hline
16 & 794.33 & -0.05855 & 0.1086 & 0.7579 \\ 
\hline
17 & 1000.00 & -0.0833 & 0.1196 & 0.7795 \\ 
\hline
18 & 1258.93 & -0.08619 & 0.09809 & 0.7935 \\ 
\hline
19 & 1584.89 & -0.08311 & 0.08867 & 0.8005 \\ 
\hline
20 & 1995.26 & -0.09317 & 0.08655 & 0.8077 \\ 
\hline
21 & 2511.89 & -0.1231 & 0.09174 & 0.7957 \\ 
\hline
22 & 3162.28 & -0.0991 & 0.06587 & 0.7985 \\ 
\hline
23 & 3981.07 & -0.06462 & 0.0427 & 0.8006 \\ 
\hline
24 & 5011.87 & -0.03127 & 0.02038 & 0.7939 \\ 
\hline
25 & 6309.57 & 0.0325 & 0.01033 & 0.8055 \\ 
\hline
26 & 7943.28 & 0.02386 & 0.009707 & 0.8062 \\ 
\hline
27 & 10000.00 & 0.07546 & 0.02013 & 0.659 \\ 
\hline
28 & 12589.25 & 0.156 & 0.03743 & 0.7598 \\ 
\hline
29 & 15848.93 & 0.2225 & 0.06813 & 0.645 \\ 
\hline
30 & 19952.62 & 0.3174 & 0.1277 & 0.05588 \\ 
\hline
\else
\fi
\end{tabular}%
\end{table*}%
%
%
\clearpage
\subsubsection{Fan noise}
\begin{figure}[!ht]
	\ifarXiv
\centerline{\epsfig{figure=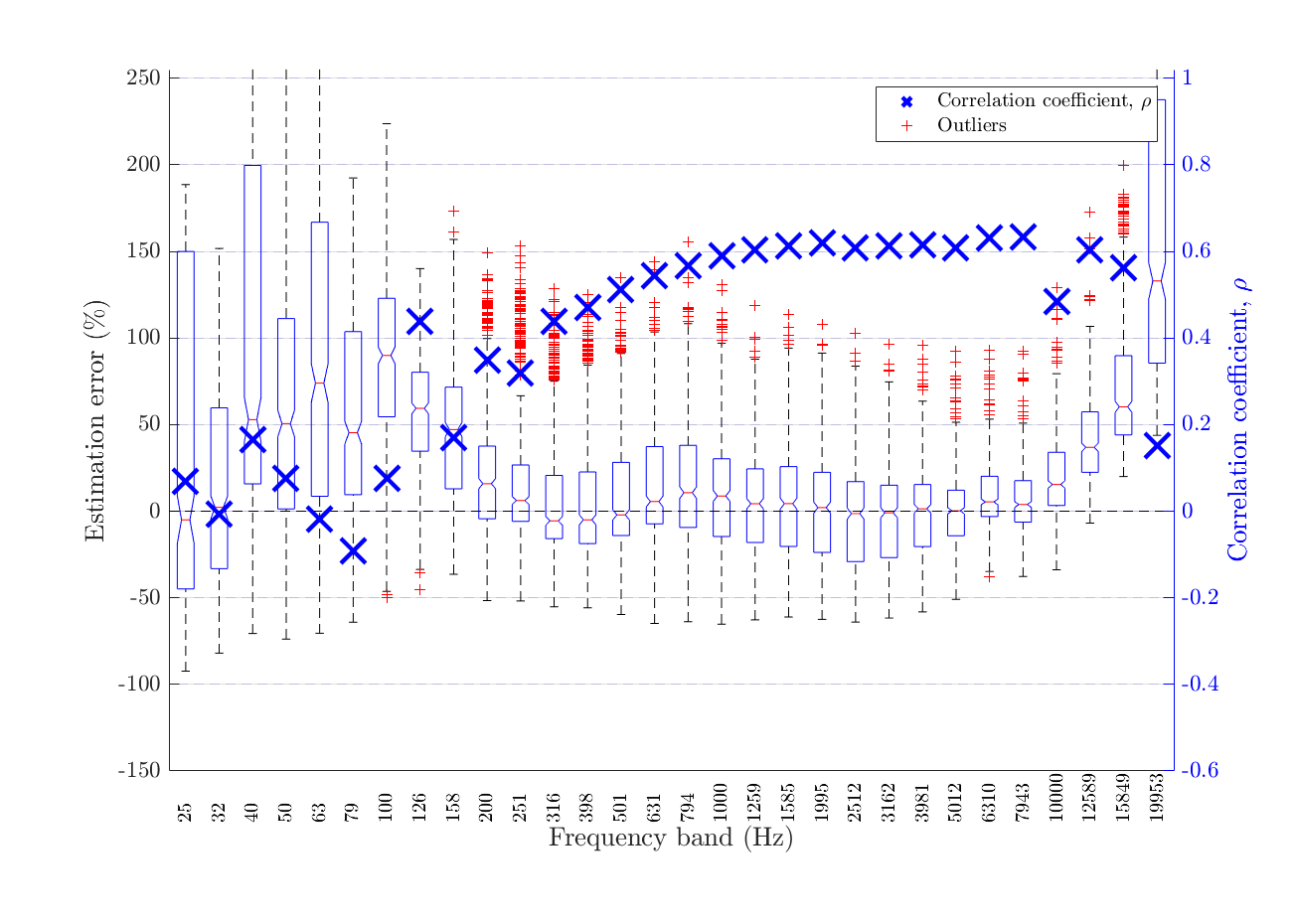,
	width=\figWidthACETR,viewport=45 10 765 530,clip}}%
	\else
	\centerline{\epsfig{figure=FigsACE/ana_eval_gt_partic_results_combined_Phase3_TR_P3S_T60_Perc_All_SNR_Fan_sub_Frequency-dependent-RTE.png,
	width=\figWidthACETR,viewport=45 10 765 530,clip}}%
	\fi
	\caption{{Frequency-dependent \ac{T60} estimation error in fan noise for all \acp{SNR} for algorithm Model-based \ac{SB} RTE~\cite{Lollmann2015}}}%
\label{fig:ACE_T60_Sub_Fan_Lollman}%
\end{figure}%
\begin{table*}[!htb]\small
\caption{Frequency-dependent \ac{T60} estimation error in fan noise for all \acp{SNR} for algorithm Model-based \ac{SB} RTE~\cite{Lollmann2015}}
\vspace{5mm} 
\centering
\begin{tabular}{crrrl}%
\hline%
Freq. band
& Centre Freq. (Hz)
& Bias
& \acs{MSE}
& $\PearsonCC$

\\
\hline%
\hline%
\ifarXiv
 1 & 25.12 & -0.5453 & 2.973 & 0.06903 \\ 
\hline
 2 & 31.62 & -0.08427 & 0.5741 & -0.006387 \\ 
\hline
 3 & 39.81 & 0.3657 & 0.5685 & 0.1654 \\ 
\hline
 4 & 50.12 & 0.3503 & 0.4786 & 0.07602 \\ 
\hline
 5 & 63.10 & 0.4203 & 0.5074 & -0.01927 \\ 
\hline
 6 & 79.43 & 0.3445 & 0.3058 & -0.0928 \\ 
\hline
 7 & 100.00 & 0.4956 & 0.332 & 0.07602 \\ 
\hline
 8 & 125.89 & 0.3763 & 0.1928 & 0.4387 \\ 
\hline
 9 & 158.49 & 0.3124 & 0.1692 & 0.17 \\ 
\hline
10 & 199.53 & 0.1865 & 0.1155 & 0.3485 \\ 
\hline
11 & 251.19 & 0.1267 & 0.1149 & 0.3184 \\ 
\hline
12 & 316.23 & 0.08556 & 0.09114 & 0.4388 \\ 
\hline
13 & 398.11 & 0.06364 & 0.09533 & 0.4702 \\ 
\hline
14 & 501.19 & 0.04794 & 0.1091 & 0.5121 \\ 
\hline
15 & 630.96 & 0.03658 & 0.1335 & 0.5428 \\ 
\hline
16 & 794.33 & 0.01523 & 0.1391 & 0.5671 \\ 
\hline
17 & 1000.00 & -0.01617 & 0.1454 & 0.589 \\ 
\hline
18 & 1258.93 & -0.02493 & 0.1198 & 0.6031 \\ 
\hline
19 & 1584.89 & -0.02688 & 0.1085 & 0.6124 \\ 
\hline
20 & 1995.26 & -0.04114 & 0.1042 & 0.6203 \\ 
\hline
21 & 2511.89 & -0.07453 & 0.106 & 0.6074 \\ 
\hline
22 & 3162.28 & -0.05326 & 0.07968 & 0.6119 \\ 
\hline
23 & 3981.07 & -0.02097 & 0.05682 & 0.6155 \\ 
\hline
24 & 5011.87 & 0.01065 & 0.03318 & 0.6086 \\ 
\hline
25 & 6309.57 & 0.07309 & 0.0232 & 0.6303 \\ 
\hline
26 & 7943.28 & 0.06342 & 0.02147 & 0.6341 \\ 
\hline
27 & 10000.00 & 0.1142 & 0.03443 & 0.4842 \\ 
\hline
28 & 12589.25 & 0.1941 & 0.05513 & 0.6027 \\ 
\hline
29 & 15848.93 & 0.2602 & 0.08826 & 0.563 \\ 
\hline
30 & 19952.62 & 0.3547 & 0.1522 & 0.1521 \\ 
\hline
\else
\fi
\end{tabular}%
\end{table*}%
%
%
%
%
\clearpage
\subsection{Frequency-dependent \ac{T60} estimation results by noise type and \ac{SNR}}
\subsubsection{Ambient noise at \dBel{18}}
\begin{figure}[!ht]
	\ifarXiv
\centerline{\epsfig{figure=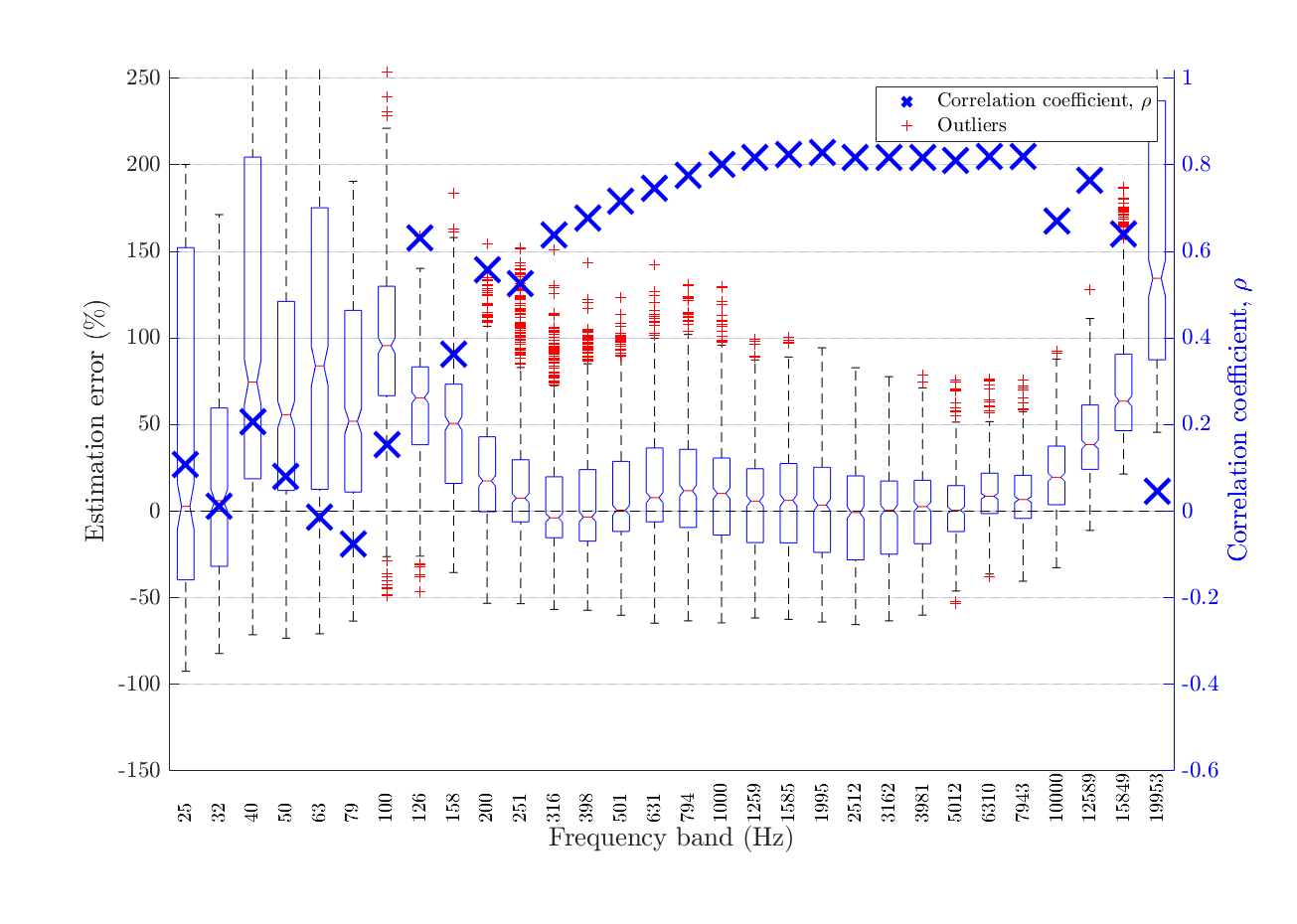,
	width=\figWidthACETR,viewport=45 10 765 530,clip}}%
	\else
	\centerline{\epsfig{figure=FigsACE/ana_eval_gt_partic_results_combined_Phase3_TR_P3S_T60_Perc_18dB_SNR_Ambient_sub_Frequency-dependent-RTE.png,
	width=\figWidthACETR,viewport=45 10 765 530,clip}}%
	\fi
	\caption{{Frequency-dependent \ac{T60} estimation error in ambient noise at \dBel{18} \ac{SNR} for algorithm Model-based \ac{SB} RTE~\cite{Lollmann2015}}}%
\label{fig:ACE_T60_Sub_Ambient_18dB_SNR_Lollman}%
\end{figure}%
\begin{table*}[!htb]\small
\caption{Frequency-dependent \ac{T60} estimation error in ambient noise at \dBel{18} \ac{SNR} for algorithm Model-based \ac{SB} RTE~\cite{Lollmann2015}}
\vspace{5mm} 
\centering
\begin{tabular}{crrrl}%
\hline%
Freq. band
& Centre Freq. (Hz)
& Bias
& \acs{MSE}
& $\PearsonCC$

\\
\hline%
\hline%
\ifarXiv
 1 & 25.12 & -0.492 & 2.884 & 0.109 \\ 
\hline
 2 & 31.62 & 0.0215 & 0.54 & 0.01075 \\ 
\hline
 3 & 39.81 & 0.498 & 0.6376 & 0.2061 \\ 
\hline
 4 & 50.12 & 0.481 & 0.5514 & 0.08078 \\ 
\hline
 5 & 63.10 & 0.5267 & 0.5733 & -0.01295 \\ 
\hline
 6 & 79.43 & 0.4138 & 0.3301 & -0.07604 \\ 
\hline
 7 & 100.00 & 0.5253 & 0.3417 & 0.1527 \\ 
\hline
 8 & 125.89 & 0.3714 & 0.1678 & 0.6304 \\ 
\hline
 9 & 158.49 & 0.2818 & 0.1326 & 0.3616 \\ 
\hline
10 & 199.53 & 0.1397 & 0.07238 & 0.5571 \\ 
\hline
11 & 251.19 & 0.07161 & 0.07183 & 0.5249 \\ 
\hline
12 & 316.23 & 0.02794 & 0.05357 & 0.6385 \\ 
\hline
13 & 398.11 & 0.007045 & 0.05709 & 0.6765 \\ 
\hline
14 & 501.19 & -0.005716 & 0.07074 & 0.7149 \\ 
\hline
15 & 630.96 & -0.0133 & 0.09399 & 0.7461 \\ 
\hline
16 & 794.33 & -0.03069 & 0.1015 & 0.7762 \\ 
\hline
17 & 1000.00 & -0.0583 & 0.1106 & 0.8008 \\ 
\hline
18 & 1258.93 & -0.06361 & 0.08942 & 0.8165 \\ 
\hline
19 & 1584.89 & -0.06253 & 0.08054 & 0.8231 \\ 
\hline
20 & 1995.26 & -0.07417 & 0.07826 & 0.8294 \\ 
\hline
21 & 2511.89 & -0.1054 & 0.08277 & 0.8158 \\ 
\hline
22 & 3162.28 & -0.08224 & 0.05875 & 0.8173 \\ 
\hline
23 & 3981.07 & -0.04842 & 0.03784 & 0.8173 \\ 
\hline
24 & 5011.87 & -0.01555 & 0.01819 & 0.8093 \\ 
\hline
25 & 6309.57 & 0.04791 & 0.01181 & 0.8193 \\ 
\hline
26 & 7943.28 & 0.03906 & 0.01096 & 0.8198 \\ 
\hline
27 & 10000.00 & 0.09053 & 0.02404 & 0.6691 \\ 
\hline
28 & 12589.25 & 0.171 & 0.04441 & 0.7651 \\ 
\hline
29 & 15848.93 & 0.2375 & 0.07786 & 0.6411 \\ 
\hline
30 & 19952.62 & 0.3324 & 0.1408 & 0.04633 \\ 
\hline
\else
\fi
\end{tabular}%
\end{table*}%
%
%
\clearpage
\subsubsection{Ambient noise at \dBel{12}}
\begin{figure}[!ht]
	\ifarXiv
\centerline{\epsfig{figure=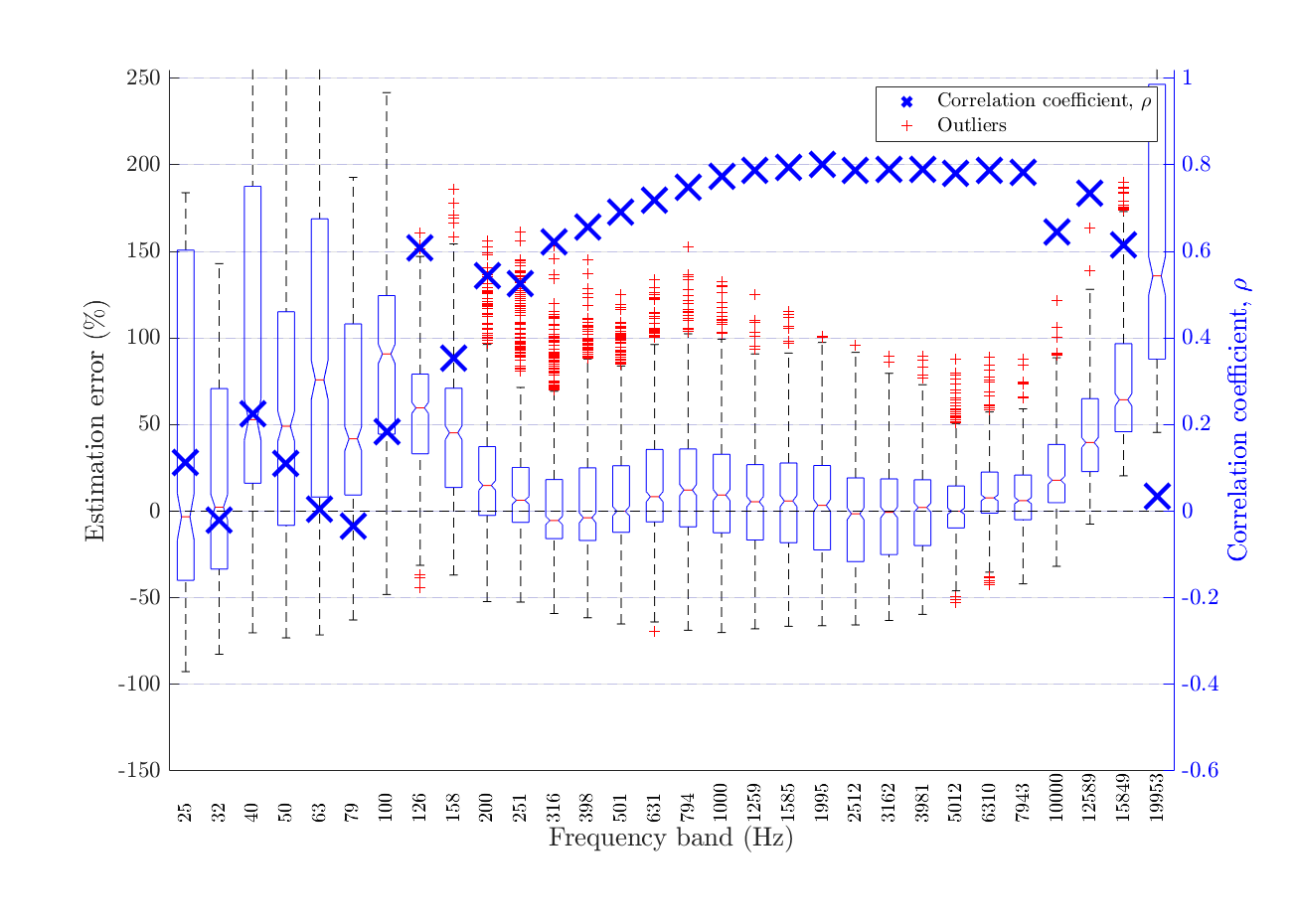, width=\figWidthACETR,viewport=45 10 765 530,clip}}%
	\else
	\centerline{\epsfig{figure=FigsACE/ana_eval_gt_partic_results_combined_Phase3_TR_P3S_T60_Perc_12dB_SNR_Ambient_sub_Frequency-dependent-RTE.png, width=\figWidthACETR,viewport=45 10 765 530,clip}}%
	\fi
	\caption{{Frequency-dependent \ac{T60} estimation error in ambient noise at \dBel{12} \ac{SNR} for algorithm Model-based \ac{SB} RTE~\cite{Lollmann2015}}}%
\label{fig:ACE_T60_Sub_Ambient_12dB_SNR_Lollman}%
\end{figure}%
\begin{table*}[!htb]\small
\caption{Frequency-dependent \ac{T60} estimation error in ambient noise at \dBel{12} \ac{SNR} for algorithm Model-based \ac{SB} RTE~\cite{Lollmann2015}}
\vspace{5mm} 
\centering
\begin{tabular}{crrrl}%
\hline%
Freq. band
& Centre Freq. (Hz)
& Bias
& \acs{MSE}
& $\PearsonCC$

\\
\hline%
\hline%
\ifarXiv
 1 & 25.12 & -0.5143 & 2.904 & 0.1135 \\ 
\hline
 2 & 31.62 & -0.01651 & 0.5708 & -0.01967 \\ 
\hline
 3 & 39.81 & 0.4505 & 0.6029 & 0.2257 \\ 
\hline
 4 & 50.12 & 0.4308 & 0.5244 & 0.1093 \\ 
\hline
 5 & 63.10 & 0.4796 & 0.5502 & 0.004409 \\ 
\hline
 6 & 79.43 & 0.3734 & 0.315 & -0.03412 \\ 
\hline
 7 & 100.00 & 0.4931 & 0.3183 & 0.1834 \\ 
\hline
 8 & 125.89 & 0.3471 & 0.1556 & 0.6079 \\ 
\hline
 9 & 158.49 & 0.2642 & 0.1249 & 0.3536 \\ 
\hline
10 & 199.53 & 0.1272 & 0.06996 & 0.5449 \\ 
\hline
11 & 251.19 & 0.06291 & 0.06977 & 0.5247 \\ 
\hline
12 & 316.23 & 0.02203 & 0.05508 & 0.6214 \\ 
\hline
13 & 398.11 & 0.003237 & 0.06022 & 0.6552 \\ 
\hline
14 & 501.19 & -0.007883 & 0.07524 & 0.6904 \\ 
\hline
15 & 630.96 & -0.01415 & 0.09982 & 0.7174 \\ 
\hline
16 & 794.33 & -0.03047 & 0.1069 & 0.7491 \\ 
\hline
17 & 1000.00 & -0.05721 & 0.1159 & 0.7725 \\ 
\hline
18 & 1258.93 & -0.06186 & 0.09411 & 0.7872 \\ 
\hline
19 & 1584.89 & -0.06028 & 0.08475 & 0.7943 \\ 
\hline
20 & 1995.26 & -0.07157 & 0.08228 & 0.8002 \\ 
\hline
21 & 2511.89 & -0.1025 & 0.08625 & 0.7882 \\ 
\hline
22 & 3162.28 & -0.0793 & 0.06193 & 0.7889 \\ 
\hline
23 & 3981.07 & -0.04544 & 0.04072 & 0.7884 \\ 
\hline
24 & 5011.87 & -0.01257 & 0.02039 & 0.78 \\ 
\hline
25 & 6309.57 & 0.05084 & 0.01353 & 0.7865 \\ 
\hline
26 & 7943.28 & 0.04192 & 0.01275 & 0.7837 \\ 
\hline
27 & 10000.00 & 0.09331 & 0.0251 & 0.6454 \\ 
\hline
28 & 12589.25 & 0.1737 & 0.04568 & 0.7334 \\ 
\hline
29 & 15848.93 & 0.2401 & 0.07893 & 0.6148 \\ 
\hline
30 & 19952.62 & 0.335 & 0.1421 & 0.03508 \\ 
\hline
\else
\fi
\end{tabular}%
\end{table*}%
%
%
\clearpage
\subsubsection{Ambient noise at \dBel{-1}}
\begin{figure}[!ht]
	\ifarXiv
\centerline{\epsfig{figure=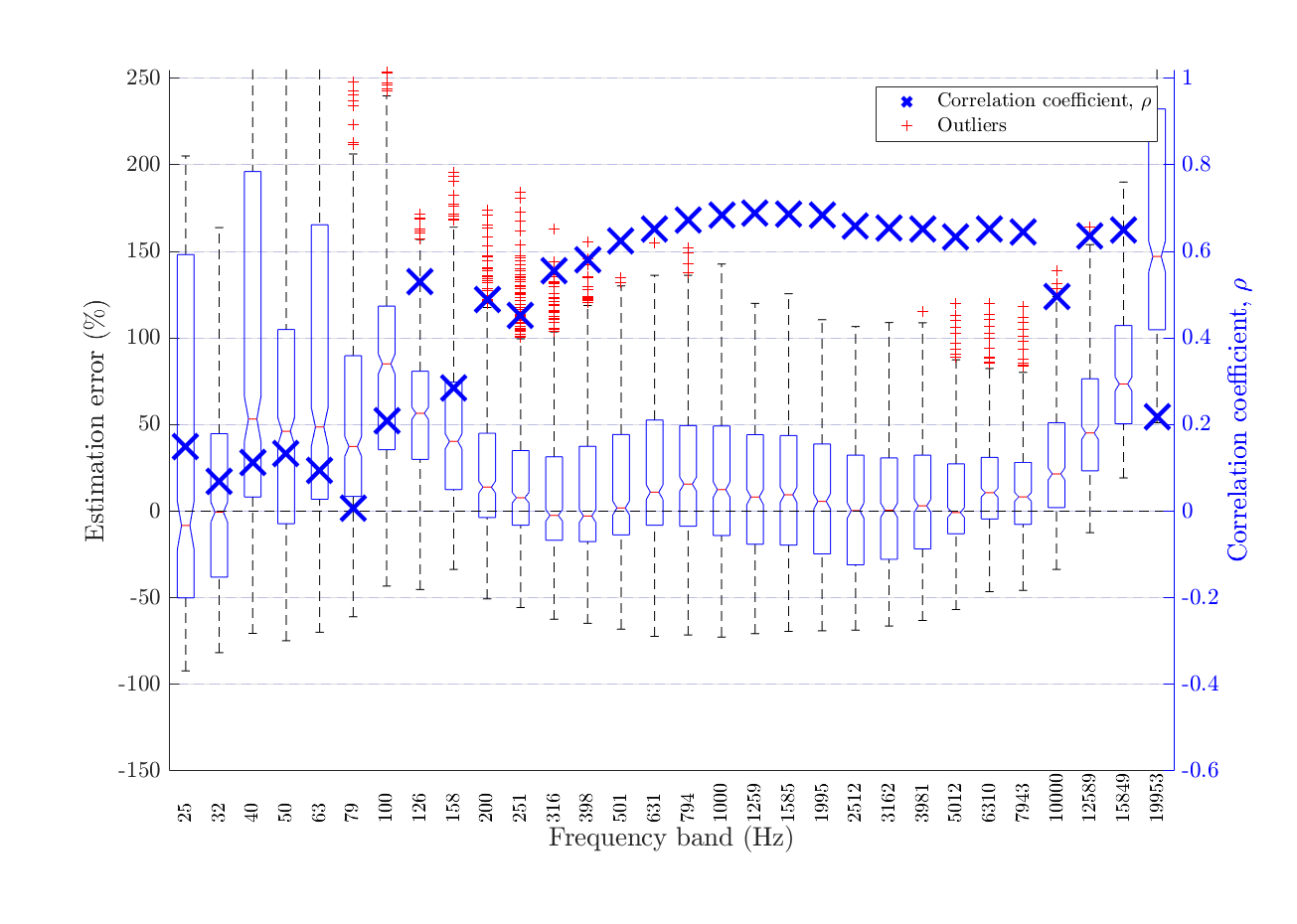, width=\figWidthACETR,viewport=45 10 765 530,clip}}%
	\else
	\centerline{\epsfig{figure=FigsACE/ana_eval_gt_partic_results_combined_Phase3_TR_P3S_T60_Perc_-1dB_SNR_Ambient_sub_Frequency-dependent-RTE.png,	width=\figWidthACETR,viewport=45 10 765 530,clip}}%
	\fi
	\caption{{Frequency-dependent \ac{T60} estimation error in ambient noise at \dBel{-1} \ac{SNR} for algorithm Model-based \ac{SB} RTE~\cite{Lollmann2015}}}%
\label{fig:ACE_T60_Sub_Ambient_-1dB_SNR_Lollman}%
\end{figure}%
\begin{table*}[!htb]\small
\caption{Frequency-dependent \ac{T60} estimation error in ambient noise at \dBel{-1} \ac{SNR} for algorithm Model-based \ac{SB} RTE~\cite{Lollmann2015}}
\vspace{5mm} 
\centering
\begin{tabular}{crrrl}%
\hline%
Freq. band
& Centre Freq. (Hz)
& Bias
& \acs{MSE}
& $\PearsonCC$

\\
\hline%
\hline%
\ifarXiv
 1 & 25.12 & -0.529 & 2.889 & 0.1493 \\ 
\hline
 2 & 31.62 & -0.05149 & 0.5329 & 0.06865 \\ 
\hline
 3 & 39.81 & 0.4044 & 0.6307 & 0.1121 \\ 
\hline
 4 & 50.12 & 0.3837 & 0.4894 & 0.134 \\ 
\hline
 5 & 63.10 & 0.4396 & 0.4971 & 0.094 \\ 
\hline
 6 & 79.43 & 0.3455 & 0.2986 & 0.006754 \\ 
\hline
 7 & 100.00 & 0.4785 & 0.3081 & 0.2094 \\ 
\hline
 8 & 125.89 & 0.3444 & 0.1628 & 0.5294 \\ 
\hline
 9 & 158.49 & 0.2706 & 0.1351 & 0.2846 \\ 
\hline
10 & 199.53 & 0.1397 & 0.08085 & 0.4884 \\ 
\hline
11 & 251.19 & 0.079 & 0.08276 & 0.4521 \\ 
\hline
12 & 316.23 & 0.03987 & 0.06589 & 0.5556 \\ 
\hline
13 & 398.11 & 0.02168 & 0.07236 & 0.5817 \\ 
\hline
14 & 501.19 & 0.01055 & 0.087 & 0.6233 \\ 
\hline
15 & 630.96 & 0.003987 & 0.112 & 0.6525 \\ 
\hline
16 & 794.33 & -0.01273 & 0.1209 & 0.6713 \\ 
\hline
17 & 1000.00 & -0.03988 & 0.131 & 0.6837 \\ 
\hline
18 & 1258.93 & -0.04487 & 0.1089 & 0.6873 \\ 
\hline
19 & 1584.89 & -0.04358 & 0.09981 & 0.6859 \\ 
\hline
20 & 1995.26 & -0.0551 & 0.09755 & 0.6836 \\ 
\hline
21 & 2511.89 & -0.08624 & 0.1018 & 0.6576 \\ 
\hline
22 & 3162.28 & -0.06312 & 0.07626 & 0.6543 \\ 
\hline
23 & 3981.07 & -0.02934 & 0.0538 & 0.6517 \\ 
\hline
24 & 5011.87 & 0.003485 & 0.03111 & 0.634 \\ 
\hline
25 & 6309.57 & 0.06688 & 0.02107 & 0.6511 \\ 
\hline
26 & 7943.28 & 0.05796 & 0.02 & 0.6456 \\ 
\hline
27 & 10000.00 & 0.1094 & 0.03256 & 0.4948 \\ 
\hline
28 & 12589.25 & 0.1897 & 0.05224 & 0.6363 \\ 
\hline
29 & 15848.93 & 0.2562 & 0.08457 & 0.6484 \\ 
\hline
30 & 19952.62 & 0.3511 & 0.148 & 0.2188 \\ 
\hline
\else
\fi
\end{tabular}%
\end{table*}%
%
%
%
\clearpage
\subsubsection{Babble noise at \dBel{18}}
\begin{figure}[!ht]
	\ifarXiv
\centerline{\epsfig{figure=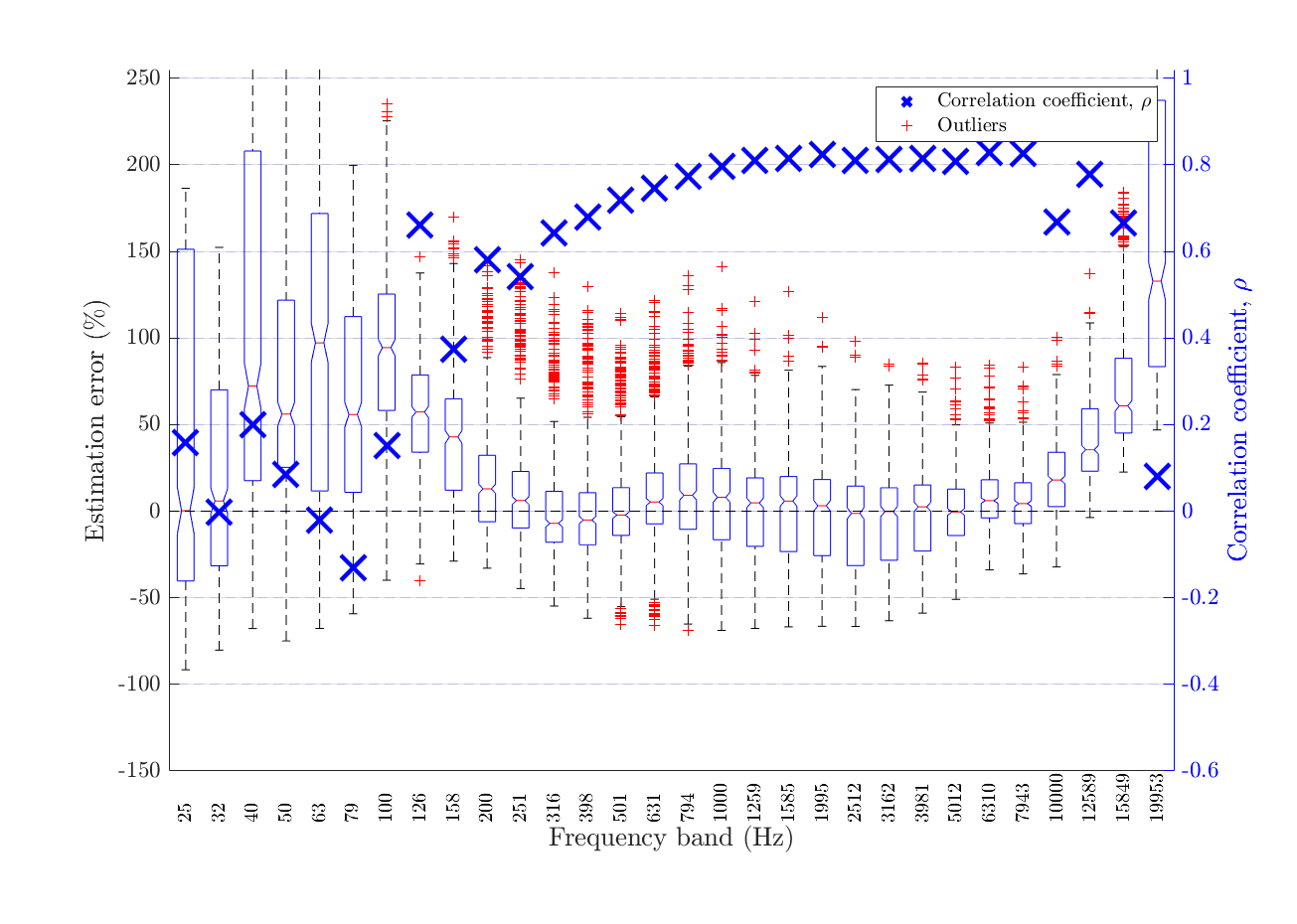,
	width=\figWidthACETR,viewport=45 10 765 530,clip}}%
	\else
	\centerline{\epsfig{figure=FigsACE/ana_eval_gt_partic_results_combined_Phase3_TR_P3S_T60_Perc_18dB_SNR_Babble_sub_Frequency-dependent-RTE.png,
	width=\figWidthACETR,viewport=45 10 765 530,clip}}%
	\fi
	\caption{{Frequency-dependent \ac{T60} estimation error in babble noise at \dBel{18} \ac{SNR} for algorithm Model-based \ac{SB} RTE~\cite{Lollmann2015}}}%
\label{fig:ACE_T60_Sub_Babble_18dB_SNR_Lollman}%
\end{figure}%
\begin{table*}[!htb]\small
\caption{Frequency-dependent \ac{T60} estimation error in babble noise at \dBel{18} \ac{SNR} for algorithm Model-based \ac{SB} RTE~\cite{Lollmann2015}}
\vspace{5mm} 
\centering
\begin{tabular}{crrrl}%
\hline%
Freq. band
& Centre Freq. (Hz)
& Bias
& \acs{MSE}
& $\PearsonCC$

\\
\hline%
\hline%
\ifarXiv
 1 & 25.12 & -0.4858 & 2.839 & 0.1578 \\ 
\hline
 2 & 31.62 & 0.0403 & 0.5332 & -0.003286 \\ 
\hline
 3 & 39.81 & 0.5213 & 0.6535 & 0.1986 \\ 
\hline
 4 & 50.12 & 0.5001 & 0.5587 & 0.0844 \\ 
\hline
 5 & 63.10 & 0.5351 & 0.579 & -0.02045 \\ 
\hline
 6 & 79.43 & 0.4082 & 0.332 & -0.1301 \\ 
\hline
 7 & 100.00 & 0.5059 & 0.321 & 0.1525 \\ 
\hline
 8 & 125.89 & 0.341 & 0.1427 & 0.6613 \\ 
\hline
 9 & 158.49 & 0.2442 & 0.1083 & 0.3732 \\ 
\hline
10 & 199.53 & 0.09864 & 0.05667 & 0.5816 \\ 
\hline
11 & 251.19 & 0.03031 & 0.0623 & 0.5417 \\ 
\hline
12 & 316.23 & -0.01143 & 0.05043 & 0.6434 \\ 
\hline
13 & 398.11 & -0.02912 & 0.05685 & 0.6797 \\ 
\hline
14 & 501.19 & -0.0381 & 0.07234 & 0.7176 \\ 
\hline
15 & 630.96 & -0.04182 & 0.09694 & 0.7467 \\ 
\hline
16 & 794.33 & -0.05557 & 0.1059 & 0.7736 \\ 
\hline
17 & 1000.00 & -0.07991 & 0.1163 & 0.7966 \\ 
\hline
18 & 1258.93 & -0.08241 & 0.09489 & 0.8098 \\ 
\hline
19 & 1584.89 & -0.07899 & 0.08569 & 0.8149 \\ 
\hline
20 & 1995.26 & -0.08874 & 0.08324 & 0.8227 \\ 
\hline
21 & 2511.89 & -0.1184 & 0.08826 & 0.809 \\ 
\hline
22 & 3162.28 & -0.09418 & 0.06279 & 0.8121 \\ 
\hline
23 & 3981.07 & -0.0595 & 0.04034 & 0.8135 \\ 
\hline
24 & 5011.87 & -0.026 & 0.01891 & 0.8081 \\ 
\hline
25 & 6309.57 & 0.03792 & 0.009949 & 0.8278 \\ 
\hline
26 & 7943.28 & 0.02939 & 0.009371 & 0.8265 \\ 
\hline
27 & 10000.00 & 0.08108 & 0.02127 & 0.6687 \\ 
\hline
28 & 12589.25 & 0.1617 & 0.0396 & 0.7775 \\ 
\hline
29 & 15848.93 & 0.2283 & 0.07146 & 0.6648 \\ 
\hline
30 & 19952.62 & 0.3233 & 0.1322 & 0.08033 \\ 
\hline
\else
\fi
\end{tabular}%
\end{table*}%
%
%
\clearpage
\subsubsection{Babble noise at \dBel{12}}
\begin{figure}[!ht]
	\ifarXiv
\centerline{\epsfig{figure=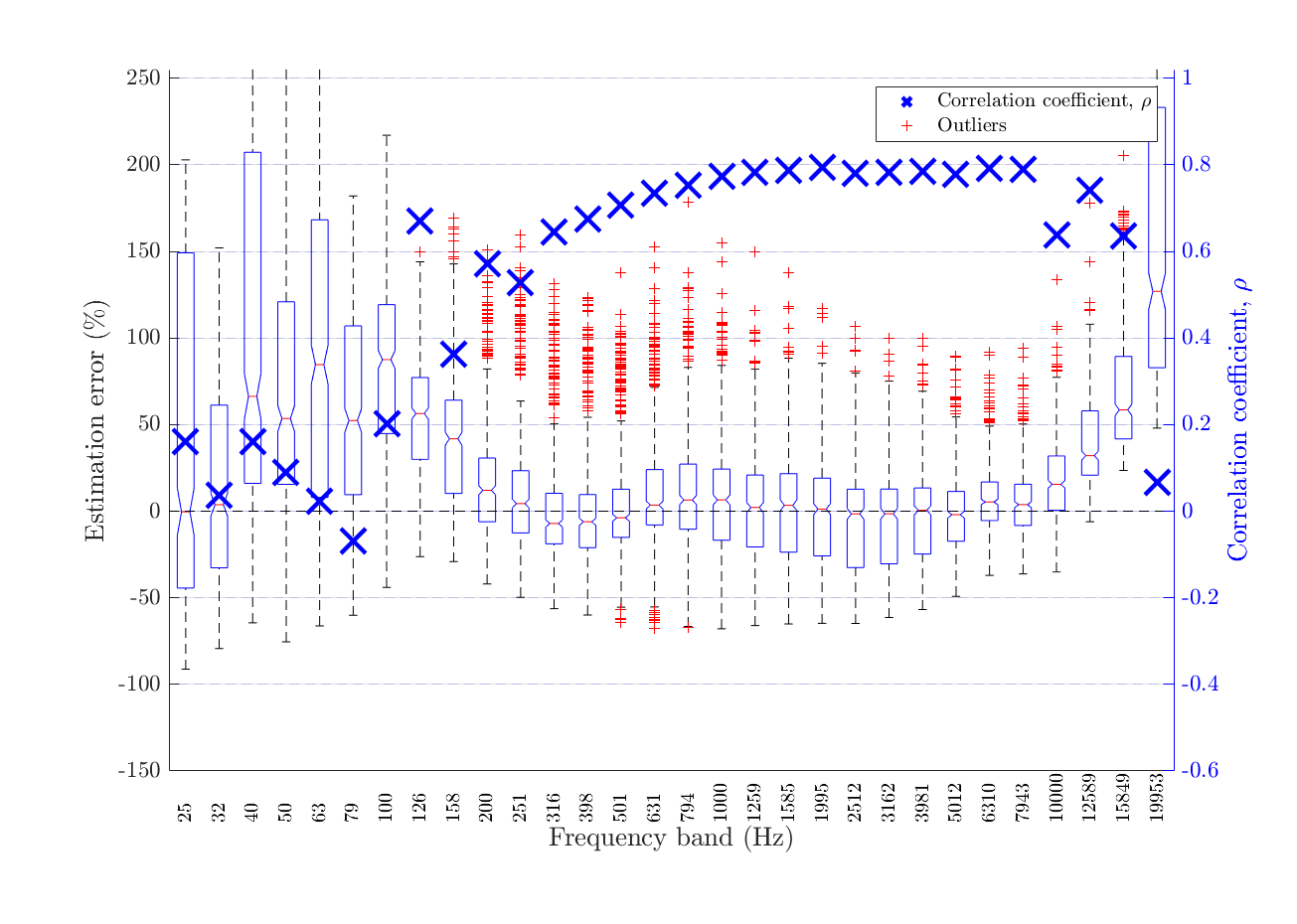, width=\figWidthACETR,viewport=45 10 765 530,clip}}%
	\else
	\centerline{\epsfig{figure=FigsACE/ana_eval_gt_partic_results_combined_Phase3_TR_P3S_T60_Perc_12dB_SNR_Babble_sub_Frequency-dependent-RTE.png, width=\figWidthACETR,viewport=45 10 765 530,clip}}%
	\fi
	\caption{{Frequency-dependent \ac{T60} estimation error in babble noise at \dBel{12} \ac{SNR} for algorithm Model-based \ac{SB} RTE~\cite{Lollmann2015}}}%
\label{fig:ACE_T60_Sub_Babble_12dB_SNR_Lollman}%
\end{figure}%
\begin{table*}[!htb]\small
\caption{Frequency-dependent \ac{T60} estimation error in babble noise at \dBel{12} \ac{SNR} for algorithm Model-based \ac{SB} RTE~\cite{Lollmann2015}}
\vspace{5mm} 
\centering
\begin{tabular}{crrrl}%
\hline%
Freq. band
& Centre Freq. (Hz)
& Bias
& \acs{MSE}
& $\PearsonCC$

\\
\hline%
\hline%
\ifarXiv
 1 & 25.12 & -0.5075 & 2.858 & 0.1617 \\ 
\hline
 2 & 31.62 & 0.008911 & 0.5158 & 0.03594 \\ 
\hline
 3 & 39.81 & 0.485 & 0.638 & 0.1617 \\ 
\hline
 4 & 50.12 & 0.464 & 0.5279 & 0.08925 \\ 
\hline
 5 & 63.10 & 0.5033 & 0.5386 & 0.02186 \\ 
\hline
 6 & 79.43 & 0.3832 & 0.3057 & -0.0682 \\ 
\hline
 7 & 100.00 & 0.4881 & 0.3005 & 0.2029 \\ 
\hline
 8 & 125.89 & 0.3295 & 0.1347 & 0.6701 \\ 
\hline
 9 & 158.49 & 0.2374 & 0.1067 & 0.3634 \\ 
\hline
10 & 199.53 & 0.09501 & 0.05799 & 0.571 \\ 
\hline
11 & 251.19 & 0.02834 & 0.06502 & 0.5271 \\ 
\hline
12 & 316.23 & -0.01278 & 0.05061 & 0.6459 \\ 
\hline
13 & 398.11 & -0.0305 & 0.05799 & 0.6737 \\ 
\hline
14 & 501.19 & -0.03988 & 0.07403 & 0.7077 \\ 
\hline
15 & 630.96 & -0.04417 & 0.0994 & 0.7336 \\ 
\hline
16 & 794.33 & -0.05855 & 0.1101 & 0.7527 \\ 
\hline
17 & 1000.00 & -0.08353 & 0.1217 & 0.7726 \\ 
\hline
18 & 1258.93 & -0.08663 & 0.1008 & 0.7823 \\ 
\hline
19 & 1584.89 & -0.08376 & 0.09157 & 0.7878 \\ 
\hline
20 & 1995.26 & -0.09401 & 0.08975 & 0.7933 \\ 
\hline
21 & 2511.89 & -0.1241 & 0.09515 & 0.78 \\ 
\hline
22 & 3162.28 & -0.1003 & 0.06887 & 0.7823 \\ 
\hline
23 & 3981.07 & -0.06593 & 0.04512 & 0.7845 \\ 
\hline
24 & 5011.87 & -0.03272 & 0.02183 & 0.7776 \\ 
\hline
25 & 6309.57 & 0.03094 & 0.01047 & 0.7916 \\ 
\hline
26 & 7943.28 & 0.0222 & 0.009941 & 0.79 \\ 
\hline
27 & 10000.00 & 0.07371 & 0.01981 & 0.6373 \\ 
\hline
28 & 12589.25 & 0.1542 & 0.03626 & 0.7413 \\ 
\hline
29 & 15848.93 & 0.2206 & 0.06609 & 0.6362 \\ 
\hline
30 & 19952.62 & 0.3155 & 0.1247 & 0.06677 \\ 
\hline
\else
\fi
\end{tabular}%
\end{table*}%
%
%
\clearpage
\subsubsection{Babble noise at \dBel{-1}}
\begin{figure}[!ht]
	\ifarXiv
\centerline{\epsfig{figure=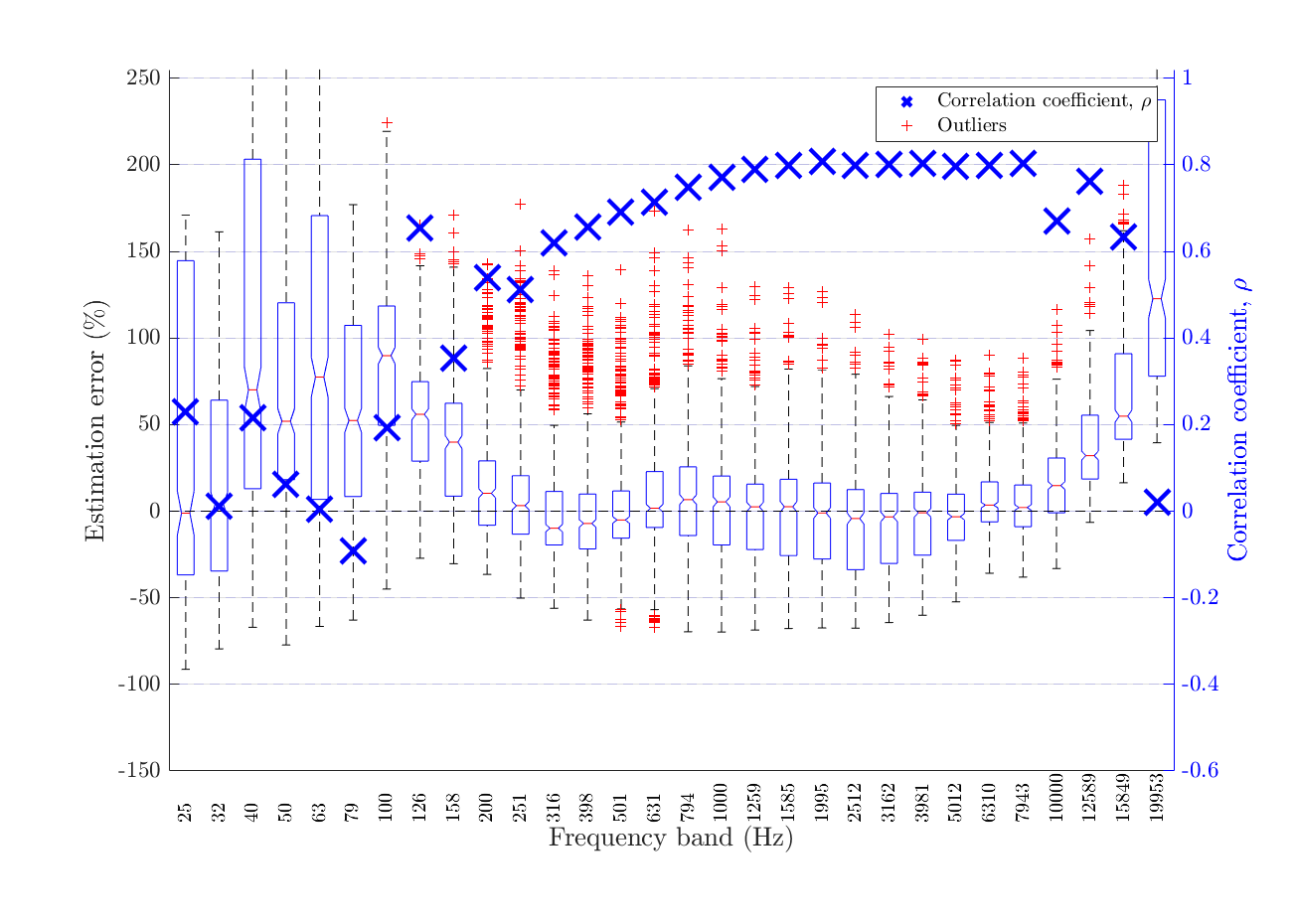, 	width=\figWidthACETR,viewport=45 10 765 530,clip}}%
	\else
	\centerline{\epsfig{figure=FigsACE/ana_eval_gt_partic_results_combined_Phase3_TR_P3S_T60_Perc_-1dB_SNR_Babble_sub_Frequency-dependent-RTE.png, 	width=\figWidthACETR,viewport=45 10 765 530,clip}}%
	\fi
	\caption{{Frequency-dependent \ac{T60} estimation error in babble noise at \dBel{-1} \ac{SNR} for algorithm Model-based \ac{SB} RTE~\cite{Lollmann2015}}}%
\label{fig:ACE_T60_Sub_Babble_-1dB_SNR_Lollman}%
\end{figure}%
\begin{table*}[!htb]\small
\caption{Frequency-dependent \ac{T60} estimation error in babble noise at \dBel{-1} \ac{SNR} for algorithm Model-based \ac{SB} RTE~\cite{Lollmann2015}}
\vspace{5mm} 
\centering
\begin{tabular}{crrrl}%
\hline%
Freq. band
& Centre Freq. (Hz)
& Bias
& \acs{MSE}
& $\PearsonCC$

\\
\hline%
\hline%
\ifarXiv
 1 & 25.12 & -0.5131 & 2.813 & 0.2285 \\ 
\hline
 2 & 31.62 & 0.0009175 & 0.5312 & 0.01062 \\ 
\hline
 3 & 39.81 & 0.4756 & 0.6085 & 0.2152 \\ 
\hline
 4 & 50.12 & 0.4544 & 0.5349 & 0.0614 \\ 
\hline
 5 & 63.10 & 0.4943 & 0.5416 & 0.003491 \\ 
\hline
 6 & 79.43 & 0.3754 & 0.31 & -0.09157 \\ 
\hline
 7 & 100.00 & 0.4817 & 0.2997 & 0.1922 \\ 
\hline
 8 & 125.89 & 0.3243 & 0.1349 & 0.6534 \\ 
\hline
 9 & 158.49 & 0.2332 & 0.1087 & 0.3537 \\ 
\hline
10 & 199.53 & 0.09138 & 0.0629 & 0.5397 \\ 
\hline
11 & 251.19 & 0.02506 & 0.0688 & 0.511 \\ 
\hline
12 & 316.23 & -0.01588 & 0.05542 & 0.6203 \\ 
\hline
13 & 398.11 & -0.03353 & 0.06141 & 0.6558 \\ 
\hline
14 & 501.19 & -0.04289 & 0.07666 & 0.6918 \\ 
\hline
15 & 630.96 & -0.04717 & 0.1023 & 0.7134 \\ 
\hline
16 & 794.33 & -0.06153 & 0.1098 & 0.7479 \\ 
\hline
17 & 1000.00 & -0.08647 & 0.1207 & 0.77 \\ 
\hline
18 & 1258.93 & -0.08953 & 0.09858 & 0.7887 \\ 
\hline
19 & 1584.89 & -0.08659 & 0.08876 & 0.7991 \\ 
\hline
20 & 1995.26 & -0.09676 & 0.08667 & 0.8074 \\ 
\hline
21 & 2511.89 & -0.1268 & 0.0918 & 0.7984 \\ 
\hline
22 & 3162.28 & -0.1028 & 0.06595 & 0.8014 \\ 
\hline
23 & 3981.07 & -0.06841 & 0.04263 & 0.8042 \\ 
\hline
24 & 5011.87 & -0.0351 & 0.02041 & 0.7964 \\ 
\hline
25 & 6309.57 & 0.02865 & 0.01057 & 0.7977 \\ 
\hline
26 & 7943.28 & 0.01999 & 0.009809 & 0.8025 \\ 
\hline
27 & 10000.00 & 0.07158 & 0.01931 & 0.6711 \\ 
\hline
28 & 12589.25 & 0.1521 & 0.03644 & 0.7608 \\ 
\hline
29 & 15848.93 & 0.2186 & 0.06684 & 0.6344 \\ 
\hline
30 & 19952.62 & 0.3136 & 0.1261 & 0.02066 \\ 
\hline
\else
\fi
\end{tabular}%
\end{table*}%
%
%
\clearpage
\subsubsection{Fan noise at \dBel{18}}
\begin{figure}[!ht]
	\ifarXiv
\centerline{\epsfig{figure=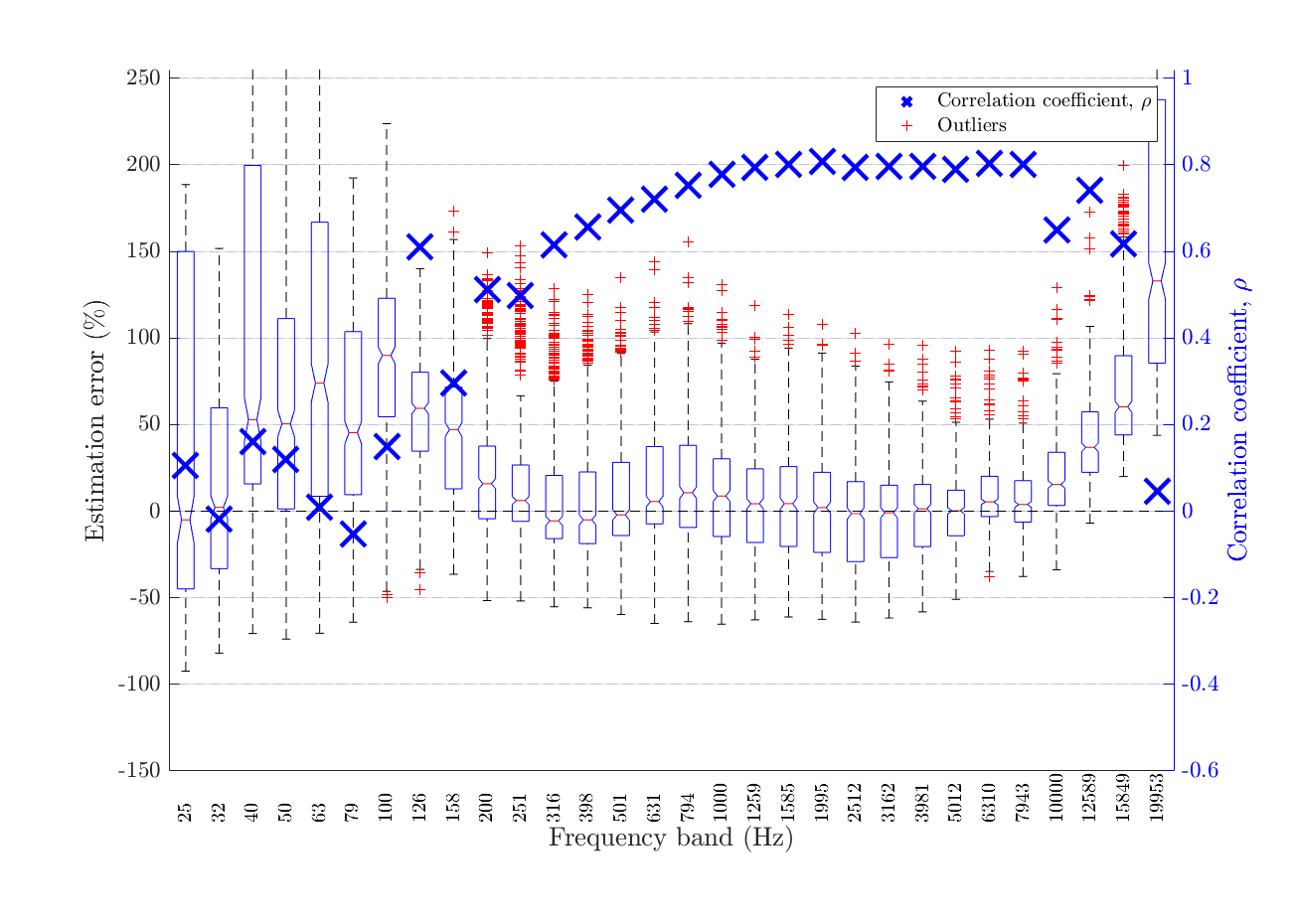,
	width=\figWidthACETR,viewport=45 10 765 530,clip}}%
	\else
	\centerline{\epsfig{figure=FigsACE/ana_eval_gt_partic_results_combined_Phase3_TR_P3S_T60_Perc_18dB_SNR_Fan_sub_Frequency-dependent-RTE.png,
	width=\figWidthACETR,viewport=45 10 765 530,clip}}%
	\fi
	\caption{{Frequency-dependent \ac{T60} estimation error in fan noise at \dBel{18} \ac{SNR} for algorithm Model-based \ac{SB} RTE~\cite{Lollmann2015}}}%
\label{fig:ACE_T60_Sub_Fan_18dB_SNR_Lollman}%
\end{figure}%
\begin{table*}[!htb]\small
\caption{Frequency-dependent \ac{T60} estimation error in fan noise at \dBel{18} \ac{SNR} for algorithm Model-based \ac{SB} RTE~\cite{Lollmann2015}}
\vspace{5mm} 
\centering
\begin{tabular}{crrrl}%
\hline%
Freq. band
& Centre Freq. (Hz)
& Bias
& \acs{MSE}
& $\PearsonCC$

\\
\hline%
\hline%
\ifarXiv
 1 & 25.12 & -0.5218 & 2.916 & 0.106 \\ 
\hline
 2 & 31.62 & -0.02202 & 0.5605 & -0.01754 \\ 
\hline
 3 & 39.81 & 0.4467 & 0.6222 & 0.16 \\ 
\hline
 4 & 50.12 & 0.4284 & 0.5049 & 0.1203 \\ 
\hline
 5 & 63.10 & 0.4782 & 0.5333 & 0.00933 \\ 
\hline
 6 & 79.43 & 0.3726 & 0.3054 & -0.05196 \\ 
\hline
 7 & 100.00 & 0.4924 & 0.3136 & 0.1494 \\ 
\hline
 8 & 125.89 & 0.3462 & 0.1522 & 0.6111 \\ 
\hline
 9 & 158.49 & 0.2626 & 0.1258 & 0.2967 \\ 
\hline
10 & 199.53 & 0.1248 & 0.07269 & 0.5116 \\ 
\hline
11 & 251.19 & 0.05942 & 0.07299 & 0.4974 \\ 
\hline
12 & 316.23 & 0.01739 & 0.05579 & 0.6146 \\ 
\hline
13 & 398.11 & -0.002582 & 0.05993 & 0.6565 \\ 
\hline
14 & 501.19 & -0.01483 & 0.07473 & 0.695 \\ 
\hline
15 & 630.96 & -0.02213 & 0.1003 & 0.7195 \\ 
\hline
16 & 794.33 & -0.03937 & 0.1077 & 0.7533 \\ 
\hline
17 & 1000.00 & -0.06689 & 0.1177 & 0.7774 \\ 
\hline
18 & 1258.93 & -0.07216 & 0.09601 & 0.7929 \\ 
\hline
19 & 1584.89 & -0.07106 & 0.08678 & 0.7998 \\ 
\hline
20 & 1995.26 & -0.0827 & 0.08452 & 0.8068 \\ 
\hline
21 & 2511.89 & -0.1139 & 0.08919 & 0.7947 \\ 
\hline
22 & 3162.28 & -0.09082 & 0.06407 & 0.796 \\ 
\hline
23 & 3981.07 & -0.05704 & 0.04181 & 0.796 \\ 
\hline
24 & 5011.87 & -0.0242 & 0.02017 & 0.79 \\ 
\hline
25 & 6309.57 & 0.03922 & 0.01118 & 0.8042 \\ 
\hline
26 & 7943.28 & 0.03034 & 0.01064 & 0.801 \\ 
\hline
27 & 10000.00 & 0.08177 & 0.02217 & 0.6486 \\ 
\hline
28 & 12589.25 & 0.1622 & 0.04067 & 0.7418 \\ 
\hline
29 & 15848.93 & 0.2287 & 0.07238 & 0.6166 \\ 
\hline
30 & 19952.62 & 0.3236 & 0.133 & 0.04615 \\ 
\hline
\else
\fi
\end{tabular}%
\end{table*}%
%
%
\clearpage
\subsubsection{Fan noise at \dBel{12}}
\begin{figure}[!ht]
	\ifarXiv
\centerline{\epsfig{figure=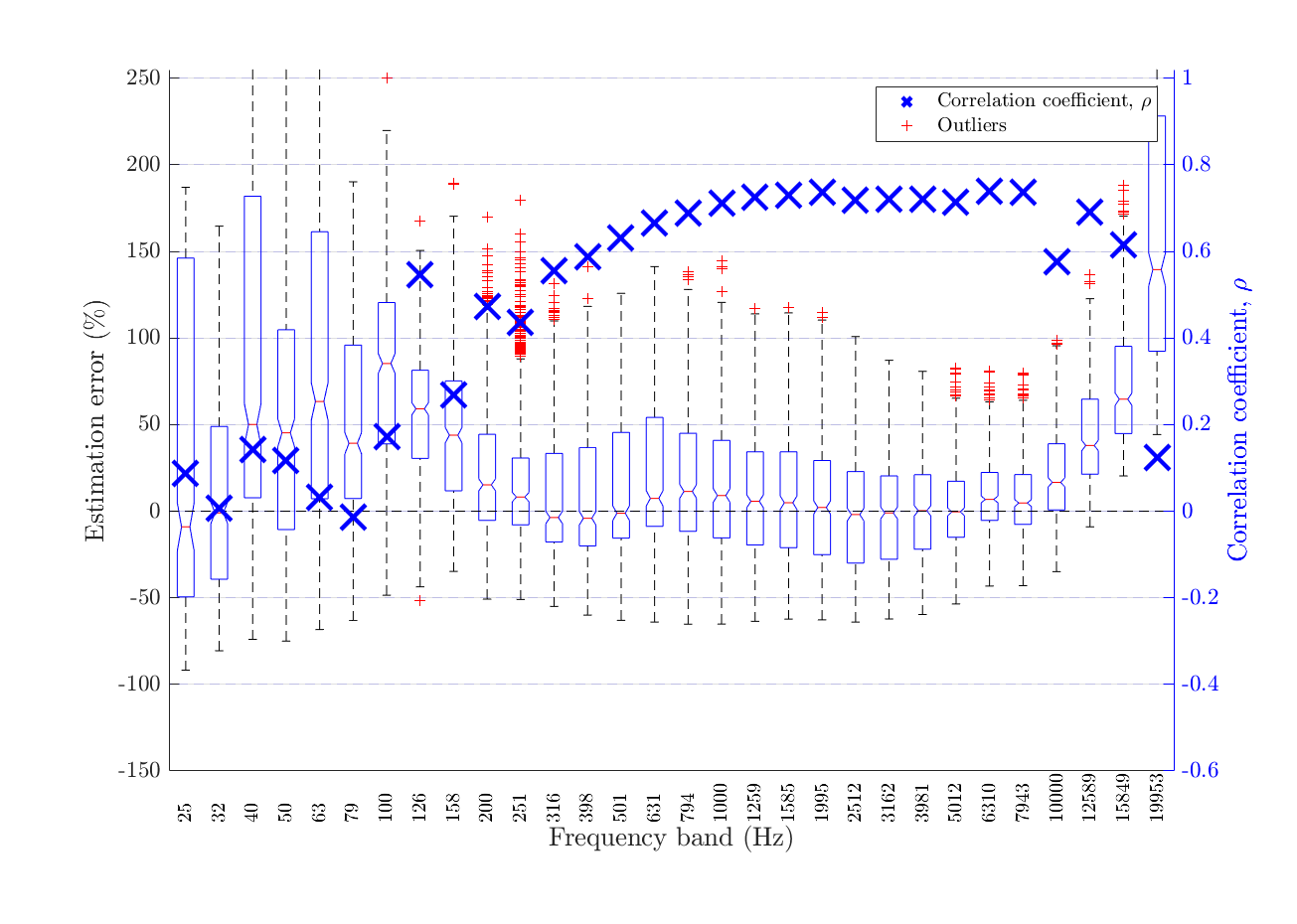, width=\figWidthACETR,viewport=45 10 765 530,clip}}%
	\else
	\centerline{\epsfig{figure=FigsACE/ana_eval_gt_partic_results_combined_Phase3_TR_P3S_T60_Perc_12dB_SNR_Fan_sub_Frequency-dependent-RTE.png, width=\figWidthACETR,viewport=45 10 765 530,clip}}%
	\fi
	\caption{{Frequency-dependent \ac{T60} estimation error in fan noise at \dBel{12} \ac{SNR} for algorithm Model-based \ac{SB} RTE~\cite{Lollmann2015}}}%
\label{fig:ACE_T60_Sub_Fan_12dB_SNR_Lollman}%
\end{figure}%
\begin{table*}[!htb]\small
\caption{Frequency-dependent \ac{T60} estimation error in fan noise at \dBel{12} \ac{SNR} for algorithm Model-based \ac{SB} RTE~\cite{Lollmann2015}}
\vspace{5mm} 
\centering
\begin{tabular}{crrrl}%
\hline%
Freq. band
& Centre Freq. (Hz)
& Bias
& \acs{MSE}
& $\PearsonCC$

\\
\hline%
\hline%
\ifarXiv
 1 & 25.12 & -0.5521 & 2.964 & 0.0869 \\ 
\hline
 2 & 31.62 & -0.07531 & 0.5598 & 0.006393 \\ 
\hline
 3 & 39.81 & 0.3818 & 0.5859 & 0.143 \\ 
\hline
 4 & 50.12 & 0.3642 & 0.4662 & 0.1176 \\ 
\hline
 5 & 63.10 & 0.4246 & 0.4877 & 0.03276 \\ 
\hline
 6 & 79.43 & 0.3351 & 0.2795 & -0.0127 \\ 
\hline
 7 & 100.00 & 0.4722 & 0.2956 & 0.1716 \\ 
\hline
 8 & 125.89 & 0.341 & 0.1547 & 0.5455 \\ 
\hline
 9 & 158.49 & 0.2686 & 0.1315 & 0.2693 \\ 
\hline
10 & 199.53 & 0.1377 & 0.08082 & 0.4728 \\ 
\hline
11 & 251.19 & 0.07589 & 0.08383 & 0.4362 \\ 
\hline
12 & 316.23 & 0.03487 & 0.06505 & 0.5561 \\ 
\hline
13 & 398.11 & 0.01437 & 0.07111 & 0.5864 \\ 
\hline
14 & 501.19 & 0.0007776 & 0.08572 & 0.6304 \\ 
\hline
15 & 630.96 & -0.008209 & 0.1101 & 0.666 \\ 
\hline
16 & 794.33 & -0.02718 & 0.1191 & 0.6892 \\ 
\hline
17 & 1000.00 & -0.05634 & 0.1287 & 0.7125 \\ 
\hline
18 & 1258.93 & -0.06308 & 0.1062 & 0.7247 \\ 
\hline
19 & 1584.89 & -0.06325 & 0.09676 & 0.7294 \\ 
\hline
20 & 1995.26 & -0.07597 & 0.09419 & 0.7359 \\ 
\hline
21 & 2511.89 & -0.1081 & 0.09894 & 0.7186 \\ 
\hline
22 & 3162.28 & -0.08569 & 0.07265 & 0.7208 \\ 
\hline
23 & 3981.07 & -0.05249 & 0.04929 & 0.7209 \\ 
\hline
24 & 5011.87 & -0.02011 & 0.02566 & 0.7138 \\ 
\hline
25 & 6309.57 & 0.04296 & 0.01406 & 0.7384 \\ 
\hline
26 & 7943.28 & 0.03379 & 0.01331 & 0.7363 \\ 
\hline
27 & 10000.00 & 0.08501 & 0.02439 & 0.5759 \\ 
\hline
28 & 12589.25 & 0.1653 & 0.04171 & 0.6908 \\ 
\hline
29 & 15848.93 & 0.2316 & 0.07233 & 0.6144 \\ 
\hline
30 & 19952.62 & 0.3264 & 0.132 & 0.124 \\ 
\hline
\else
\fi
\end{tabular}%
\end{table*}%
%
%
\clearpage
\subsubsection{Fan noise at \dBel{-1}}
\begin{figure}[!ht]
	\ifarXiv
\centerline{\epsfig{figure=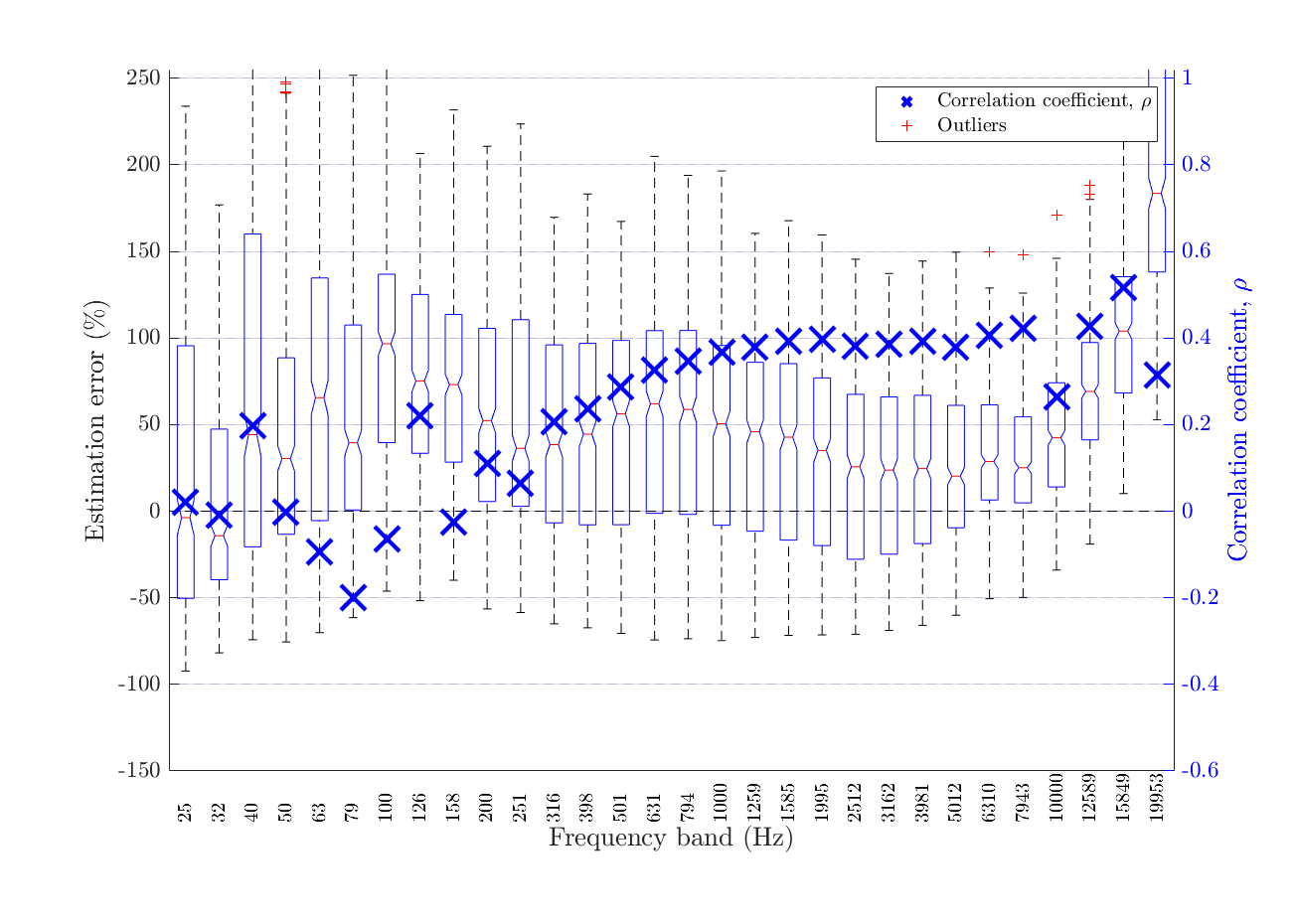, 	width=\figWidthACETR,viewport=45 10 765 530,clip}}%
	\else
	\centerline{\epsfig{figure=FigsACE/ana_eval_gt_partic_results_combined_Phase3_TR_P3S_T60_Perc_-1dB_SNR_Fan_sub_Frequency-dependent-RTE.png, 	width=\figWidthACETR,viewport=45 10 765 530,clip}}%
	\fi
	\caption{{Frequency-dependent \ac{T60} estimation error in fan noise at \dBel{-1} \ac{SNR} for algorithm Model-based \ac{SB} RTE~\cite{Lollmann2015}}}%
\label{fig:ACE_T60_Sub_Fan_-1dB_SNR_Lollman}%
\end{figure}%
\begin{table*}[!htb]\small
\caption{Frequency-dependent \ac{T60} estimation error in fan noise at \dBel{-1} \ac{SNR} for algorithm Model-based \ac{SB} RTE~\cite{Lollmann2015}}
\vspace{5mm} 
\centering
\begin{tabular}{crrrl}%
\hline%
Freq. band
& Centre Freq. (Hz)
& Bias
& \acs{MSE}
& $\PearsonCC$

\\
\hline%
\hline%
\ifarXiv
 1 & 25.12 & -0.5619 & 3.04 & 0.01991 \\ 
\hline
 2 & 31.62 & -0.1555 & 0.6021 & -0.008327 \\ 
\hline
 3 & 39.81 & 0.2686 & 0.4974 & 0.1984 \\ 
\hline
 4 & 50.12 & 0.2583 & 0.4648 & -0.002017 \\ 
\hline
 5 & 63.10 & 0.3581 & 0.5013 & -0.09391 \\ 
\hline
 6 & 79.43 & 0.3256 & 0.3326 & -0.2 \\ 
\hline
 7 & 100.00 & 0.5222 & 0.3869 & -0.06485 \\ 
\hline
 8 & 125.89 & 0.4418 & 0.2714 & 0.221 \\ 
\hline
 9 & 158.49 & 0.4061 & 0.2503 & -0.02438 \\ 
\hline
10 & 199.53 & 0.2971 & 0.193 & 0.1108 \\ 
\hline
11 & 251.19 & 0.2447 & 0.1878 & 0.06494 \\ 
\hline
12 & 316.23 & 0.2044 & 0.1526 & 0.2076 \\ 
\hline
13 & 398.11 & 0.1791 & 0.155 & 0.2356 \\ 
\hline
14 & 501.19 & 0.1579 & 0.167 & 0.2878 \\ 
\hline
15 & 630.96 & 0.1401 & 0.19 & 0.3268 \\ 
\hline
16 & 794.33 & 0.1122 & 0.1905 & 0.3465 \\ 
\hline
17 & 1000.00 & 0.07473 & 0.1897 & 0.3665 \\ 
\hline
18 & 1258.93 & 0.06045 & 0.1571 & 0.3794 \\ 
\hline
19 & 1584.89 & 0.05367 & 0.1418 & 0.3925 \\ 
\hline
20 & 1995.26 & 0.03527 & 0.134 & 0.398 \\ 
\hline
21 & 2511.89 & -0.001626 & 0.1299 & 0.3801 \\ 
\hline
22 & 3162.28 & 0.01673 & 0.1023 & 0.3858 \\ 
\hline
23 & 3981.07 & 0.04662 & 0.07934 & 0.3927 \\ 
\hline
24 & 5011.87 & 0.07628 & 0.05372 & 0.3796 \\ 
\hline
25 & 6309.57 & 0.1371 & 0.04437 & 0.406 \\ 
\hline
26 & 7943.28 & 0.1261 & 0.04046 & 0.4213 \\ 
\hline
27 & 10000.00 & 0.1758 & 0.05674 & 0.265 \\ 
\hline
28 & 12589.25 & 0.2549 & 0.08301 & 0.4278 \\ 
\hline
29 & 15848.93 & 0.3202 & 0.1201 & 0.5153 \\ 
\hline
30 & 19952.62 & 0.4142 & 0.1916 & 0.3153 \\ 
\hline
\else
\fi
\end{tabular}%
\end{table*}%
%
%
%
\clearpage
\section{\ac{DRR} estimation results}
\subsection{Fullband \ac{DRR} estimation results by noise type}
\subsubsection{Ambient noise}
\begin{figure}[!ht]
	\ifarXiv
\centerline{\epsfig{figure=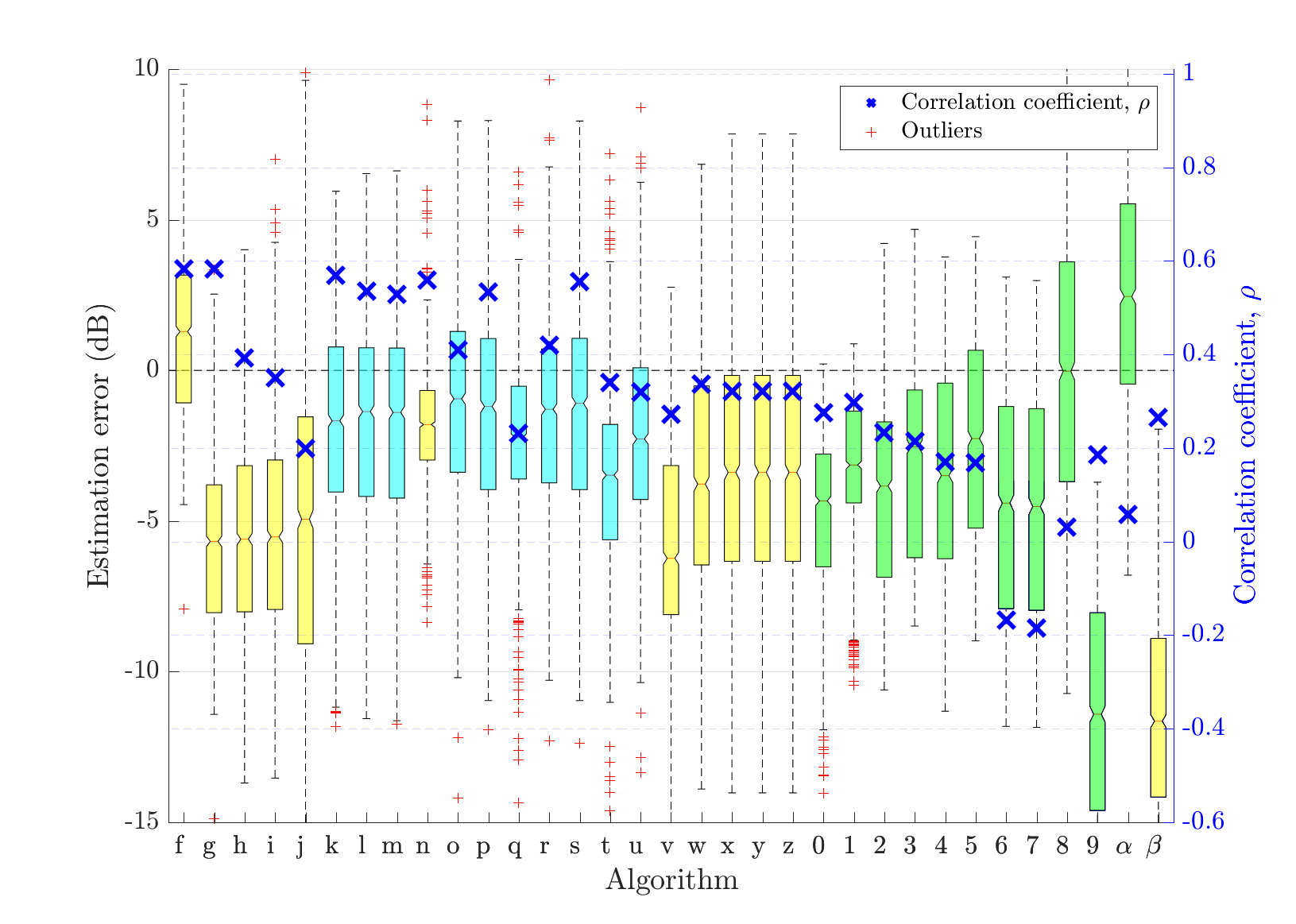,
	width=\figWidthACETR,viewport=45 10 765 530,clip}}%
	\else
	\centerline{\epsfig{figure=FigsACE/ana_eval_gt_partic_results_combined_Phase3_All_WASPAA_P3_DRR_dB_A2_Ambient.png,
	width=\figWidthACETR,viewport=45 10 765 530,clip}}%
	\fi
	\caption{Fullband {\ac{DRR} estimation error in ambient noise for all \acp{SNR}}}%
\label{fig:ACE_DRR_Ambient}%
\end{figure}%
%
\acused{PSD}%
\begin{table*}[!ht]\small
\caption{\ac{DRR} estimation algorithm performance in ambient noise for all \acp{SNR}}
\vspace{5mm} 
\centering
\begin{tabular}{cllllllll}%
\hline%
Ref.
& Algorithm
& Class
& Mic. Config.
& Bias
& \acs{MSE}
& $\PearsonCC$
& \ac{RTF}
\\
\hline
\hline
\ifarXiv
f & PSD est. in beamspace, bias comp.~\cite{Hioka2015} & \ac{ABC} & Mobile & 1.21 & 8.32 & 0.583 & 0.757\\ 
\hline
g & PSD est. in beamspace (Raw)~\cite{Hioka2015} & \ac{ABC} & Mobile & -5.76 & 40 & 0.583 & 3.15\\ 
\hline
h & PSD est. in beamspace v2~\cite{Hioka2015} & \ac{ABC} & Mobile & -5.46 & 40.1 & 0.393 & 0.844\\ 
\hline
i & PSD est. by twin BF~\cite{Hioka2012} & \ac{ABC} & Mobile & -5.34 & 39.9 & 0.351 & 0.614\\ 
\hline
j & Spatial Covariance in matrix mode~\cite{Hioka2011} & \ac{ABC} & Mobile & -5.17 & 65 & 0.2 & 0.627\\ 
\hline
k & NIRAv2~\cite{Parada2015} & \ac{MLMF} & Single & -1.68 & 14 & 0.568 & 0.897$^\dagger$\\ 
\hline
l & NIRAv3~\cite{Parada2015} & \ac{MLMF} & Single & -1.8 & 15 & 0.536 & 0.897$^\dagger$\\ 
\hline
m & NIRAv1~\cite{Parada2015} & \ac{MLMF} & Single & -1.84 & 15.3 & 0.528 & 0.897$^\dagger$\\ 
\hline
n & Particle velocity~\cite{Chen2015} & \ac{ABC} & EM32 & -1.85 & 6.62 & 0.559 & 0.134\\ 
\hline
o & Multi-layer perceptron~\cite{Xiong2015} & \ac{MLMF} & Single & -1.12 & 16.2 & 0.409 & 0.0578$^\ddagger$\\ 
\hline
p & Multi-layer perceptron P2~\cite{Xiong2015} & \ac{MLMF} & Single & -1.42 & 15.4 & 0.533 & 0.0578$^\ddagger$\\ 
\hline
q & Multi-layer perceptron P2~\cite{Xiong2015} & \ac{MLMF} & Chromebook & -2.58 & 14.6 & 0.231 & 0.0589$^\ddagger$\\ 
\hline
r & Multi-layer perceptron P2~\cite{Xiong2015} & \ac{MLMF} & Mobile & -1.36 & 13.6 & 0.42 & 0.0557$^\ddagger$\\ 
\hline
s & Multi-layer perceptron P2~\cite{Xiong2015} & \ac{MLMF} & Crucif & -1.33 & 14.6 & 0.555 & 0.0569$^\ddagger$\\ 
\hline
t & Multi-layer perceptron P2~\cite{Xiong2015} & \ac{MLMF} & Lin8Ch & -3.57 & 24.7 & 0.34 & 0.062$^\ddagger$\\ 
\hline
u & Multi-layer perceptron P2~\cite{Xiong2015} & \ac{MLMF} & EM32 & -2.02 & 13.7 & 0.32 & 0.0578$^\ddagger$\\ 
\hline
v & \ac{DENBE} no noise reduction~\cite{Eaton2015} & \ac{ABC} & Chromebook & -5.78 & 46.8 & 0.272 & 0.0323\\ 
\hline
w & \ac{DENBE} spectral subtraction~\cite{Eaton2015c} & \ac{ABC} & Chromebook & -3.53 & 25.5 & 0.337 & 0.0602\\ 
\hline
x & \ac{DENBE} spec. sub. Gerkmann~\cite{Eaton2015} & \ac{ABC} & Chromebook & -3.24 & 24.5 & 0.321 & 0.0474\\ 
\hline
y & \ac{DENBE} filtered subbands~\cite{Eaton2015c} & \ac{ABC} & Chromebook & -3.24 & 24.5 & 0.321 & 0.775\\ 
\hline
z & \ac{DENBE} FFT derived subbands~\cite{Eaton2015c} & \ac{ABC} & Chromebook & -3.24 & 24.5 & 0.321 & 0.0449\\ 
\hline
0 & \ac{NOSRMR} {\sectMidSent} 2.2.~\cite{Senoussaoui2015} & \ac{SFM} & Chromebook & -4.75 & 29.5 & 0.276 & 1.04\\ 
\hline
1 & \ac{OSRMR} {\sectMidSent} 2.2.~\cite{Senoussaoui2015} & \ac{SFM} & Chromebook & -3.23 & 15.6 & 0.298 & 0.831\\ 
\hline
2 & \ac{NOSRMR} {\sectMidSent} 2.2.~\cite{Senoussaoui2015} & \ac{SFM} & Mobile & -3.96 & 25.9 & 0.233 & 1.59\\ 
\hline
3 & \ac{OSRMR} {\sectMidSent} 2.2.~\cite{Senoussaoui2015} & \ac{SFM} & Mobile & -2.78 & 17.7 & 0.215 & 1.26\\ 
\hline
4 & \ac{NOSRMR} {\sectMidSent} 2.2.~\cite{Senoussaoui2015} & \ac{SFM} & Crucif & -3.55 & 25.3 & 0.171 & 2.63\\ 
\hline
5 & \ac{OSRMR} {\sectMidSent} 2.2.~\cite{Senoussaoui2015} & \ac{SFM} & Crucif & -2.39 & 18 & 0.169 & 2.09\\ 
\hline
6 & \ac{NOSRMR} {\sectMidSent} 2.2.~\cite{Senoussaoui2015} & \ac{SFM} & Single & -4.19 & 34.2 & -0.168 & 0.543\\ 
\hline
7 & \ac{OSRMR} {\sectMidSent} 2.2.~\cite{Senoussaoui2015} & \ac{SFM} & Single & -4.28 & 34.9 & -0.185 & 0.446\\ 
\hline
8 & Per acoust. band SRMR {\sectMidSent} 2.5.~\cite{Senoussaoui2015} & \ac{SFM} & Single & -0.0744 & 22.1 & 0.0317 & 0.58\\ 
\hline
9 & Temporal dynamics~\cite{Falk2009} & \ac{SFM} & Single & -11.2 & 142 & 0.185 & 0.0819\\ 
\hline
$\alpha$ & QA Reverb~\cite{Prego2015} & \ac{SFM} & Single & 2.41 & 23 & 0.0583 & 0.391\\ 
\hline
$\beta$ & Blind est. of coherent-to-diffuse energy ratio~\cite{Jeub2011} & \ac{ABC} & Chromebook & -11.4 & 146 & 0.266 & 0.019\\ 
\hline

\else

\fi
\end{tabular}
\end{table*}
\clearpage
%
\acused{RTF}%
\acused{NSRMR}%
\acused{SRMR}%
\acused{SDDSA-G}%
\clearpage
\subsubsection{Babble noise}
\begin{figure}[!ht]
	\ifarXiv
\centerline{\epsfig{figure=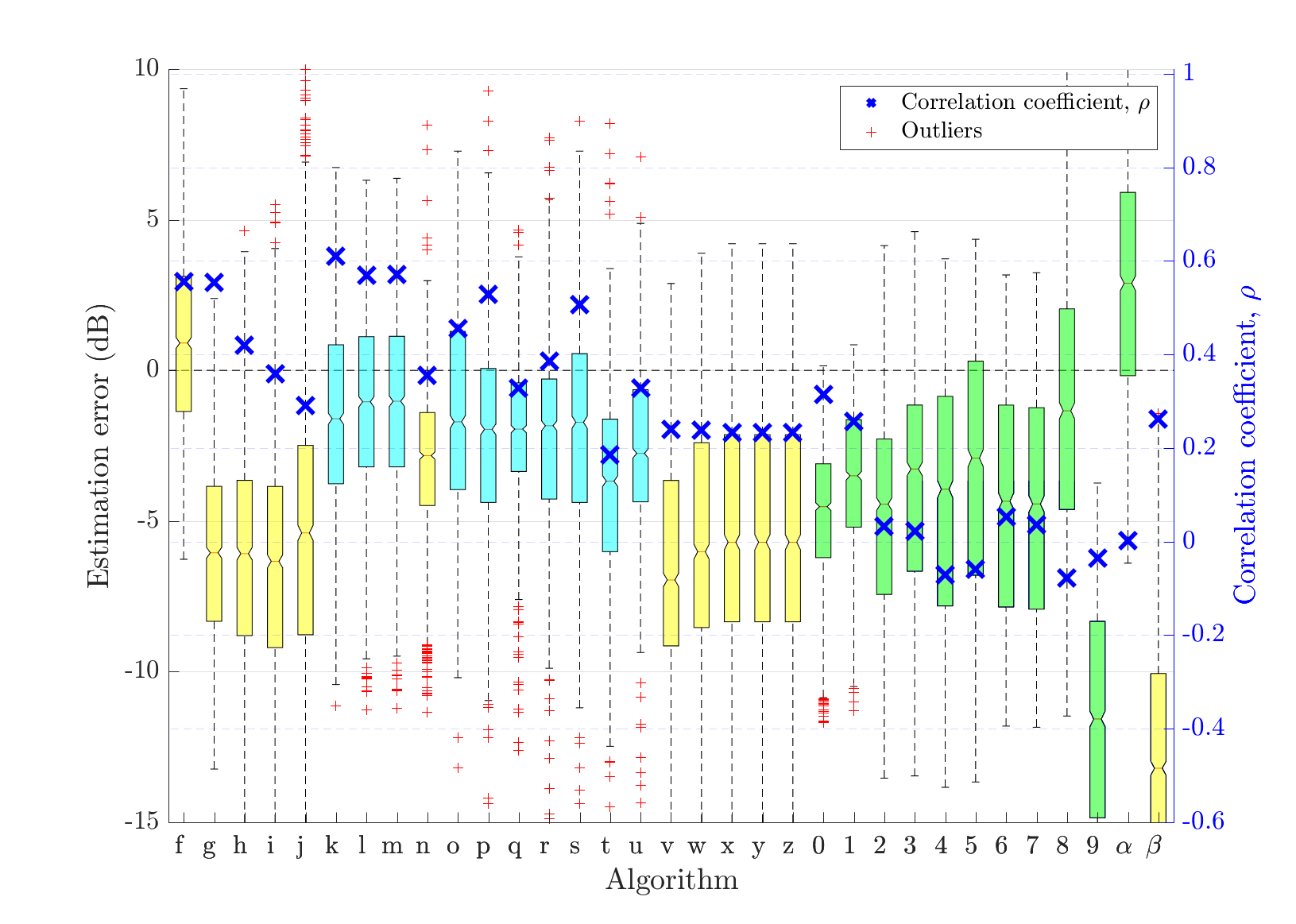,
	width=\figWidthACETR,viewport=45 10 765 530,clip}}%
	\else
	\centerline{\epsfig{figure=FigsACE/ana_eval_gt_partic_results_combined_Phase3_All_WASPAA_P3_DRR_dB_A2_Babble.png,
	width=\figWidthACETR,viewport=45 10 765 530,clip}}%
	\fi
	\caption{Fullband {\ac{DRR} estimation error in babble noise for all \acp{SNR}}}%
\label{fig:ACE_DRR_Babble}%
\end{figure}%
%
\acused{PSD}%
\begin{table*}[!ht]\small
\caption{\ac{DRR} estimation algorithm performance in babble noise for all \acp{SNR}}
\vspace{5mm} 
\centering
\begin{tabular}{cllllllll}%
\hline%
Ref.
& Algorithm
& Class
& Mic. Config.
& Bias
& \acs{MSE}
& $\PearsonCC$
& \ac{RTF}
\\
\hline
\hline
\ifarXiv
f & PSD est. in beamspace, bias comp.~\cite{Hioka2015} & \ac{ABC} & Mobile & 0.839 & 8.2 & 0.555 & 0.757\\ 
\hline
g & PSD est. in beamspace (Raw)~\cite{Hioka2015} & \ac{ABC} & Mobile & -6.13 & 45 & 0.555 & 3.17\\ 
\hline
h & PSD est. in beamspace v2~\cite{Hioka2015} & \ac{ABC} & Mobile & -6.1 & 48.4 & 0.42 & 0.843\\ 
\hline
i & PSD est. by twin BF~\cite{Hioka2012} & \ac{ABC} & Mobile & -6.38 & 54.6 & 0.358 & 0.615\\ 
\hline
j & Spatial Covariance in matrix mode~\cite{Hioka2011} & \ac{ABC} & Mobile & -5.6 & 57 & 0.29 & 0.627\\ 
\hline
k & NIRAv2~\cite{Parada2015} & \ac{MLMF} & Single & -1.66 & 13.2 & 0.61 & 0.906$^\dagger$\\ 
\hline
l & NIRAv3~\cite{Parada2015} & \ac{MLMF} & Single & -1.17 & 12.7 & 0.57 & 0.906$^\dagger$\\ 
\hline
m & NIRAv1~\cite{Parada2015} & \ac{MLMF} & Single & -1.14 & 12.6 & 0.571 & 0.906$^\dagger$\\ 
\hline
n & Particle velocity~\cite{Chen2015} & \ac{ABC} & EM32 & -3.13 & 16.2 & 0.356 & 0.134\\ 
\hline
o & Multi-layer perceptron~\cite{Xiong2015} & \ac{MLMF} & Single & -1.53 & 15.7 & 0.455 & 0.0579$^\ddagger$\\ 
\hline
p & Multi-layer perceptron P2~\cite{Xiong2015} & \ac{MLMF} & Single & -1.95 & 17 & 0.528 & 0.0579$^\ddagger$\\ 
\hline
q & Multi-layer perceptron P2~\cite{Xiong2015} & \ac{MLMF} & Chromebook & -2.31 & 13 & 0.328 & 0.0588$^\ddagger$\\ 
\hline
r & Multi-layer perceptron P2~\cite{Xiong2015} & \ac{MLMF} & Mobile & -2.25 & 17.3 & 0.386 & 0.0555$^\ddagger$\\ 
\hline
s & Multi-layer perceptron P2~\cite{Xiong2015} & \ac{MLMF} & Crucif & -1.93 & 17.2 & 0.506 & 0.057$^\ddagger$\\ 
\hline
t & Multi-layer perceptron P2~\cite{Xiong2015} & \ac{MLMF} & Lin8Ch & -3.75 & 28.4 & 0.185 & 0.0618$^\ddagger$\\ 
\hline
u & Multi-layer perceptron P2~\cite{Xiong2015} & \ac{MLMF} & EM32 & -2.6 & 16.4 & 0.329 & 0.0576$^\ddagger$\\ 
\hline
v & \ac{DENBE} no noise reduction~\cite{Eaton2015} & \ac{ABC} & Chromebook & -6.59 & 59.3 & 0.24 & 0.0323\\ 
\hline
w & \ac{DENBE} spectral subtraction~\cite{Eaton2015c} & \ac{ABC} & Chromebook & -5.74 & 50 & 0.237 & 0.0577\\ 
\hline
x & \ac{DENBE} spec. sub. Gerkmann~\cite{Eaton2015} & \ac{ABC} & Chromebook & -5.5 & 47.6 & 0.232 & 0.0476\\ 
\hline
y & \ac{DENBE} filtered subbands~\cite{Eaton2015c} & \ac{ABC} & Chromebook & -5.5 & 47.6 & 0.232 & 0.778\\ 
\hline
z & \ac{DENBE} FFT derived subbands~\cite{Eaton2015c} & \ac{ABC} & Chromebook & -5.5 & 47.6 & 0.232 & 0.0448\\ 
\hline
0 & \ac{NOSRMR} {\sectMidSent} 2.2.~\cite{Senoussaoui2015} & \ac{SFM} & Chromebook & -4.72 & 27.7 & 0.315 & 1.04\\ 
\hline
1 & \ac{OSRMR} {\sectMidSent} 2.2.~\cite{Senoussaoui2015} & \ac{SFM} & Chromebook & -3.68 & 19.5 & 0.257 & 0.833\\ 
\hline
2 & \ac{NOSRMR} {\sectMidSent} 2.2.~\cite{Senoussaoui2015} & \ac{SFM} & Mobile & -4.71 & 35.3 & 0.0325 & 1.58\\ 
\hline
3 & \ac{OSRMR} {\sectMidSent} 2.2.~\cite{Senoussaoui2015} & \ac{SFM} & Mobile & -3.73 & 27.2 & 0.0231 & 1.26\\ 
\hline
4 & \ac{NOSRMR} {\sectMidSent} 2.2.~\cite{Senoussaoui2015} & \ac{SFM} & Crucif & -4.29 & 34.6 & -0.0707 & 2.63\\ 
\hline
5 & \ac{OSRMR} {\sectMidSent} 2.2.~\cite{Senoussaoui2015} & \ac{SFM} & Crucif & -3.31 & 27.1 & -0.0591 & 2.1\\ 
\hline
6 & \ac{NOSRMR} {\sectMidSent} 2.2.~\cite{Senoussaoui2015} & \ac{SFM} & Single & -4.14 & 33.6 & 0.0538 & 0.534\\ 
\hline
7 & \ac{OSRMR} {\sectMidSent} 2.2.~\cite{Senoussaoui2015} & \ac{SFM} & Single & -4.21 & 34.2 & 0.0352 & 0.444\\ 
\hline
8 & Per acoust. band SRMR {\sectMidSent} 2.5.~\cite{Senoussaoui2015} & \ac{SFM} & Single & -1.3 & 22.5 & -0.0786 & 0.579\\ 
\hline
9 & Temporal dynamics~\cite{Falk2009} & \ac{SFM} & Single & -11.6 & 152 & -0.0352 & 0.0823\\ 
\hline
$\alpha$ & QA Reverb~\cite{Prego2015} & \ac{SFM} & Single & 2.79 & 25.5 & 0.00216 & 0.392\\ 
\hline
$\beta$ & Blind est. of coherent-to-diffuse energy ratio~\cite{Jeub2011} & \ac{ABC} & Chromebook & -12.8 & 179 & 0.261 & 0.019\\ 
\hline

\else

\fi
\end{tabular}
\end{table*}
\clearpage
\subsubsection{Fan noise}
\begin{figure}[!ht]
	\ifarXiv
\centerline{\epsfig{figure=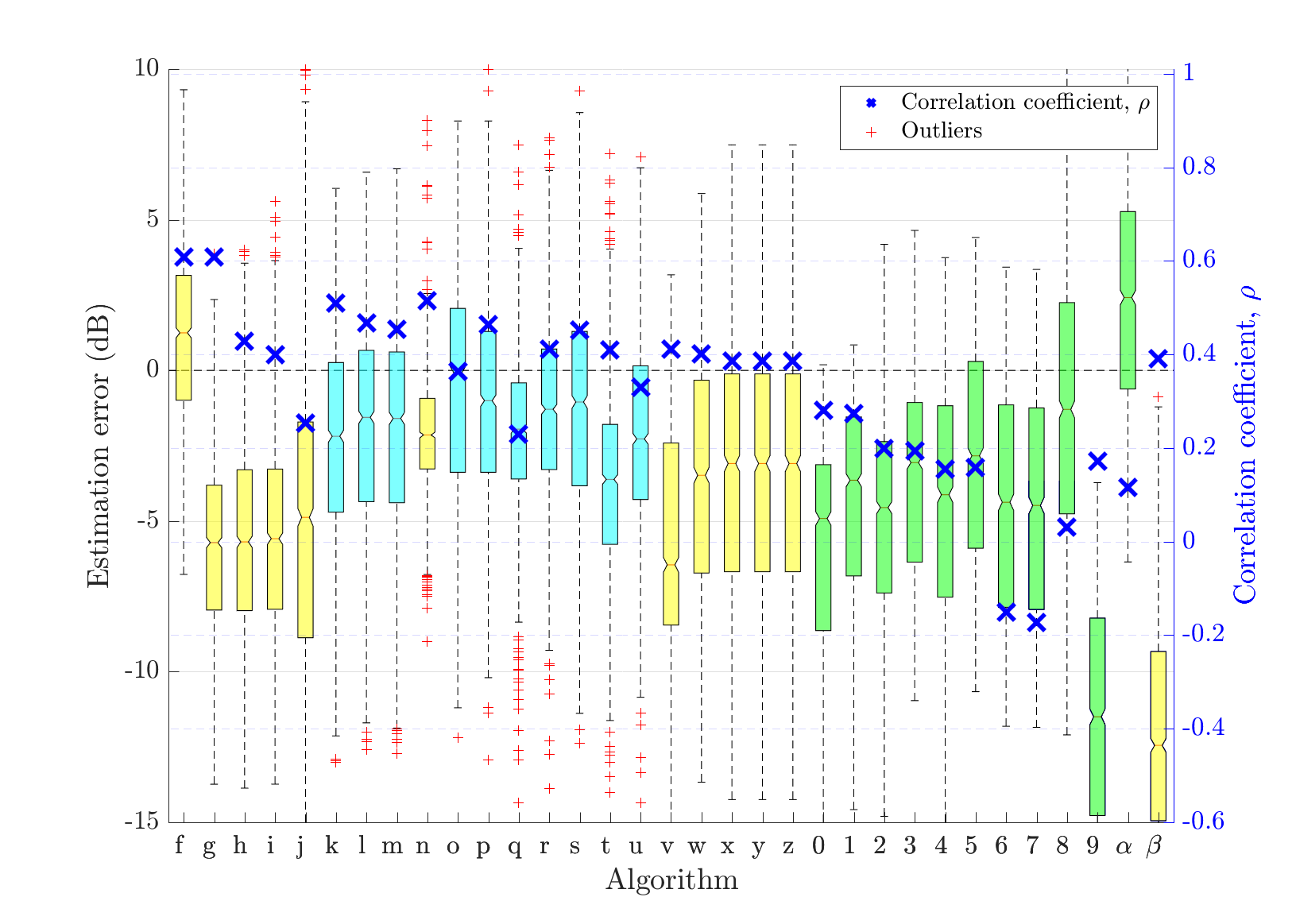,
	width=\figWidthACETR,viewport=45 10 765 530,clip}}%
	\else
	\centerline{\epsfig{figure=FigsACE/ana_eval_gt_partic_results_combined_Phase3_All_WASPAA_P3_DRR_dB_A2_Fan.png,
	width=\figWidthACETR,viewport=45 10 765 530,clip}}%
	\fi
	\caption{Fullband {\ac{DRR} estimation error in fan noise for all \acp{SNR}}}%
\label{fig:ACE_DRR_Fan}%
\end{figure}%
%
\acused{PSD}%
\begin{table*}[!ht]\small
\caption{\ac{DRR} estimation algorithm performance in fan noise for all \acp{SNR}}
\vspace{5mm} 
\centering
\begin{tabular}{cllllllll}%
\hline%
Ref.
& Algorithm
& Class
& Mic. Config.
& Bias
& \acs{MSE}
& $\PearsonCC$
& \ac{RTF}
\\
\hline
\hline
\ifarXiv
f & PSD est. in beamspace, bias comp.~\cite{Hioka2015} & \ac{ABC} & Mobile & 1.16 & 7.89 & 0.608 & 0.757\\ 
\hline
g & PSD est. in beamspace (Raw)~\cite{Hioka2015} & \ac{ABC} & Mobile & -5.8 & 40.2 & 0.608 & 3.18\\ 
\hline
h & PSD est. in beamspace v2~\cite{Hioka2015} & \ac{ABC} & Mobile & -5.54 & 40.4 & 0.428 & 0.844\\ 
\hline
i & PSD est. by twin BF~\cite{Hioka2012} & \ac{ABC} & Mobile & -5.42 & 40 & 0.4 & 0.613\\ 
\hline
j & Spatial Covariance in matrix mode~\cite{Hioka2011} & \ac{ABC} & Mobile & -5.33 & 61.4 & 0.254 & 0.627\\ 
\hline
k & NIRAv2~\cite{Parada2015} & \ac{MLMF} & Single & -2.23 & 17.2 & 0.511 & 0.895$^\dagger$\\ 
\hline
l & NIRAv3~\cite{Parada2015} & \ac{MLMF} & Single & -1.88 & 16.5 & 0.467 & 0.895$^\dagger$\\ 
\hline
m & NIRAv1~\cite{Parada2015} & \ac{MLMF} & Single & -1.93 & 16.9 & 0.455 & 0.895$^\dagger$\\ 
\hline
n & Particle velocity~\cite{Chen2015} & \ac{ABC} & EM32 & -2.15 & 8.28 & 0.515 & 0.134\\ 
\hline
o & Multi-layer perceptron~\cite{Xiong2015} & \ac{MLMF} & Single & -0.773 & 15.9 & 0.363 & 0.0578$^\ddagger$\\ 
\hline
p & Multi-layer perceptron P2~\cite{Xiong2015} & \ac{MLMF} & Single & -1.2 & 15.9 & 0.465 & 0.0578$^\ddagger$\\ 
\hline
q & Multi-layer perceptron P2~\cite{Xiong2015} & \ac{MLMF} & Chromebook & -2.41 & 13.4 & 0.23 & 0.059$^\ddagger$\\ 
\hline
r & Multi-layer perceptron P2~\cite{Xiong2015} & \ac{MLMF} & Mobile & -1.39 & 13.9 & 0.412 & 0.0555$^\ddagger$\\ 
\hline
s & Multi-layer perceptron P2~\cite{Xiong2015} & \ac{MLMF} & Crucif & -1.24 & 16.3 & 0.451 & 0.0569$^\ddagger$\\ 
\hline
t & Multi-layer perceptron P2~\cite{Xiong2015} & \ac{MLMF} & Lin8Ch & -3.62 & 24.2 & 0.41 & 0.0617$^\ddagger$\\ 
\hline
u & Multi-layer perceptron P2~\cite{Xiong2015} & \ac{MLMF} & EM32 & -2.04 & 13.9 & 0.33 & 0.0574$^\ddagger$\\ 
\hline
v & \ac{DENBE} no noise reduction~\cite{Eaton2015} & \ac{ABC} & Chromebook & -5.77 & 47.4 & 0.411 & 0.0322\\ 
\hline
w & \ac{DENBE} spectral subtraction~\cite{Eaton2015c} & \ac{ABC} & Chromebook & -3.48 & 26.9 & 0.401 & 0.0588\\ 
\hline
x & \ac{DENBE} spec. sub. Gerkmann~\cite{Eaton2015} & \ac{ABC} & Chromebook & -3.27 & 26.4 & 0.386 & 0.048\\ 
\hline
y & \ac{DENBE} filtered subbands~\cite{Eaton2015c} & \ac{ABC} & Chromebook & -3.27 & 26.4 & 0.386 & 0.774\\ 
\hline
z & \ac{DENBE} FFT derived subbands~\cite{Eaton2015c} & \ac{ABC} & Chromebook & -3.27 & 26.4 & 0.386 & 0.0452\\ 
\hline
0 & \ac{NOSRMR} {\sectMidSent} 2.2.~\cite{Senoussaoui2015} & \ac{SFM} & Chromebook & -5.82 & 45.6 & 0.281 & 1.03\\ 
\hline
1 & \ac{OSRMR} {\sectMidSent} 2.2.~\cite{Senoussaoui2015} & \ac{SFM} & Chromebook & -4.21 & 26.8 & 0.275 & 0.824\\ 
\hline
2 & \ac{NOSRMR} {\sectMidSent} 2.2.~\cite{Senoussaoui2015} & \ac{SFM} & Mobile & -4.74 & 34.9 & 0.199 & 1.58\\ 
\hline
3 & \ac{OSRMR} {\sectMidSent} 2.2.~\cite{Senoussaoui2015} & \ac{SFM} & Mobile & -3.33 & 21.8 & 0.193 & 1.26\\ 
\hline
4 & \ac{NOSRMR} {\sectMidSent} 2.2.~\cite{Senoussaoui2015} & \ac{SFM} & Crucif & -4.3 & 33.2 & 0.155 & 2.61\\ 
\hline
5 & \ac{OSRMR} {\sectMidSent} 2.2.~\cite{Senoussaoui2015} & \ac{SFM} & Crucif & -2.93 & 21.6 & 0.158 & 2.08\\ 
\hline
6 & \ac{NOSRMR} {\sectMidSent} 2.2.~\cite{Senoussaoui2015} & \ac{SFM} & Single & -4.14 & 33.8 & -0.151 & 0.543\\ 
\hline
7 & \ac{OSRMR} {\sectMidSent} 2.2.~\cite{Senoussaoui2015} & \ac{SFM} & Single & -4.24 & 34.6 & -0.173 & 0.447\\ 
\hline
8 & Per acoust. band SRMR {\sectMidSent} 2.5.~\cite{Senoussaoui2015} & \ac{SFM} & Single & -1.33 & 23.7 & 0.0307 & 0.576\\ 
\hline
9 & Temporal dynamics~\cite{Falk2009} & \ac{SFM} & Single & -11.4 & 147 & 0.173 & 0.0818\\ 
\hline
$\alpha$ & QA Reverb~\cite{Prego2015} & \ac{SFM} & Single & 2.34 & 22.1 & 0.116 & 0.391\\ 
\hline
$\beta$ & Blind est. of coherent-to-diffuse energy ratio~\cite{Jeub2011} & \ac{ABC} & Chromebook & -12 & 160 & 0.391 & 0.019\\ 
\hline

\else

\fi
\end{tabular}
\end{table*}
\clearpage
\subsection{Fullband \ac{DRR} estimation results by noise type and \ac{SNR}}
\subsubsection{Ambient noise at \dBel{18} \ac{SNR}}
\begin{figure}[!ht]
	\ifarXiv
\centerline{\epsfig{figure=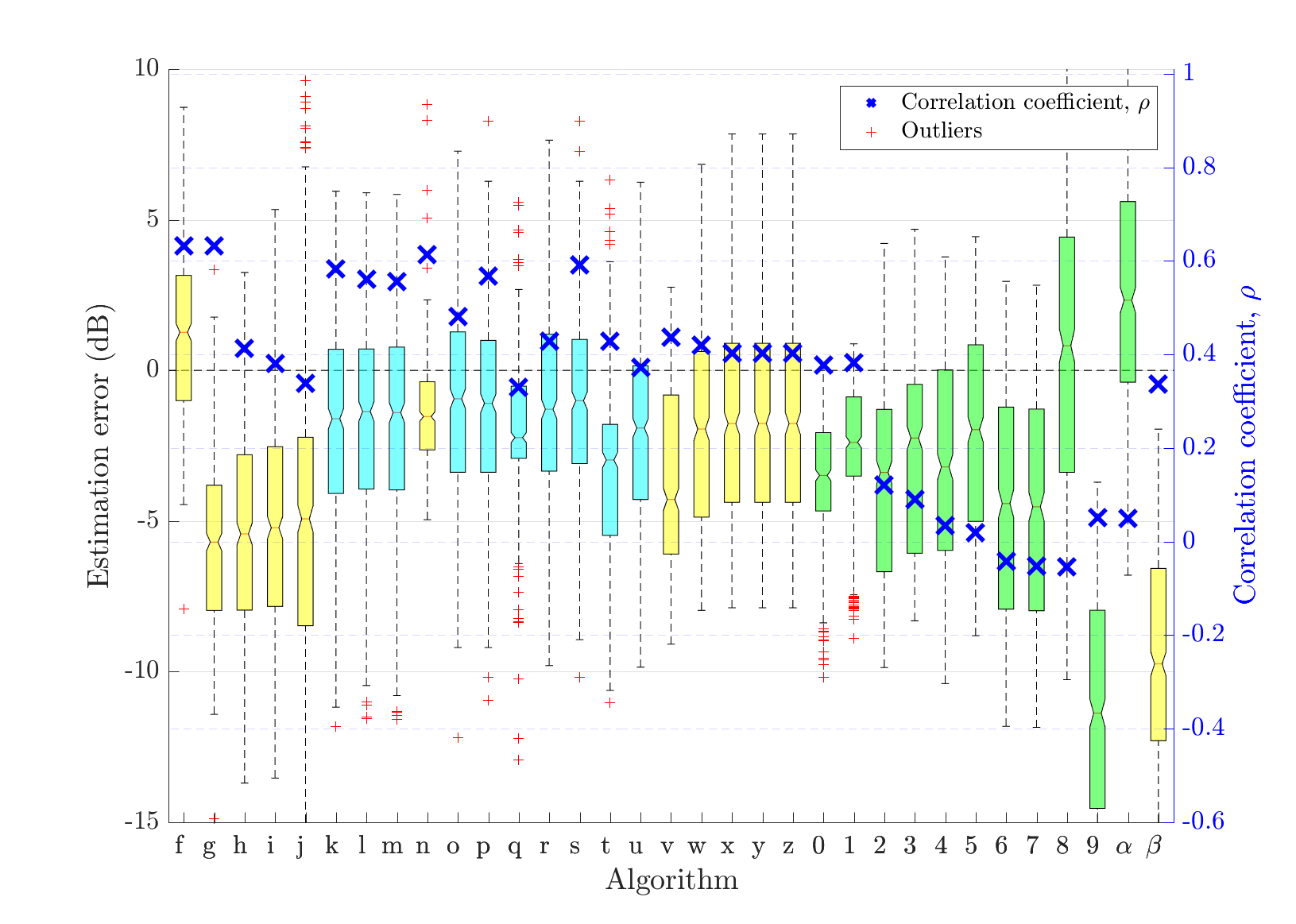,
	width=\figWidthACETR,viewport=45 10 765 530,clip}}%
	\else
	\centerline{\epsfig{figure=FigsACE/ana_eval_gt_partic_results_combined_Phase3_All_WASPAA_P3_DRR_dB_H_Ambient.png,
	width=\figWidthACETR,viewport=45 10 765 530,clip}}%
	\fi
	\caption{Fullband {\ac{DRR} estimation error in ambient noise at \dBel{18} \ac{SNR}}}%
\label{fig:ACE_DRR_Ambient_H}%
\end{figure}%
\begin{table*}[!ht]\small
\caption{\ac{DRR} estimation algorithm performance in ambient noise at \dBel{18} \ac{SNR}}
\vspace{5mm} 
\centering
\begin{tabular}{clllllll}%
\hline%
Ref.
& Algorithm
& Class
& Mic. Config.
& Bias
& MSE
&  $\PearsonCC$
& \ac{RTF}
\\
\hline
\hline
\ifarXiv
f & PSD est. in beamspace, bias comp.~\cite{Hioka2015} & \ac{ABC} & Mobile & 1.14 & 7.81 & 0.632 & 0.757\\ 
\hline
g & PSD est. in beamspace (Raw)~\cite{Hioka2015} & \ac{ABC} & Mobile & -5.82 & 40.4 & 0.632 & 3.15\\ 
\hline
h & PSD est. in beamspace v2~\cite{Hioka2015} & \ac{ABC} & Mobile & -5.37 & 40.7 & 0.413 & 0.844\\ 
\hline
i & PSD est. by twin BF~\cite{Hioka2012} & \ac{ABC} & Mobile & -5.11 & 39.2 & 0.381 & 0.614\\ 
\hline
j & Spatial Covariance in matrix mode~\cite{Hioka2011} & \ac{ABC} & Mobile & -5.22 & 53 & 0.339 & 0.627\\ 
\hline
k & NIRAv2~\cite{Parada2015} & \ac{MLMF} & Single & -1.73 & 13.9 & 0.582 & 0.897$^\dagger$\\ 
\hline
l & NIRAv3~\cite{Parada2015} & \ac{MLMF} & Single & -1.81 & 14.6 & 0.561 & 0.897$^\dagger$\\ 
\hline
m & NIRAv1~\cite{Parada2015} & \ac{MLMF} & Single & -1.84 & 14.8 & 0.557 & 0.897$^\dagger$\\ 
\hline
n & Particle velocity~\cite{Chen2015} & \ac{ABC} & EM32 & -1.44 & 4.89 & 0.613 & 0.134\\ 
\hline
o & Multi-layer perceptron~\cite{Xiong2015} & \ac{MLMF} & Single & -1.14 & 15.4 & 0.48 & 0.0578$^\ddagger$\\ 
\hline
p & Multi-layer perceptron P2~\cite{Xiong2015} & \ac{MLMF} & Single & -1.29 & 14.3 & 0.567 & 0.0578$^\ddagger$\\ 
\hline
q & Multi-layer perceptron P2~\cite{Xiong2015} & \ac{MLMF} & Chromebook & -2.28 & 11.6 & 0.331 & 0.0589$^\ddagger$\\ 
\hline
r & Multi-layer perceptron P2~\cite{Xiong2015} & \ac{MLMF} & Mobile & -1.17 & 12.8 & 0.428 & 0.0557$^\ddagger$\\ 
\hline
s & Multi-layer perceptron P2~\cite{Xiong2015} & \ac{MLMF} & Crucif & -1.16 & 13.5 & 0.592 & 0.0569$^\ddagger$\\ 
\hline
t & Multi-layer perceptron P2~\cite{Xiong2015} & \ac{MLMF} & Lin8Ch & -3.37 & 21.1 & 0.428 & 0.062$^\ddagger$\\ 
\hline
u & Multi-layer perceptron P2~\cite{Xiong2015} & \ac{MLMF} & EM32 & -1.93 & 11.9 & 0.373 & 0.0578$^\ddagger$\\ 
\hline
v & \ac{DENBE} no noise reduction~\cite{Eaton2015} & \ac{ABC} & Chromebook & -3.51 & 21.4 & 0.437 & 0.0323\\ 
\hline
w & \ac{DENBE} spectral subtraction~\cite{Eaton2015c} & \ac{ABC} & Chromebook & -1.91 & 14.2 & 0.42 & 0.0602\\ 
\hline
x & \ac{DENBE} spec. sub. Gerkmann~\cite{Eaton2015} & \ac{ABC} & Chromebook & -1.58 & 13.8 & 0.403 & 0.0474\\ 
\hline
y & \ac{DENBE} filtered subbands~\cite{Eaton2015c} & \ac{ABC} & Chromebook & -1.58 & 13.8 & 0.403 & 0.775\\ 
\hline
z & \ac{DENBE} FFT derived subbands~\cite{Eaton2015c} & \ac{ABC} & Chromebook & -1.58 & 13.8 & 0.403 & 0.0449\\ 
\hline
0 & \ac{NOSRMR} {\sectMidSent} 2.2.~\cite{Senoussaoui2015} & \ac{SFM} & Chromebook & -3.66 & 17.8 & 0.377 & 1.04\\ 
\hline
1 & \ac{OSRMR} {\sectMidSent} 2.2.~\cite{Senoussaoui2015} & \ac{SFM} & Chromebook & -2.51 & 10.7 & 0.382 & 0.831\\ 
\hline
2 & \ac{NOSRMR} {\sectMidSent} 2.2.~\cite{Senoussaoui2015} & \ac{SFM} & Mobile & -3.53 & 23.2 & 0.121 & 1.59\\ 
\hline
3 & \ac{OSRMR} {\sectMidSent} 2.2.~\cite{Senoussaoui2015} & \ac{SFM} & Mobile & -2.53 & 16.8 & 0.0908 & 1.26\\ 
\hline
4 & \ac{NOSRMR} {\sectMidSent} 2.2.~\cite{Senoussaoui2015} & \ac{SFM} & Crucif & -3.16 & 23.3 & 0.0351 & 2.63\\ 
\hline
5 & \ac{OSRMR} {\sectMidSent} 2.2.~\cite{Senoussaoui2015} & \ac{SFM} & Crucif & -2.15 & 17.5 & 0.0185 & 2.09\\ 
\hline
6 & \ac{NOSRMR} {\sectMidSent} 2.2.~\cite{Senoussaoui2015} & \ac{SFM} & Single & -4.22 & 34.3 & -0.0427 & 0.543\\ 
\hline
7 & \ac{OSRMR} {\sectMidSent} 2.2.~\cite{Senoussaoui2015} & \ac{SFM} & Single & -4.3 & 35 & -0.0515 & 0.446\\ 
\hline
8 & Per acoust. band SRMR {\sectMidSent} 2.5.~\cite{Senoussaoui2015} & \ac{SFM} & Single & 0.511 & 24.1 & -0.0548 & 0.58\\ 
\hline
9 & Temporal dynamics~\cite{Falk2009} & \ac{SFM} & Single & -11.1 & 140 & 0.0515 & 0.0819\\ 
\hline
$\alpha$ & QA Reverb~\cite{Prego2015} & \ac{SFM} & Single & 2.41 & 23.5 & 0.0488 & 0.391\\ 
\hline
$\beta$ & Blind est. of coherent-to-diffuse energy ratio~\cite{Jeub2011} & \ac{ABC} & Chromebook & -9.71 & 109 & 0.337 & 0.019\\ 
\hline

\else

\fi
\end{tabular}
\end{table*}
\clearpage
\subsubsection{Ambient noise at \dBel{12} \ac{SNR}}
\begin{figure}[!ht]
	\ifarXiv
\centerline{\epsfig{figure=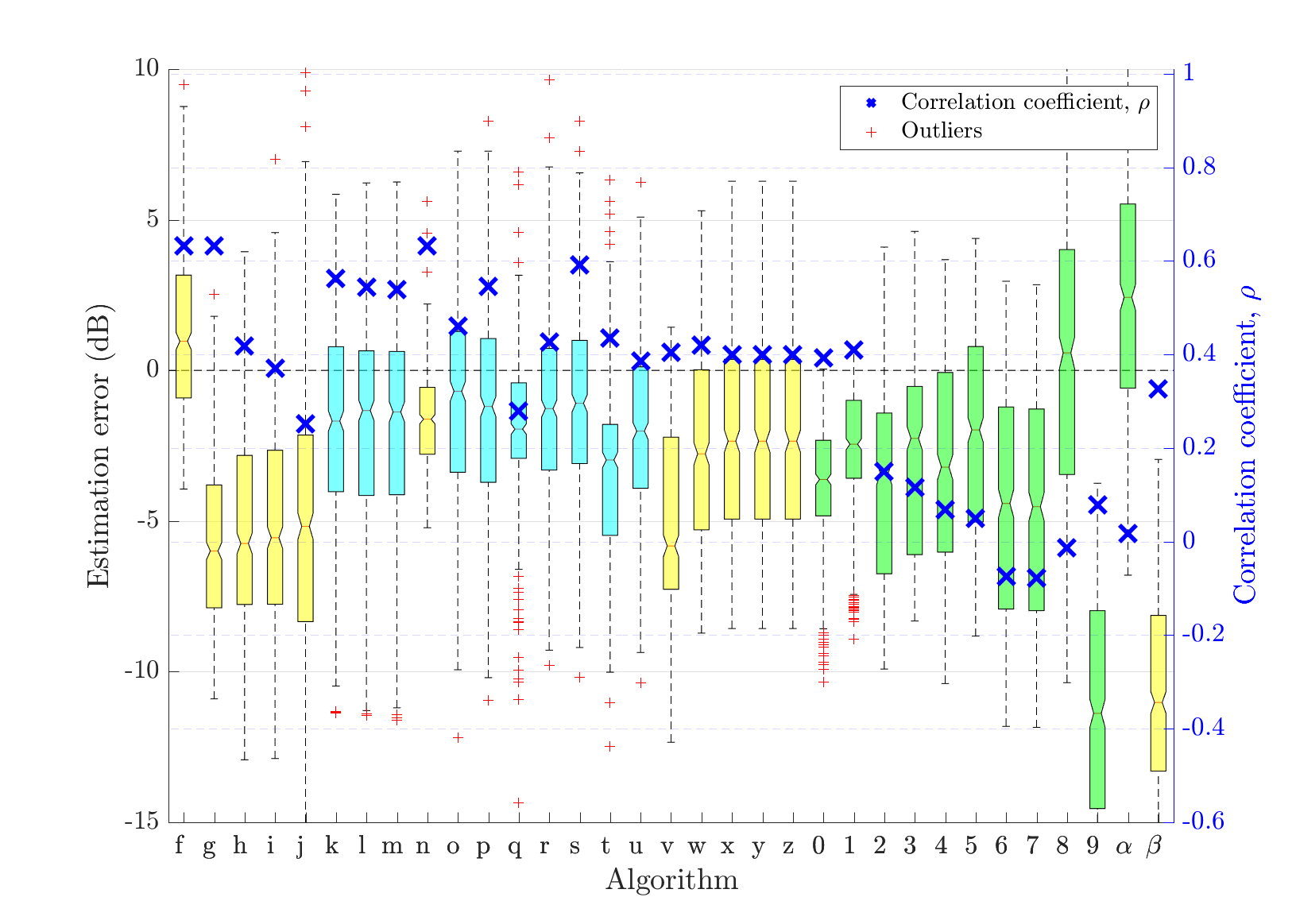,
	width=\figWidthACETR,viewport=45 10 765 530,clip}}%
	\else
	\centerline{\epsfig{figure=FigsACE/ana_eval_gt_partic_results_combined_Phase3_All_WASPAA_P3_DRR_dB_M_Ambient.png,
	width=\figWidthACETR,viewport=45 10 765 530,clip}}%
	\fi
	\caption{Fullband {\ac{DRR} estimation error in ambient noise at \dBel{12} \ac{SNR}}}%
\label{fig:ACE_DRR_Ambient_M}%
\end{figure}%
\begin{table*}[!ht]\small
\caption{\ac{DRR} estimation algorithm performance in ambient noise at \dBel{12} \ac{SNR}}
\vspace{5mm} 
\centering
\begin{tabular}{clllllll}%
\hline%
Ref.
& Algorithm
& Class
& Mic. Config.
& Bias
& MSE
&  $\PearsonCC$
& \ac{RTF}
\\
\hline
\hline
\ifarXiv
f & PSD est. in beamspace, bias comp.~\cite{Hioka2015} & \ac{ABC} & Mobile & 1.18 & 7.63 & 0.632 & 0.757\\ 
\hline
g & PSD est. in beamspace (Raw)~\cite{Hioka2015} & \ac{ABC} & Mobile & -5.79 & 39.8 & 0.632 & 3.15\\ 
\hline
h & PSD est. in beamspace v2~\cite{Hioka2015} & \ac{ABC} & Mobile & -5.38 & 39.2 & 0.419 & 0.844\\ 
\hline
i & PSD est. by twin BF~\cite{Hioka2012} & \ac{ABC} & Mobile & -5.16 & 38.1 & 0.37 & 0.614\\ 
\hline
j & Spatial Covariance in matrix mode~\cite{Hioka2011} & \ac{ABC} & Mobile & -5.11 & 52.6 & 0.251 & 0.627\\ 
\hline
k & NIRAv2~\cite{Parada2015} & \ac{MLMF} & Single & -1.71 & 14.2 & 0.562 & 0.897$^\dagger$\\ 
\hline
l & NIRAv3~\cite{Parada2015} & \ac{MLMF} & Single & -1.82 & 14.9 & 0.543 & 0.897$^\dagger$\\ 
\hline
m & NIRAv1~\cite{Parada2015} & \ac{MLMF} & Single & -1.84 & 15.1 & 0.539 & 0.897$^\dagger$\\ 
\hline
n & Particle velocity~\cite{Chen2015} & \ac{ABC} & EM32 & -1.64 & 5.2 & 0.632 & 0.134\\ 
\hline
o & Multi-layer perceptron~\cite{Xiong2015} & \ac{MLMF} & Single & -1.06 & 15.2 & 0.46 & 0.0578$^\ddagger$\\ 
\hline
p & Multi-layer perceptron P2~\cite{Xiong2015} & \ac{MLMF} & Single & -1.33 & 15 & 0.545 & 0.0578$^\ddagger$\\ 
\hline
q & Multi-layer perceptron P2~\cite{Xiong2015} & \ac{MLMF} & Chromebook & -2.27 & 12.2 & 0.279 & 0.0589$^\ddagger$\\ 
\hline
r & Multi-layer perceptron P2~\cite{Xiong2015} & \ac{MLMF} & Mobile & -1.1 & 12.8 & 0.426 & 0.0557$^\ddagger$\\ 
\hline
s & Multi-layer perceptron P2~\cite{Xiong2015} & \ac{MLMF} & Crucif & -1.15 & 13.2 & 0.592 & 0.0569$^\ddagger$\\ 
\hline
t & Multi-layer perceptron P2~\cite{Xiong2015} & \ac{MLMF} & Lin8Ch & -3.37 & 21.5 & 0.435 & 0.062$^\ddagger$\\ 
\hline
u & Multi-layer perceptron P2~\cite{Xiong2015} & \ac{MLMF} & EM32 & -1.88 & 11.8 & 0.386 & 0.0578$^\ddagger$\\ 
\hline
v & \ac{DENBE} no noise reduction~\cite{Eaton2015} & \ac{ABC} & Chromebook & -4.96 & 33.2 & 0.404 & 0.0323\\ 
\hline
w & \ac{DENBE} spectral subtraction~\cite{Eaton2015c} & \ac{ABC} & Chromebook & -2.68 & 16.6 & 0.42 & 0.0602\\ 
\hline
x & \ac{DENBE} spec. sub. Gerkmann~\cite{Eaton2015} & \ac{ABC} & Chromebook & -2.28 & 15.2 & 0.399 & 0.0474\\ 
\hline
y & \ac{DENBE} filtered subbands~\cite{Eaton2015c} & \ac{ABC} & Chromebook & -2.28 & 15.2 & 0.399 & 0.775\\ 
\hline
z & \ac{DENBE} FFT derived subbands~\cite{Eaton2015c} & \ac{ABC} & Chromebook & -2.28 & 15.2 & 0.399 & 0.0449\\ 
\hline
0 & \ac{NOSRMR} {\sectMidSent} 2.2.~\cite{Senoussaoui2015} & \ac{SFM} & Chromebook & -3.83 & 19 & 0.392 & 1.04\\ 
\hline
1 & \ac{OSRMR} {\sectMidSent} 2.2.~\cite{Senoussaoui2015} & \ac{SFM} & Chromebook & -2.62 & 11.2 & 0.41 & 0.831\\ 
\hline
2 & \ac{NOSRMR} {\sectMidSent} 2.2.~\cite{Senoussaoui2015} & \ac{SFM} & Mobile & -3.6 & 23.4 & 0.15 & 1.59\\ 
\hline
3 & \ac{OSRMR} {\sectMidSent} 2.2.~\cite{Senoussaoui2015} & \ac{SFM} & Mobile & -2.56 & 16.9 & 0.116 & 1.26\\ 
\hline
4 & \ac{NOSRMR} {\sectMidSent} 2.2.~\cite{Senoussaoui2015} & \ac{SFM} & Crucif & -3.22 & 23.5 & 0.0688 & 2.63\\ 
\hline
5 & \ac{OSRMR} {\sectMidSent} 2.2.~\cite{Senoussaoui2015} & \ac{SFM} & Crucif & -2.18 & 17.5 & 0.0496 & 2.09\\ 
\hline
6 & \ac{NOSRMR} {\sectMidSent} 2.2.~\cite{Senoussaoui2015} & \ac{SFM} & Single & -4.22 & 34.3 & -0.0748 & 0.543\\ 
\hline
7 & \ac{OSRMR} {\sectMidSent} 2.2.~\cite{Senoussaoui2015} & \ac{SFM} & Single & -4.29 & 35 & -0.0777 & 0.446\\ 
\hline
8 & Per acoust. band SRMR {\sectMidSent} 2.5.~\cite{Senoussaoui2015} & \ac{SFM} & Single & 0.283 & 22.8 & -0.0139 & 0.58\\ 
\hline
9 & Temporal dynamics~\cite{Falk2009} & \ac{SFM} & Single & -11.1 & 140 & 0.0777 & 0.0819\\ 
\hline
$\alpha$ & QA Reverb~\cite{Prego2015} & \ac{SFM} & Single & 2.37 & 23.5 & 0.0171 & 0.391\\ 
\hline
$\beta$ & Blind est. of coherent-to-diffuse energy ratio~\cite{Jeub2011} & \ac{ABC} & Chromebook & -10.9 & 131 & 0.327 & 0.019\\ 
\hline

\else

\fi
\end{tabular}
\end{table*}
\clearpage
\subsubsection{Ambient noise at \dBel{-1} \ac{SNR}}
\begin{figure}[!ht]
	\ifarXiv
\centerline{\epsfig{figure=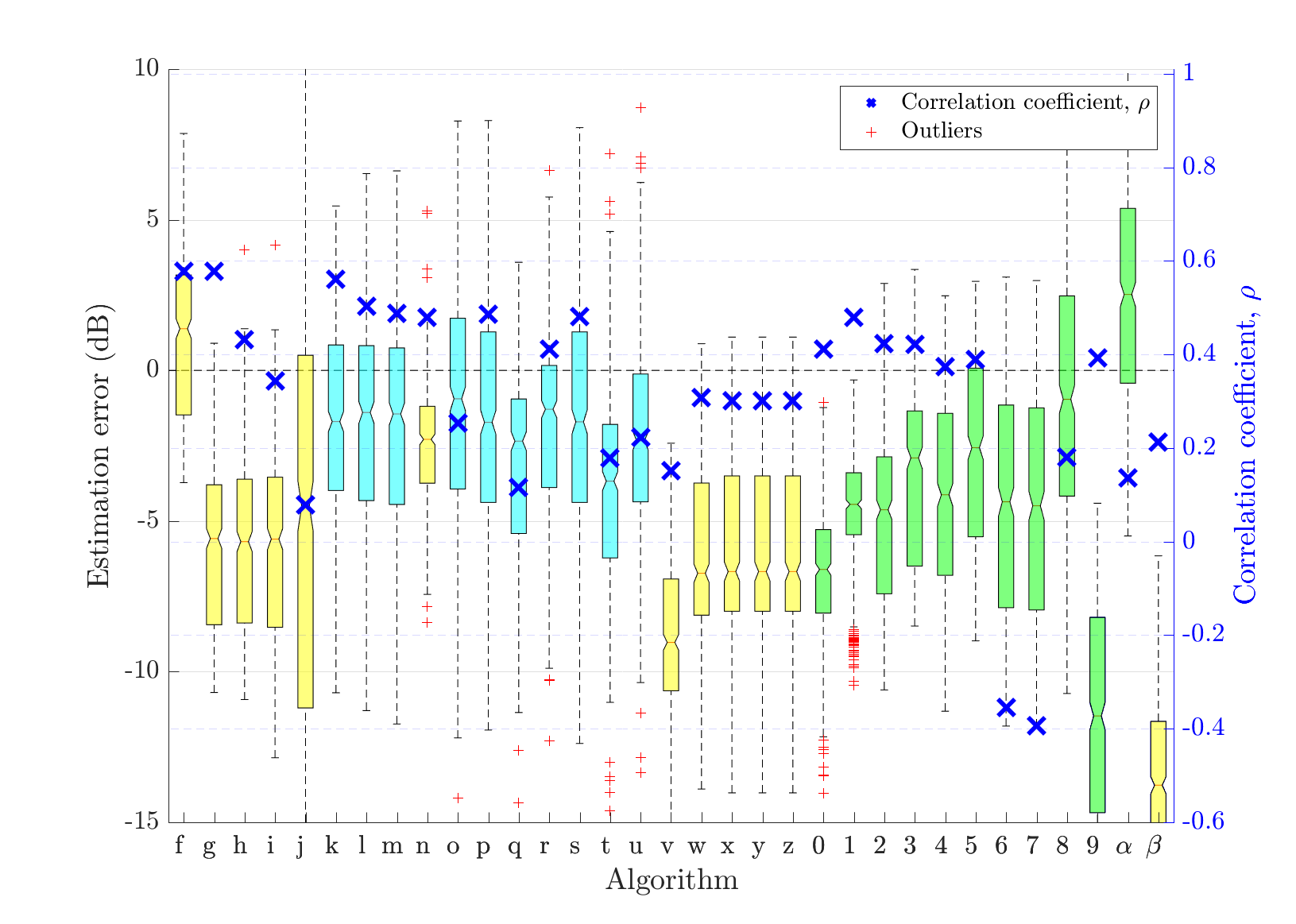,
	width=\figWidthACETR,viewport=45 10 765 530,clip}}%
	\else
	\centerline{\epsfig{figure=FigsACE/ana_eval_gt_partic_results_combined_Phase3_All_WASPAA_P3_DRR_dB_L_Ambient.png,
	width=\figWidthACETR,viewport=45 10 765 530,clip}}%
	\fi
	\caption{Fullband {\ac{DRR} estimation error in ambient noise at \dBel{-1} \ac{SNR}}}%
\label{fig:ACE_DRR_Ambient_L}%
\end{figure}%
\begin{table*}[!ht]\small
\caption{\ac{DRR} estimation algorithm performance in ambient noise at \dBel{-1} \ac{SNR}}
\vspace{5mm} 
\centering
\begin{tabular}{clllllll}%
\hline%
Ref.
& Algorithm
& Class
& Mic. Config.
& Bias
& MSE
&  $\PearsonCC$
& \ac{RTF}
\\
\hline
\hline
\ifarXiv
f & PSD est. in beamspace, bias comp.~\cite{Hioka2015} & \ac{ABC} & Mobile & 1.3 & 9.51 & 0.578 & 0.757\\ 
\hline
g & PSD est. in beamspace (Raw)~\cite{Hioka2015} & \ac{ABC} & Mobile & -5.67 & 39.9 & 0.578 & 3.15\\ 
\hline
h & PSD est. in beamspace v2~\cite{Hioka2015} & \ac{ABC} & Mobile & -5.63 & 40.2 & 0.431 & 0.844\\ 
\hline
i & PSD est. by twin BF~\cite{Hioka2012} & \ac{ABC} & Mobile & -5.76 & 42.4 & 0.344 & 0.614\\ 
\hline
j & Spatial Covariance in matrix mode~\cite{Hioka2011} & \ac{ABC} & Mobile & -5.17 & 89.4 & 0.0787 & 0.627\\ 
\hline
k & NIRAv2~\cite{Parada2015} & \ac{MLMF} & Single & -1.6 & 13.9 & 0.561 & 0.897$^\dagger$\\ 
\hline
l & NIRAv3~\cite{Parada2015} & \ac{MLMF} & Single & -1.78 & 15.5 & 0.503 & 0.897$^\dagger$\\ 
\hline
m & NIRAv1~\cite{Parada2015} & \ac{MLMF} & Single & -1.86 & 16.1 & 0.488 & 0.897$^\dagger$\\ 
\hline
n & Particle velocity~\cite{Chen2015} & \ac{ABC} & EM32 & -2.48 & 9.77 & 0.479 & 0.134\\ 
\hline
o & Multi-layer perceptron~\cite{Xiong2015} & \ac{MLMF} & Single & -1.15 & 17.9 & 0.253 & 0.0578$^\ddagger$\\ 
\hline
p & Multi-layer perceptron P2~\cite{Xiong2015} & \ac{MLMF} & Single & -1.63 & 17 & 0.486 & 0.0578$^\ddagger$\\ 
\hline
q & Multi-layer perceptron P2~\cite{Xiong2015} & \ac{MLMF} & Chromebook & -3.17 & 20 & 0.116 & 0.0589$^\ddagger$\\ 
\hline
r & Multi-layer perceptron P2~\cite{Xiong2015} & \ac{MLMF} & Mobile & -1.81 & 15.2 & 0.412 & 0.0557$^\ddagger$\\ 
\hline
s & Multi-layer perceptron P2~\cite{Xiong2015} & \ac{MLMF} & Crucif & -1.66 & 17.1 & 0.481 & 0.0569$^\ddagger$\\ 
\hline
t & Multi-layer perceptron P2~\cite{Xiong2015} & \ac{MLMF} & Lin8Ch & -3.97 & 31.4 & 0.178 & 0.062$^\ddagger$\\ 
\hline
u & Multi-layer perceptron P2~\cite{Xiong2015} & \ac{MLMF} & EM32 & -2.23 & 17.4 & 0.223 & 0.0578$^\ddagger$\\ 
\hline
v & \ac{DENBE} no noise reduction~\cite{Eaton2015} & \ac{ABC} & Chromebook & -8.85 & 85.7 & 0.152 & 0.0323\\ 
\hline
w & \ac{DENBE} spectral subtraction~\cite{Eaton2015c} & \ac{ABC} & Chromebook & -5.99 & 45.8 & 0.308 & 0.0602\\ 
\hline
x & \ac{DENBE} spec. sub. Gerkmann~\cite{Eaton2015} & \ac{ABC} & Chromebook & -5.88 & 44.5 & 0.302 & 0.0474\\ 
\hline
y & \ac{DENBE} filtered subbands~\cite{Eaton2015c} & \ac{ABC} & Chromebook & -5.88 & 44.5 & 0.302 & 0.775\\ 
\hline
z & \ac{DENBE} FFT derived subbands~\cite{Eaton2015c} & \ac{ABC} & Chromebook & -5.88 & 44.5 & 0.302 & 0.0449\\ 
\hline
0 & \ac{NOSRMR} {\sectMidSent} 2.2.~\cite{Senoussaoui2015} & \ac{SFM} & Chromebook & -6.77 & 51.5 & 0.411 & 1.04\\ 
\hline
1 & \ac{OSRMR} {\sectMidSent} 2.2.~\cite{Senoussaoui2015} & \ac{SFM} & Chromebook & -4.55 & 24.8 & 0.479 & 0.831\\ 
\hline
2 & \ac{NOSRMR} {\sectMidSent} 2.2.~\cite{Senoussaoui2015} & \ac{SFM} & Mobile & -4.74 & 31 & 0.423 & 1.59\\ 
\hline
3 & \ac{OSRMR} {\sectMidSent} 2.2.~\cite{Senoussaoui2015} & \ac{SFM} & Mobile & -3.26 & 19.3 & 0.422 & 1.26\\ 
\hline
4 & \ac{NOSRMR} {\sectMidSent} 2.2.~\cite{Senoussaoui2015} & \ac{SFM} & Crucif & -4.28 & 29.2 & 0.374 & 2.63\\ 
\hline
5 & \ac{OSRMR} {\sectMidSent} 2.2.~\cite{Senoussaoui2015} & \ac{SFM} & Crucif & -2.84 & 19 & 0.389 & 2.09\\ 
\hline
6 & \ac{NOSRMR} {\sectMidSent} 2.2.~\cite{Senoussaoui2015} & \ac{SFM} & Single & -4.14 & 34 & -0.355 & 0.543\\ 
\hline
7 & \ac{OSRMR} {\sectMidSent} 2.2.~\cite{Senoussaoui2015} & \ac{SFM} & Single & -4.25 & 34.8 & -0.393 & 0.446\\ 
\hline
8 & Per acoust. band SRMR {\sectMidSent} 2.5.~\cite{Senoussaoui2015} & \ac{SFM} & Single & -1.02 & 19.3 & 0.181 & 0.58\\ 
\hline
9 & Temporal dynamics~\cite{Falk2009} & \ac{SFM} & Single & -11.4 & 145 & 0.393 & 0.0819\\ 
\hline
$\alpha$ & QA Reverb~\cite{Prego2015} & \ac{SFM} & Single & 2.43 & 22.2 & 0.137 & 0.391\\ 
\hline
$\beta$ & Blind est. of coherent-to-diffuse energy ratio~\cite{Jeub2011} & \ac{ABC} & Chromebook & -13.8 & 199 & 0.212 & 0.019\\ 
\hline

\else

\fi
\end{tabular}
\end{table*}
\clearpage
\subsubsection{Babble noise at \dBel{18} \ac{SNR}}
\begin{figure}[!ht]
	\ifarXiv
\centerline{\epsfig{figure=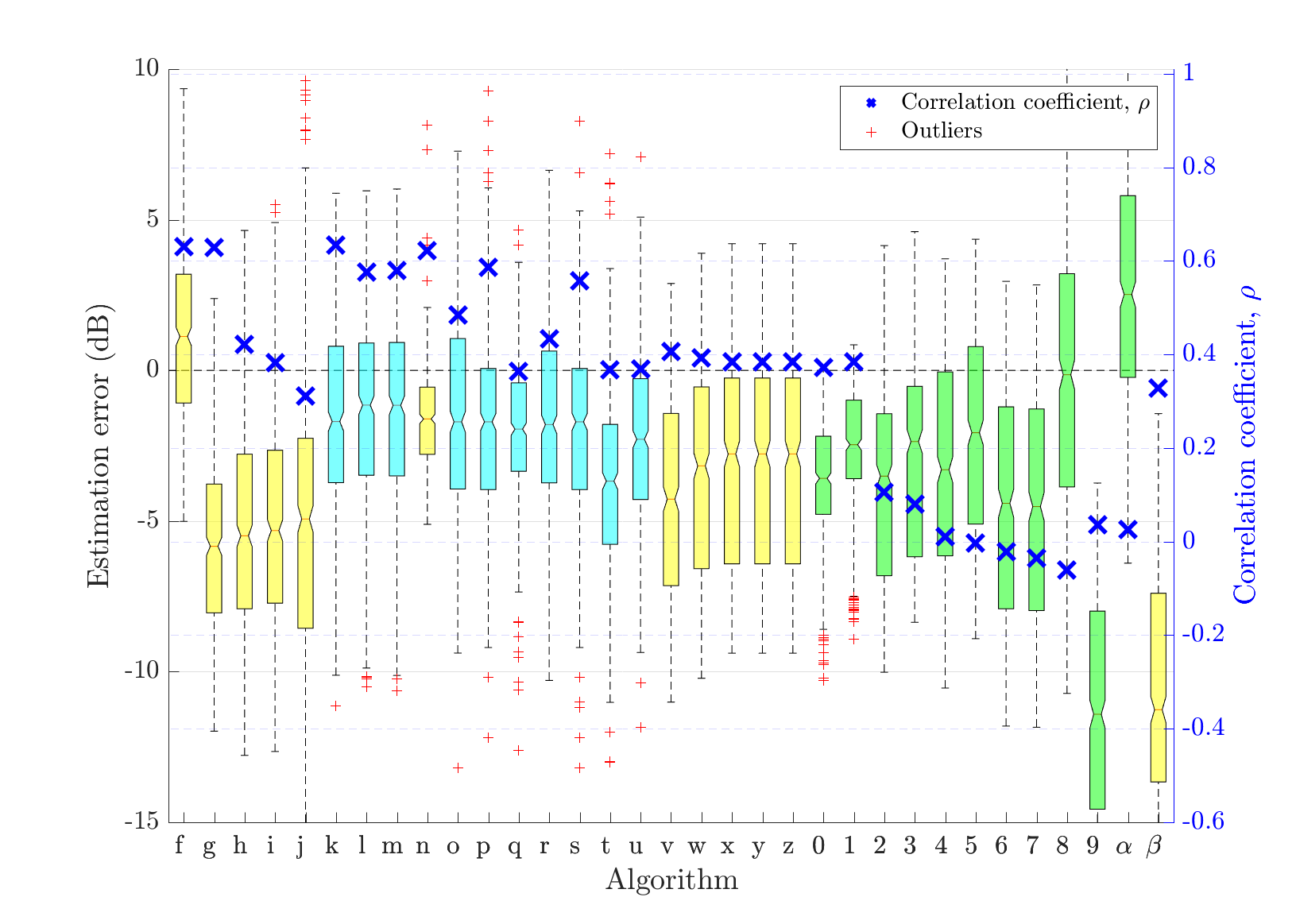,
	width=\figWidthACETR,viewport=45 10 765 530,clip}}%
	\else
	\centerline{\epsfig{figure=FigsACE/ana_eval_gt_partic_results_combined_Phase3_All_WASPAA_P3_DRR_dB_H_Babble.png,
	width=\figWidthACETR,viewport=45 10 765 530,clip}}%
	\fi
	\caption{Fullband {\ac{DRR} estimation error in babble noise at \dBel{18} \ac{SNR}}}%
\label{fig:ACE_DRR_Babble_H}%
\end{figure}%
\begin{table*}[!ht]\small
\caption{\ac{DRR} estimation algorithm performance in babble noise at \dBel{18} \ac{SNR}}
\vspace{5mm} 
\centering
\begin{tabular}{clllllll}%
\hline%
Ref.
& Algorithm
& Class
& Mic. Config.
& Bias
& MSE
&  $\PearsonCC$
& \ac{RTF}
\\
\hline
\hline
\ifarXiv
f & PSD est. in beamspace, bias comp.~\cite{Hioka2015} & \ac{ABC} & Mobile & 1.12 & 7.96 & 0.631 & 0.757\\ 
\hline
g & PSD est. in beamspace (Raw)~\cite{Hioka2015} & \ac{ABC} & Mobile & -5.85 & 40.9 & 0.629 & 3.17\\ 
\hline
h & PSD est. in beamspace v2~\cite{Hioka2015} & \ac{ABC} & Mobile & -5.38 & 41.2 & 0.422 & 0.843\\ 
\hline
i & PSD est. by twin BF~\cite{Hioka2012} & \ac{ABC} & Mobile & -5.13 & 40 & 0.382 & 0.615\\ 
\hline
j & Spatial Covariance in matrix mode~\cite{Hioka2011} & \ac{ABC} & Mobile & -5.09 & 51.7 & 0.312 & 0.627\\ 
\hline
k & NIRAv2~\cite{Parada2015} & \ac{MLMF} & Single & -1.67 & 12.8 & 0.633 & 0.906$^\dagger$\\ 
\hline
l & NIRAv3~\cite{Parada2015} & \ac{MLMF} & Single & -1.35 & 13 & 0.576 & 0.906$^\dagger$\\ 
\hline
m & NIRAv1~\cite{Parada2015} & \ac{MLMF} & Single & -1.35 & 12.9 & 0.579 & 0.906$^\dagger$\\ 
\hline
n & Particle velocity~\cite{Chen2015} & \ac{ABC} & EM32 & -1.62 & 5.28 & 0.623 & 0.134\\ 
\hline
o & Multi-layer perceptron~\cite{Xiong2015} & \ac{MLMF} & Single & -1.54 & 15.5 & 0.485 & 0.0579$^\ddagger$\\ 
\hline
p & Multi-layer perceptron P2~\cite{Xiong2015} & \ac{MLMF} & Single & -1.6 & 14.1 & 0.586 & 0.0579$^\ddagger$\\ 
\hline
q & Multi-layer perceptron P2~\cite{Xiong2015} & \ac{MLMF} & Chromebook & -2.16 & 11.1 & 0.363 & 0.0588$^\ddagger$\\ 
\hline
r & Multi-layer perceptron P2~\cite{Xiong2015} & \ac{MLMF} & Mobile & -1.76 & 13.8 & 0.434 & 0.0555$^\ddagger$\\ 
\hline
s & Multi-layer perceptron P2~\cite{Xiong2015} & \ac{MLMF} & Crucif & -1.62 & 14.6 & 0.557 & 0.057$^\ddagger$\\ 
\hline
t & Multi-layer perceptron P2~\cite{Xiong2015} & \ac{MLMF} & Lin8Ch & -3.56 & 23.3 & 0.368 & 0.0618$^\ddagger$\\ 
\hline
u & Multi-layer perceptron P2~\cite{Xiong2015} & \ac{MLMF} & EM32 & -2.2 & 13.5 & 0.369 & 0.0576$^\ddagger$\\ 
\hline
v & \ac{DENBE} no noise reduction~\cite{Eaton2015} & \ac{ABC} & Chromebook & -4.11 & 27.4 & 0.406 & 0.0323\\ 
\hline
w & \ac{DENBE} spectral subtraction~\cite{Eaton2015c} & \ac{ABC} & Chromebook & -3.27 & 22 & 0.393 & 0.0577\\ 
\hline
x & \ac{DENBE} spec. sub. Gerkmann~\cite{Eaton2015} & \ac{ABC} & Chromebook & -3.02 & 20.6 & 0.385 & 0.0476\\ 
\hline
y & \ac{DENBE} filtered subbands~\cite{Eaton2015c} & \ac{ABC} & Chromebook & -3.02 & 20.6 & 0.385 & 0.778\\ 
\hline
z & \ac{DENBE} FFT derived subbands~\cite{Eaton2015c} & \ac{ABC} & Chromebook & -3.02 & 20.6 & 0.385 & 0.0448\\ 
\hline
0 & \ac{NOSRMR} {\sectMidSent} 2.2.~\cite{Senoussaoui2015} & \ac{SFM} & Chromebook & -3.76 & 18.5 & 0.373 & 1.04\\ 
\hline
1 & \ac{OSRMR} {\sectMidSent} 2.2.~\cite{Senoussaoui2015} & \ac{SFM} & Chromebook & -2.6 & 11.1 & 0.384 & 0.833\\ 
\hline
2 & \ac{NOSRMR} {\sectMidSent} 2.2.~\cite{Senoussaoui2015} & \ac{SFM} & Mobile & -3.66 & 24.2 & 0.106 & 1.58\\ 
\hline
3 & \ac{OSRMR} {\sectMidSent} 2.2.~\cite{Senoussaoui2015} & \ac{SFM} & Mobile & -2.62 & 17.4 & 0.0805 & 1.26\\ 
\hline
4 & \ac{NOSRMR} {\sectMidSent} 2.2.~\cite{Senoussaoui2015} & \ac{SFM} & Crucif & -3.28 & 24.3 & 0.0103 & 2.63\\ 
\hline
5 & \ac{OSRMR} {\sectMidSent} 2.2.~\cite{Senoussaoui2015} & \ac{SFM} & Crucif & -2.24 & 18 & -0.00234 & 2.1\\ 
\hline
6 & \ac{NOSRMR} {\sectMidSent} 2.2.~\cite{Senoussaoui2015} & \ac{SFM} & Single & -4.21 & 34.3 & -0.0213 & 0.534\\ 
\hline
7 & \ac{OSRMR} {\sectMidSent} 2.2.~\cite{Senoussaoui2015} & \ac{SFM} & Single & -4.29 & 34.9 & -0.0354 & 0.444\\ 
\hline
8 & Per acoust. band SRMR {\sectMidSent} 2.5.~\cite{Senoussaoui2015} & \ac{SFM} & Single & -0.337 & 21.5 & -0.0605 & 0.579\\ 
\hline
9 & Temporal dynamics~\cite{Falk2009} & \ac{SFM} & Single & -11.1 & 141 & 0.0354 & 0.0823\\ 
\hline
$\alpha$ & QA Reverb~\cite{Prego2015} & \ac{SFM} & Single & 2.63 & 24.8 & 0.0256 & 0.392\\ 
\hline
$\beta$ & Blind est. of coherent-to-diffuse energy ratio~\cite{Jeub2011} & \ac{ABC} & Chromebook & -10.8 & 133 & 0.329 & 0.019\\ 
\hline

\else

\fi
\end{tabular}
\end{table*}
\clearpage
\subsubsection{Babble noise at \dBel{12} \ac{SNR}}
\begin{figure}[!ht]
	\ifarXiv
\centerline{\epsfig{figure=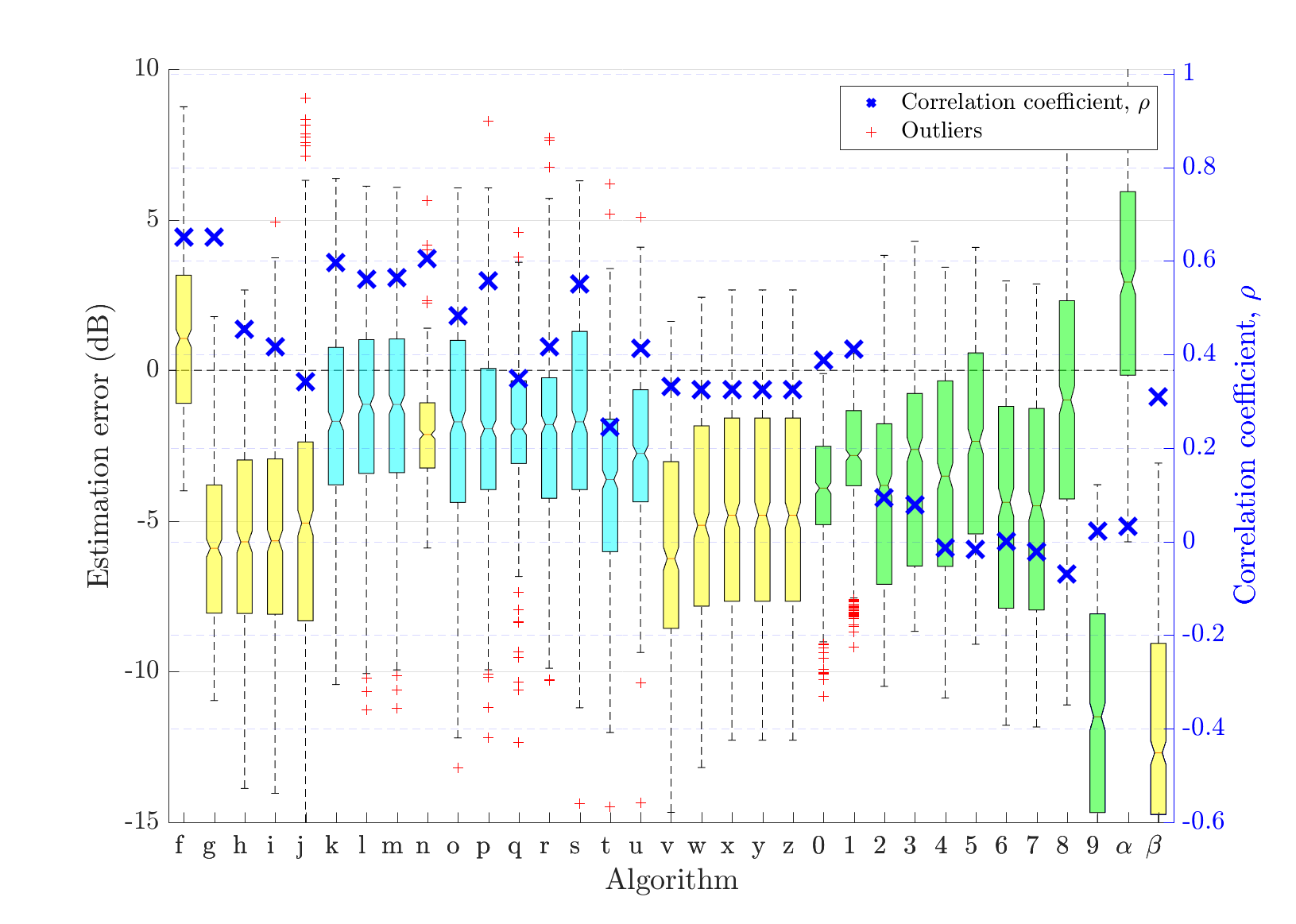,
	width=\figWidthACETR,viewport=45 10 765 530,clip}}%
	\else
	\centerline{\epsfig{figure=FigsACE/ana_eval_gt_partic_results_combined_Phase3_All_WASPAA_P3_DRR_dB_M_Babble.png,
	width=\figWidthACETR,viewport=45 10 765 530,clip}}%
	\fi
	\caption{Fullband {\ac{DRR} estimation error in babble noise at \dBel{12} \ac{SNR}}}%
\label{fig:ACE_DRR_Babble_M}%
\end{figure}%
\begin{table*}[!ht]\small
\caption{\ac{DRR} estimation algorithm performance in babble noise at \dBel{12} \ac{SNR}}
\vspace{5mm} 
\centering
\begin{tabular}{clllllll}%
\hline%
Ref.
& Algorithm
& Class
& Mic. Config.
& Bias
& MSE
&  $\PearsonCC$
& \ac{RTF}
\\
\hline
\hline
\ifarXiv
f & PSD est. in beamspace, bias comp.~\cite{Hioka2015} & \ac{ABC} & Mobile & 1.11 & 7.41 & 0.651 & 0.757\\ 
\hline
g & PSD est. in beamspace (Raw)~\cite{Hioka2015} & \ac{ABC} & Mobile & -5.86 & 40.5 & 0.651 & 3.17\\ 
\hline
h & PSD est. in beamspace v2~\cite{Hioka2015} & \ac{ABC} & Mobile & -5.49 & 40.9 & 0.454 & 0.843\\ 
\hline
i & PSD est. by twin BF~\cite{Hioka2012} & \ac{ABC} & Mobile & -5.4 & 40.8 & 0.416 & 0.615\\ 
\hline
j & Spatial Covariance in matrix mode~\cite{Hioka2011} & \ac{ABC} & Mobile & -5.2 & 48.4 & 0.342 & 0.627\\ 
\hline
k & NIRAv2~\cite{Parada2015} & \ac{MLMF} & Single & -1.73 & 13.7 & 0.596 & 0.906$^\dagger$\\ 
\hline
l & NIRAv3~\cite{Parada2015} & \ac{MLMF} & Single & -1.26 & 13 & 0.561 & 0.906$^\dagger$\\ 
\hline
m & NIRAv1~\cite{Parada2015} & \ac{MLMF} & Single & -1.23 & 12.9 & 0.564 & 0.906$^\dagger$\\ 
\hline
n & Particle velocity~\cite{Chen2015} & \ac{ABC} & EM32 & -2.15 & 7.25 & 0.604 & 0.134\\ 
\hline
o & Multi-layer perceptron~\cite{Xiong2015} & \ac{MLMF} & Single & -1.57 & 15.4 & 0.483 & 0.0579$^\ddagger$\\ 
\hline
p & Multi-layer perceptron P2~\cite{Xiong2015} & \ac{MLMF} & Single & -1.8 & 15.5 & 0.558 & 0.0579$^\ddagger$\\ 
\hline
q & Multi-layer perceptron P2~\cite{Xiong2015} & \ac{MLMF} & Chromebook & -2.11 & 11.3 & 0.349 & 0.0588$^\ddagger$\\ 
\hline
r & Multi-layer perceptron P2~\cite{Xiong2015} & \ac{MLMF} & Mobile & -2.01 & 15.6 & 0.416 & 0.0555$^\ddagger$\\ 
\hline
s & Multi-layer perceptron P2~\cite{Xiong2015} & \ac{MLMF} & Crucif & -1.64 & 14.8 & 0.551 & 0.057$^\ddagger$\\ 
\hline
t & Multi-layer perceptron P2~\cite{Xiong2015} & \ac{MLMF} & Lin8Ch & -3.64 & 25.8 & 0.246 & 0.0618$^\ddagger$\\ 
\hline
u & Multi-layer perceptron P2~\cite{Xiong2015} & \ac{MLMF} & EM32 & -2.51 & 14.1 & 0.413 & 0.0576$^\ddagger$\\ 
\hline
v & \ac{DENBE} no noise reduction~\cite{Eaton2015} & \ac{ABC} & Chromebook & -5.91 & 46 & 0.331 & 0.0323\\ 
\hline
w & \ac{DENBE} spectral subtraction~\cite{Eaton2015c} & \ac{ABC} & Chromebook & -4.9 & 35.7 & 0.325 & 0.0577\\ 
\hline
x & \ac{DENBE} spec. sub. Gerkmann~\cite{Eaton2015} & \ac{ABC} & Chromebook & -4.62 & 33 & 0.324 & 0.0476\\ 
\hline
y & \ac{DENBE} filtered subbands~\cite{Eaton2015c} & \ac{ABC} & Chromebook & -4.62 & 33 & 0.324 & 0.778\\ 
\hline
z & \ac{DENBE} FFT derived subbands~\cite{Eaton2015c} & \ac{ABC} & Chromebook & -4.62 & 33 & 0.324 & 0.0448\\ 
\hline
0 & \ac{NOSRMR} {\sectMidSent} 2.2.~\cite{Senoussaoui2015} & \ac{SFM} & Chromebook & -4.09 & 21.1 & 0.388 & 1.04\\ 
\hline
1 & \ac{OSRMR} {\sectMidSent} 2.2.~\cite{Senoussaoui2015} & \ac{SFM} & Chromebook & -2.88 & 12.6 & 0.411 & 0.833\\ 
\hline
2 & \ac{NOSRMR} {\sectMidSent} 2.2.~\cite{Senoussaoui2015} & \ac{SFM} & Mobile & -4 & 27 & 0.0937 & 1.58\\ 
\hline
3 & \ac{OSRMR} {\sectMidSent} 2.2.~\cite{Senoussaoui2015} & \ac{SFM} & Mobile & -2.88 & 18.8 & 0.0782 & 1.26\\ 
\hline
4 & \ac{NOSRMR} {\sectMidSent} 2.2.~\cite{Senoussaoui2015} & \ac{SFM} & Crucif & -3.59 & 26.7 & -0.014 & 2.63\\ 
\hline
5 & \ac{OSRMR} {\sectMidSent} 2.2.~\cite{Senoussaoui2015} & \ac{SFM} & Crucif & -2.49 & 19.3 & -0.0169 & 2.1\\ 
\hline
6 & \ac{NOSRMR} {\sectMidSent} 2.2.~\cite{Senoussaoui2015} & \ac{SFM} & Single & -4.19 & 34 & -0.000435 & 0.534\\ 
\hline
7 & \ac{OSRMR} {\sectMidSent} 2.2.~\cite{Senoussaoui2015} & \ac{SFM} & Single & -4.27 & 34.8 & -0.0218 & 0.444\\ 
\hline
8 & Per acoust. band SRMR {\sectMidSent} 2.5.~\cite{Senoussaoui2015} & \ac{SFM} & Single & -1.12 & 21.2 & -0.0696 & 0.579\\ 
\hline
9 & Temporal dynamics~\cite{Falk2009} & \ac{SFM} & Single & -11.2 & 143 & 0.0218 & 0.0823\\ 
\hline
$\alpha$ & QA Reverb~\cite{Prego2015} & \ac{SFM} & Single & 2.8 & 25.5 & 0.0333 & 0.392\\ 
\hline
$\beta$ & Blind est. of coherent-to-diffuse energy ratio~\cite{Jeub2011} & \ac{ABC} & Chromebook & -12.2 & 163 & 0.31 & 0.019\\ 
\hline

\else

\fi
\end{tabular}
\end{table*}
\clearpage
\subsubsection{Babble noise at \dBel{-1} \ac{SNR}}
\begin{figure}[!ht]
	\ifarXiv
\centerline{\epsfig{figure=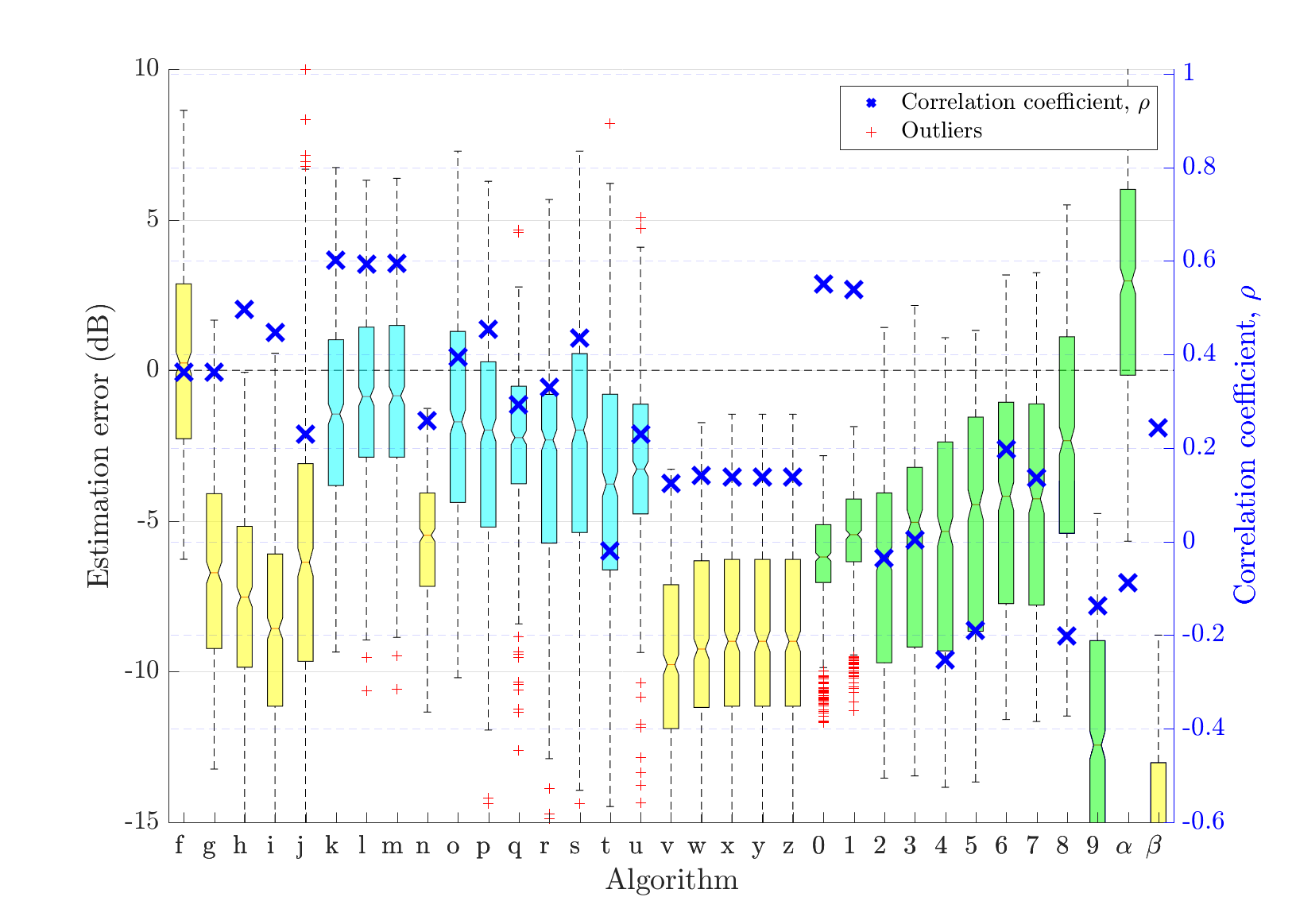,
	width=\figWidthACETR,viewport=45 10 765 530,clip}}%
	\else
	\centerline{\epsfig{figure=FigsACE/ana_eval_gt_partic_results_combined_Phase3_All_WASPAA_P3_DRR_dB_L_Babble.png,
	width=\figWidthACETR,viewport=45 10 765 530,clip}}%
	\fi
	\caption{Fullband {\ac{DRR} estimation error in babble noise at \dBel{-1} \ac{SNR}}}%
\label{fig:ACE_DRR_Babble_L}%
\end{figure}%
\begin{table*}[!ht]\small
\caption{\ac{DRR} estimation algorithm performance in babble noise at \dBel{-1} \ac{SNR}}
\vspace{5mm} 
\centering
\begin{tabular}{clllllll}%
\hline%
Ref.
& Algorithm
& Class
& Mic. Config.
& Bias
& MSE
&  $\PearsonCC$
& \ac{RTF}
\\
\hline
\hline
\ifarXiv
f & PSD est. in beamspace, bias comp.~\cite{Hioka2015} & \ac{ABC} & Mobile & 0.289 & 9.23 & 0.362 & 0.757\\ 
\hline
g & PSD est. in beamspace (Raw)~\cite{Hioka2015} & \ac{ABC} & Mobile & -6.67 & 53.7 & 0.362 & 3.17\\ 
\hline
h & PSD est. in beamspace v2~\cite{Hioka2015} & \ac{ABC} & Mobile & -7.42 & 63.1 & 0.496 & 0.843\\ 
\hline
i & PSD est. by twin BF~\cite{Hioka2012} & \ac{ABC} & Mobile & -8.61 & 83.1 & 0.447 & 0.615\\ 
\hline
j & Spatial Covariance in matrix mode~\cite{Hioka2011} & \ac{ABC} & Mobile & -6.51 & 71 & 0.23 & 0.627\\ 
\hline
k & NIRAv2~\cite{Parada2015} & \ac{MLMF} & Single & -1.58 & 13.1 & 0.601 & 0.906$^\dagger$\\ 
\hline
l & NIRAv3~\cite{Parada2015} & \ac{MLMF} & Single & -0.885 & 12.1 & 0.594 & 0.906$^\dagger$\\ 
\hline
m & NIRAv1~\cite{Parada2015} & \ac{MLMF} & Single & -0.848 & 12.1 & 0.595 & 0.906$^\dagger$\\ 
\hline
n & Particle velocity~\cite{Chen2015} & \ac{ABC} & EM32 & -5.63 & 36.2 & 0.259 & 0.134\\ 
\hline
o & Multi-layer perceptron~\cite{Xiong2015} & \ac{MLMF} & Single & -1.46 & 16.1 & 0.395 & 0.0579$^\ddagger$\\ 
\hline
p & Multi-layer perceptron P2~\cite{Xiong2015} & \ac{MLMF} & Single & -2.45 & 21.5 & 0.454 & 0.0579$^\ddagger$\\ 
\hline
q & Multi-layer perceptron P2~\cite{Xiong2015} & \ac{MLMF} & Chromebook & -2.66 & 16.5 & 0.293 & 0.0588$^\ddagger$\\ 
\hline
r & Multi-layer perceptron P2~\cite{Xiong2015} & \ac{MLMF} & Mobile & -2.99 & 22.6 & 0.331 & 0.0555$^\ddagger$\\ 
\hline
s & Multi-layer perceptron P2~\cite{Xiong2015} & \ac{MLMF} & Crucif & -2.54 & 22.2 & 0.434 & 0.057$^\ddagger$\\ 
\hline
t & Multi-layer perceptron P2~\cite{Xiong2015} & \ac{MLMF} & Lin8Ch & -4.04 & 36 & -0.0201 & 0.0618$^\ddagger$\\ 
\hline
u & Multi-layer perceptron P2~\cite{Xiong2015} & \ac{MLMF} & EM32 & -3.07 & 21.5 & 0.23 & 0.0576$^\ddagger$\\ 
\hline
v & \ac{DENBE} no noise reduction~\cite{Eaton2015} & \ac{ABC} & Chromebook & -9.74 & 105 & 0.124 & 0.0323\\ 
\hline
w & \ac{DENBE} spectral subtraction~\cite{Eaton2015c} & \ac{ABC} & Chromebook & -9.03 & 92.2 & 0.142 & 0.0577\\ 
\hline
x & \ac{DENBE} spec. sub. Gerkmann~\cite{Eaton2015} & \ac{ABC} & Chromebook & -8.87 & 89.1 & 0.137 & 0.0476\\ 
\hline
y & \ac{DENBE} filtered subbands~\cite{Eaton2015c} & \ac{ABC} & Chromebook & -8.87 & 89.1 & 0.137 & 0.778\\ 
\hline
z & \ac{DENBE} FFT derived subbands~\cite{Eaton2015c} & \ac{ABC} & Chromebook & -8.87 & 89.1 & 0.137 & 0.0448\\ 
\hline
0 & \ac{NOSRMR} {\sectMidSent} 2.2.~\cite{Senoussaoui2015} & \ac{SFM} & Chromebook & -6.33 & 43.6 & 0.551 & 1.04\\ 
\hline
1 & \ac{OSRMR} {\sectMidSent} 2.2.~\cite{Senoussaoui2015} & \ac{SFM} & Chromebook & -5.57 & 34.7 & 0.539 & 0.833\\ 
\hline
2 & \ac{NOSRMR} {\sectMidSent} 2.2.~\cite{Senoussaoui2015} & \ac{SFM} & Mobile & -6.47 & 54.7 & -0.0348 & 1.58\\ 
\hline
3 & \ac{OSRMR} {\sectMidSent} 2.2.~\cite{Senoussaoui2015} & \ac{SFM} & Mobile & -5.69 & 45.3 & 0.00456 & 1.26\\ 
\hline
4 & \ac{NOSRMR} {\sectMidSent} 2.2.~\cite{Senoussaoui2015} & \ac{SFM} & Crucif & -5.99 & 52.9 & -0.253 & 2.63\\ 
\hline
5 & \ac{OSRMR} {\sectMidSent} 2.2.~\cite{Senoussaoui2015} & \ac{SFM} & Crucif & -5.21 & 44.1 & -0.19 & 2.1\\ 
\hline
6 & \ac{NOSRMR} {\sectMidSent} 2.2.~\cite{Senoussaoui2015} & \ac{SFM} & Single & -4.02 & 32.5 & 0.197 & 0.534\\ 
\hline
7 & \ac{OSRMR} {\sectMidSent} 2.2.~\cite{Senoussaoui2015} & \ac{SFM} & Single & -4.07 & 32.9 & 0.136 & 0.444\\ 
\hline
8 & Per acoust. band SRMR {\sectMidSent} 2.5.~\cite{Senoussaoui2015} & \ac{SFM} & Single & -2.45 & 24.9 & -0.201 & 0.579\\ 
\hline
9 & Temporal dynamics~\cite{Falk2009} & \ac{SFM} & Single & -12.4 & 172 & -0.136 & 0.0823\\ 
\hline
$\alpha$ & QA Reverb~\cite{Prego2015} & \ac{SFM} & Single & 2.96 & 26.3 & -0.0876 & 0.392\\ 
\hline
$\beta$ & Blind est. of coherent-to-diffuse energy ratio~\cite{Jeub2011} & \ac{ABC} & Chromebook & -15.3 & 241 & 0.244 & 0.019\\ 
\hline

\else

\fi
\end{tabular}
\end{table*}
\clearpage
\subsubsection{Fan noise at \dBel{18} \ac{SNR}}
\begin{figure}[!ht]
	\ifarXiv
\centerline{\epsfig{figure=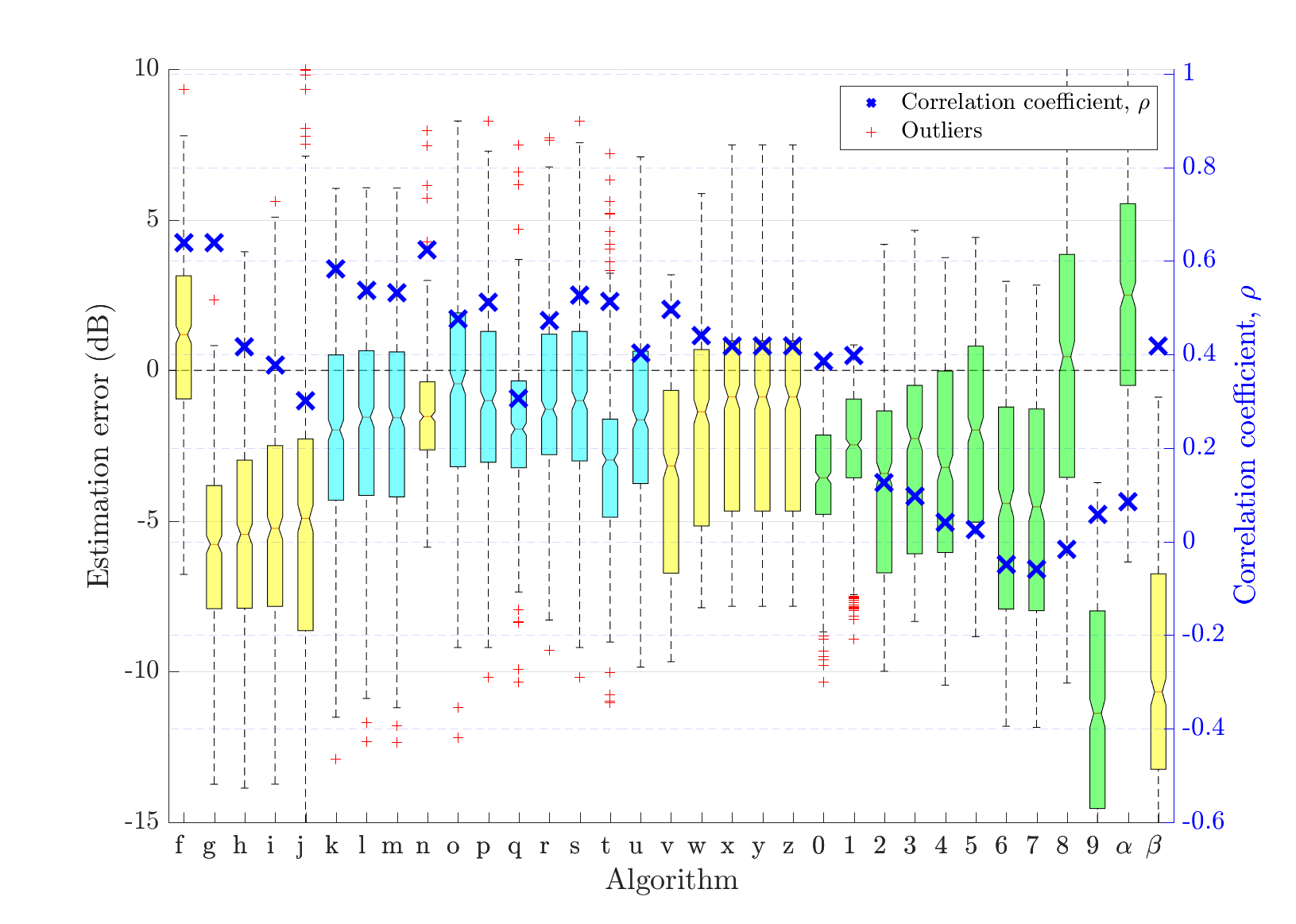,
	width=\figWidthACETR,viewport=45 10 765 530,clip}}%
	\else
	\centerline{\epsfig{figure=FigsACE/ana_eval_gt_partic_results_combined_Phase3_All_WASPAA_P3_DRR_dB_H_Fan.png,
	width=\figWidthACETR,viewport=45 10 765 530,clip}}%
	\fi
	\caption{Fullband {\ac{DRR} estimation error in fan noise at \dBel{18} \ac{SNR}}}%
\label{fig:ACE_DRR_Fan_H}%
\end{figure}%
\begin{table*}[!ht]\small
\caption{\ac{DRR} estimation algorithm performance in fan noise at \dBel{18} \ac{SNR}}
\vspace{5mm} 
\centering
\begin{tabular}{clllllll}%
\hline%
Ref.
& Algorithm
& Class
& Mic. Config.
& Bias
& MSE
&  $\PearsonCC$
& \ac{RTF}
\\
\hline
\hline
\ifarXiv
f & PSD est. in beamspace, bias comp.~\cite{Hioka2015} & \ac{ABC} & Mobile & 1.13 & 7.67 & 0.638 & 0.757\\ 
\hline
g & PSD est. in beamspace (Raw)~\cite{Hioka2015} & \ac{ABC} & Mobile & -5.83 & 40.4 & 0.639 & 3.18\\ 
\hline
h & PSD est. in beamspace v2~\cite{Hioka2015} & \ac{ABC} & Mobile & -5.38 & 40.9 & 0.416 & 0.844\\ 
\hline
i & PSD est. by twin BF~\cite{Hioka2012} & \ac{ABC} & Mobile & -5.09 & 39.4 & 0.378 & 0.613\\ 
\hline
j & Spatial Covariance in matrix mode~\cite{Hioka2011} & \ac{ABC} & Mobile & -5.18 & 52.6 & 0.301 & 0.627\\ 
\hline
k & NIRAv2~\cite{Parada2015} & \ac{MLMF} & Single & -1.97 & 14.8 & 0.583 & 0.895$^\dagger$\\ 
\hline
l & NIRAv3~\cite{Parada2015} & \ac{MLMF} & Single & -1.85 & 15.1 & 0.537 & 0.895$^\dagger$\\ 
\hline
m & NIRAv1~\cite{Parada2015} & \ac{MLMF} & Single & -1.86 & 15.3 & 0.531 & 0.895$^\dagger$\\ 
\hline
n & Particle velocity~\cite{Chen2015} & \ac{ABC} & EM32 & -1.45 & 4.9 & 0.624 & 0.134\\ 
\hline
o & Multi-layer perceptron~\cite{Xiong2015} & \ac{MLMF} & Single & -0.829 & 14.4 & 0.476 & 0.0578$^\ddagger$\\ 
\hline
p & Multi-layer perceptron P2~\cite{Xiong2015} & \ac{MLMF} & Single & -0.883 & 14.3 & 0.512 & 0.0578$^\ddagger$\\ 
\hline
q & Multi-layer perceptron P2~\cite{Xiong2015} & \ac{MLMF} & Chromebook & -2.09 & 10.8 & 0.306 & 0.059$^\ddagger$\\ 
\hline
r & Multi-layer perceptron P2~\cite{Xiong2015} & \ac{MLMF} & Mobile & -0.997 & 11.7 & 0.472 & 0.0555$^\ddagger$\\ 
\hline
s & Multi-layer perceptron P2~\cite{Xiong2015} & \ac{MLMF} & Crucif & -0.985 & 14.3 & 0.526 & 0.0569$^\ddagger$\\ 
\hline
t & Multi-layer perceptron P2~\cite{Xiong2015} & \ac{MLMF} & Lin8Ch & -3.19 & 19.3 & 0.514 & 0.0617$^\ddagger$\\ 
\hline
u & Multi-layer perceptron P2~\cite{Xiong2015} & \ac{MLMF} & EM32 & -1.61 & 11.2 & 0.403 & 0.0574$^\ddagger$\\ 
\hline
v & \ac{DENBE} no noise reduction~\cite{Eaton2015} & \ac{ABC} & Chromebook & -3.49 & 22.8 & 0.497 & 0.0322\\ 
\hline
w & \ac{DENBE} spectral subtraction~\cite{Eaton2015c} & \ac{ABC} & Chromebook & -1.83 & 15.5 & 0.439 & 0.0588\\ 
\hline
x & \ac{DENBE} spec. sub. Gerkmann~\cite{Eaton2015} & \ac{ABC} & Chromebook & -1.55 & 15.2 & 0.418 & 0.048\\ 
\hline
y & \ac{DENBE} filtered subbands~\cite{Eaton2015c} & \ac{ABC} & Chromebook & -1.55 & 15.2 & 0.418 & 0.774\\ 
\hline
z & \ac{DENBE} FFT derived subbands~\cite{Eaton2015c} & \ac{ABC} & Chromebook & -1.55 & 15.2 & 0.418 & 0.0452\\ 
\hline
0 & \ac{NOSRMR} {\sectMidSent} 2.2.~\cite{Senoussaoui2015} & \ac{SFM} & Chromebook & -3.72 & 18.2 & 0.386 & 1.03\\ 
\hline
1 & \ac{OSRMR} {\sectMidSent} 2.2.~\cite{Senoussaoui2015} & \ac{SFM} & Chromebook & -2.56 & 10.9 & 0.398 & 0.824\\ 
\hline
2 & \ac{NOSRMR} {\sectMidSent} 2.2.~\cite{Senoussaoui2015} & \ac{SFM} & Mobile & -3.58 & 23.4 & 0.127 & 1.58\\ 
\hline
3 & \ac{OSRMR} {\sectMidSent} 2.2.~\cite{Senoussaoui2015} & \ac{SFM} & Mobile & -2.55 & 16.9 & 0.0964 & 1.26\\ 
\hline
4 & \ac{NOSRMR} {\sectMidSent} 2.2.~\cite{Senoussaoui2015} & \ac{SFM} & Crucif & -3.2 & 23.5 & 0.0412 & 2.61\\ 
\hline
5 & \ac{OSRMR} {\sectMidSent} 2.2.~\cite{Senoussaoui2015} & \ac{SFM} & Crucif & -2.17 & 17.6 & 0.0261 & 2.08\\ 
\hline
6 & \ac{NOSRMR} {\sectMidSent} 2.2.~\cite{Senoussaoui2015} & \ac{SFM} & Single & -4.22 & 34.3 & -0.0485 & 0.543\\ 
\hline
7 & \ac{OSRMR} {\sectMidSent} 2.2.~\cite{Senoussaoui2015} & \ac{SFM} & Single & -4.3 & 35 & -0.0587 & 0.447\\ 
\hline
8 & Per acoust. band SRMR {\sectMidSent} 2.5.~\cite{Senoussaoui2015} & \ac{SFM} & Single & 0.136 & 22.2 & -0.0161 & 0.576\\ 
\hline
9 & Temporal dynamics~\cite{Falk2009} & \ac{SFM} & Single & -11.1 & 140 & 0.0587 & 0.0818\\ 
\hline
$\alpha$ & QA Reverb~\cite{Prego2015} & \ac{SFM} & Single & 2.39 & 22.9 & 0.086 & 0.391\\ 
\hline
$\beta$ & Blind est. of coherent-to-diffuse energy ratio~\cite{Jeub2011} & \ac{ABC} & Chromebook & -10.2 & 120 & 0.419 & 0.019\\ 
\hline

\else

\fi
\end{tabular}
\end{table*}
\clearpage
\subsubsection{Fan noise at \dBel{12} \ac{SNR}}
\begin{figure}[!ht]
	\ifarXiv
\centerline{\epsfig{figure=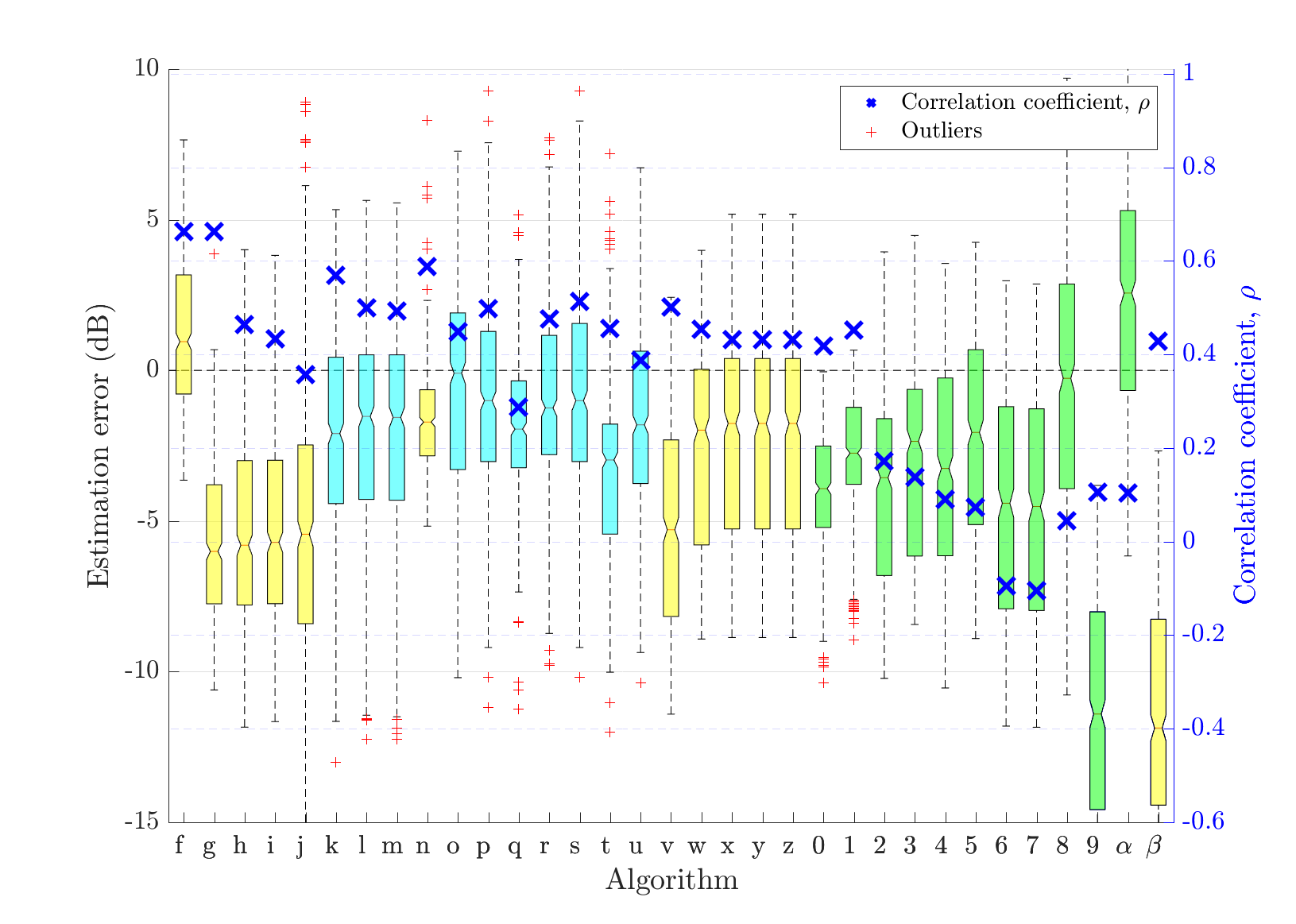,
	width=\figWidthACETR,viewport=45 10 765 530,clip}}%
	\else
	\centerline{\epsfig{figure=FigsACE/ana_eval_gt_partic_results_combined_Phase3_All_WASPAA_P3_DRR_dB_M_Fan.png,
	width=\figWidthACETR,viewport=45 10 765 530,clip}}%
	\fi
	\caption{Fullband {\ac{DRR} estimation error in fan noise at \dBel{12} \ac{SNR}}}%
\label{fig:ACE_DRR_Fan_M}%
\end{figure}%
\begin{table*}[!ht]\small
\caption{\ac{DRR} estimation algorithm performance in fan noise at \dBel{12} \ac{SNR}}
\vspace{5mm} 
\centering
\begin{tabular}{clllllll}%
\hline%
Ref.
& Algorithm
& Class
& Mic. Config.
& Bias
& MSE
&  $\PearsonCC$
& \ac{RTF}
\\
\hline
\hline
\ifarXiv
f & PSD est. in beamspace, bias comp.~\cite{Hioka2015} & \ac{ABC} & Mobile & 1.15 & 7.15 & 0.662 & 0.757\\ 
\hline
g & PSD est. in beamspace (Raw)~\cite{Hioka2015} & \ac{ABC} & Mobile & -5.81 & 39.6 & 0.662 & 3.18\\ 
\hline
h & PSD est. in beamspace v2~\cite{Hioka2015} & \ac{ABC} & Mobile & -5.48 & 39.7 & 0.464 & 0.844\\ 
\hline
i & PSD est. by twin BF~\cite{Hioka2012} & \ac{ABC} & Mobile & -5.3 & 38.6 & 0.434 & 0.613\\ 
\hline
j & Spatial Covariance in matrix mode~\cite{Hioka2011} & \ac{ABC} & Mobile & -5.4 & 50.5 & 0.357 & 0.627\\ 
\hline
k & NIRAv2~\cite{Parada2015} & \ac{MLMF} & Single & -2.04 & 15.3 & 0.57 & 0.895$^\dagger$\\ 
\hline
l & NIRAv3~\cite{Parada2015} & \ac{MLMF} & Single & -1.92 & 16.1 & 0.5 & 0.895$^\dagger$\\ 
\hline
m & NIRAv1~\cite{Parada2015} & \ac{MLMF} & Single & -1.93 & 16.3 & 0.493 & 0.895$^\dagger$\\ 
\hline
n & Particle velocity~\cite{Chen2015} & \ac{ABC} & EM32 & -1.67 & 5.69 & 0.589 & 0.134\\ 
\hline
o & Multi-layer perceptron~\cite{Xiong2015} & \ac{MLMF} & Single & -0.743 & 14.4 & 0.449 & 0.0578$^\ddagger$\\ 
\hline
p & Multi-layer perceptron P2~\cite{Xiong2015} & \ac{MLMF} & Single & -0.941 & 14.7 & 0.498 & 0.0578$^\ddagger$\\ 
\hline
q & Multi-layer perceptron P2~\cite{Xiong2015} & \ac{MLMF} & Chromebook & -2.06 & 10.1 & 0.288 & 0.059$^\ddagger$\\ 
\hline
r & Multi-layer perceptron P2~\cite{Xiong2015} & \ac{MLMF} & Mobile & -0.905 & 11.7 & 0.476 & 0.0555$^\ddagger$\\ 
\hline
s & Multi-layer perceptron P2~\cite{Xiong2015} & \ac{MLMF} & Crucif & -0.893 & 14 & 0.513 & 0.0569$^\ddagger$\\ 
\hline
t & Multi-layer perceptron P2~\cite{Xiong2015} & \ac{MLMF} & Lin8Ch & -3.22 & 20.5 & 0.456 & 0.0617$^\ddagger$\\ 
\hline
u & Multi-layer perceptron P2~\cite{Xiong2015} & \ac{MLMF} & EM32 & -1.65 & 11.8 & 0.388 & 0.0574$^\ddagger$\\ 
\hline
v & \ac{DENBE} no noise reduction~\cite{Eaton2015} & \ac{ABC} & Chromebook & -5.06 & 36.3 & 0.501 & 0.0322\\ 
\hline
w & \ac{DENBE} spectral subtraction~\cite{Eaton2015c} & \ac{ABC} & Chromebook & -2.58 & 17.6 & 0.455 & 0.0588\\ 
\hline
x & \ac{DENBE} spec. sub. Gerkmann~\cite{Eaton2015} & \ac{ABC} & Chromebook & -2.21 & 16.4 & 0.432 & 0.048\\ 
\hline
y & \ac{DENBE} filtered subbands~\cite{Eaton2015c} & \ac{ABC} & Chromebook & -2.21 & 16.4 & 0.432 & 0.774\\ 
\hline
z & \ac{DENBE} FFT derived subbands~\cite{Eaton2015c} & \ac{ABC} & Chromebook & -2.21 & 16.4 & 0.432 & 0.0452\\ 
\hline
0 & \ac{NOSRMR} {\sectMidSent} 2.2.~\cite{Senoussaoui2015} & \ac{SFM} & Chromebook & -4.07 & 20.8 & 0.418 & 1.03\\ 
\hline
1 & \ac{OSRMR} {\sectMidSent} 2.2.~\cite{Senoussaoui2015} & \ac{SFM} & Chromebook & -2.78 & 11.9 & 0.452 & 0.824\\ 
\hline
2 & \ac{NOSRMR} {\sectMidSent} 2.2.~\cite{Senoussaoui2015} & \ac{SFM} & Mobile & -3.74 & 24.4 & 0.172 & 1.58\\ 
\hline
3 & \ac{OSRMR} {\sectMidSent} 2.2.~\cite{Senoussaoui2015} & \ac{SFM} & Mobile & -2.65 & 17.3 & 0.137 & 1.26\\ 
\hline
4 & \ac{NOSRMR} {\sectMidSent} 2.2.~\cite{Senoussaoui2015} & \ac{SFM} & Crucif & -3.36 & 24.3 & 0.09 & 2.61\\ 
\hline
5 & \ac{OSRMR} {\sectMidSent} 2.2.~\cite{Senoussaoui2015} & \ac{SFM} & Crucif & -2.27 & 17.8 & 0.0731 & 2.08\\ 
\hline
6 & \ac{NOSRMR} {\sectMidSent} 2.2.~\cite{Senoussaoui2015} & \ac{SFM} & Single & -4.21 & 34.3 & -0.0954 & 0.543\\ 
\hline
7 & \ac{OSRMR} {\sectMidSent} 2.2.~\cite{Senoussaoui2015} & \ac{SFM} & Single & -4.29 & 34.9 & -0.105 & 0.447\\ 
\hline
8 & Per acoust. band SRMR {\sectMidSent} 2.5.~\cite{Senoussaoui2015} & \ac{SFM} & Single & -0.669 & 20.3 & 0.0451 & 0.576\\ 
\hline
9 & Temporal dynamics~\cite{Falk2009} & \ac{SFM} & Single & -11.2 & 141 & 0.105 & 0.0818\\ 
\hline
$\alpha$ & QA Reverb~\cite{Prego2015} & \ac{SFM} & Single & 2.41 & 22.7 & 0.104 & 0.391\\ 
\hline
$\beta$ & Blind est. of coherent-to-diffuse energy ratio~\cite{Jeub2011} & \ac{ABC} & Chromebook & -11.5 & 147 & 0.428 & 0.019\\ 
\hline

\else

\fi
\end{tabular}
\end{table*}
\clearpage
\subsubsection{Fan noise at \dBel{-1} \ac{SNR}}
\begin{figure}[!ht]
	\ifarXiv
\centerline{\epsfig{figure=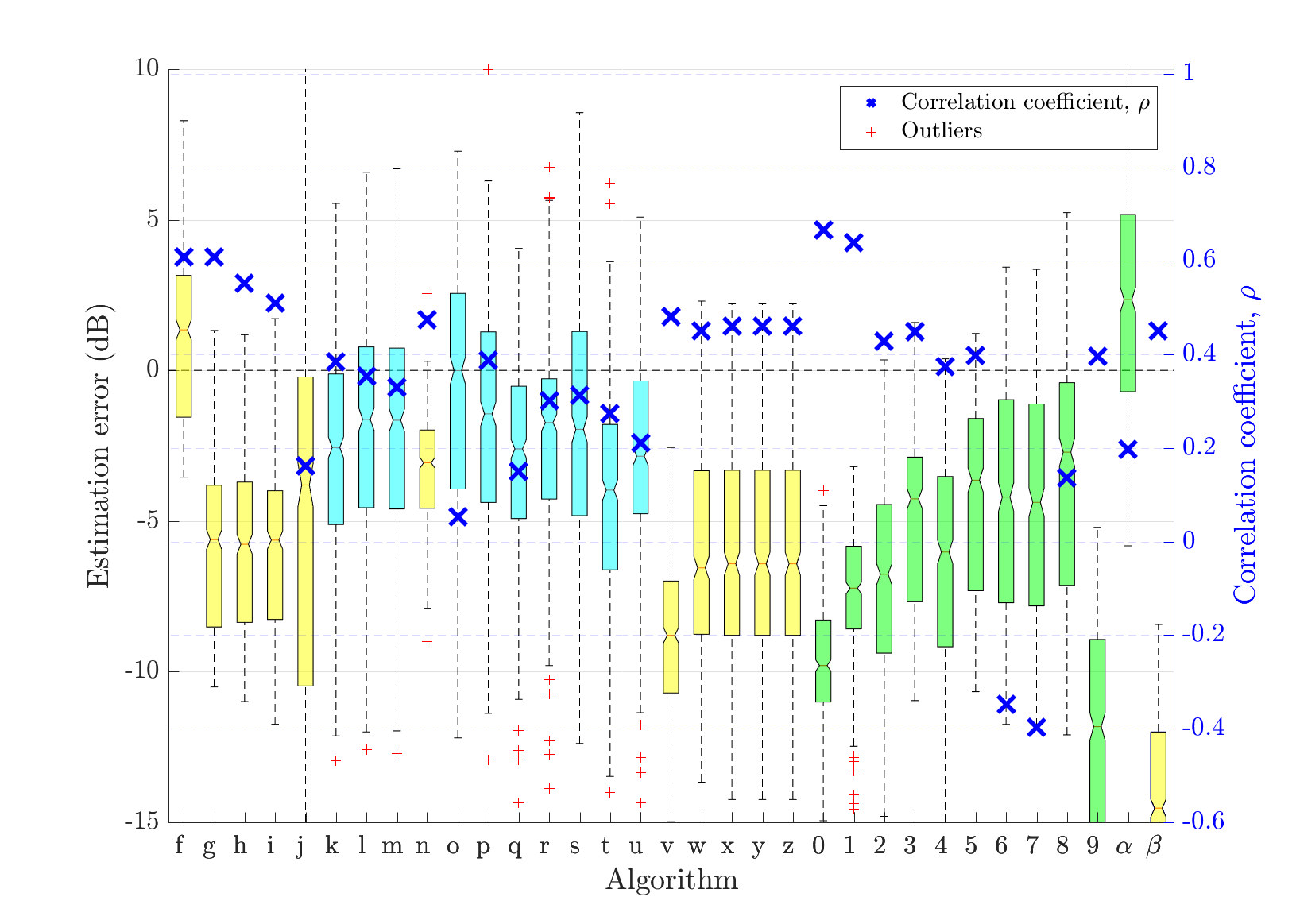,
	width=\figWidthACETR,viewport=45 10 765 530,clip}}%
	\else
	\centerline{\epsfig{figure=FigsACE/ana_eval_gt_partic_results_combined_Phase3_All_WASPAA_P3_DRR_dB_L_Fan.png,
	width=\figWidthACETR,viewport=45 10 765 530,clip}}%
	\fi
	\caption{Fullband {\ac{DRR} estimation error in fan noise at \dBel{-1} \ac{SNR}}}%
\label{fig:ACE_DRR_Fan_L}%
\end{figure}%
\begin{table*}[!ht]\small
\caption{\ac{DRR} estimation algorithm performance in fan noise at \dBel{-1} \ac{SNR}}
\vspace{5mm} 
\centering
\begin{tabular}{clllllll}%
\hline%
Ref.
& Algorithm
& Class
& Mic. Config.
& Bias
& MSE
&  $\PearsonCC$
& \ac{RTF}
\\
\hline
\hline
\ifarXiv
f & PSD est. in beamspace, bias comp.~\cite{Hioka2015} & \ac{ABC} & Mobile & 1.2 & 8.85 & 0.609 & 0.757\\ 
\hline
g & PSD est. in beamspace (Raw)~\cite{Hioka2015} & \ac{ABC} & Mobile & -5.76 & 40.6 & 0.609 & 3.18\\ 
\hline
h & PSD est. in beamspace v2~\cite{Hioka2015} & \ac{ABC} & Mobile & -5.75 & 40.6 & 0.553 & 0.844\\ 
\hline
i & PSD est. by twin BF~\cite{Hioka2012} & \ac{ABC} & Mobile & -5.86 & 42.1 & 0.51 & 0.613\\ 
\hline
j & Spatial Covariance in matrix mode~\cite{Hioka2011} & \ac{ABC} & Mobile & -5.4 & 81.2 & 0.161 & 0.627\\ 
\hline
k & NIRAv2~\cite{Parada2015} & \ac{MLMF} & Single & -2.67 & 21.5 & 0.384 & 0.895$^\dagger$\\ 
\hline
l & NIRAv3~\cite{Parada2015} & \ac{MLMF} & Single & -1.89 & 18.3 & 0.354 & 0.895$^\dagger$\\ 
\hline
m & NIRAv1~\cite{Parada2015} & \ac{MLMF} & Single & -1.98 & 19.1 & 0.33 & 0.895$^\dagger$\\ 
\hline
n & Particle velocity~\cite{Chen2015} & \ac{ABC} & EM32 & -3.32 & 14.3 & 0.474 & 0.134\\ 
\hline
o & Multi-layer perceptron~\cite{Xiong2015} & \ac{MLMF} & Single & -0.747 & 18.7 & 0.053 & 0.0578$^\ddagger$\\ 
\hline
p & Multi-layer perceptron P2~\cite{Xiong2015} & \ac{MLMF} & Single & -1.77 & 18.7 & 0.387 & 0.0578$^\ddagger$\\ 
\hline
q & Multi-layer perceptron P2~\cite{Xiong2015} & \ac{MLMF} & Chromebook & -3.08 & 19.2 & 0.149 & 0.059$^\ddagger$\\ 
\hline
r & Multi-layer perceptron P2~\cite{Xiong2015} & \ac{MLMF} & Mobile & -2.26 & 18.3 & 0.302 & 0.0555$^\ddagger$\\ 
\hline
s & Multi-layer perceptron P2~\cite{Xiong2015} & \ac{MLMF} & Crucif & -1.83 & 20.6 & 0.312 & 0.0569$^\ddagger$\\ 
\hline
t & Multi-layer perceptron P2~\cite{Xiong2015} & \ac{MLMF} & Lin8Ch & -4.44 & 32.6 & 0.274 & 0.0617$^\ddagger$\\ 
\hline
u & Multi-layer perceptron P2~\cite{Xiong2015} & \ac{MLMF} & EM32 & -2.86 & 18.6 & 0.212 & 0.0574$^\ddagger$\\ 
\hline
v & \ac{DENBE} no noise reduction~\cite{Eaton2015} & \ac{ABC} & Chromebook & -8.75 & 83 & 0.481 & 0.0322\\ 
\hline
w & \ac{DENBE} spectral subtraction~\cite{Eaton2015c} & \ac{ABC} & Chromebook & -6.05 & 47.6 & 0.45 & 0.0588\\ 
\hline
x & \ac{DENBE} spec. sub. Gerkmann~\cite{Eaton2015} & \ac{ABC} & Chromebook & -6.05 & 47.5 & 0.46 & 0.048\\ 
\hline
y & \ac{DENBE} filtered subbands~\cite{Eaton2015c} & \ac{ABC} & Chromebook & -6.05 & 47.5 & 0.46 & 0.774\\ 
\hline
z & \ac{DENBE} FFT derived subbands~\cite{Eaton2015c} & \ac{ABC} & Chromebook & -6.05 & 47.5 & 0.46 & 0.0452\\ 
\hline
0 & \ac{NOSRMR} {\sectMidSent} 2.2.~\cite{Senoussaoui2015} & \ac{SFM} & Chromebook & -9.66 & 97.8 & 0.667 & 1.03\\ 
\hline
1 & \ac{OSRMR} {\sectMidSent} 2.2.~\cite{Senoussaoui2015} & \ac{SFM} & Chromebook & -7.29 & 57.5 & 0.639 & 0.824\\ 
\hline
2 & \ac{NOSRMR} {\sectMidSent} 2.2.~\cite{Senoussaoui2015} & \ac{SFM} & Mobile & -6.9 & 57 & 0.429 & 1.58\\ 
\hline
3 & \ac{OSRMR} {\sectMidSent} 2.2.~\cite{Senoussaoui2015} & \ac{SFM} & Mobile & -4.78 & 31.1 & 0.449 & 1.26\\ 
\hline
4 & \ac{NOSRMR} {\sectMidSent} 2.2.~\cite{Senoussaoui2015} & \ac{SFM} & Crucif & -6.34 & 52 & 0.374 & 2.61\\ 
\hline
5 & \ac{OSRMR} {\sectMidSent} 2.2.~\cite{Senoussaoui2015} & \ac{SFM} & Crucif & -4.35 & 29.5 & 0.398 & 2.08\\ 
\hline
6 & \ac{NOSRMR} {\sectMidSent} 2.2.~\cite{Senoussaoui2015} & \ac{SFM} & Single & -3.99 & 32.9 & -0.347 & 0.543\\ 
\hline
7 & \ac{OSRMR} {\sectMidSent} 2.2.~\cite{Senoussaoui2015} & \ac{SFM} & Single & -4.13 & 34 & -0.397 & 0.447\\ 
\hline
8 & Per acoust. band SRMR {\sectMidSent} 2.5.~\cite{Senoussaoui2015} & \ac{SFM} & Single & -3.44 & 28.6 & 0.136 & 0.576\\ 
\hline
9 & Temporal dynamics~\cite{Falk2009} & \ac{SFM} & Single & -12.1 & 160 & 0.397 & 0.0818\\ 
\hline
$\alpha$ & QA Reverb~\cite{Prego2015} & \ac{SFM} & Single & 2.22 & 20.8 & 0.198 & 0.391\\ 
\hline
$\beta$ & Blind est. of coherent-to-diffuse energy ratio~\cite{Jeub2011} & \ac{ABC} & Chromebook & -14.3 & 211 & 0.451 & 0.019\\ 
\hline

\else

\fi
\end{tabular}
\end{table*}
\clearpage
%
%
\subsection{Frequency-dependent \ac{DRR} estimation results}
\begin{figure}[!ht]
	\ifarXiv
\centerline{\epsfig{figure=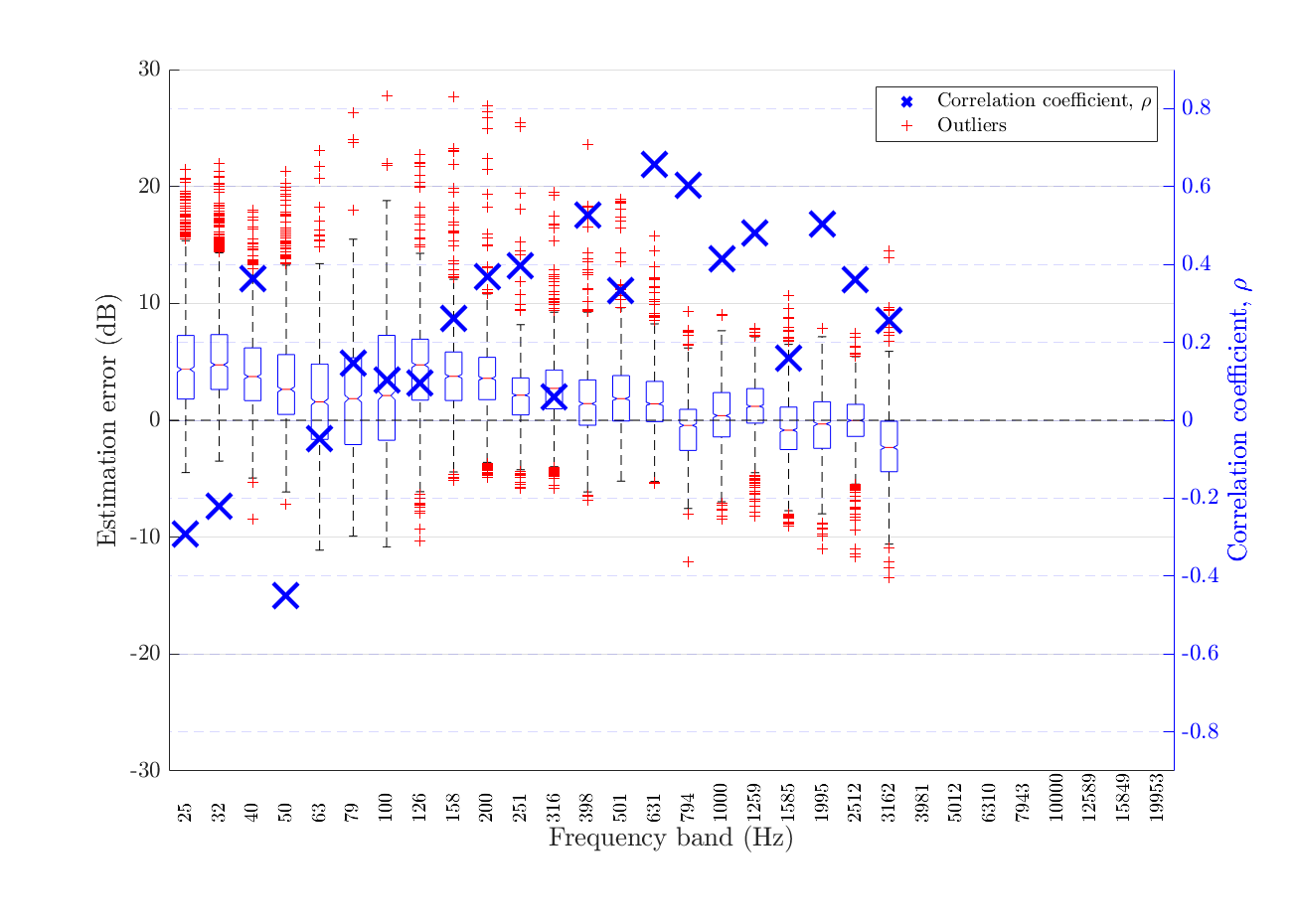,
	width=\figWidthACETR,viewport=45 10 765 530,clip}}%
	\else
	\centerline{\epsfig{figure=FigsACE/ana_eval_gt_partic_results_combined_Phase3_TR_P3S_DRR_dB_All_SNR_All_Noises_sub_Velocity.png,
	width=\figWidthACETR,viewport=45 10 765 530,clip}}%
	\fi
	\caption{{Frequency-dependent \ac{DRR} estimation error in all noises for all \acp{SNR} for algorithm Particle Velocity~\cite{Chen2015}}}%
\label{fig:ACE_DRR_Sub_All_Velocity}%
\end{figure}%
\begin{table*}[!ht]\small
\caption{Frequency-dependent \ac{DRR} estimation error in all noises for all \acp{SNR} for algorithm Particle Velocity~\cite{Chen2015}}
\vspace{5mm} 
\centering
\begin{tabular}{crrrl}%
\hline%
Freq. band
& Centre Freq. (Hz)
& Bias
& \acs{MSE}
& $\PearsonCC$

\\
\hline
\hline
\ifarXiv
 1 & 25.12 & 5.915 & 59.49 & -0.2928 \\ 
\hline
 2 & 31.62 & 6.399 & 62.41 & -0.2196 \\ 
\hline
 3 & 39.81 & 4.915 & 40.5 & 0.3622 \\ 
\hline
 4 & 50.12 & 3.624 & 37.37 & -0.4494 \\ 
\hline
 5 & 63.10 & 1.311 & 23.91 & -0.04796 \\ 
\hline
 6 & 79.43 & 0.6641 & 20.28 & 0.1454 \\ 
\hline
 7 & 100.00 & 1.154 & 25.77 & 0.104 \\ 
\hline
 8 & 125.89 & 1.867 & 27.15 & 0.09463 \\ 
\hline
 9 & 158.49 & 2.251 & 20.78 & 0.2602 \\ 
\hline
10 & 199.53 & 2.809 & 18.57 & 0.3691 \\ 
\hline
11 & 251.19 & 1.455 & 11.06 & 0.3971 \\ 
\hline
12 & 316.23 & 1.66 & 14.86 & 0.05926 \\ 
\hline
13 & 398.11 & 1.008 & 12.1 & 0.5263 \\ 
\hline
14 & 501.19 & 1.512 & 14.1 & 0.334 \\ 
\hline
15 & 630.96 & 1.12 & 11.15 & 0.6573 \\ 
\hline
16 & 794.33 & -1.25 & 11.12 & 0.6027 \\ 
\hline
17 & 1000.00 & -0.2177 & 10.31 & 0.4151 \\ 
\hline
18 & 1258.93 & 0.6023 & 9.464 & 0.4811 \\ 
\hline
19 & 1584.89 & -1.002 & 10.28 & 0.1604 \\ 
\hline
20 & 1995.26 & -0.8029 & 8.668 & 0.5033 \\ 
\hline
21 & 2511.89 & -0.7828 & 8.903 & 0.3599 \\ 
\hline
22 & 3162.28 & -2.95 & 22.64 & 0.2575 \\ 
\hline

\else

\fi
\end{tabular}
\end{table*}
%
%
%
\begin{figure}[!ht]
	\ifarXiv
\centerline{\epsfig{figure=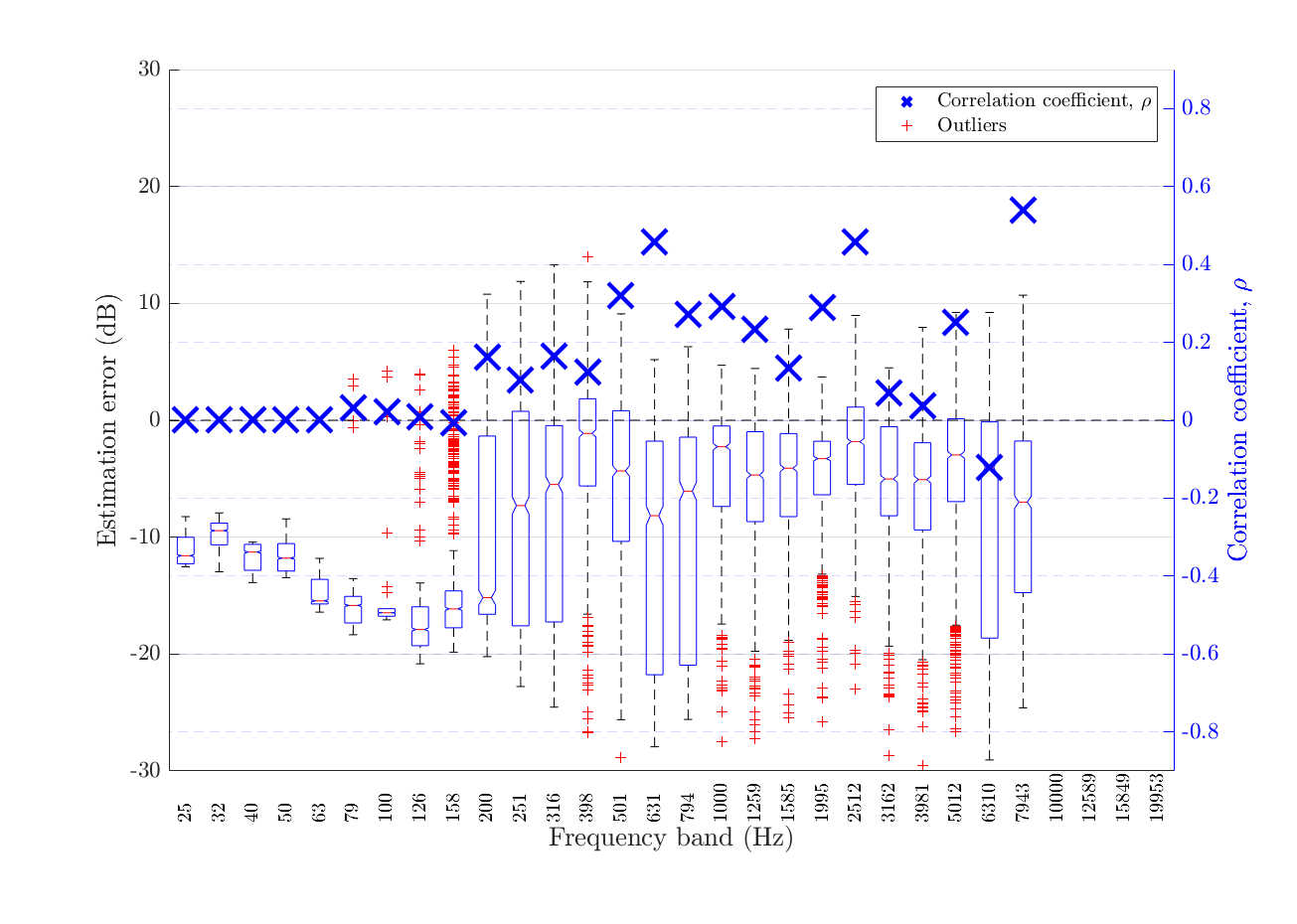,
	width=\figWidthACETR,viewport=45 10 765 530,clip}}%
	\else
	\centerline{\epsfig{figure=FigsACE/ana_eval_gt_partic_results_combined_Phase3_TR_P3S_DRR_dB_All_SNR_All_Noises_sub_ICASSP_2015_DRR_2-ch_Gerkmann_NR_FFT_subband.png,
	width=\figWidthACETR,viewport=45 10 765 530,clip}}%
	\fi
	\caption{{Frequency-dependent \ac{DRR} estimation error in all noises for all \acp{SNR} for algorithm \ac{DENBE} with FFT derived subbands~\cite{Eaton2015c}}}%
\label{fig:ACE_DRR_Sub_All_Eaton_FFT}%
\end{figure}%
\begin{table*}[!ht]\small
\caption{Frequency-dependent \ac{DRR} estimation error in all noises for all \acp{SNR} for algorithm \ac{DENBE} with FFT derived subbands~\cite{Eaton2015c}}
\vspace{5mm} 
\centering
\begin{tabular}{crrrl}%
\hline%
Freq. band
& Centre Freq. (Hz)
& Bias
& \acs{MSE}
& $\PearsonCC$

\\
\hline
\hline
\ifarXiv
 1 & 25.12 & -11.07 & 124.4 & 0 \\ 
\hline
 2 & 31.62 & -9.93 & 100.8 & 0 \\ 
\hline
 3 & 39.81 & -11.69 & 138.1 & 0 \\ 
\hline
 4 & 50.12 & -11.55 & 135.4 & 0 \\ 
\hline
 5 & 63.10 & -14.82 & 221.5 & 0 \\ 
\hline
 6 & 79.43 & -16.07 & 260.2 & 0.0318 \\ 
\hline
 7 & 100.00 & -16.16 & 262.3 & 0.02187 \\ 
\hline
 8 & 125.89 & -17.68 & 317.1 & 0.009427 \\ 
\hline
 9 & 158.49 & -16.39 & 276.4 & -0.005167 \\ 
\hline
10 & 199.53 & -13.75 & 231.9 & 0.1624 \\ 
\hline
11 & 251.19 & -12.72 & 233 & 0.1023 \\ 
\hline
12 & 316.23 & -14.05 & 275.1 & 0.1637 \\ 
\hline
13 & 398.11 & -10.56 & 228.8 & 0.1242 \\ 
\hline
14 & 501.19 & -11.52 & 237.1 & 0.3193 \\ 
\hline
15 & 630.96 & -14.88 & 312.2 & 0.4584 \\ 
\hline
16 & 794.33 & -13.78 & 282.8 & 0.2725 \\ 
\hline
17 & 1000.00 & -10.74 & 211.8 & 0.2908 \\ 
\hline
18 & 1258.93 & -10.4 & 187.5 & 0.2331 \\ 
\hline
19 & 1584.89 & -8.162 & 140.1 & 0.1336 \\ 
\hline
20 & 1995.26 & -7.311 & 102.7 & 0.2892 \\ 
\hline
21 & 2511.89 & -5.217 & 74.37 & 0.4572 \\ 
\hline
22 & 3162.28 & -6.642 & 82.46 & 0.07081 \\ 
\hline
23 & 3981.07 & -7.008 & 81.94 & 0.03811 \\ 
\hline
24 & 5011.87 & -6.333 & 95.62 & 0.2504 \\ 
\hline
25 & 6309.57 & -9.64 & 207.7 & -0.1221 \\ 
\hline
26 & 7943.28 & -9.647 & 166.5 & 0.5396 \\ 
\hline

\else

\fi
\end{tabular}
\end{table*}
%
%
%
%
\begin{figure}[!ht]
	\ifarXiv
\centerline{\epsfig{figure=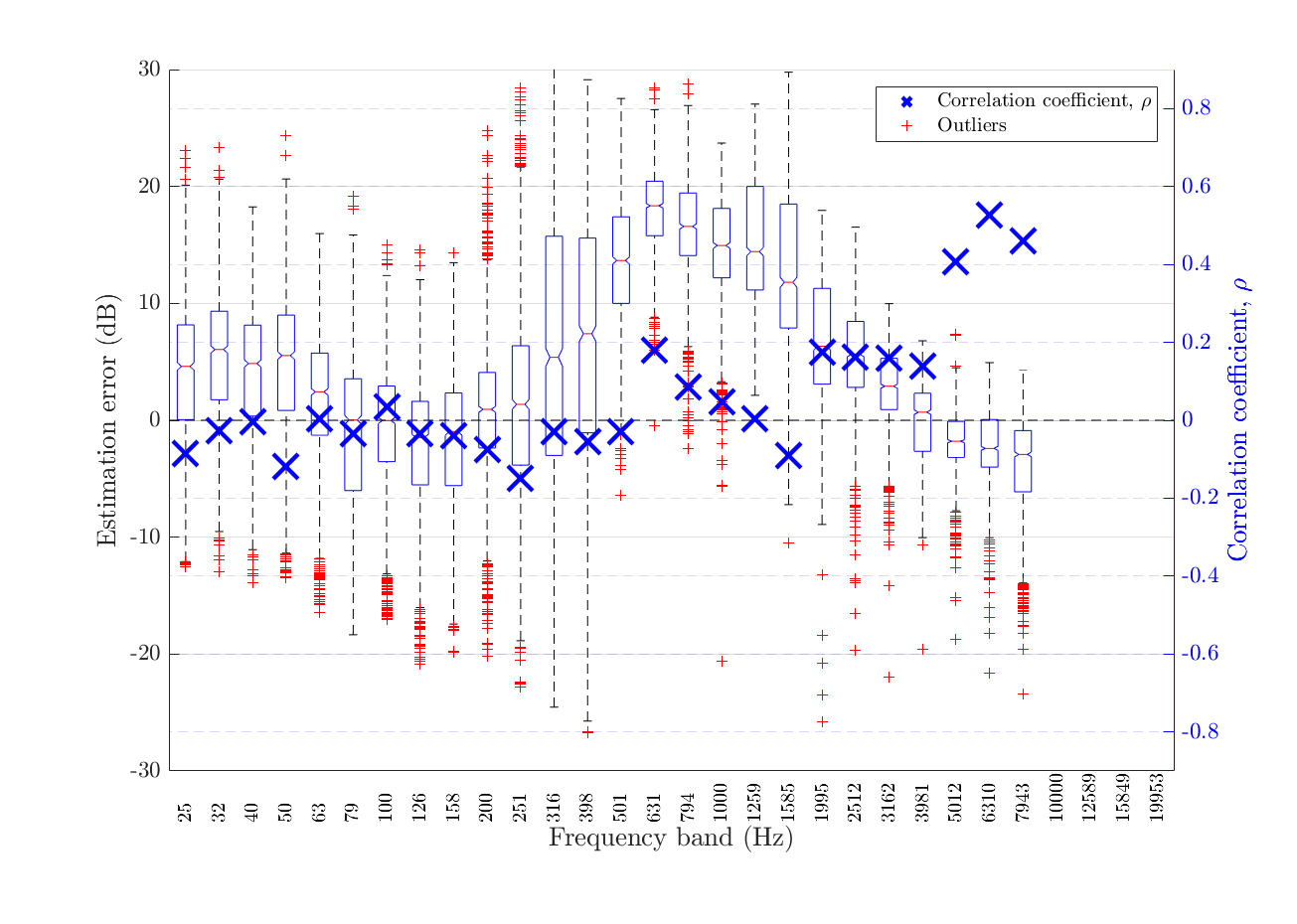,
	width=\figWidthACETR,viewport=45 10 765 530,clip}}%
	\else
	\centerline{\epsfig{figure=FigsACE/ana_eval_gt_partic_results_combined_Phase3_TR_P3S_DRR_dB_All_SNR_All_Noises_sub_ICASSP_2015_DRR_2-ch_Gerkmann_NR_filtered_subband.png,
	width=\figWidthACETR,viewport=45 10 765 530,clip}}%
	\fi
	\caption{{Frequency-dependent \ac{DRR} estimation error in all noises for all \acp{SNR} for algorithm \ac{DENBE} with filtered subbands~\cite{Eaton2015c}}}%
\label{fig:ACE_DRR_Sub_All_Eaton_filt}%
\end{figure}%
\begin{table*}[!ht]\small
\caption{Frequency-dependent \ac{DRR} estimation error in all noises for all \acp{SNR} for algorithm \ac{DENBE} with filtered subbands~\cite{Eaton2015c}}
\vspace{5mm} 
\centering
\begin{tabular}{crrrl}%
\hline%
Freq. band
& Centre Freq. (Hz)
& Bias
& \acs{MSE}
& $\PearsonCC$

\\
\hline
\hline
\ifarXiv
 1 & 25.12 & 2.367 & 60.71 & -0.08501 \\ 
\hline
 2 & 31.62 & 3.618 & 64.55 & -0.0276 \\ 
\hline
 3 & 39.81 & 2.117 & 52.78 & -0.004386 \\ 
\hline
 4 & 50.12 & 2.7 & 61.79 & -0.1188 \\ 
\hline
 5 & 63.10 & -0.01441 & 47.99 & 0.004362 \\ 
\hline
 6 & 79.43 & -2.464 & 67.34 & -0.03461 \\ 
\hline
 7 & 100.00 & -2.482 & 50.57 & 0.03417 \\ 
\hline
 8 & 125.89 & -3.691 & 58.41 & -0.03403 \\ 
\hline
 9 & 158.49 & -3.092 & 59.07 & -0.03996 \\ 
\hline
10 & 199.53 & -1.593 & 51.98 & -0.07489 \\ 
\hline
11 & 251.19 & -1.831 & 81.92 & -0.1493 \\ 
\hline
12 & 316.23 & 1.369 & 128 & -0.0298 \\ 
\hline
13 & 398.11 & 1.402 & 121.7 & -0.0544 \\ 
\hline
14 & 501.19 & 6.935 & 135.3 & -0.02971 \\ 
\hline
15 & 630.96 & 12.77 & 215.3 & 0.179 \\ 
\hline
16 & 794.33 & 11.48 & 195.7 & 0.08491 \\ 
\hline
17 & 1000.00 & 9.2 & 157.1 & 0.04596 \\ 
\hline
18 & 1258.93 & 11.13 & 184.6 & 0.003646 \\ 
\hline
19 & 1584.89 & 8.951 & 153.7 & -0.0894 \\ 
\hline
20 & 1995.26 & 3.492 & 68.87 & 0.1737 \\ 
\hline
21 & 2511.89 & 2.02 & 57.47 & 0.1612 \\ 
\hline
22 & 3162.28 & -0.5084 & 40.61 & 0.16 \\ 
\hline
23 & 3981.07 & -2.036 & 27 & 0.1399 \\ 
\hline
24 & 5011.87 & -3.013 & 25.41 & 0.4063 \\ 
\hline
25 & 6309.57 & -3.702 & 30.21 & 0.5259 \\ 
\hline
26 & 7943.28 & -5.567 & 56.88 & 0.4599 \\ 
\hline

\else

\fi
\end{tabular}
\end{table*}
%
%
%
\clearpage
\subsection{Frequency-dependent \ac{DRR} estimation results by noise type}
\subsubsection{Ambient noise}
\begin{figure}[!ht]
	\ifarXiv
\centerline{\epsfig{figure=
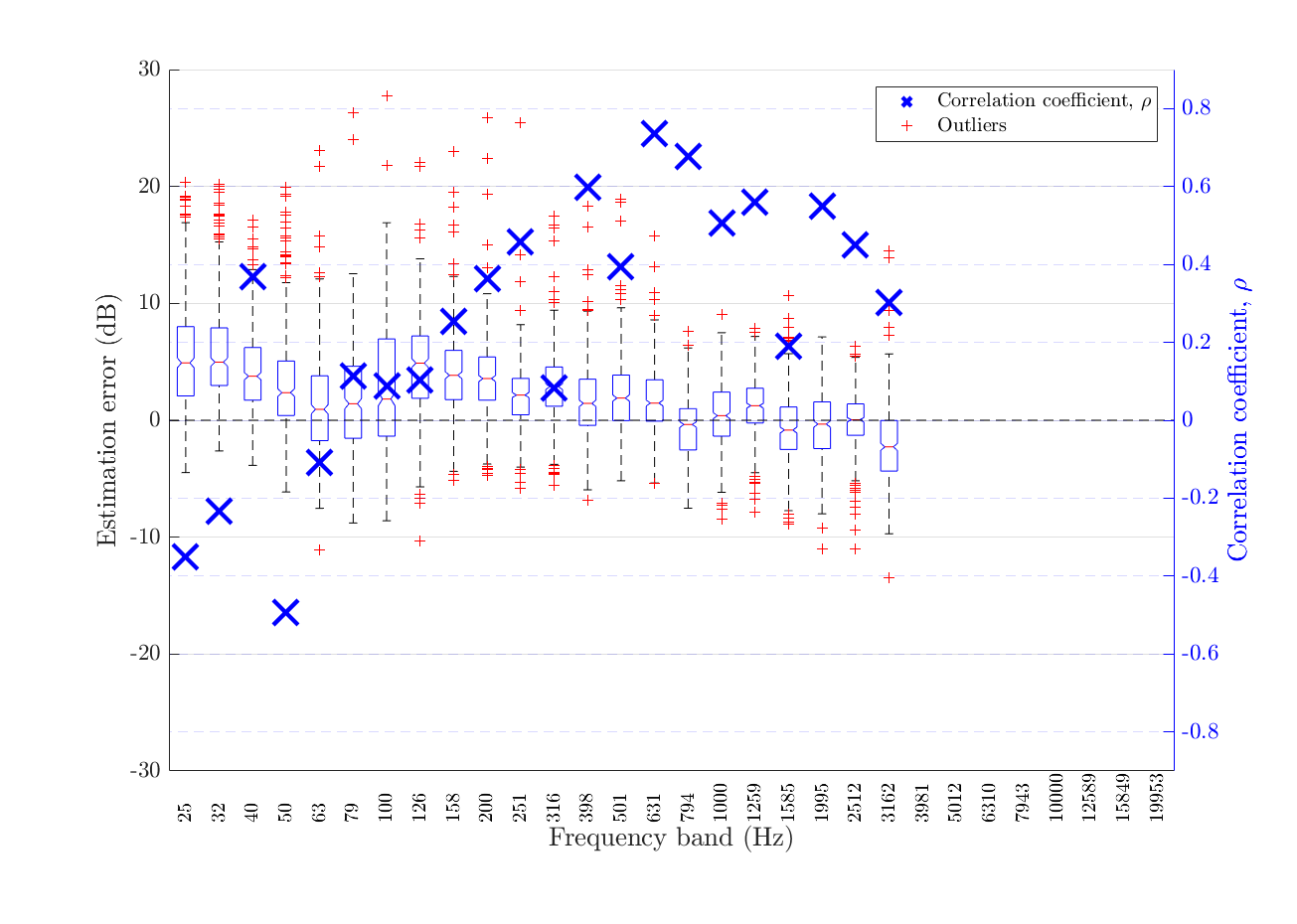,
	width=\figWidthACETR,viewport=45 10 765 530,clip}}%
	\else
	\centerline{\epsfig{figure=FigsACE/ana_eval_gt_partic_results_combined_Phase3_TR_P3S_DRR_dB_All_SNR_Ambient_sub_Velocity.png,
	width=\figWidthACETR,viewport=45 10 765 530,clip}}%
	\fi
	\caption{{
	Frequency-dependent \ac{DRR} estimation error in ambient noise for all \acp{SNR} for algorithm Particle Velocity~\cite{Chen2015}
	}}%
\label{fig:ACE_DRR_Sub_Ambient_Velocity}%
\end{figure}%
\begin{table*}[!ht]\small
\caption{
Frequency-dependent \ac{DRR} estimation error in ambient noise for all \acp{SNR} for algorithm Particle Velocity~\cite{Chen2015}
}
\vspace{5mm} 
\centering
\begin{tabular}{crrrl}%
\hline%
Freq. band
& Centre Freq. (Hz)
& Bias
& \acs{MSE}
& $\PearsonCC$

\\
\hline
\hline
\ifarXiv
 1 & 25.12 & 6.214 & 63.74 & -0.3506 \\ 
\hline
 2 & 31.62 & 6.689 & 66.26 & -0.2323 \\ 
\hline
 3 & 39.81 & 5.141 & 42.68 & 0.3684 \\ 
\hline
 4 & 50.12 & 3.689 & 37.53 & -0.4925 \\ 
\hline
 5 & 63.10 & 1.167 & 21.71 & -0.109 \\ 
\hline
 6 & 79.43 & 0.511 & 17.14 & 0.1127 \\ 
\hline
 7 & 100.00 & 1.343 & 23.59 & 0.08844 \\ 
\hline
 8 & 125.89 & 2.361 & 26.85 & 0.103 \\ 
\hline
 9 & 158.49 & 2.723 & 21.22 & 0.2526 \\ 
\hline
10 & 199.53 & 3.108 & 19.59 & 0.3639 \\ 
\hline
11 & 251.19 & 1.817 & 11.37 & 0.4577 \\ 
\hline
12 & 316.23 & 2.511 & 15.81 & 0.0833 \\ 
\hline
13 & 398.11 & 1.649 & 11.81 & 0.5972 \\ 
\hline
14 & 501.19 & 2.16 & 15.08 & 0.3946 \\ 
\hline
15 & 630.96 & 1.818 & 10.86 & 0.7346 \\ 
\hline
16 & 794.33 & -0.6569 & 7.836 & 0.6764 \\ 
\hline
17 & 1000.00 & 0.4372 & 7.952 & 0.5051 \\ 
\hline
18 & 1258.93 & 1.151 & 8.026 & 0.5595 \\ 
\hline
19 & 1584.89 & -0.7626 & 9.359 & 0.1898 \\ 
\hline
20 & 1995.26 & -0.4096 & 7.21 & 0.55 \\ 
\hline
21 & 2511.89 & -0.2113 & 6.052 & 0.451 \\ 
\hline
22 & 3162.28 & -2.222 & 16.39 & 0.302 \\ 
\hline

\else

\fi
\end{tabular}
\end{table*}
%
%
%
\begin{figure}[!ht]
	\ifarXiv
\centerline{\epsfig{figure=
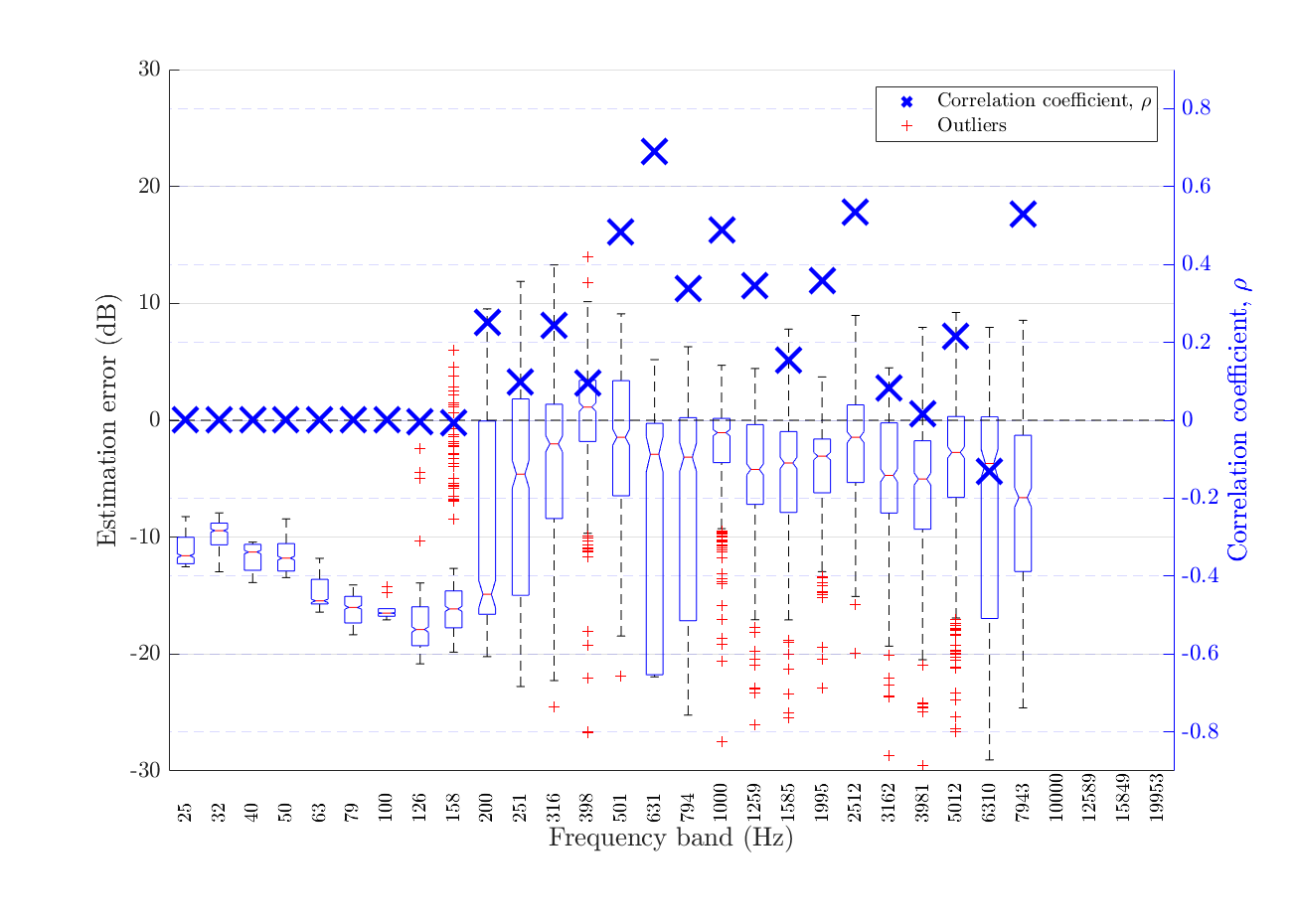,
	width=\figWidthACETR,viewport=45 10 765 530,clip}}%
	\else
	\centerline{\epsfig{figure=FigsACE/ana_eval_gt_partic_results_combined_Phase3_TR_P3S_DRR_dB_All_SNR_Ambient_sub_ICASSP_2015_DRR_2-ch_Gerkmann_NR_FFT_subband.png,
	width=\figWidthACETR,viewport=45 10 765 530,clip}}%
	\fi
	\caption{{
	Frequency-dependent \ac{DRR} estimation error in ambient noise for all \acp{SNR} for algorithm \ac{DENBE} with FFT derived subbands~\cite{Eaton2015c}
	}}%
\label{fig:ACE_DRR_Sub_Ambient_Eaton_FFT}%
\end{figure}%
\begin{table*}[!ht]\small
\caption{
	Frequency-dependent \ac{DRR} estimation error in ambient noise for all \acp{SNR} for algorithm \ac{DENBE} with FFT derived subbands~\cite{Eaton2015c}
}
\vspace{5mm} 
\centering
\begin{tabular}{crrrl}%
\hline%
Freq. band
& Centre Freq. (Hz)
& Bias
& \acs{MSE}
& $\PearsonCC$

\\
\hline
\hline
\ifarXiv
 1 & 25.12 & -11.07 & 124.4 & 0 \\ 
\hline
 2 & 31.62 & -9.93 & 100.8 & 0 \\ 
\hline
 3 & 39.81 & -11.69 & 138.1 & 0 \\ 
\hline
 4 & 50.12 & -11.55 & 135.4 & 0 \\ 
\hline
 5 & 63.10 & -14.82 & 221.5 & 0 \\ 
\hline
 6 & 79.43 & -16.09 & 260.6 & 0 \\ 
\hline
 7 & 100.00 & -16.18 & 262.6 & 0 \\ 
\hline
 8 & 125.89 & -17.71 & 317.6 & -0.002839 \\ 
\hline
 9 & 158.49 & -16.36 & 276.1 & -0.006109 \\ 
\hline
10 & 199.53 & -13.14 & 222 & 0.2499 \\ 
\hline
11 & 251.19 & -11.33 & 212.9 & 0.09857 \\ 
\hline
12 & 316.23 & -11.3 & 218 & 0.2435 \\ 
\hline
13 & 398.11 & -5.595 & 135.9 & 0.0947 \\ 
\hline
14 & 501.19 & -5.816 & 112.4 & 0.4825 \\ 
\hline
15 & 630.96 & -10.64 & 209.3 & 0.6889 \\ 
\hline
16 & 794.33 & -9.63 & 187.8 & 0.339 \\ 
\hline
17 & 1000.00 & -6.313 & 104.9 & 0.4895 \\ 
\hline
18 & 1258.93 & -7.227 & 107.4 & 0.3443 \\ 
\hline
19 & 1584.89 & -5.902 & 86.14 & 0.1545 \\ 
\hline
20 & 1995.26 & -5.517 & 58.78 & 0.359 \\ 
\hline
21 & 2511.89 & -3.762 & 45.75 & 0.5346 \\ 
\hline
22 & 3162.28 & -5.89 & 69.48 & 0.0819 \\ 
\hline
23 & 3981.07 & -6.871 & 82 & 0.01629 \\ 
\hline
24 & 5011.87 & -5.961 & 90.28 & 0.2161 \\ 
\hline
25 & 6309.57 & -9.023 & 193.2 & -0.1301 \\ 
\hline
26 & 7943.28 & -8.983 & 151.1 & 0.529 \\ 
\hline

\else

\fi
\end{tabular}
\end{table*}
%
%
%
\begin{figure}[!ht]
	\ifarXiv
\centerline{\epsfig{figure=
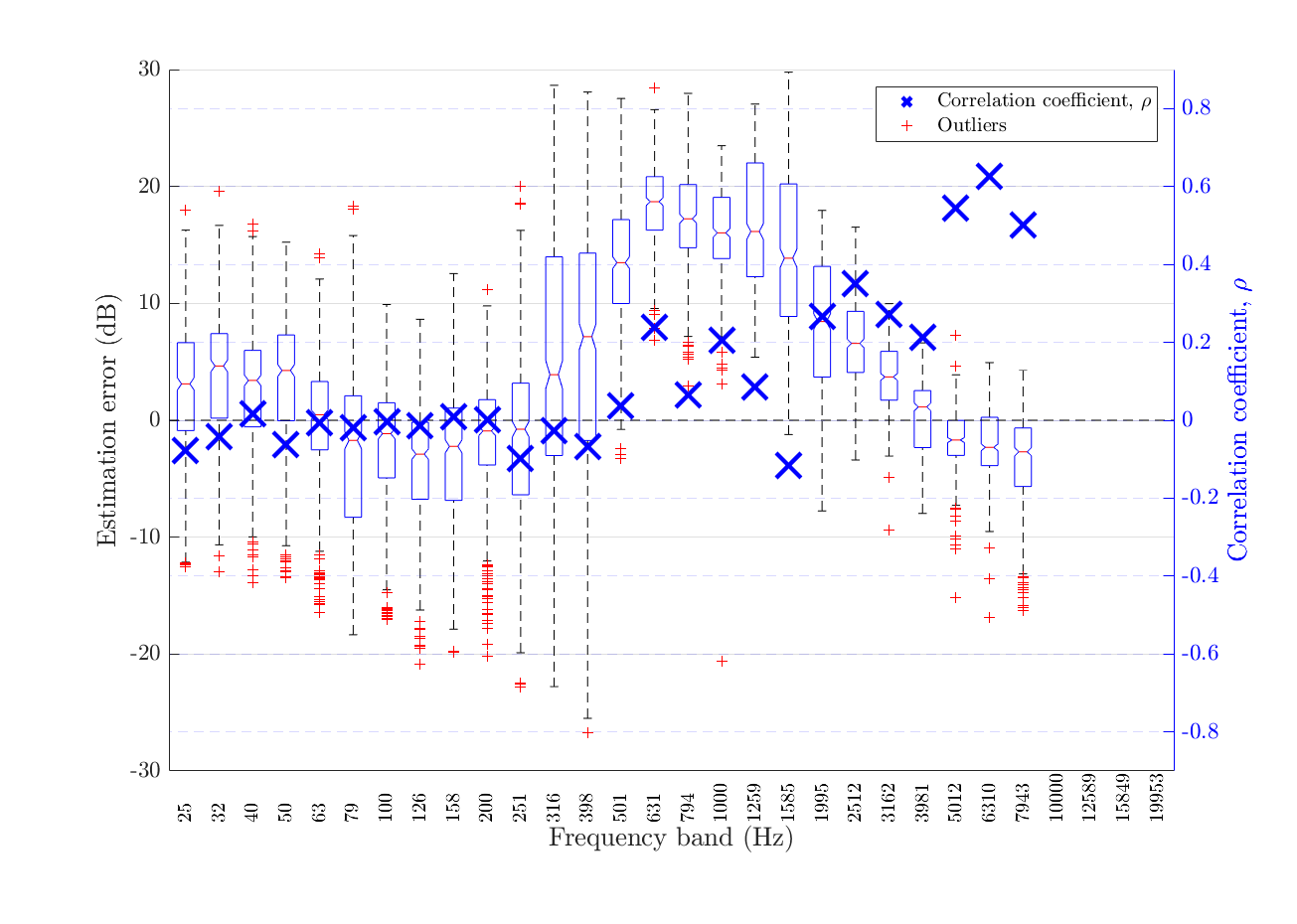,
	width=\figWidthACETR,viewport=45 10 765 530,clip}}%
	\else
	\centerline{\epsfig{figure=
	FigsACE/ana_eval_gt_partic_results_combined_Phase3_TR_P3S_DRR_dB_All_SNR_Ambient_sub_ICASSP_2015_DRR_2-ch_Gerkmann_NR_filtered_subband.png,
	width=\figWidthACETR,viewport=45 10 765 530,clip}}%
	\fi
	\caption{{
	Frequency-dependent \ac{DRR} estimation error in ambient noise for all \acp{SNR} for algorithm \ac{DENBE} with filtered subbands~\cite{Eaton2015c}
	}}%
\label{fig:ACE_DRR_Sub_Ambient_Eaton_filt}%
\end{figure}%
\begin{table*}[!ht]\small
\caption{
	Frequency-dependent \ac{DRR} estimation error in ambient noise for all \acp{SNR} for algorithm \ac{DENBE} with filtered subbands~\cite{Eaton2015c}
}
\vspace{5mm} 
\centering
\begin{tabular}{crrrl}%
\hline%
Freq. band
& Centre Freq. (Hz)
& Bias
& \acs{MSE}
& $\PearsonCC$

\\
\hline
\hline
\ifarXiv
 1 & 25.12 & 1.66 & 54.97 & -0.07702 \\ 
\hline
 2 & 31.62 & 2.713 & 53.7 & -0.04108 \\ 
\hline
 3 & 39.81 & 1.356 & 42.18 & 0.01744 \\ 
\hline
 4 & 50.12 & 1.582 & 50.25 & -0.06225 \\ 
\hline
 5 & 63.10 & -1.442 & 43.29 & -0.006312 \\ 
\hline
 6 & 79.43 & -3.108 & 68.75 & -0.01959 \\ 
\hline
 7 & 100.00 & -3.231 & 49.12 & -0.003524 \\ 
\hline
 8 & 125.89 & -4.595 & 60.49 & -0.01394 \\ 
\hline
 9 & 158.49 & -3.882 & 60.76 & 0.008567 \\ 
\hline
10 & 199.53 & -2.874 & 44.48 & 0.002132 \\ 
\hline
11 & 251.19 & -3.767 & 69.47 & -0.0971 \\ 
\hline
12 & 316.23 & 0.3985 & 112.6 & -0.02558 \\ 
\hline
13 & 398.11 & 0.7077 & 118.2 & -0.06832 \\ 
\hline
14 & 501.19 & 6.848 & 131.9 & 0.03671 \\ 
\hline
15 & 630.96 & 13.44 & 229.6 & 0.2388 \\ 
\hline
16 & 794.33 & 12.93 & 227.2 & 0.0649 \\ 
\hline
17 & 1000.00 & 12.07 & 193.2 & 0.2053 \\ 
\hline
18 & 1258.93 & 14.78 & 247.4 & 0.08421 \\ 
\hline
19 & 1584.89 & 12.73 & 207.5 & -0.1166 \\ 
\hline
20 & 1995.26 & 6.81 & 80.34 & 0.2655 \\ 
\hline
21 & 2511.89 & 5.048 & 51.05 & 0.3499 \\ 
\hline
22 & 3162.28 & 1.967 & 24.94 & 0.2718 \\ 
\hline
23 & 3981.07 & -0.7185 & 14.42 & 0.2117 \\ 
\hline
24 & 5011.87 & -2.156 & 14.58 & 0.544 \\ 
\hline
25 & 6309.57 & -2.896 & 19.87 & 0.6265 \\ 
\hline
26 & 7943.28 & -4.729 & 43.81 & 0.5001 \\ 
\hline

\else

\fi
\end{tabular}
\end{table*}
%
%
%
\clearpage
\subsubsection{Babble noise}
\begin{figure}[!ht]
	\ifarXiv
\centerline{\epsfig{figure=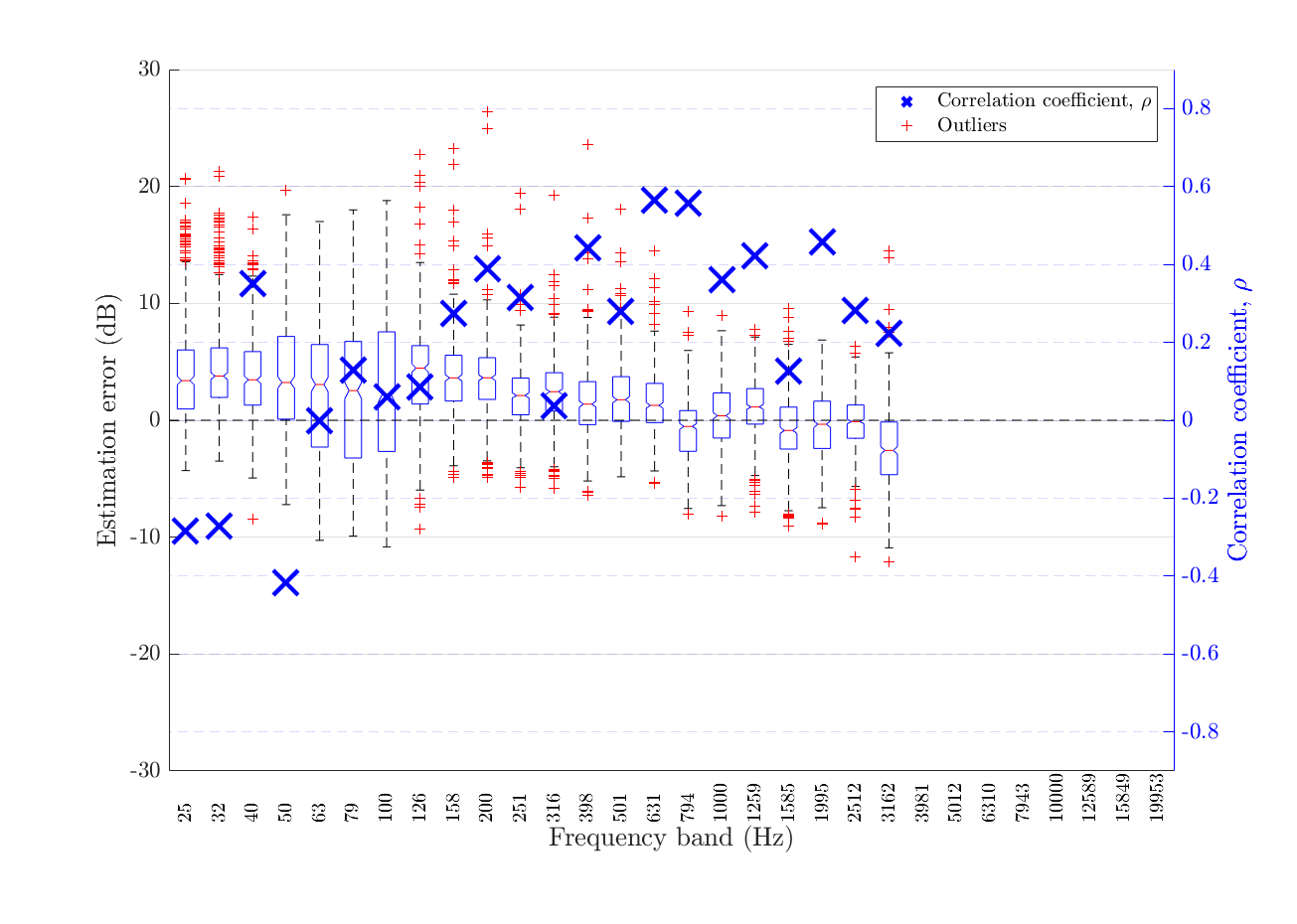,
	width=\figWidthACETR,viewport=45 10 765 530,clip}}%
	\else
	\centerline{\epsfig{figure=FigsACE/ana_eval_gt_partic_results_combined_Phase3_TR_P3S_DRR_dB_All_SNR_Babble_sub_Velocity.png,
	width=\figWidthACETR,viewport=45 10 765 530,clip}}%
	\fi
	\caption{{Frequency-dependent \ac{DRR} estimation error in babble noise for all \acp{SNR} for algorithm Particle Velocity~\cite{Chen2015}}}%
\label{fig:ACE_DRR_Sub_Babble_Velocity}%
\end{figure}%
\begin{table*}[!ht]\small
\caption{
Frequency-dependent \ac{DRR} estimation error in babble noise for all \acp{SNR} for algorithm Particle Velocity~\cite{Chen2015}
}
\vspace{5mm} 
\centering
\begin{tabular}{crrrl}%
\hline%
Freq. band
& Centre Freq. (Hz)
& Bias
& \acs{MSE}
& $\PearsonCC$

\\
\hline
\hline
\ifarXiv
 1 & 25.12 & 5.283 & 53.02 & -0.2852 \\ 
\hline
 2 & 31.62 & 5.764 & 55.5 & -0.2719 \\ 
\hline
 3 & 39.81 & 4.42 & 36.07 & 0.35 \\ 
\hline
 4 & 50.12 & 3.432 & 37.01 & -0.4171 \\ 
\hline
 5 & 63.10 & 1.334 & 27.61 & -0.0008789 \\ 
\hline
 6 & 79.43 & 0.4524 & 25.49 & 0.1279 \\ 
\hline
 7 & 100.00 & 0.4457 & 29.46 & 0.0596 \\ 
\hline
 8 & 125.89 & 0.9477 & 27.44 & 0.08608 \\ 
\hline
 9 & 158.49 & 1.447 & 19.41 & 0.2753 \\ 
\hline
10 & 199.53 & 2.22 & 16.05 & 0.3898 \\ 
\hline
11 & 251.19 & 0.7781 & 10.82 & 0.3151 \\ 
\hline
12 & 316.23 & 0.4684 & 14.71 & 0.03585 \\ 
\hline
13 & 398.11 & 0.1692 & 13.43 & 0.4431 \\ 
\hline
14 & 501.19 & 0.6781 & 13.49 & 0.2792 \\ 
\hline
15 & 630.96 & 0.1452 & 12.36 & 0.5645 \\ 
\hline
16 & 794.33 & -1.864 & 14.5 & 0.5561 \\ 
\hline
17 & 1000.00 & -0.8126 & 12.97 & 0.3599 \\ 
\hline
18 & 1258.93 & -0.05593 & 11.58 & 0.4217 \\ 
\hline
19 & 1584.89 & -1.384 & 12.04 & 0.1265 \\ 
\hline
20 & 1995.26 & -1.305 & 10.75 & 0.4583 \\ 
\hline
21 & 2511.89 & -1.619 & 13.53 & 0.2812 \\ 
\hline
22 & 3162.28 & -3.902 & 31.41 & 0.2219 \\ 
\hline

\else

\fi
\end{tabular}
\end{table*}
%
%
%
\begin{figure}[!ht]
	\ifarXiv
\centerline{\epsfig{figure=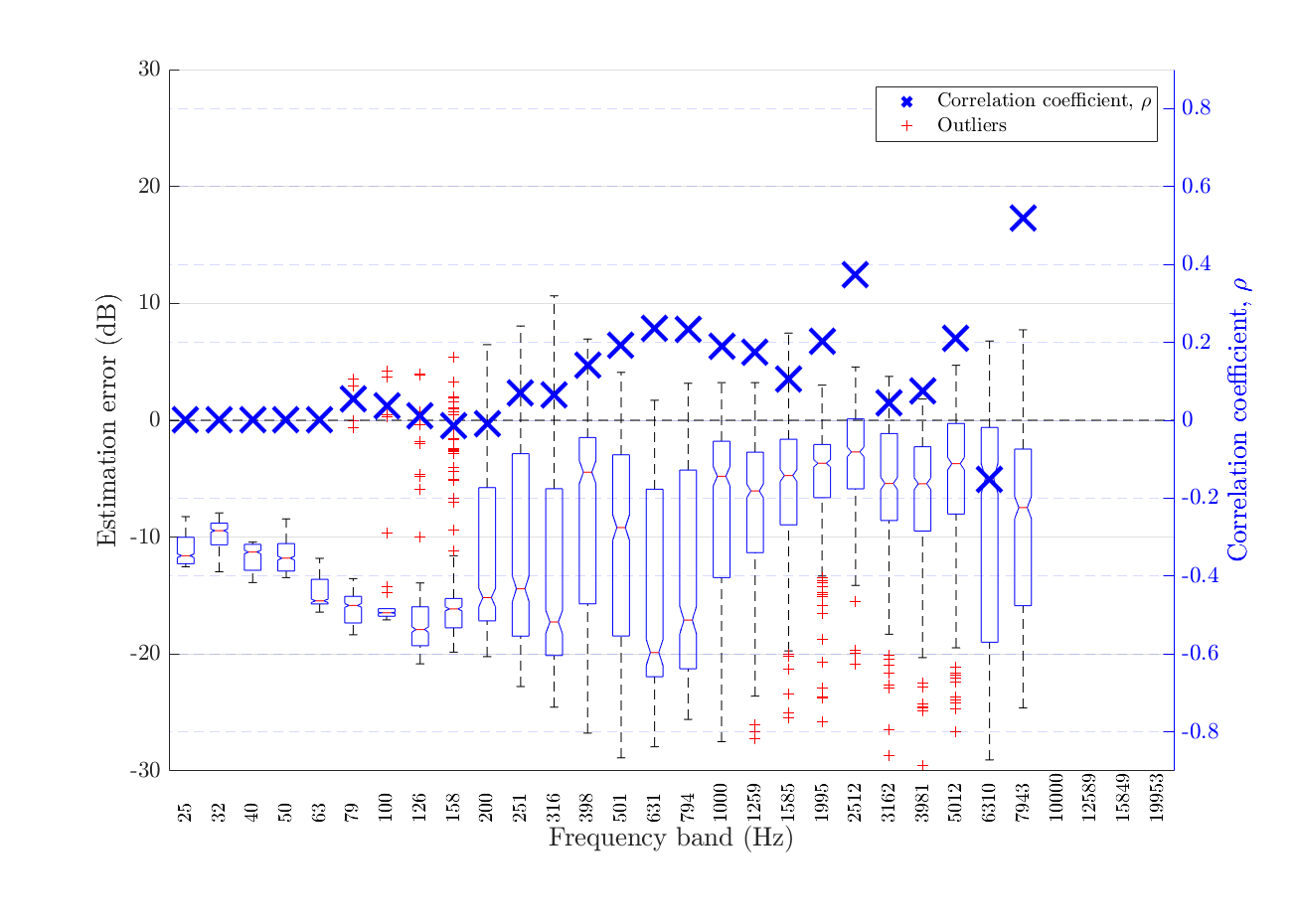,
	width=\figWidthACETR,viewport=45 10 765 530,clip}}%
	\else
	\centerline{\epsfig{figure=FigsACE/ana_eval_gt_partic_results_combined_Phase3_TR_P3S_DRR_dB_All_SNR_Babble_sub_ICASSP_2015_DRR_2-ch_Gerkmann_NR_FFT_subband.png,
	width=\figWidthACETR,viewport=45 10 765 530,clip}}%
	\fi
	\caption{{Frequency-dependent \ac{DRR} estimation error in babble noise for all \acp{SNR} for algorithm \ac{DENBE} with FFT derived subbands~\cite{Eaton2015c}}}%
\label{fig:ACE_DRR_Sub_Babble_Eaton_FFT}%
\end{figure}%
\begin{table*}[!ht]\small
\caption{
	Frequency-dependent \ac{DRR} estimation error in babble noise for all \acp{SNR} for algorithm \ac{DENBE} with FFT derived subbands~\cite{Eaton2015c}
}
\vspace{5mm} 
\centering
\begin{tabular}{crrrl}%
\hline%
Freq. band
& Centre Freq. (Hz)
& Bias
& \acs{MSE}
& $\PearsonCC$

\\
\hline
\hline
\ifarXiv
 1 & 25.12 & -11.07 & 124.4 & 0 \\ 
\hline
 2 & 31.62 & -9.93 & 100.8 & 0 \\ 
\hline
 3 & 39.81 & -11.69 & 138.1 & 0 \\ 
\hline
 4 & 50.12 & -11.55 & 135.4 & 0 \\ 
\hline
 5 & 63.10 & -14.82 & 221.5 & 0 \\ 
\hline
 6 & 79.43 & -16.03 & 259.6 & 0.05514 \\ 
\hline
 7 & 100.00 & -16.12 & 261.6 & 0.03793 \\ 
\hline
 8 & 125.89 & -17.62 & 316 & 0.01081 \\ 
\hline
 9 & 158.49 & -16.47 & 278 & -0.01475 \\ 
\hline
10 & 199.53 & -14.94 & 252.4 & -0.009532 \\ 
\hline
11 & 251.19 & -15.14 & 272.3 & 0.07024 \\ 
\hline
12 & 316.23 & -17.67 & 353.7 & 0.06498 \\ 
\hline
13 & 398.11 & -16.26 & 345.6 & 0.1415 \\ 
\hline
14 & 501.19 & -17.53 & 378.2 & 0.1929 \\ 
\hline
15 & 630.96 & -19.54 & 428.4 & 0.237 \\ 
\hline
16 & 794.33 & -18.05 & 382.3 & 0.2328 \\ 
\hline
17 & 1000.00 & -15.23 & 321.6 & 0.1888 \\ 
\hline
18 & 1258.93 & -14.58 & 290.9 & 0.1744 \\ 
\hline
19 & 1584.89 & -11.62 & 222.4 & 0.1068 \\ 
\hline
20 & 1995.26 & -10.08 & 173.3 & 0.2017 \\ 
\hline
21 & 2511.89 & -7.693 & 123.8 & 0.3742 \\ 
\hline
22 & 3162.28 & -8.071 & 109.7 & 0.04468 \\ 
\hline
23 & 3981.07 & -7.607 & 88.77 & 0.07405 \\ 
\hline
24 & 5011.87 & -7.238 & 104.3 & 0.2093 \\ 
\hline
25 & 6309.57 & -11.01 & 231.9 & -0.1509 \\ 
\hline
26 & 7943.28 & -11.06 & 193.3 & 0.519 \\ 
\hline

\else

\fi
\end{tabular}
\end{table*}
%
%
%
\begin{figure}[!ht]
	\ifarXiv
\centerline{\epsfig{figure=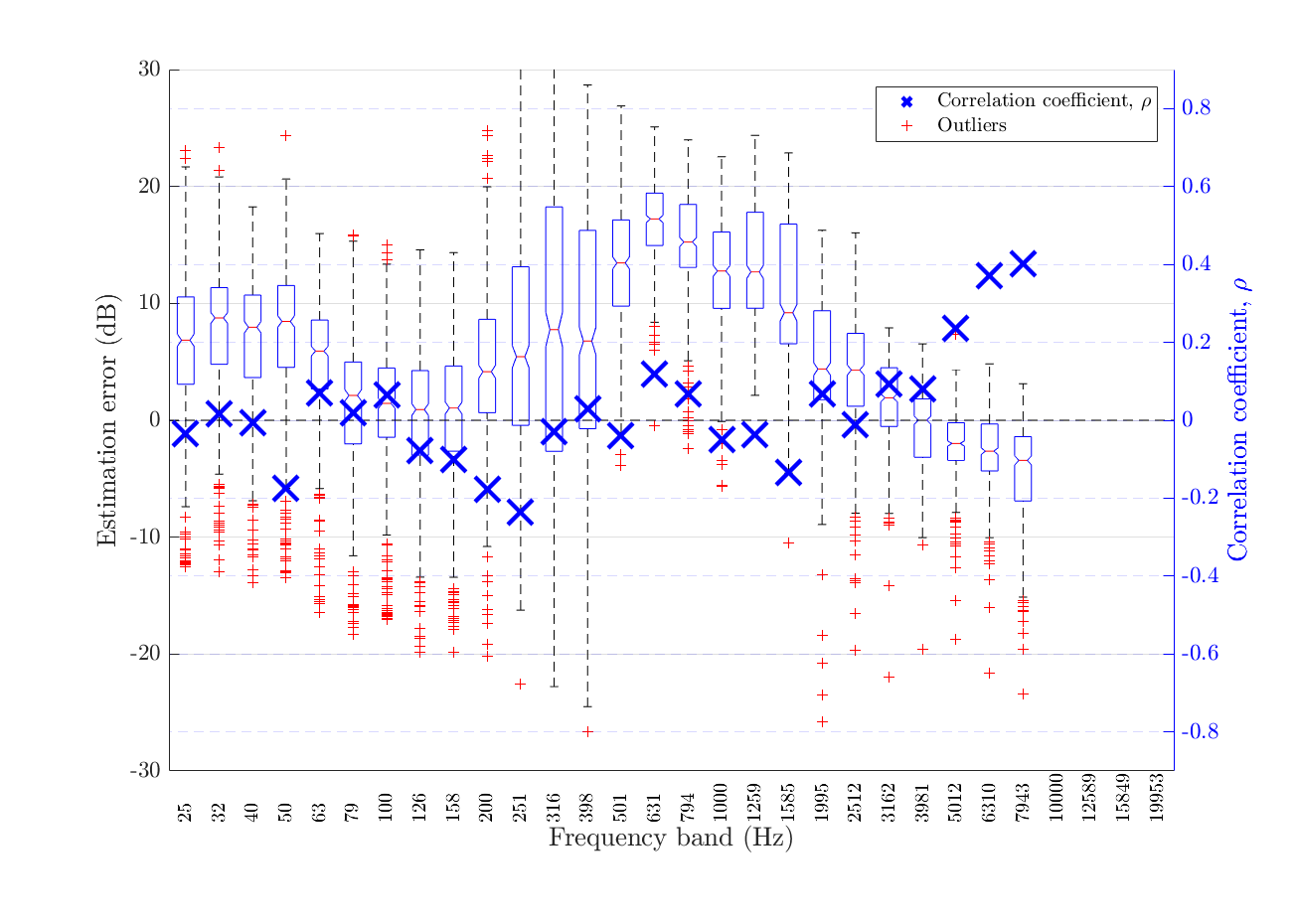,
	width=\figWidthACETR,viewport=45 10 765 530,clip}}%
	\else
	\centerline{\epsfig{figure=FigsACE/ana_eval_gt_partic_results_combined_Phase3_TR_P3S_DRR_dB_All_SNR_Babble_sub_ICASSP_2015_DRR_2-ch_Gerkmann_NR_filtered_subband.png,
	width=\figWidthACETR,viewport=45 10 765 530,clip}}%
	\fi
	\caption{{Frequency-dependent \ac{DRR} estimation error in babble noise for all \acp{SNR} for algorithm \ac{DENBE} with filtered subbands~\cite{Eaton2015c}}}%
\label{fig:ACE_DRR_Sub_Babble_Eaton_filt}%
\end{figure}%
\begin{table*}[!ht]\small
\caption{
	Frequency-dependent \ac{DRR} estimation error in babble noise for all \acp{SNR} for algorithm \ac{DENBE} with filtered subbands~\cite{Eaton2015c}
}
\vspace{5mm} 
\centering
\begin{tabular}{crrrl}%
\hline%
Freq. band
& Centre Freq. (Hz)
& Bias
& \acs{MSE}
& $\PearsonCC$

\\
\hline
\hline
\ifarXiv
 1 & 25.12 & 3.769 & 68.3 & -0.03378 \\ 
\hline
 2 & 31.62 & 5.437 & 76.85 & 0.01647 \\ 
\hline
 3 & 39.81 & 4.093 & 62.42 & -0.00727 \\ 
\hline
 4 & 50.12 & 5.104 & 76.03 & -0.1741 \\ 
\hline
 5 & 63.10 & 2.715 & 46.94 & 0.06954 \\ 
\hline
 6 & 79.43 & -1.091 & 55.54 & 0.01963 \\ 
\hline
 7 & 100.00 & -1.016 & 42.39 & 0.06502 \\ 
\hline
 8 & 125.89 & -1.576 & 39.75 & -0.07832 \\ 
\hline
 9 & 158.49 & -1.467 & 47.49 & -0.09999 \\ 
\hline
10 & 199.53 & 1.03 & 52.19 & -0.1781 \\ 
\hline
11 & 251.19 & 1.692 & 94.37 & -0.2364 \\ 
\hline
12 & 316.23 & 2.913 & 147.8 & -0.03055 \\ 
\hline
13 & 398.11 & 1.472 & 123.6 & 0.02826 \\ 
\hline
14 & 501.19 & 6.685 & 135.6 & -0.03896 \\ 
\hline
15 & 630.96 & 10.94 & 179.4 & 0.1174 \\ 
\hline
16 & 794.33 & 8.818 & 151.9 & 0.06773 \\ 
\hline
17 & 1000.00 & 5.442 & 119.5 & -0.05079 \\ 
\hline
18 & 1258.93 & 6.697 & 124.4 & -0.03634 \\ 
\hline
19 & 1584.89 & 4.199 & 105 & -0.133 \\ 
\hline
20 & 1995.26 & -0.5712 & 68.04 & 0.06788 \\ 
\hline
21 & 2511.89 & -1.244 & 73.39 & -0.01247 \\ 
\hline
22 & 3162.28 & -3.337 & 63.95 & 0.09364 \\ 
\hline
23 & 3981.07 & -4.13 & 49.48 & 0.07909 \\ 
\hline
24 & 5011.87 & -4.5 & 46.33 & 0.2359 \\ 
\hline
25 & 6309.57 & -4.977 & 47.57 & 0.3699 \\ 
\hline
26 & 7943.28 & -7.132 & 82.14 & 0.4012 \\ 
\hline

\else

\fi
\end{tabular}
\end{table*}
%
%
%
%
%
\clearpage
\subsubsection{Fan noise}
\begin{figure}[!ht]
	\ifarXiv
\centerline{\epsfig{figure=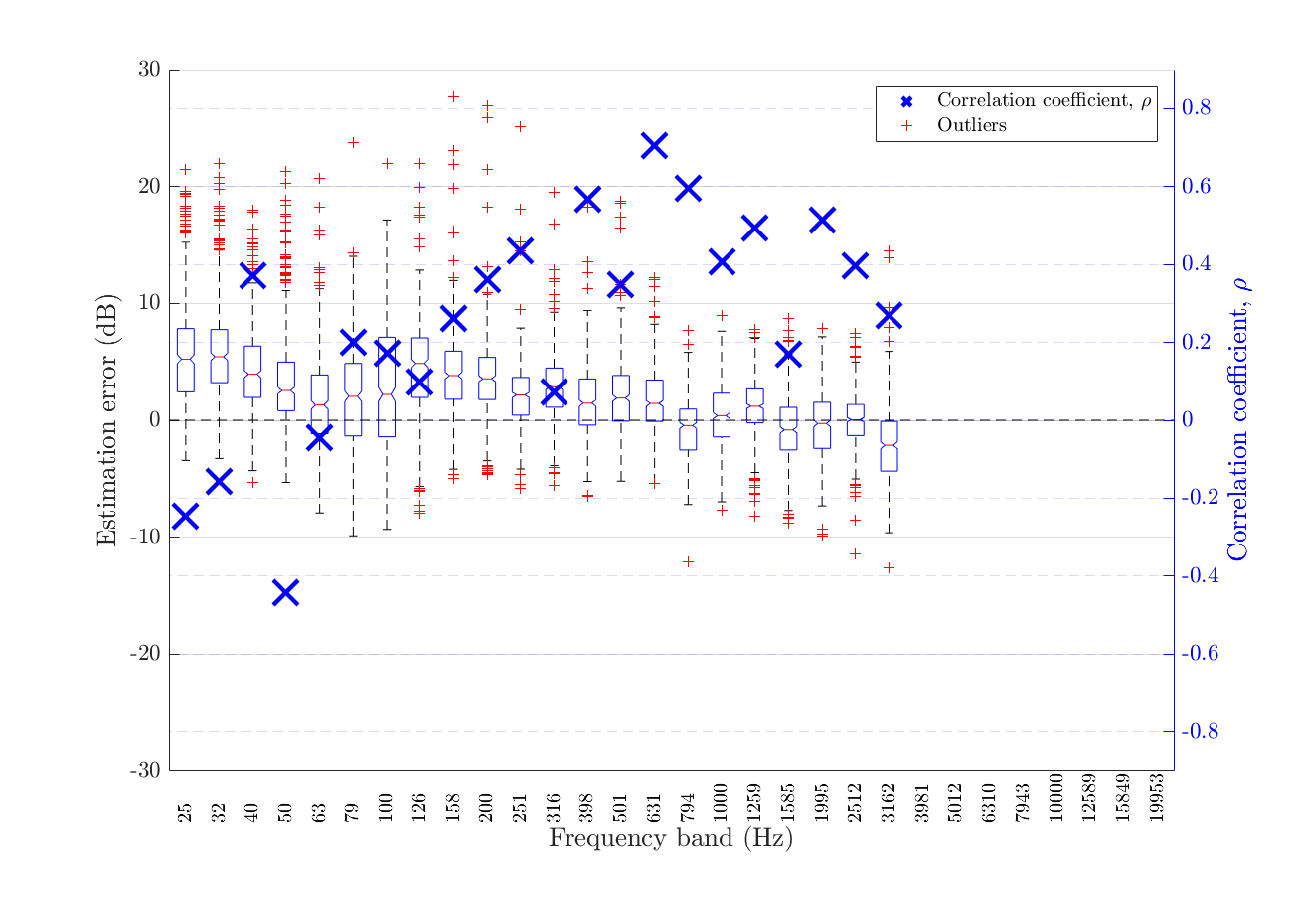,
	width=\figWidthACETR,viewport=45 10 765 530,clip}}%
	\else
	\centerline{\epsfig{figure=FigsACE/ana_eval_gt_partic_results_combined_Phase3_TR_P3S_DRR_dB_All_SNR_Fan_sub_Velocity.png,
	width=\figWidthACETR,viewport=45 10 765 530,clip}}%
	\fi
	\caption{{Frequency-dependent \ac{DRR} estimation error in fan noise for all \acp{SNR} for algorithm Particle Velocity~\cite{Chen2015}}}%
\label{fig:ACE_DRR_Sub_Fan_Velocity}%
\end{figure}%
\begin{table*}[!ht]\small
\caption{
Frequency-dependent \ac{DRR} estimation error in fan noise for all \acp{SNR} for algorithm Particle Velocity~\cite{Chen2015}
}
\vspace{5mm} 
\centering
\begin{tabular}{crrrl}%
\hline%
Freq. band
& Centre Freq. (Hz)
& Bias
& \acs{MSE}
& $\PearsonCC$

\\
\hline
\hline
\ifarXiv
 1 & 25.12 & 6.247 & 61.73 & -0.2465 \\ 
\hline
 2 & 31.62 & 6.743 & 65.47 & -0.1562 \\ 
\hline
 3 & 39.81 & 5.183 & 42.73 & 0.3718 \\ 
\hline
 4 & 50.12 & 3.751 & 37.56 & -0.4414 \\ 
\hline
 5 & 63.10 & 1.432 & 22.4 & -0.04497 \\ 
\hline
 6 & 79.43 & 1.029 & 18.2 & 0.2009 \\ 
\hline
 7 & 100.00 & 1.672 & 24.28 & 0.172 \\ 
\hline
 8 & 125.89 & 2.291 & 27.15 & 0.0991 \\ 
\hline
 9 & 158.49 & 2.585 & 21.72 & 0.2602 \\ 
\hline
10 & 199.53 & 3.098 & 20.07 & 0.3609 \\ 
\hline
11 & 251.19 & 1.77 & 11.01 & 0.4359 \\ 
\hline
12 & 316.23 & 2 & 14.07 & 0.07199 \\ 
\hline
13 & 398.11 & 1.204 & 11.06 & 0.5675 \\ 
\hline
14 & 501.19 & 1.697 & 13.72 & 0.3477 \\ 
\hline
15 & 630.96 & 1.398 & 10.22 & 0.7052 \\ 
\hline
16 & 794.33 & -1.228 & 11.02 & 0.5956 \\ 
\hline
17 & 1000.00 & -0.2778 & 10.02 & 0.4073 \\ 
\hline
18 & 1258.93 & 0.7117 & 8.782 & 0.4946 \\ 
\hline
19 & 1584.89 & -0.8601 & 9.44 & 0.1695 \\ 
\hline
20 & 1995.26 & -0.6946 & 8.04 & 0.515 \\ 
\hline
21 & 2511.89 & -0.5183 & 7.125 & 0.3956 \\ 
\hline
22 & 3162.28 & -2.725 & 20.13 & 0.2696 \\ 
\hline

\else

\fi
\end{tabular}
\end{table*}
%
%
%
\begin{figure}[!ht]
	\ifarXiv
\centerline{\epsfig{figure=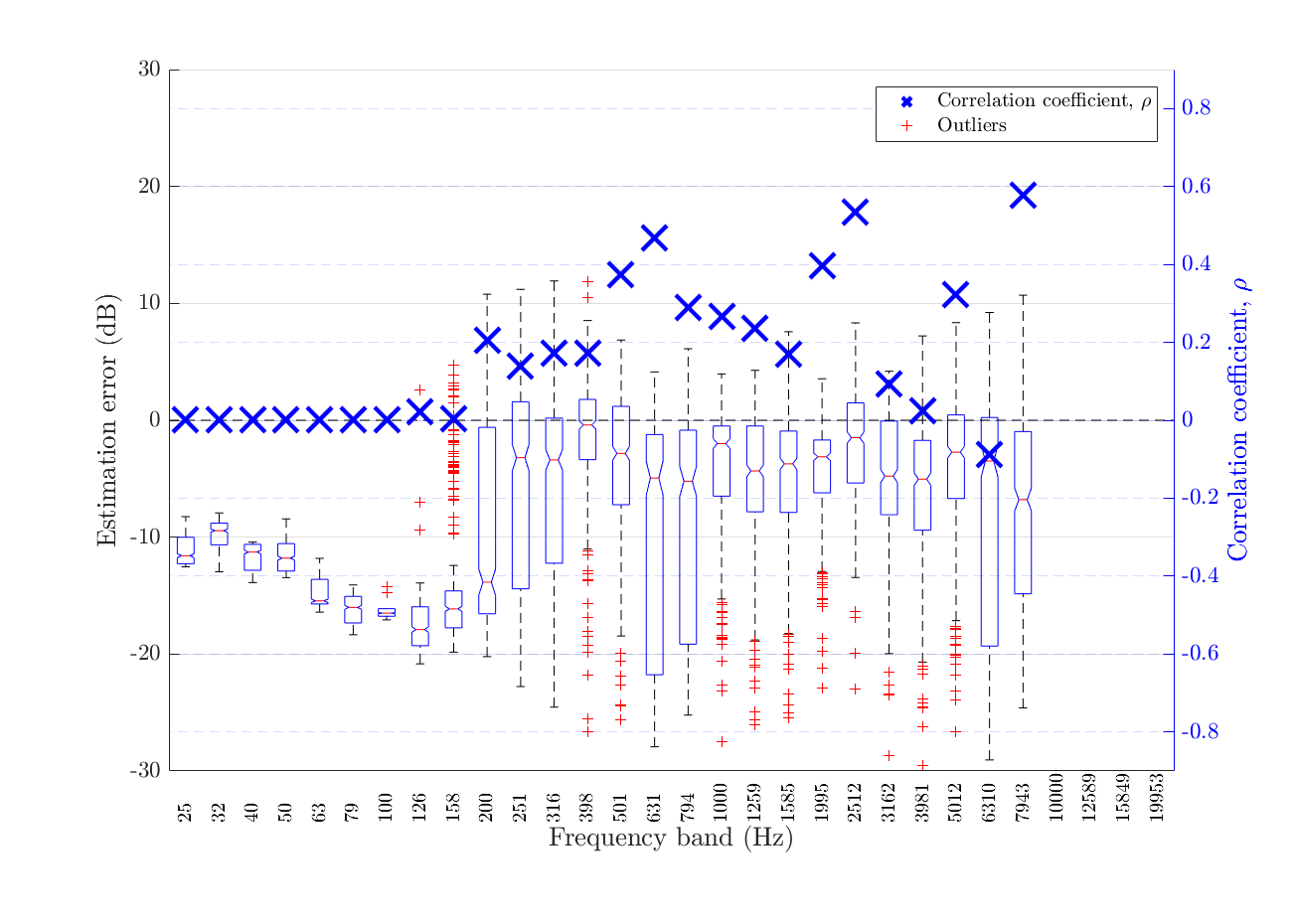,
	width=\figWidthACETR,viewport=45 10 765 530,clip}}%
	\else
	\centerline{\epsfig{figure=FigsACE/ana_eval_gt_partic_results_combined_Phase3_TR_P3S_DRR_dB_All_SNR_Fan_sub_ICASSP_2015_DRR_2-ch_Gerkmann_NR_FFT_subband.png,
	width=\figWidthACETR,viewport=45 10 765 530,clip}}%
	\fi
	\caption{{Frequency-dependent \ac{DRR} estimation error in fan noise for all \acp{SNR} for algorithm \ac{DENBE} with FFT derived subbands~\cite{Eaton2015c}}}%
\label{fig:ACE_DRR_Sub_Fan_Eaton_FFT}%
\end{figure}%
\begin{table*}[!ht]\small
\caption{
	Frequency-dependent \ac{DRR} estimation error in fan noise for all \acp{SNR} for algorithm \ac{DENBE} with FFT derived subbands~\cite{Eaton2015c}
}
\vspace{5mm} 
\centering
\begin{tabular}{crrrl}%
\hline%
Freq. band
& Centre Freq. (Hz)
& Bias
& \acs{MSE}
& $\PearsonCC$

\\
\hline
\hline
\ifarXiv
 1 & 25.12 & -11.07 & 124.4 & 0 \\ 
\hline
 2 & 31.62 & -9.93 & 100.8 & 0 \\ 
\hline
 3 & 39.81 & -11.69 & 138.1 & 0 \\ 
\hline
 4 & 50.12 & -11.55 & 135.4 & 0 \\ 
\hline
 5 & 63.10 & -14.82 & 221.5 & 0 \\ 
\hline
 6 & 79.43 & -16.09 & 260.6 & 0 \\ 
\hline
 7 & 100.00 & -16.18 & 262.6 & 0 \\ 
\hline
 8 & 125.89 & -17.71 & 317.7 & 0.02099 \\ 
\hline
 9 & 158.49 & -16.32 & 275.2 & 0.003448 \\ 
\hline
10 & 199.53 & -13.16 & 221.2 & 0.2047 \\ 
\hline
11 & 251.19 & -11.67 & 213.8 & 0.1388 \\ 
\hline
12 & 316.23 & -13.19 & 253.5 & 0.1732 \\ 
\hline
13 & 398.11 & -9.816 & 205 & 0.1723 \\ 
\hline
14 & 501.19 & -11.2 & 220.7 & 0.3728 \\ 
\hline
15 & 630.96 & -14.46 & 298.9 & 0.4677 \\ 
\hline
16 & 794.33 & -13.67 & 278.3 & 0.2903 \\ 
\hline
17 & 1000.00 & -10.67 & 209 & 0.2675 \\ 
\hline
18 & 1258.93 & -9.385 & 164.2 & 0.236 \\ 
\hline
19 & 1584.89 & -6.966 & 111.6 & 0.1705 \\ 
\hline
20 & 1995.26 & -6.339 & 76 & 0.3956 \\ 
\hline
21 & 2511.89 & -4.197 & 53.6 & 0.5345 \\ 
\hline
22 & 3162.28 & -5.964 & 68.26 & 0.09377 \\ 
\hline
23 & 3981.07 & -6.548 & 75.06 & 0.02546 \\ 
\hline
24 & 5011.87 & -5.8 & 92.28 & 0.322 \\ 
\hline
25 & 6309.57 & -8.89 & 198 & -0.08916 \\ 
\hline
26 & 7943.28 & -8.902 & 155 & 0.5775 \\ 
\hline

\else

\fi
\end{tabular}
\end{table*}
%
%
%
\begin{figure}[!ht]
	\ifarXiv
\centerline{\epsfig{figure=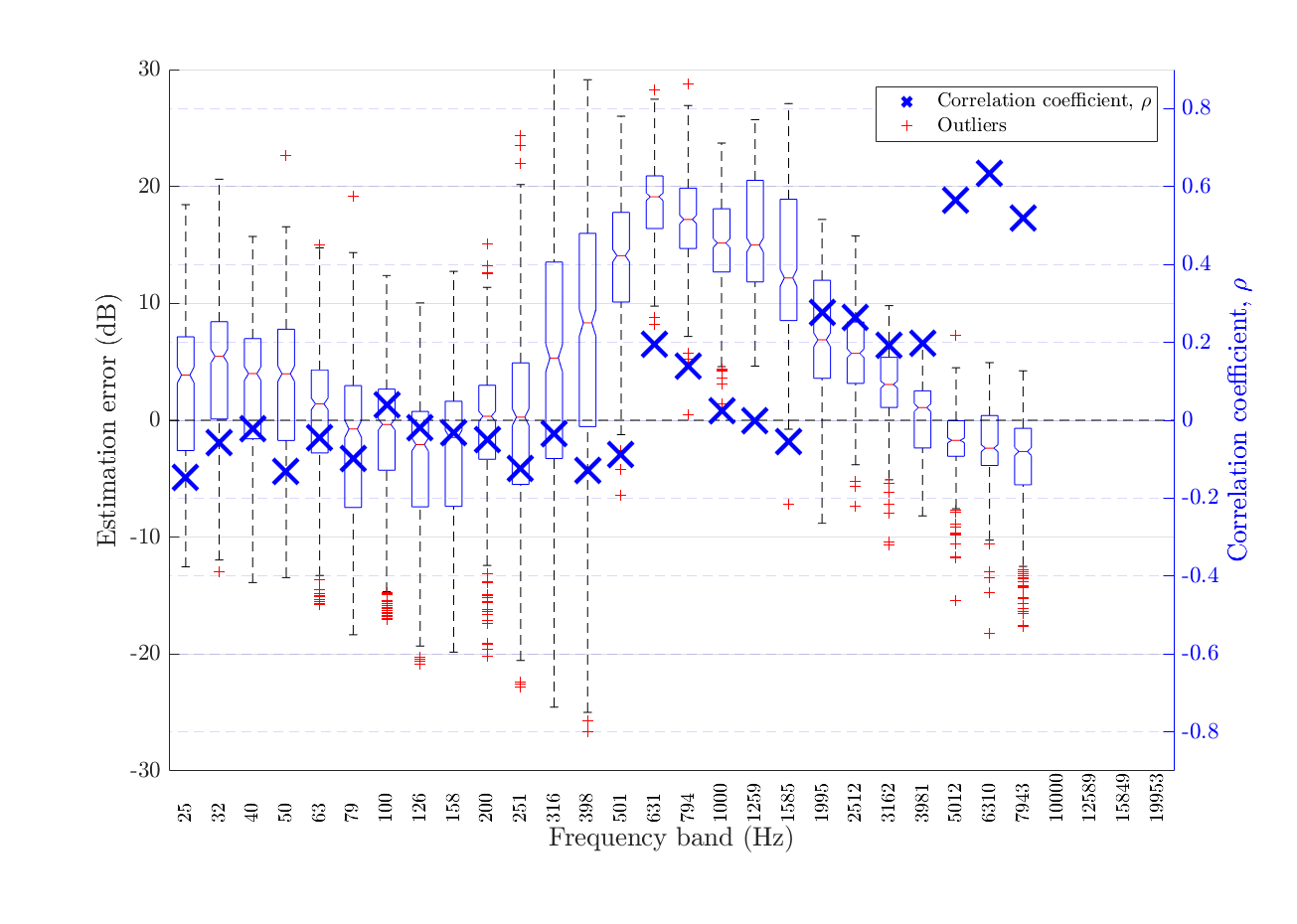,
	width=\figWidthACETR,viewport=45 10 765 530,clip}}%
	\else
	\centerline{\epsfig{figure=FigsACE/ana_eval_gt_partic_results_combined_Phase3_TR_P3S_DRR_dB_All_SNR_Fan_sub_ICASSP_2015_DRR_2-ch_Gerkmann_NR_filtered_subband.png,
	width=\figWidthACETR,viewport=45 10 765 530,clip}}%
	\fi
	\caption{{Frequency-dependent \ac{DRR} estimation error in fan noise for all \acp{SNR} for algorithm \ac{DENBE} with filtered subbands~\cite{Eaton2015c}}}%
\label{fig:ACE_DRR_Sub_Fan_Eaton_filt}%
\end{figure}%
\begin{table*}[!ht]\small
\caption{
	Frequency-dependent \ac{DRR} estimation error in fan noise for all \acp{SNR} for algorithm \ac{DENBE} with filtered subbands~\cite{Eaton2015c}
}
\vspace{5mm} 
\centering
\begin{tabular}{crrrl}%
\hline%
Freq. band
& Centre Freq. (Hz)
& Bias
& \acs{MSE}
& $\PearsonCC$

\\
\hline
\hline
\ifarXiv
 1 & 25.12 & 1.672 & 58.87 & -0.1467 \\ 
\hline
 2 & 31.62 & 2.705 & 63.12 & -0.05782 \\ 
\hline
 3 & 39.81 & 0.9042 & 53.73 & -0.0211 \\ 
\hline
 4 & 50.12 & 1.413 & 59.08 & -0.1302 \\ 
\hline
 5 & 63.10 & -1.317 & 53.74 & -0.04346 \\ 
\hline
 6 & 79.43 & -3.195 & 77.74 & -0.09915 \\ 
\hline
 7 & 100.00 & -3.199 & 60.21 & 0.04029 \\ 
\hline
 8 & 125.89 & -4.903 & 74.99 & -0.01795 \\ 
\hline
 9 & 158.49 & -3.928 & 68.95 & -0.0327 \\ 
\hline
10 & 199.53 & -2.936 & 59.26 & -0.04891 \\ 
\hline
11 & 251.19 & -3.417 & 81.91 & -0.124 \\ 
\hline
12 & 316.23 & 0.7962 & 123.5 & -0.0336 \\ 
\hline
13 & 398.11 & 2.026 & 123.4 & -0.1288 \\ 
\hline
14 & 501.19 & 7.271 & 138.6 & -0.08914 \\ 
\hline
15 & 630.96 & 13.94 & 237 & 0.196 \\ 
\hline
16 & 794.33 & 12.71 & 208 & 0.1394 \\ 
\hline
17 & 1000.00 & 10.09 & 158.5 & 0.02309 \\ 
\hline
18 & 1258.93 & 11.91 & 182 & -0.001627 \\ 
\hline
19 & 1584.89 & 9.922 & 148.6 & -0.05414 \\ 
\hline
20 & 1995.26 & 4.237 & 58.22 & 0.2764 \\ 
\hline
21 & 2511.89 & 2.256 & 47.98 & 0.2649 \\ 
\hline
22 & 3162.28 & -0.1546 & 32.93 & 0.1915 \\ 
\hline
23 & 3981.07 & -1.258 & 17.11 & 0.1974 \\ 
\hline
24 & 5011.87 & -2.383 & 15.33 & 0.5647 \\ 
\hline
25 & 6309.57 & -3.232 & 23.2 & 0.6342 \\ 
\hline
26 & 7943.28 & -4.841 & 44.68 & 0.5195 \\ 
\hline

\else

\fi
\end{tabular}
\end{table*}
%
%
%
\clearpage
\subsection{Frequency-dependent \ac{DRR} estimation results by noise type and \ac{SNR}}
\subsubsection{Ambient noise at \dBel{18}}
\begin{figure}[!ht]
	\ifarXiv
\centerline{\epsfig{figure=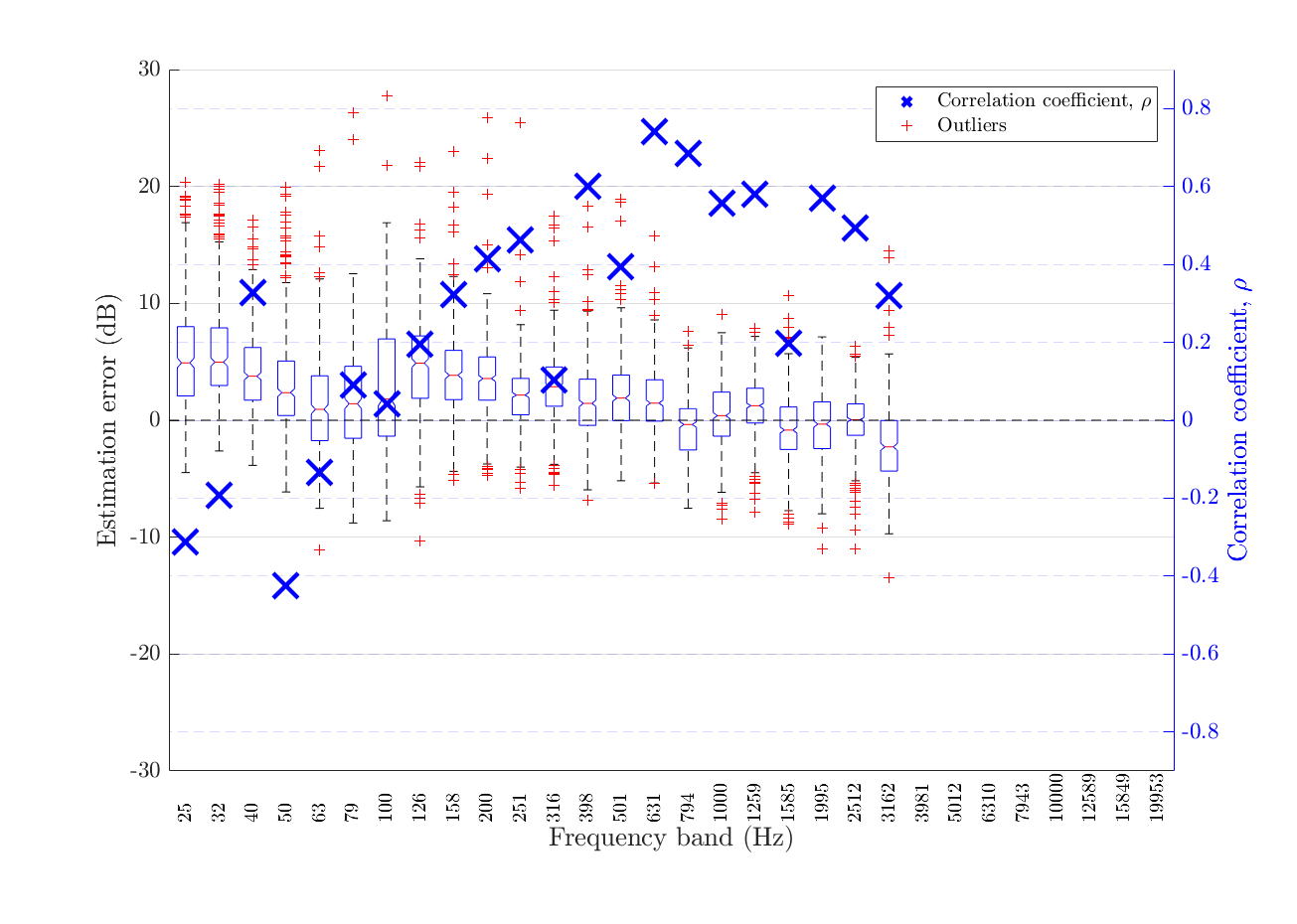,
	width=\figWidthACETR,viewport=45 10 765 530,clip}}%
	\else
	\centerline{\epsfig{figure=FigsACE/ana_eval_gt_partic_results_combined_Phase3_TR_P3S_DRR_dB_18dB_SNR_Ambient_sub_Velocity.png,
	width=\figWidthACETR,viewport=45 10 765 530,clip}}%
	\fi
	\caption{{
	Frequency-dependent \ac{DRR} estimation error in ambient noise at \dBel{18} \ac{SNR} for algorithm Particle Velocity~\cite{Chen2015}
	}}%
\label{fig:ACE_DRR_Sub_Ambient_18dB_Velocity}%
\end{figure}%
\begin{table*}[!ht]\small
\caption{
	Frequency-dependent \ac{DRR} estimation error in ambient noise at \dBel{18} \ac{SNR} for algorithm Particle Velocity~\cite{Chen2015}
}
\vspace{5mm} 
\centering
\begin{tabular}{crrrl}%
\hline%
Freq. band
& Centre Freq. (Hz)
& Bias
& \acs{MSE}
& $\PearsonCC$

\\
\hline
\hline
\ifarXiv
 1 & 25.12 & 5.406 & 50.56 & -0.3128 \\ 
\hline
 2 & 31.62 & 5.76 & 50.42 & -0.1923 \\ 
\hline
 3 & 39.81 & 4.176 & 30.1 & 0.3286 \\ 
\hline
 4 & 50.12 & 3.073 & 26.77 & -0.4243 \\ 
\hline
 5 & 63.10 & 1.351 & 19.74 & -0.1328 \\ 
\hline
 6 & 79.43 & 1.491 & 21.83 & 0.08933 \\ 
\hline
 7 & 100.00 & 2.656 & 35.8 & 0.04144 \\ 
\hline
 8 & 125.89 & 4.562 & 38.54 & 0.1947 \\ 
\hline
 9 & 158.49 & 4.027 & 27.7 & 0.3228 \\ 
\hline
10 & 199.53 & 3.599 & 22.6 & 0.4139 \\ 
\hline
11 & 251.19 & 2.115 & 11.63 & 0.4632 \\ 
\hline
12 & 316.23 & 2.923 & 16.9 & 0.1038 \\ 
\hline
13 & 398.11 & 1.741 & 11.87 & 0.6015 \\ 
\hline
14 & 501.19 & 2.165 & 14.94 & 0.3948 \\ 
\hline
15 & 630.96 & 1.808 & 10.65 & 0.7419 \\ 
\hline
16 & 794.33 & -0.5607 & 7.33 & 0.6849 \\ 
\hline
17 & 1000.00 & 0.5146 & 7.047 & 0.5579 \\ 
\hline
18 & 1258.93 & 1.212 & 7.82 & 0.5805 \\ 
\hline
19 & 1584.89 & -0.7134 & 9.087 & 0.1987 \\ 
\hline
20 & 1995.26 & -0.4129 & 6.879 & 0.5693 \\ 
\hline
21 & 2511.89 & -0.07927 & 5.183 & 0.4921 \\ 
\hline
22 & 3162.28 & -2.098 & 15.07 & 0.3196 \\ 
\hline

\else

\fi
\end{tabular}
\end{table*}
%
%
%
\begin{figure}[!ht]
	\ifarXiv
\centerline{\epsfig{figure=
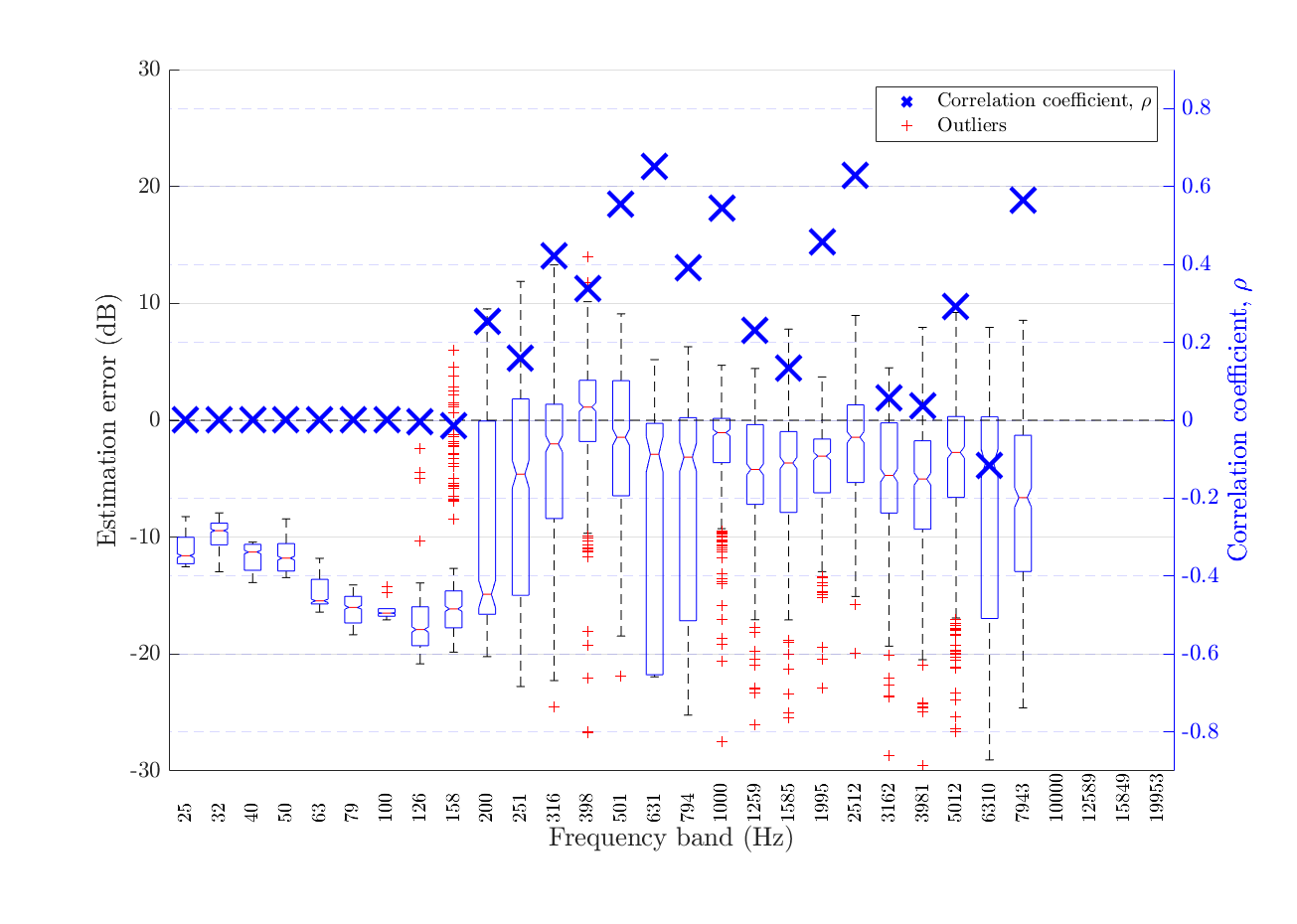,
	width=\figWidthACETR,viewport=45 10 765 530,clip}}%
	\else
	\centerline{\epsfig{figure=FigsACE/ana_eval_gt_partic_results_combined_Phase3_TR_P3S_DRR_dB_18dB_SNR_Ambient_sub_ICASSP_2015_DRR_2-ch_Gerkmann_NR_FFT_subband.png,
	width=\figWidthACETR,viewport=45 10 765 530,clip}}%
	\fi
	\caption{{
	Frequency-dependent \ac{DRR} estimation error in ambient noise  at \dBel{18} \ac{SNR} for algorithm \ac{DENBE} with FFT derived subbands~\cite{Eaton2015c}
	}}%
\label{fig:ACE_DRR_Sub_Ambient_18dB_Eaton_FFT}%
\end{figure}%
\begin{table*}[!ht]\small
\caption{
	Frequency-dependent \ac{DRR} estimation error in ambient noise  at \dBel{18} \ac{SNR} for algorithm \ac{DENBE} with FFT derived subbands~\cite{Eaton2015c}
}
\vspace{5mm} 
\centering
\begin{tabular}{crrrl}%
\hline%
Freq. band
& Centre Freq. (Hz)
& Bias
& \acs{MSE}
& $\PearsonCC$

\\
\hline
\hline
\ifarXiv
 1 & 25.12 & -11.07 & 124.4 & 0 \\ 
\hline
 2 & 31.62 & -9.93 & 100.8 & 0 \\ 
\hline
 3 & 39.81 & -11.69 & 138.1 & 0 \\ 
\hline
 4 & 50.12 & -11.55 & 135.4 & 0 \\ 
\hline
 5 & 63.10 & -14.82 & 221.5 & 0 \\ 
\hline
 6 & 79.43 & -16.09 & 260.6 & 0 \\ 
\hline
 7 & 100.00 & -16.18 & 262.6 & 0 \\ 
\hline
 8 & 125.89 & -17.64 & 316.2 & -0.004931 \\ 
\hline
 9 & 158.49 & -15.71 & 264.6 & -0.01359 \\ 
\hline
10 & 199.53 & -9.331 & 164.4 & 0.2525 \\ 
\hline
11 & 251.19 & -6.262 & 131.6 & 0.1594 \\ 
\hline
12 & 316.23 & -4.357 & 78.91 & 0.4217 \\ 
\hline
13 & 398.11 & 0.1566 & 32.37 & 0.3379 \\ 
\hline
14 & 501.19 & -2.666 & 60.08 & 0.5553 \\ 
\hline
15 & 630.96 & -8.696 & 173.6 & 0.6512 \\ 
\hline
16 & 794.33 & -6.646 & 126.9 & 0.391 \\ 
\hline
17 & 1000.00 & -2.447 & 27.52 & 0.5453 \\ 
\hline
18 & 1258.93 & -4.887 & 58.83 & 0.2303 \\ 
\hline
19 & 1584.89 & -4.395 & 55.2 & 0.1331 \\ 
\hline
20 & 1995.26 & -4.28 & 33.51 & 0.4573 \\ 
\hline
21 & 2511.89 & -2.04 & 20.88 & 0.6292 \\ 
\hline
22 & 3162.28 & -5.056 & 58.67 & 0.05753 \\ 
\hline
23 & 3981.07 & -6.156 & 71.91 & 0.0363 \\ 
\hline
24 & 5011.87 & -5.07 & 88.46 & 0.2912 \\ 
\hline
25 & 6309.57 & -8.196 & 196.5 & -0.1169 \\ 
\hline
26 & 7943.28 & -8.268 & 154.6 & 0.5657 \\ 
\hline

\else

\fi
\end{tabular}
\end{table*}
%
%
%
\begin{figure}[!ht]
	\ifarXiv
\centerline{\epsfig{figure=
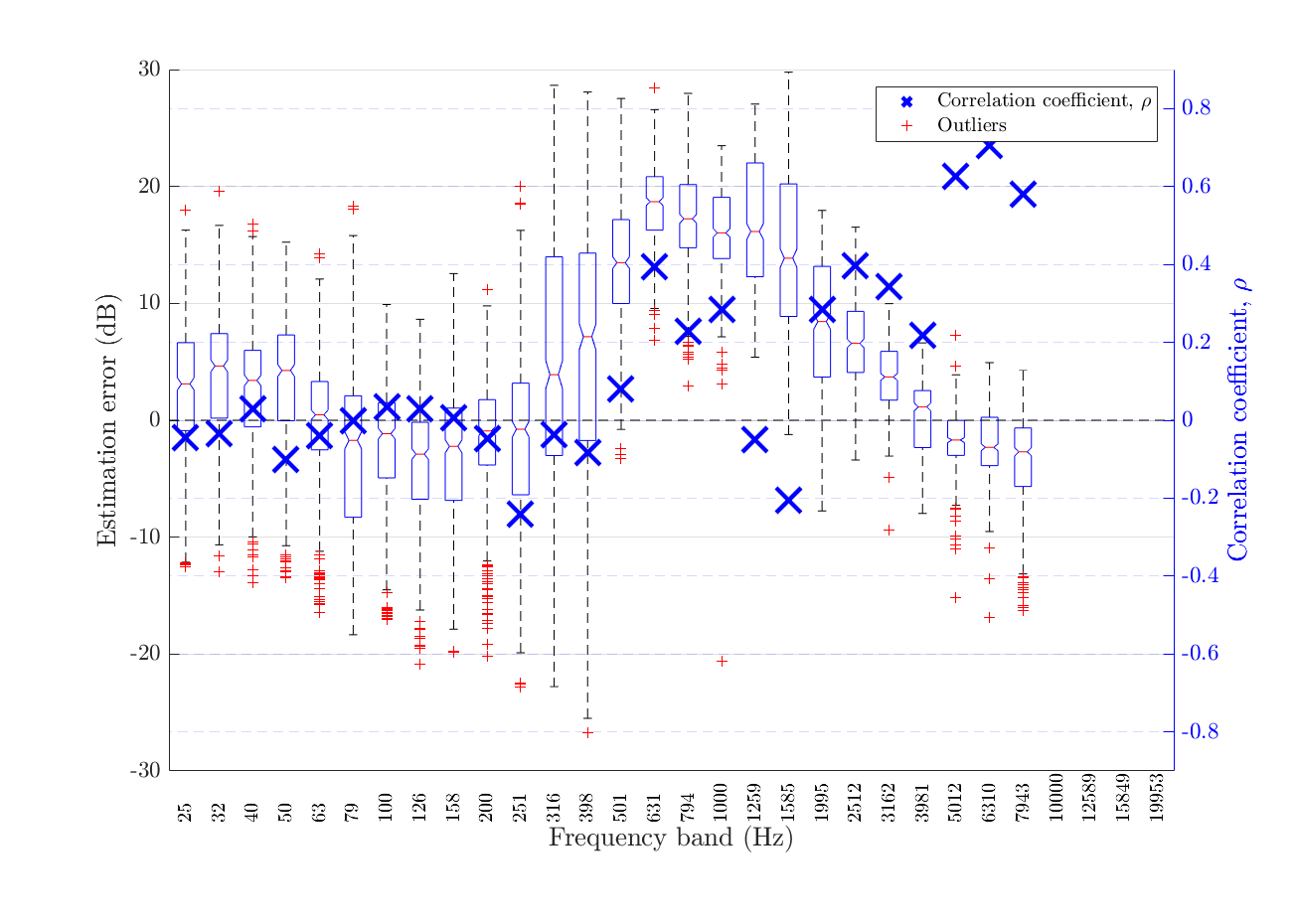,
	width=\figWidthACETR,viewport=45 10 765 530,clip}}%
	\else
	\centerline{\epsfig{figure=FigsACE/ana_eval_gt_partic_results_combined_Phase3_TR_P3S_DRR_dB_18dB_SNR_Ambient_sub_ICASSP_2015_DRR_2-ch_Gerkmann_NR_filtered_subband.png,
	width=\figWidthACETR,viewport=45 10 765 530,clip}}%
	\fi
	\caption{{
	Frequency-dependent \ac{DRR} estimation error in ambient noise  at \dBel{18} \ac{SNR} for algorithm \ac{DENBE} with filtered subbands~\cite{Eaton2015c}
	}}%
\label{fig:ACE_DRR_Sub_Ambient_18dB_Eaton_filt}%
\end{figure}%
\begin{table*}[!ht]\small
\caption{
	Frequency-dependent \ac{DRR} estimation error in ambient noise  at \dBel{18} \ac{SNR} for algorithm \ac{DENBE} with filtered subbands~\cite{Eaton2015c}
}
\vspace{5mm} 
\centering
\begin{tabular}{crrrl}%
\hline%
Freq. band
& Centre Freq. (Hz)
& Bias
& \acs{MSE}
& $\PearsonCC$

\\
\hline
\hline
\ifarXiv
 1 & 25.12 & 1.837 & 52.38 & -0.04507 \\ 
\hline
 2 & 31.62 & 2.974 & 52.09 & -0.03439 \\ 
\hline
 3 & 39.81 & 1.954 & 43.64 & 0.02952 \\ 
\hline
 4 & 50.12 & 2.518 & 51.84 & -0.1013 \\ 
\hline
 5 & 63.10 & -0.764 & 39.38 & -0.03964 \\ 
\hline
 6 & 79.43 & -3.26 & 68.83 & -0.001419 \\ 
\hline
 7 & 100.00 & -2.692 & 45.76 & 0.0342 \\ 
\hline
 8 & 125.89 & -4.244 & 56.44 & 0.0283 \\ 
\hline
 9 & 158.49 & -3.707 & 59.23 & 0.006268 \\ 
\hline
10 & 199.53 & -1.979 & 36.98 & -0.04766 \\ 
\hline
11 & 251.19 & -2.019 & 66.38 & -0.2408 \\ 
\hline
12 & 316.23 & 5.004 & 155.8 & -0.0357 \\ 
\hline
13 & 398.11 & 6.548 & 147.9 & -0.0831 \\ 
\hline
14 & 501.19 & 13.46 & 207.7 & 0.08096 \\ 
\hline
15 & 630.96 & 18.42 & 351 & 0.393 \\ 
\hline
16 & 794.33 & 17.29 & 315.8 & 0.2286 \\ 
\hline
17 & 1000.00 & 16.12 & 276.1 & 0.2841 \\ 
\hline
18 & 1258.93 & 16.81 & 311.4 & -0.04852 \\ 
\hline
19 & 1584.89 & 13.99 & 238.4 & -0.2051 \\ 
\hline
20 & 1995.26 & 8.126 & 92.09 & 0.2834 \\ 
\hline
21 & 2511.89 & 6.96 & 63.24 & 0.3958 \\ 
\hline
22 & 3162.28 & 3.66 & 20.4 & 0.3438 \\ 
\hline
23 & 3981.07 & 0.1298 & 8.961 & 0.2187 \\ 
\hline
24 & 5011.87 & -1.568 & 9.029 & 0.6269 \\ 
\hline
25 & 6309.57 & -2.109 & 12.11 & 0.7049 \\ 
\hline
26 & 7943.28 & -3.563 & 28.85 & 0.5793 \\ 
\hline

\else

\fi
\end{tabular}
\end{table*}
%
%
%
\clearpage
\subsubsection{Ambient noise at \dBel{12}}
\begin{figure}[!ht]
	\ifarXiv
\centerline{\epsfig{figure=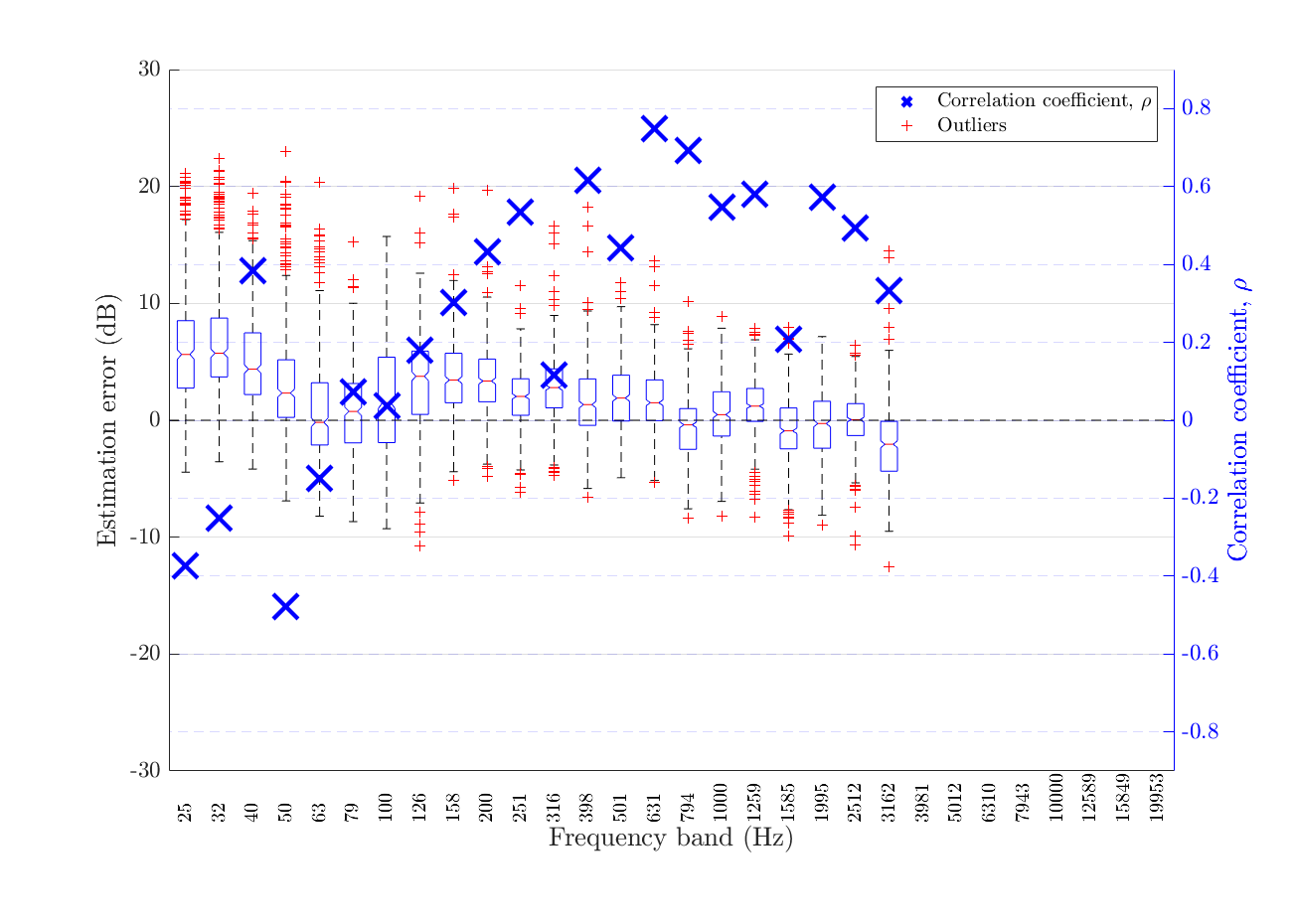,
	width=\figWidthACETR,viewport=45 10 765 530,clip}}%
	\else
	\centerline{\epsfig{figure=FigsACE/ana_eval_gt_partic_results_combined_Phase3_TR_P3S_DRR_dB_12dB_SNR_Ambient_sub_Velocity.png,
	width=\figWidthACETR,viewport=45 10 765 530,clip}}%
	\fi
	\caption{{Frequency-dependent \ac{DRR} estimation error in ambient noise at \dBel{12} \ac{SNR} for algorithm Particle Velocity~\cite{Chen2015}}}%
\label{fig:ACE_DRR_Sub_Ambient_12dB_Velocity}%
\end{figure}%
\begin{table*}[!ht]\small
\caption{
	Frequency-dependent \ac{DRR} estimation error in ambient noise at \dBel{12} \ac{SNR} for algorithm Particle Velocity~\cite{Chen2015}
}
\vspace{5mm} 
\centering
\begin{tabular}{crrrl}%
\hline%
Freq. band
& Centre Freq. (Hz)
& Bias
& \acs{MSE}
& $\PearsonCC$

\\
\hline
\hline
\ifarXiv
 1 & 25.12 & 6.359 & 65.55 & -0.3728 \\ 
\hline
 2 & 31.62 & 6.785 & 67.38 & -0.2514 \\ 
\hline
 3 & 39.81 & 5.078 & 41.35 & 0.3833 \\ 
\hline
 4 & 50.12 & 3.373 & 34.39 & -0.477 \\ 
\hline
 5 & 63.10 & 0.9007 & 19.91 & -0.148 \\ 
\hline
 6 & 79.43 & 0.7675 & 13.97 & 0.07313 \\ 
\hline
 7 & 100.00 & 1.702 & 23.86 & 0.03628 \\ 
\hline
 8 & 125.89 & 3.146 & 26.96 & 0.1796 \\ 
\hline
 9 & 158.49 & 3.641 & 23.7 & 0.3027 \\ 
\hline
10 & 199.53 & 3.398 & 19.63 & 0.4324 \\ 
\hline
11 & 251.19 & 1.981 & 9.75 & 0.5344 \\ 
\hline
12 & 316.23 & 2.793 & 15.63 & 0.1157 \\ 
\hline
13 & 398.11 & 1.74 & 11.58 & 0.615 \\ 
\hline
14 & 501.19 & 2.094 & 13.31 & 0.4419 \\ 
\hline
15 & 630.96 & 1.844 & 10.64 & 0.7478 \\ 
\hline
16 & 794.33 & -0.5516 & 7.374 & 0.6916 \\ 
\hline
17 & 1000.00 & 0.5537 & 7.12 & 0.5468 \\ 
\hline
18 & 1258.93 & 1.225 & 7.833 & 0.5796 \\ 
\hline
19 & 1584.89 & -0.7749 & 8.803 & 0.207 \\ 
\hline
20 & 1995.26 & -0.3684 & 6.802 & 0.5722 \\ 
\hline
21 & 2511.89 & -0.07648 & 5.207 & 0.4929 \\ 
\hline
22 & 3162.28 & -2.103 & 14.93 & 0.3319 \\ 
\hline

\else

\fi
\end{tabular}
\end{table*}
%
%
%
\begin{figure}[!ht]
	\ifarXiv
\centerline{\epsfig{figure=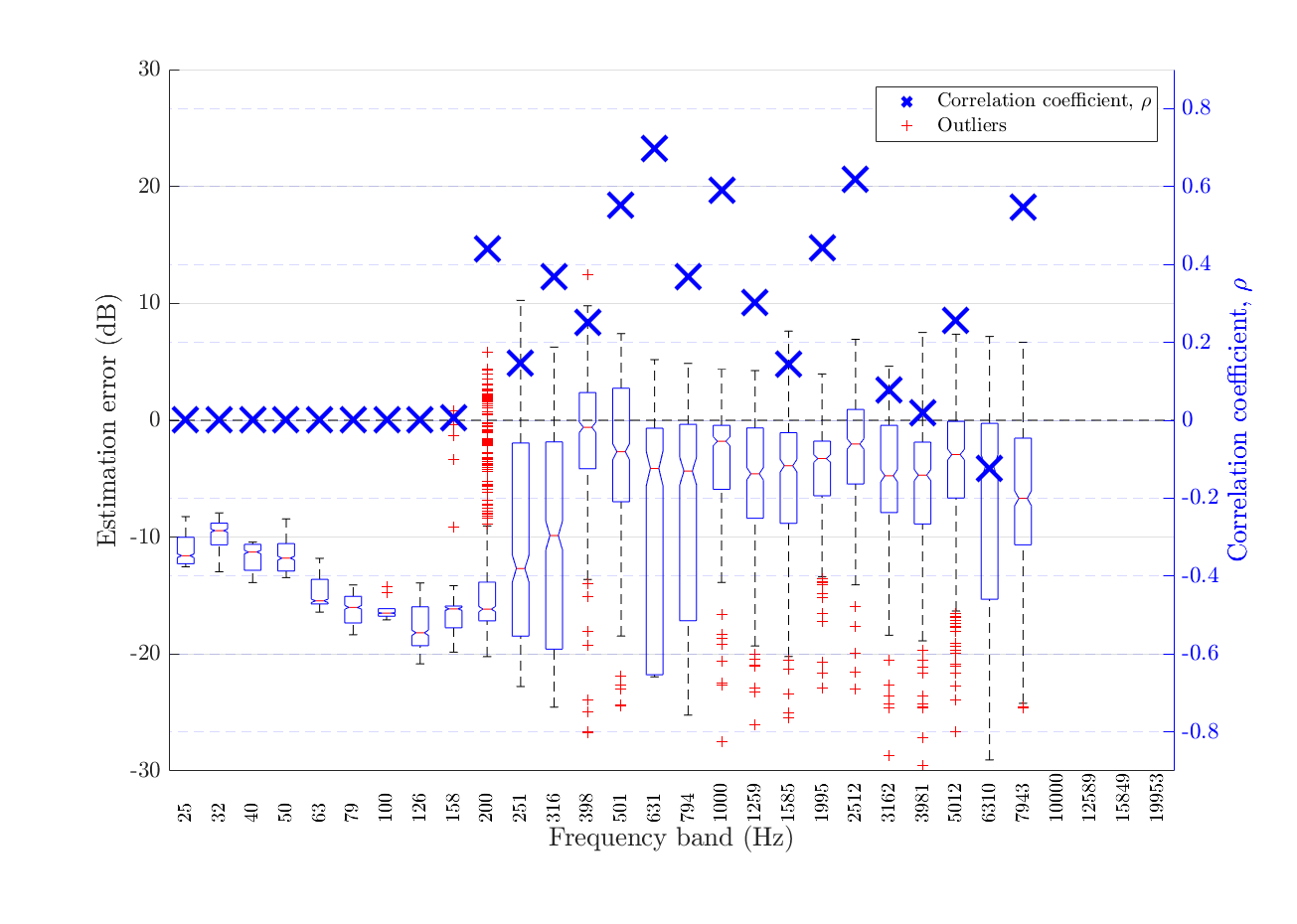,
	width=\figWidthACETR,viewport=45 10 765 530,clip}}%
	\else
	\centerline{\epsfig{figure=FigsACE/ana_eval_gt_partic_results_combined_Phase3_TR_P3S_DRR_dB_12dB_SNR_Ambient_sub_ICASSP_2015_DRR_2-ch_Gerkmann_NR_FFT_subband.png,
	width=\figWidthACETR,viewport=45 10 765 530,clip}}%
	\fi
	\caption{{Frequency-dependent \ac{DRR} estimation error in ambient noise  at \dBel{12} \ac{SNR} for algorithm \ac{DENBE} with FFT derived subbands~\cite{Eaton2015c}}}%
\label{fig:ACE_DRR_Sub_Ambient_12dB_Eaton_FFT}%
\end{figure}%
\begin{table*}[!ht]\small
\caption{
	Frequency-dependent \ac{DRR} estimation error in ambient noise  at \dBel{12} \ac{SNR} for algorithm \ac{DENBE} with FFT derived subbands~\cite{Eaton2015c}
}
\vspace{5mm} 
\centering
\begin{tabular}{crrrl}%
\hline%
Freq. band
& Centre Freq. (Hz)
& Bias
& \acs{MSE}
& $\PearsonCC$

\\
\hline
\hline
\ifarXiv
 1 & 25.12 & -11.07 & 124.4 & 0 \\ 
\hline
 2 & 31.62 & -9.93 & 100.8 & 0 \\ 
\hline
 3 & 39.81 & -11.69 & 138.1 & 0 \\ 
\hline
 4 & 50.12 & -11.55 & 135.4 & 0 \\ 
\hline
 5 & 63.10 & -14.82 & 221.5 & 0 \\ 
\hline
 6 & 79.43 & -16.09 & 260.6 & 0 \\ 
\hline
 7 & 100.00 & -16.18 & 262.6 & 0 \\ 
\hline
 8 & 125.89 & -17.74 & 318.4 & 0 \\ 
\hline
 9 & 158.49 & -16.61 & 280.5 & 0.005537 \\ 
\hline
10 & 199.53 & -13.41 & 220.1 & 0.441 \\ 
\hline
11 & 251.19 & -10.38 & 190 & 0.1459 \\ 
\hline
12 & 316.23 & -9.938 & 174.8 & 0.3679 \\ 
\hline
13 & 398.11 & -1.624 & 44.07 & 0.252 \\ 
\hline
14 & 501.19 & -3.547 & 67.21 & 0.5522 \\ 
\hline
15 & 630.96 & -9.559 & 187.6 & 0.6975 \\ 
\hline
16 & 794.33 & -7.938 & 149.4 & 0.3693 \\ 
\hline
17 & 1000.00 & -4.081 & 53.3 & 0.5898 \\ 
\hline
18 & 1258.93 & -5.65 & 72.35 & 0.3011 \\ 
\hline
19 & 1584.89 & -4.762 & 61.11 & 0.1436 \\ 
\hline
20 & 1995.26 & -4.527 & 37.28 & 0.4416 \\ 
\hline
21 & 2511.89 & -2.455 & 24.07 & 0.6174 \\ 
\hline
22 & 3162.28 & -5.193 & 61.14 & 0.07716 \\ 
\hline
23 & 3981.07 & -6.199 & 73.04 & 0.02002 \\ 
\hline
24 & 5011.87 & -5.401 & 89.86 & 0.2552 \\ 
\hline
25 & 6309.57 & -8.351 & 189.6 & -0.1232 \\ 
\hline
26 & 7943.28 & -8.355 & 147.1 & 0.548 \\ 
\hline

\else

\fi
\end{tabular}
\end{table*}
%
%
%
\begin{figure}[!ht]
	\ifarXiv
\centerline{\epsfig{figure=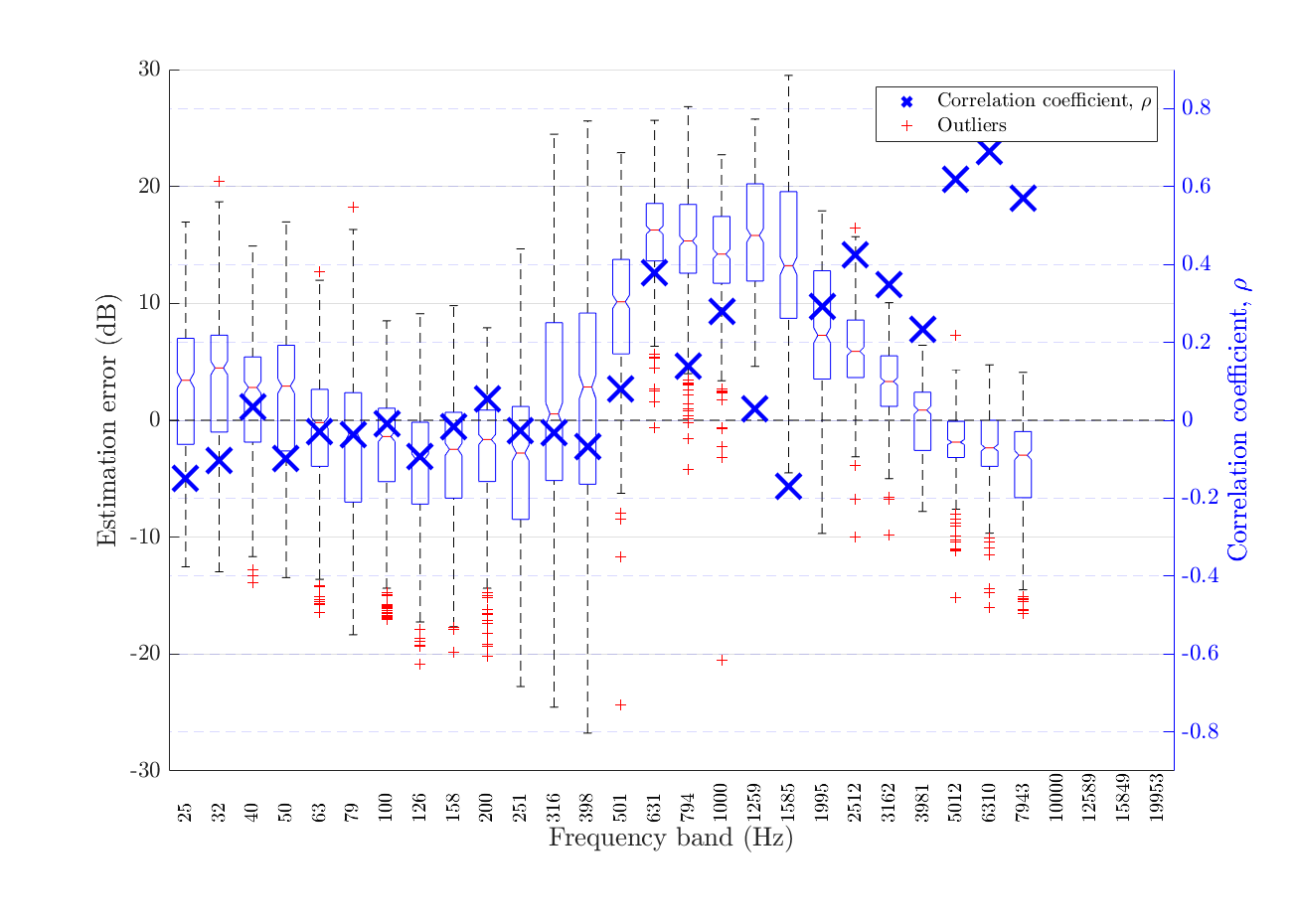,
	width=\figWidthACETR,viewport=45 10 765 530,clip}}%
	\else
	\centerline{\epsfig{figure=FigsACE/ana_eval_gt_partic_results_combined_Phase3_TR_P3S_DRR_dB_12dB_SNR_Ambient_sub_ICASSP_2015_DRR_2-ch_Gerkmann_NR_filtered_subband.png,
	width=\figWidthACETR,viewport=45 10 765 530,clip}}%
	\fi
	\caption{{Frequency-dependent \ac{DRR} estimation error in ambient noise  at \dBel{12} \ac{SNR} for algorithm \ac{DENBE} with filtered subbands~\cite{Eaton2015c}}}%
\label{fig:ACE_DRR_Sub_Ambient_12dB_Eaton_filt}%
\end{figure}%
\begin{table*}[!ht]\small
\caption{
	Frequency-dependent \ac{DRR} estimation error in ambient noise  at \dBel{12} \ac{SNR} for algorithm \ac{DENBE} with filtered subbands~\cite{Eaton2015c}
}
\vspace{5mm} 
\centering
\begin{tabular}{crrrl}%
\hline%
Freq. band
& Centre Freq. (Hz)
& Bias
& \acs{MSE}
& $\PearsonCC$

\\
\hline
\hline
\ifarXiv
 1 & 25.12 & 1.669 & 58.18 & -0.1503 \\ 
\hline
 2 & 31.62 & 2.594 & 56.2 & -0.1029 \\ 
\hline
 3 & 39.81 & 0.9571 & 42.32 & 0.03507 \\ 
\hline
 4 & 50.12 & 1.207 & 50.29 & -0.0987 \\ 
\hline
 5 & 63.10 & -1.482 & 45.9 & -0.03009 \\ 
\hline
 6 & 79.43 & -2.647 & 66.96 & -0.03601 \\ 
\hline
 7 & 100.00 & -3.125 & 49.47 & -0.008729 \\ 
\hline
 8 & 125.89 & -4.407 & 58.81 & -0.09237 \\ 
\hline
 9 & 158.49 & -3.911 & 59.85 & -0.01577 \\ 
\hline
10 & 199.53 & -3.022 & 44.1 & 0.05573 \\ 
\hline
11 & 251.19 & -4.229 & 70.15 & -0.0268 \\ 
\hline
12 & 316.23 & 1.125 & 107.4 & -0.03246 \\ 
\hline
13 & 398.11 & 2.161 & 96.84 & -0.06773 \\ 
\hline
14 & 501.19 & 9.611 & 127.5 & 0.0791 \\ 
\hline
15 & 630.96 & 15.92 & 268.8 & 0.378 \\ 
\hline
16 & 794.33 & 15.2 & 257.1 & 0.1393 \\ 
\hline
17 & 1000.00 & 14.07 & 220.2 & 0.2787 \\ 
\hline
18 & 1258.93 & 16.03 & 280.8 & 0.0291 \\ 
\hline
19 & 1584.89 & 13.58 & 225.9 & -0.1685 \\ 
\hline
20 & 1995.26 & 7.663 & 85.47 & 0.291 \\ 
\hline
21 & 2511.89 & 6.253 & 54.61 & 0.424 \\ 
\hline
22 & 3162.28 & 3.143 & 18.42 & 0.3489 \\ 
\hline
23 & 3981.07 & -0.07644 & 9.101 & 0.2326 \\ 
\hline
24 & 5011.87 & -1.73 & 9.988 & 0.6178 \\ 
\hline
25 & 6309.57 & -2.318 & 13.71 & 0.6894 \\ 
\hline
26 & 7943.28 & -3.937 & 32.1 & 0.571 \\ 
\hline

\else

\fi
\end{tabular}
\end{table*}
%
%
%
%
\clearpage
\subsubsection{Ambient noise at \dBel{-1}}
\begin{figure}[!ht]
	\ifarXiv
\centerline{\epsfig{figure=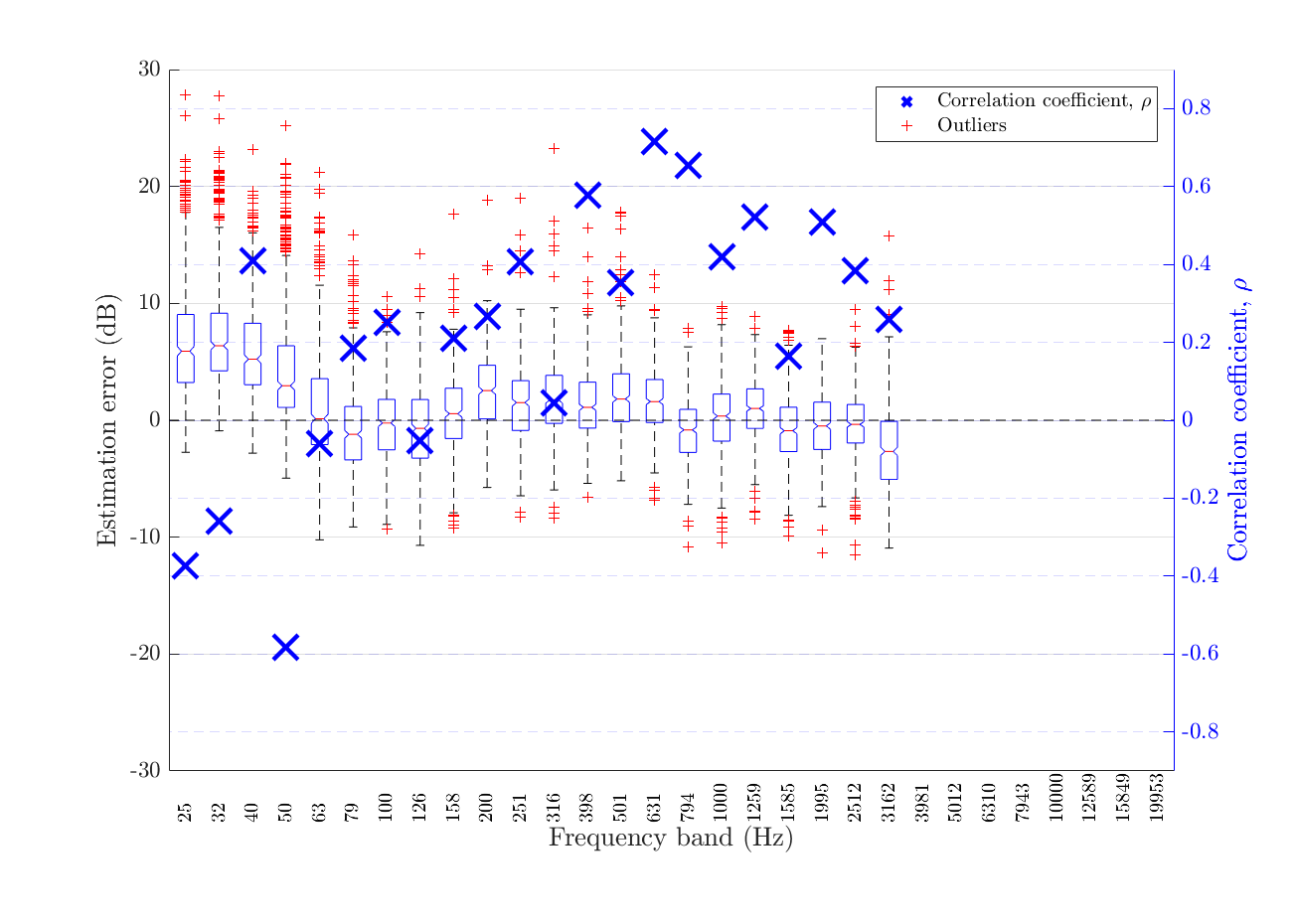,
	width=\figWidthACETR,viewport=45 10 765 530,clip}}%
	\else
	\centerline{\epsfig{figure=FigsACE/ana_eval_gt_partic_results_combined_Phase3_TR_P3S_DRR_dB_-1dB_SNR_Ambient_sub_Velocity.png,
	width=\figWidthACETR,viewport=45 10 765 530,clip}}%
	\fi
	\caption{{Frequency-dependent \ac{DRR} estimation error in ambient noise at \dBel{-1} \ac{SNR} for algorithm Particle Velocity~\cite{Chen2015}}}%
\label{fig:ACE_DRR_Sub_Ambient_-1dB_Velocity}%
\end{figure}%
\begin{table*}[!ht]\small
\caption{
	Frequency-dependent \ac{DRR} estimation error in ambient noise at \dBel{-1} \ac{SNR} for algorithm Particle Velocity~\cite{Chen2015}
}
\vspace{5mm} 
\centering
\begin{tabular}{crrrl}%
\hline%
Freq. band
& Centre Freq. (Hz)
& Bias
& \acs{MSE}
& $\PearsonCC$

\\
\hline
\hline
\ifarXiv
 1 & 25.12 & 6.877 & 75.1 & -0.3746 \\ 
\hline
 2 & 31.62 & 7.523 & 81 & -0.2588 \\ 
\hline
 3 & 39.81 & 6.169 & 56.59 & 0.4088 \\ 
\hline
 4 & 50.12 & 4.622 & 51.43 & -0.5833 \\ 
\hline
 5 & 63.10 & 1.25 & 25.48 & -0.05929 \\ 
\hline
 6 & 79.43 & -0.7251 & 15.61 & 0.1846 \\ 
\hline
 7 & 100.00 & -0.3289 & 11.11 & 0.2517 \\ 
\hline
 8 & 125.89 & -0.6238 & 15.06 & -0.05141 \\ 
\hline
 9 & 158.49 & 0.5003 & 12.24 & 0.2103 \\ 
\hline
10 & 199.53 & 2.328 & 16.54 & 0.2667 \\ 
\hline
11 & 251.19 & 1.354 & 12.72 & 0.4064 \\ 
\hline
12 & 316.23 & 1.816 & 14.92 & 0.04348 \\ 
\hline
13 & 398.11 & 1.468 & 11.99 & 0.5772 \\ 
\hline
14 & 501.19 & 2.222 & 16.97 & 0.3531 \\ 
\hline
15 & 630.96 & 1.801 & 11.3 & 0.7141 \\ 
\hline
16 & 794.33 & -0.8585 & 8.804 & 0.6549 \\ 
\hline
17 & 1000.00 & 0.2432 & 9.689 & 0.4193 \\ 
\hline
18 & 1258.93 & 1.016 & 8.424 & 0.5203 \\ 
\hline
19 & 1584.89 & -0.7994 & 10.19 & 0.1657 \\ 
\hline
20 & 1995.26 & -0.4474 & 7.949 & 0.5098 \\ 
\hline
21 & 2511.89 & -0.4781 & 7.765 & 0.3828 \\ 
\hline
22 & 3162.28 & -2.465 & 19.17 & 0.26 \\ 
\hline

\else

\fi
\end{tabular}
\end{table*}
%
%
%
\begin{figure}[!ht]
	\ifarXiv
\centerline{\epsfig{figure=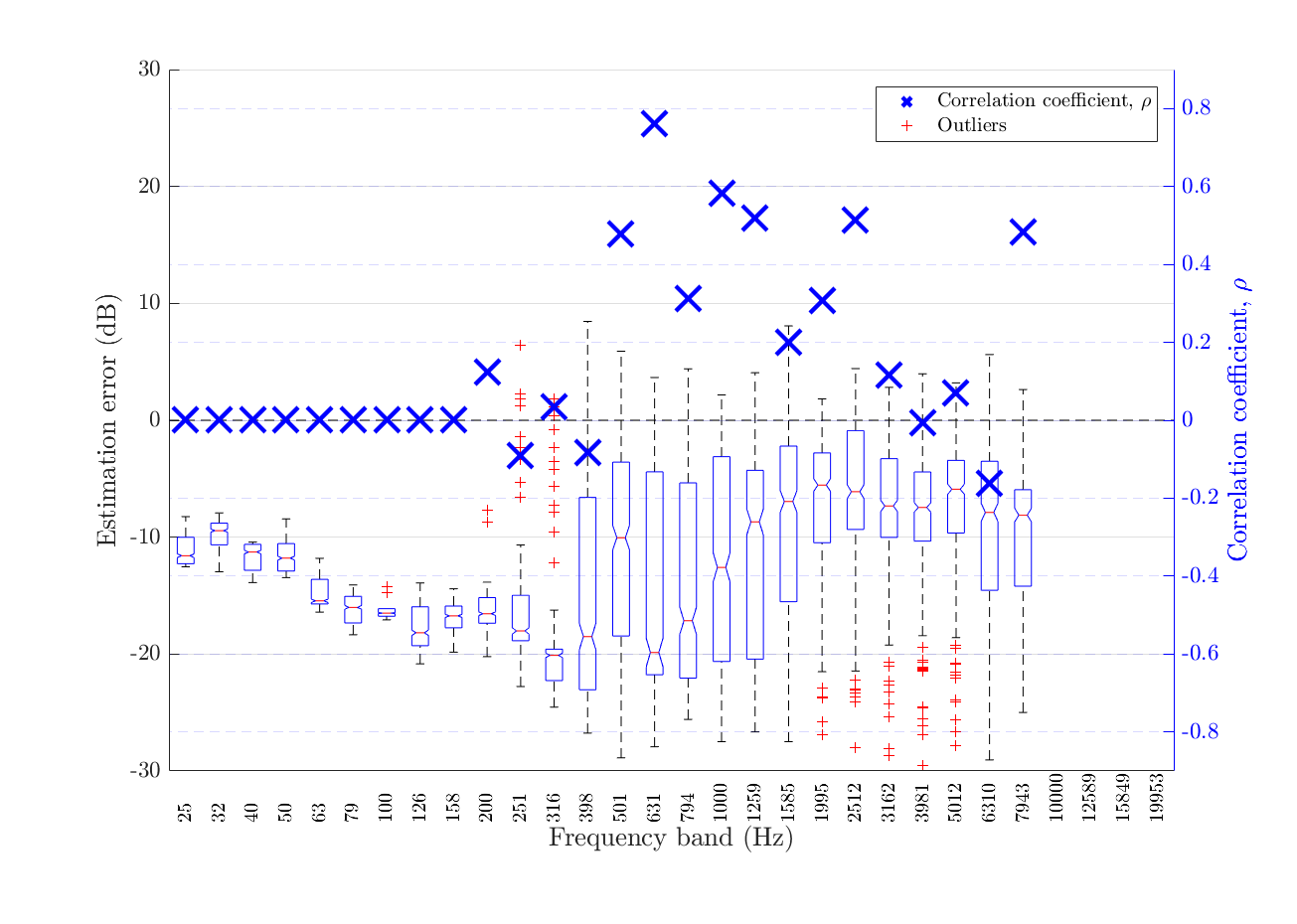,
	width=\figWidthACETR,viewport=45 10 765 530,clip}}%
	\else
	\centerline{\epsfig{figure=FigsACE/ana_eval_gt_partic_results_combined_Phase3_TR_P3S_DRR_dB_-1dB_SNR_Ambient_sub_ICASSP_2015_DRR_2-ch_Gerkmann_NR_FFT_subband.png,
	width=\figWidthACETR,viewport=45 10 765 530,clip}}%
	\fi
	\caption{{Frequency-dependent \ac{DRR} estimation error in ambient noise  at \dBel{-1} \ac{SNR} for algorithm \ac{DENBE} with FFT derived subbands~\cite{Eaton2015c}}}%
\label{fig:ACE_DRR_Sub_Ambient_-1dB_Eaton_FFT}%
\end{figure}%
\begin{table*}[!ht]\small
\caption{
	Frequency-dependent \ac{DRR} estimation error in ambient noise  at \dBel{-1} \ac{SNR} for algorithm \ac{DENBE} with FFT derived subbands~\cite{Eaton2015c}
}
\vspace{5mm} 
\centering
\begin{tabular}{crrrl}%
\hline%
Freq. band
& Centre Freq. (Hz)
& Bias
& \acs{MSE}
& $\PearsonCC$

\\
\hline
\hline
\ifarXiv
 1 & 25.12 & -11.07 & 124.4 & 0 \\ 
\hline
 2 & 31.62 & -9.93 & 100.8 & 0 \\ 
\hline
 3 & 39.81 & -11.69 & 138.1 & 0 \\ 
\hline
 4 & 50.12 & -11.55 & 135.4 & 0 \\ 
\hline
 5 & 63.10 & -14.82 & 221.5 & 0 \\ 
\hline
 6 & 79.43 & -16.09 & 260.6 & 0 \\ 
\hline
 7 & 100.00 & -16.18 & 262.6 & 0 \\ 
\hline
 8 & 125.89 & -17.74 & 318.4 & 0 \\ 
\hline
 9 & 158.49 & -16.76 & 283.2 & 0 \\ 
\hline
10 & 199.53 & -16.67 & 281.5 & 0.1229 \\ 
\hline
11 & 251.19 & -17.36 & 317 & -0.09157 \\ 
\hline
12 & 316.23 & -19.59 & 400.4 & 0.03542 \\ 
\hline
13 & 398.11 & -15.32 & 331.3 & -0.08202 \\ 
\hline
14 & 501.19 & -11.24 & 210 & 0.4774 \\ 
\hline
15 & 630.96 & -13.65 & 266.7 & 0.7611 \\ 
\hline
16 & 794.33 & -14.31 & 287.1 & 0.3133 \\ 
\hline
17 & 1000.00 & -12.41 & 234 & 0.5832 \\ 
\hline
18 & 1258.93 & -11.14 & 191 & 0.5185 \\ 
\hline
19 & 1584.89 & -8.548 & 142.1 & 0.1992 \\ 
\hline
20 & 1995.26 & -7.744 & 105.6 & 0.307 \\ 
\hline
21 & 2511.89 & -6.79 & 92.29 & 0.5136 \\ 
\hline
22 & 3162.28 & -7.42 & 88.62 & 0.1148 \\ 
\hline
23 & 3981.07 & -8.258 & 101.1 & -0.007562 \\ 
\hline
24 & 5011.87 & -7.411 & 92.54 & 0.06913 \\ 
\hline
25 & 6309.57 & -10.52 & 193.5 & -0.1615 \\ 
\hline
26 & 7943.28 & -10.32 & 151.4 & 0.4823 \\ 
\hline

\else

\fi
\end{tabular}
\end{table*}
%
%
%
\begin{figure}[!ht]
	\ifarXiv
\centerline{\epsfig{figure=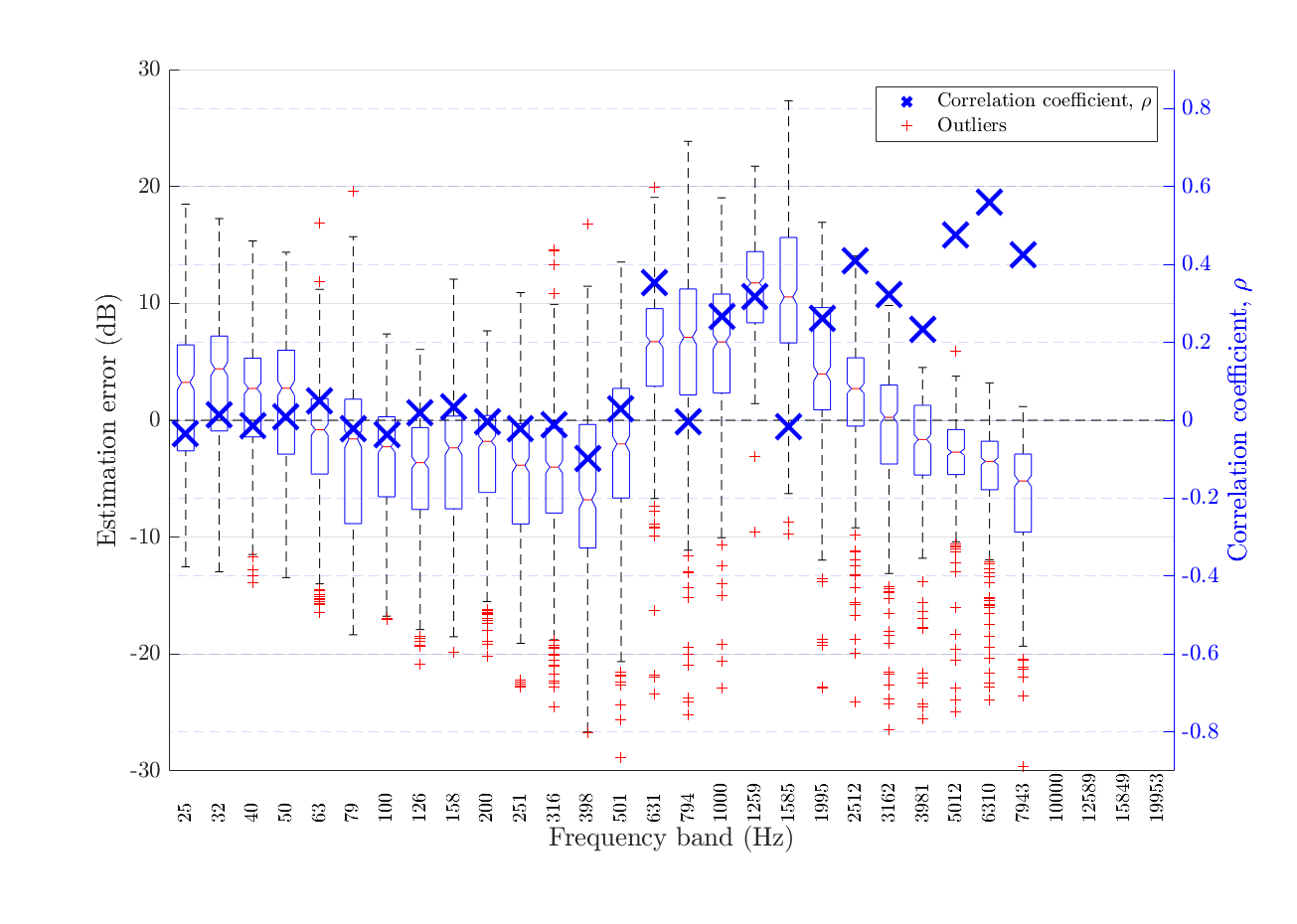,
	width=\figWidthACETR,viewport=45 10 765 530,clip}}%
	\else
	\centerline{\epsfig{figure=FigsACE/ana_eval_gt_partic_results_combined_Phase3_TR_P3S_DRR_dB_-1dB_SNR_Ambient_sub_ICASSP_2015_DRR_2-ch_Gerkmann_NR_filtered_subband.png,
	width=\figWidthACETR,viewport=45 10 765 530,clip}}%
	\fi
	\caption{{Frequency-dependent \ac{DRR} estimation error in ambient noise  at \dBel{-1} \ac{SNR} for algorithm \ac{DENBE} with filtered subbands~\cite{Eaton2015c}}}%
\label{fig:ACE_DRR_Sub_Ambient_-1dB_Eaton_filt}%
\end{figure}%
\begin{table*}[!ht]\small
\caption{
	Frequency-dependent \ac{DRR} estimation error in ambient noise  at \dBel{-1} \ac{SNR} for algorithm \ac{DENBE} with filtered subbands~\cite{Eaton2015c}
}
\vspace{5mm} 
\centering
\begin{tabular}{crrrl}%
\hline%
Freq. band
& Centre Freq. (Hz)
& Bias
& \acs{MSE}
& $\PearsonCC$

\\
\hline
\hline
\ifarXiv
 1 & 25.12 & 1.475 & 54.36 & -0.034 \\ 
\hline
 2 & 31.62 & 2.572 & 52.8 & 0.01495 \\ 
\hline
 3 & 39.81 & 1.155 & 40.57 & -0.013 \\ 
\hline
 4 & 50.12 & 1.02 & 48.62 & 0.01018 \\ 
\hline
 5 & 63.10 & -2.08 & 44.58 & 0.05004 \\ 
\hline
 6 & 79.43 & -3.417 & 70.46 & -0.02126 \\ 
\hline
 7 & 100.00 & -3.875 & 52.12 & -0.03701 \\ 
\hline
 8 & 125.89 & -5.134 & 66.23 & 0.01959 \\ 
\hline
 9 & 158.49 & -4.026 & 63.19 & 0.03421 \\ 
\hline
10 & 199.53 & -3.621 & 52.36 & -0.004437 \\ 
\hline
11 & 251.19 & -5.054 & 71.9 & -0.02234 \\ 
\hline
12 & 316.23 & -4.934 & 74.52 & -0.01068 \\ 
\hline
13 & 398.11 & -6.586 & 109.8 & -0.09895 \\ 
\hline
14 & 501.19 & -2.527 & 60.51 & 0.02878 \\ 
\hline
15 & 630.96 & 5.978 & 68.93 & 0.3528 \\ 
\hline
16 & 794.33 & 6.295 & 108.7 & -0.004775 \\ 
\hline
17 & 1000.00 & 6.022 & 83.15 & 0.2673 \\ 
\hline
18 & 1258.93 & 11.5 & 150.1 & 0.3178 \\ 
\hline
19 & 1584.89 & 10.63 & 158.3 & -0.0164 \\ 
\hline
20 & 1995.26 & 4.64 & 63.46 & 0.2619 \\ 
\hline
21 & 2511.89 & 1.932 & 35.3 & 0.41 \\ 
\hline
22 & 3162.28 & -0.9033 & 36.01 & 0.3213 \\ 
\hline
23 & 3981.07 & -2.209 & 25.19 & 0.2336 \\ 
\hline
24 & 5011.87 & -3.17 & 24.71 & 0.4751 \\ 
\hline
25 & 6309.57 & -4.261 & 33.77 & 0.5598 \\ 
\hline
26 & 7943.28 & -6.688 & 70.49 & 0.4238 \\ 
\hline

\else

\fi
\end{tabular}
\end{table*}
%
%
%
%
\clearpage
\subsubsection{Babble noise at \dBel{18}}
\begin{figure}[!ht]
	\ifarXiv
\centerline{\epsfig{figure=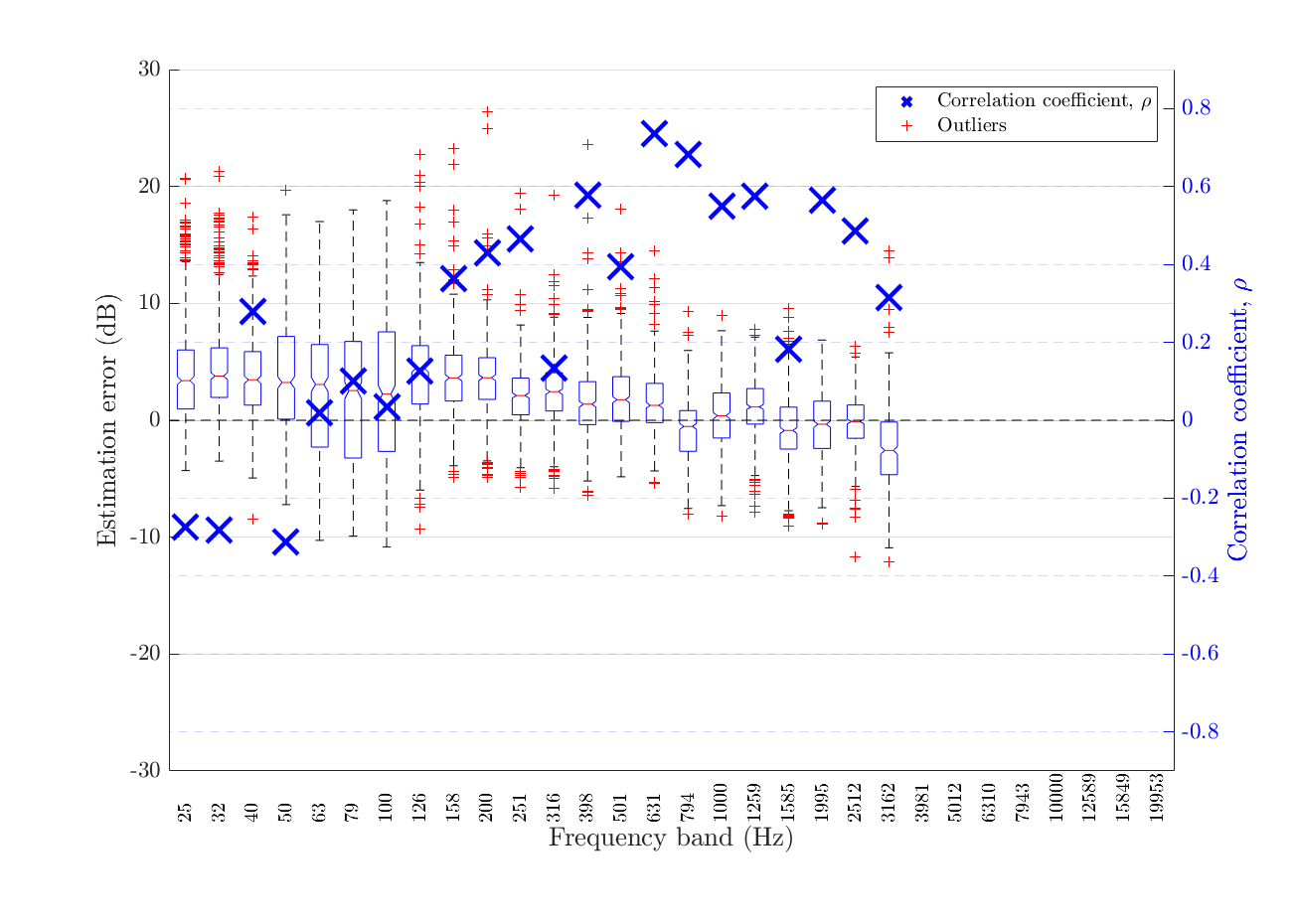,
	width=\figWidthACETR,viewport=45 10 765 530,clip}}%
	\else
	\centerline{\epsfig{figure=FigsACE/ana_eval_gt_partic_results_combined_Phase3_TR_P3S_DRR_dB_18dB_SNR_Babble_sub_Velocity.png,
	width=\figWidthACETR,viewport=45 10 765 530,clip}}%
	\fi
	\caption{{Frequency-dependent \ac{DRR} estimation error in babble noise at \dBel{18} \ac{SNR} for algorithm Particle Velocity~\cite{Chen2015}}}%
\label{fig:ACE_DRR_Sub_Babble_18dB_Velocity}%
\end{figure}%
\begin{table*}[!ht]\small
\caption{
	Frequency-dependent \ac{DRR} estimation error in babble noise at \dBel{18} \ac{SNR} for algorithm Particle Velocity~\cite{Chen2015}
}
\vspace{5mm} 
\centering
\begin{tabular}{crrrl}%
\hline%
Freq. band
& Centre Freq. (Hz)
& Bias
& \acs{MSE}
& $\PearsonCC$
\\
\hline
\hline
\ifarXiv
 1 & 25.12 & 4.045 & 36.66 & -0.2731 \\ 
\hline
 2 & 31.62 & 4.636 & 38.22 & -0.2818 \\ 
\hline
 3 & 39.81 & 3.781 & 27.1 & 0.2794 \\ 
\hline
 4 & 50.12 & 3.692 & 34.92 & -0.3127 \\ 
\hline
 5 & 63.10 & 2.471 & 33.61 & 0.02033 \\ 
\hline
 6 & 79.43 & 1.976 & 35.13 & 0.1008 \\ 
\hline
 7 & 100.00 & 2.378 & 40.53 & 0.03389 \\ 
\hline
 8 & 125.89 & 3.913 & 34.6 & 0.1254 \\ 
\hline
 9 & 158.49 & 3.809 & 25.51 & 0.3629 \\ 
\hline
10 & 199.53 & 3.569 & 22.52 & 0.429 \\ 
\hline
11 & 251.19 & 2.058 & 11.08 & 0.4648 \\ 
\hline
12 & 316.23 & 2.475 & 13.74 & 0.135 \\ 
\hline
13 & 398.11 & 1.681 & 12.31 & 0.5766 \\ 
\hline
14 & 501.19 & 2.033 & 13.76 & 0.3938 \\ 
\hline
15 & 630.96 & 1.622 & 9.909 & 0.7345 \\ 
\hline
16 & 794.33 & -0.6727 & 7.534 & 0.6814 \\ 
\hline
17 & 1000.00 & 0.4143 & 7.104 & 0.5483 \\ 
\hline
18 & 1258.93 & 1.114 & 7.761 & 0.5754 \\ 
\hline
19 & 1584.89 & -0.7544 & 9.169 & 0.1823 \\ 
\hline
20 & 1995.26 & -0.4273 & 6.97 & 0.5643 \\ 
\hline
21 & 2511.89 & -0.2249 & 5.301 & 0.4866 \\ 
\hline
22 & 3162.28 & -2.361 & 17.19 & 0.3143 \\ 
\hline

\else

\fi
\end{tabular}
\end{table*}
%
%
%
\begin{figure}[!ht]
	\ifarXiv
\centerline{\epsfig{figure=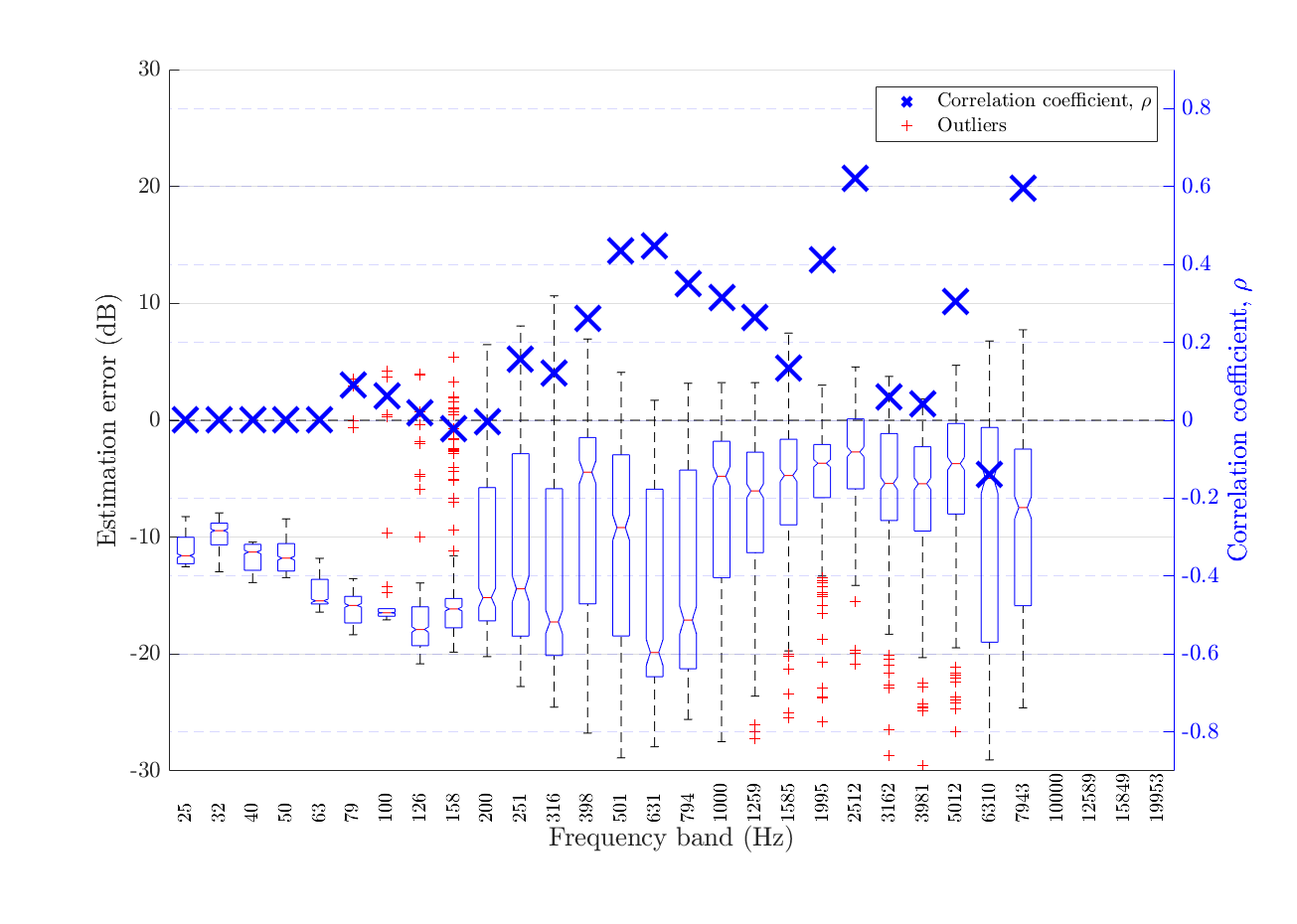,
	width=\figWidthACETR,viewport=45 10 765 530,clip}}%
	\else
	\centerline{\epsfig{figure=FigsACE/ana_eval_gt_partic_results_combined_Phase3_TR_P3S_DRR_dB_18dB_SNR_Babble_sub_ICASSP_2015_DRR_2-ch_Gerkmann_NR_FFT_subband.png,
	width=\figWidthACETR,viewport=45 10 765 530,clip}}%
	\fi
	\caption{{Frequency-dependent \ac{DRR} estimation error in babble noise  at \dBel{18} \ac{SNR} for algorithm \ac{DENBE} with FFT derived subbands~\cite{Eaton2015c}}}%
\label{fig:ACE_DRR_Sub_Babble_18dB_Eaton_FFT}%
\end{figure}%
\begin{table*}[!ht]\small
\caption{
	Frequency-dependent \ac{DRR} estimation error in babble noise  at \dBel{18} \ac{SNR} for algorithm \ac{DENBE} with FFT derived subbands~\cite{Eaton2015c}
}
\vspace{5mm} 
\centering
\begin{tabular}{crrrl}%
\hline%
Freq. band
& Centre Freq. (Hz)
& Bias
& \acs{MSE}
& $\PearsonCC$

\\
\hline
\hline
\ifarXiv
 1 & 25.12 & -11.07 & 124.4 & 0 \\ 
\hline
 2 & 31.62 & -9.93 & 100.8 & 0 \\ 
\hline
 3 & 39.81 & -11.69 & 138.1 & 0 \\ 
\hline
 4 & 50.12 & -11.55 & 135.4 & 0 \\ 
\hline
 5 & 63.10 & -14.82 & 221.5 & 0 \\ 
\hline
 6 & 79.43 & -15.93 & 258 & 0.08941 \\ 
\hline
 7 & 100.00 & -16.01 & 260.1 & 0.06148 \\ 
\hline
 8 & 125.89 & -17.43 & 312.4 & 0.01814 \\ 
\hline
 9 & 158.49 & -15.95 & 268.6 & -0.02279 \\ 
\hline
10 & 199.53 & -12.17 & 204.7 & -0.003117 \\ 
\hline
11 & 251.19 & -11.13 & 194 & 0.1571 \\ 
\hline
12 & 316.23 & -13.89 & 263.9 & 0.1212 \\ 
\hline
13 & 398.11 & -7.772 & 133.8 & 0.2609 \\ 
\hline
14 & 501.19 & -10.58 & 185.7 & 0.4345 \\ 
\hline
15 & 630.96 & -14.55 & 282.7 & 0.448 \\ 
\hline
16 & 794.33 & -13.03 & 248.9 & 0.3505 \\ 
\hline
17 & 1000.00 & -8.092 & 135.7 & 0.315 \\ 
\hline
18 & 1258.93 & -7.993 & 116.8 & 0.2627 \\ 
\hline
19 & 1584.89 & -5.421 & 70.28 & 0.1349 \\ 
\hline
20 & 1995.26 & -4.894 & 42.14 & 0.4116 \\ 
\hline
21 & 2511.89 & -3.024 & 27.58 & 0.6217 \\ 
\hline
22 & 3162.28 & -5.777 & 64.58 & 0.05932 \\ 
\hline
23 & 3981.07 & -6.629 & 74.89 & 0.0429 \\ 
\hline
24 & 5011.87 & -5.767 & 88.67 & 0.3052 \\ 
\hline
25 & 6309.57 & -9.303 & 208.7 & -0.1381 \\ 
\hline
26 & 7943.28 & -9.405 & 165.6 & 0.5956 \\ 
\hline

\else

\fi
\end{tabular}
\end{table*}
%
%
%
\begin{figure}[!ht]
	\ifarXiv
\centerline{\epsfig{figure=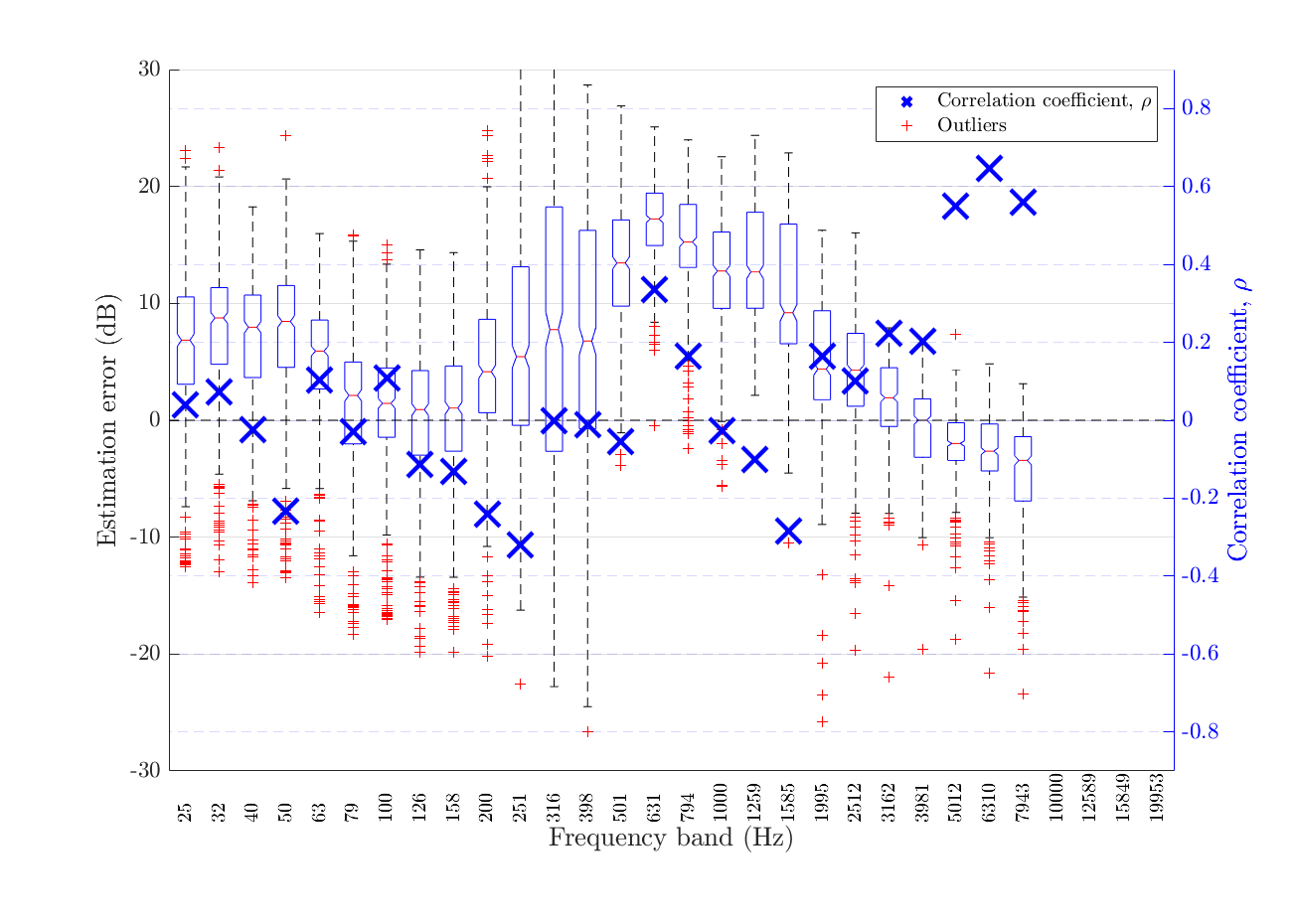,
	width=\figWidthACETR,viewport=45 10 765 530,clip}}%
	\else
	\centerline{\epsfig{figure=FigsACE/ana_eval_gt_partic_results_combined_Phase3_TR_P3S_DRR_dB_18dB_SNR_Babble_sub_ICASSP_2015_DRR_2-ch_Gerkmann_NR_filtered_subband.png,
	width=\figWidthACETR,viewport=45 10 765 530,clip}}%
	\fi
	\caption{{Frequency-dependent \ac{DRR} estimation error in babble noise  at \dBel{18} \ac{SNR} for algorithm \ac{DENBE} with filtered subbands~\cite{Eaton2015c}}}%
\label{fig:ACE_DRR_Sub_Babble_18dB_Eaton_filt}%
\end{figure}%
\begin{table*}[!ht]\small
\caption{
	Frequency-dependent \ac{DRR} estimation error in babble noise  at \dBel{18} \ac{SNR} for algorithm \ac{DENBE} with filtered subbands~\cite{Eaton2015c}
}
\vspace{5mm} 
\centering
\begin{tabular}{crrrl}%
\hline%
Freq. band
& Centre Freq. (Hz)
& Bias
& \acs{MSE}
& $\PearsonCC$

\\
\hline
\hline
\ifarXiv
 1 & 25.12 & 5.806 & 87.23 & 0.03839 \\ 
\hline
 2 & 31.62 & 7.379 & 102 & 0.07329 \\ 
\hline
 3 & 39.81 & 6.041 & 84.26 & -0.02312 \\ 
\hline
 4 & 50.12 & 6.841 & 101.5 & -0.2325 \\ 
\hline
 5 & 63.10 & 4.888 & 59.84 & 0.1023 \\ 
\hline
 6 & 79.43 & 0.7027 & 51.13 & -0.02964 \\ 
\hline
 7 & 100.00 & 0.5538 & 42.13 & 0.1081 \\ 
\hline
 8 & 125.89 & 0.3938 & 32.13 & -0.1125 \\ 
\hline
 9 & 158.49 & 0.007819 & 48.47 & -0.1312 \\ 
\hline
10 & 199.53 & 4.481 & 68.56 & -0.2401 \\ 
\hline
11 & 251.19 & 6.531 & 130.1 & -0.3197 \\ 
\hline
12 & 316.23 & 8.028 & 203.7 & -0.00221 \\ 
\hline
13 & 398.11 & 7.136 & 156.2 & -0.0126 \\ 
\hline
14 & 501.19 & 13.5 & 214 & -0.05445 \\ 
\hline
15 & 630.96 & 16.96 & 298 & 0.3353 \\ 
\hline
16 & 794.33 & 15.16 & 248.7 & 0.1636 \\ 
\hline
17 & 1000.00 & 12.61 & 184.9 & -0.027 \\ 
\hline
18 & 1258.93 & 13.49 & 205.5 & -0.09952 \\ 
\hline
19 & 1584.89 & 10.46 & 149.7 & -0.2832 \\ 
\hline
20 & 1995.26 & 5.049 & 56.02 & 0.165 \\ 
\hline
21 & 2511.89 & 4.132 & 46.24 & 0.1014 \\ 
\hline
22 & 3162.28 & 1.57 & 16.45 & 0.2225 \\ 
\hline
23 & 3981.07 & -0.7083 & 10.32 & 0.202 \\ 
\hline
24 & 5011.87 & -1.98 & 12.09 & 0.5492 \\ 
\hline
25 & 6309.57 & -2.584 & 15.93 & 0.6471 \\ 
\hline
26 & 7943.28 & -4.422 & 37.51 & 0.5584 \\ 
\hline

\else

\fi
\end{tabular}
\end{table*}
%
%
%
%
\clearpage
\subsubsection{Babble noise at \dBel{12}}
\begin{figure}[!ht]
	\ifarXiv
\centerline{\epsfig{figure=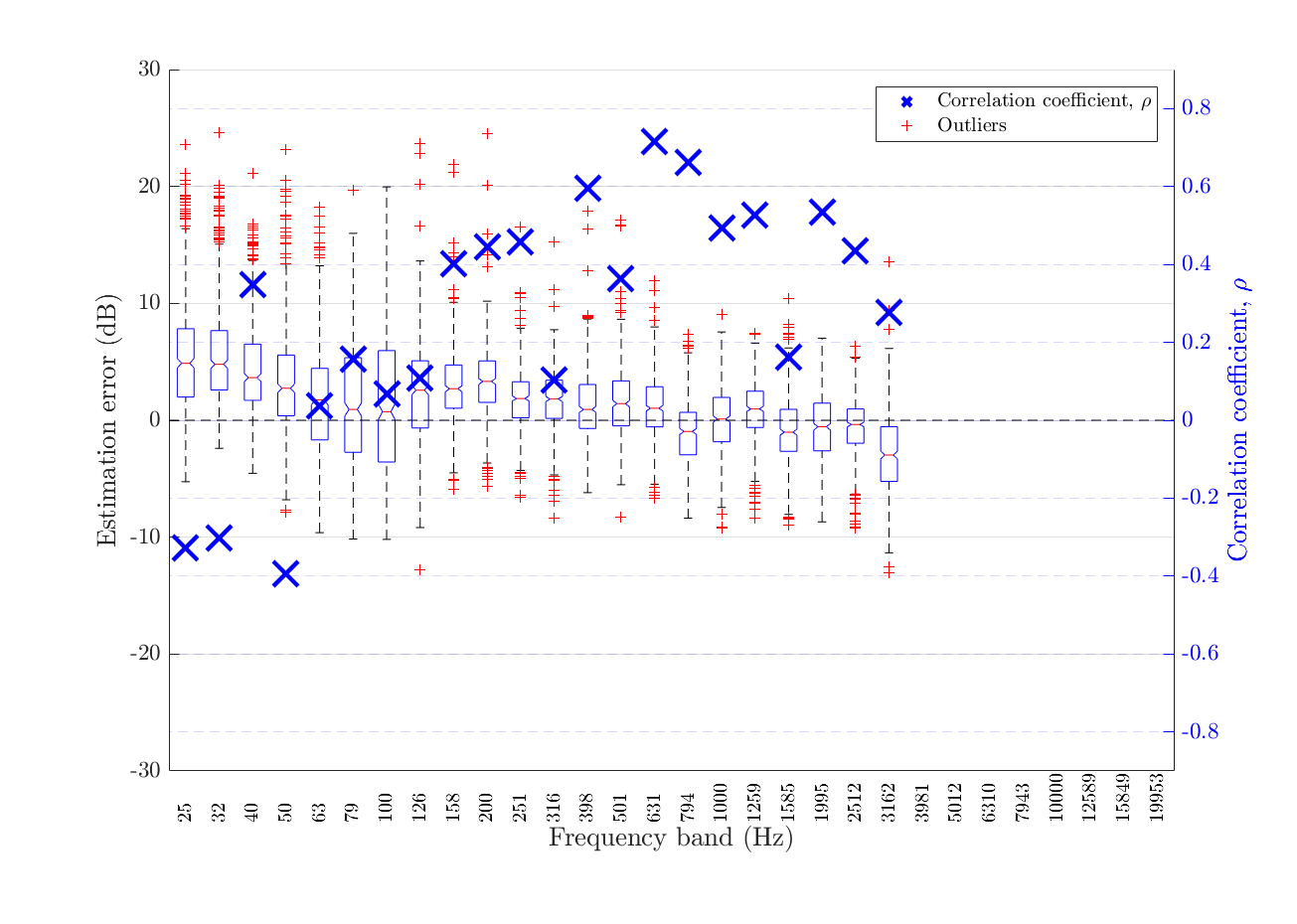,
	width=\figWidthACETR,viewport=45 10 765 530,clip}}%
	\else
	\centerline{\epsfig{figure=FigsACE/ana_eval_gt_partic_results_combined_Phase3_TR_P3S_DRR_dB_12dB_SNR_Babble_sub_Velocity.png,
	width=\figWidthACETR,viewport=45 10 765 530,clip}}%
	\fi
	\caption{{Frequency-dependent \ac{DRR} estimation error in babble noise at \dBel{12} \ac{SNR} for algorithm Particle Velocity~\cite{Chen2015}}}%
\label{fig:ACE_DRR_Sub_Babble_12dB_Velocity}%
\end{figure}%
\begin{table*}[!ht]\small
\caption{
	Frequency-dependent \ac{DRR} estimation error in babble noise at \dBel{12} \ac{SNR} for algorithm Particle Velocity~\cite{Chen2015}
}
\vspace{5mm} 
\centering
\begin{tabular}{crrrl}%
\hline%
Freq. band
& Centre Freq. (Hz)
& Bias
& \acs{MSE}
& $\PearsonCC$

\\
\hline
\hline
\ifarXiv
 1 & 25.12 & 5.482 & 55.83 & -0.3274 \\ 
\hline
 2 & 31.62 & 5.822 & 56.2 & -0.3022 \\ 
\hline
 3 & 39.81 & 4.336 & 34.14 & 0.3476 \\ 
\hline
 4 & 50.12 & 3.396 & 33.47 & -0.3935 \\ 
\hline
 5 & 63.10 & 1.658 & 24.62 & 0.03637 \\ 
\hline
 6 & 79.43 & 1.107 & 25 & 0.1574 \\ 
\hline
 7 & 100.00 & 1.103 & 32.27 & 0.06681 \\ 
\hline
 8 & 125.89 & 2.188 & 24.99 & 0.1081 \\ 
\hline
 9 & 158.49 & 2.996 & 18.72 & 0.4028 \\ 
\hline
10 & 199.53 & 3.309 & 19.73 & 0.4446 \\ 
\hline
11 & 251.19 & 1.789 & 9.746 & 0.4576 \\ 
\hline
12 & 316.23 & 1.709 & 10.74 & 0.1024 \\ 
\hline
13 & 398.11 & 1.258 & 9.83 & 0.5957 \\ 
\hline
14 & 501.19 & 1.646 & 12.84 & 0.3629 \\ 
\hline
15 & 630.96 & 1.158 & 8.498 & 0.7151 \\ 
\hline
16 & 794.33 & -1.016 & 8.382 & 0.6615 \\ 
\hline
17 & 1000.00 & 0.07851 & 7.85 & 0.493 \\ 
\hline
18 & 1258.93 & 0.8271 & 8.021 & 0.5276 \\ 
\hline
19 & 1584.89 & -0.86 & 9.62 & 0.1607 \\ 
\hline
20 & 1995.26 & -0.6218 & 7.38 & 0.5351 \\ 
\hline
21 & 2511.89 & -0.5907 & 6.343 & 0.4349 \\ 
\hline
22 & 3162.28 & -2.899 & 20.86 & 0.2771 \\ 
\hline

\else

\fi
\end{tabular}
\end{table*}
%
%
%
\begin{figure}[!ht]
	\ifarXiv
\centerline{\epsfig{figure=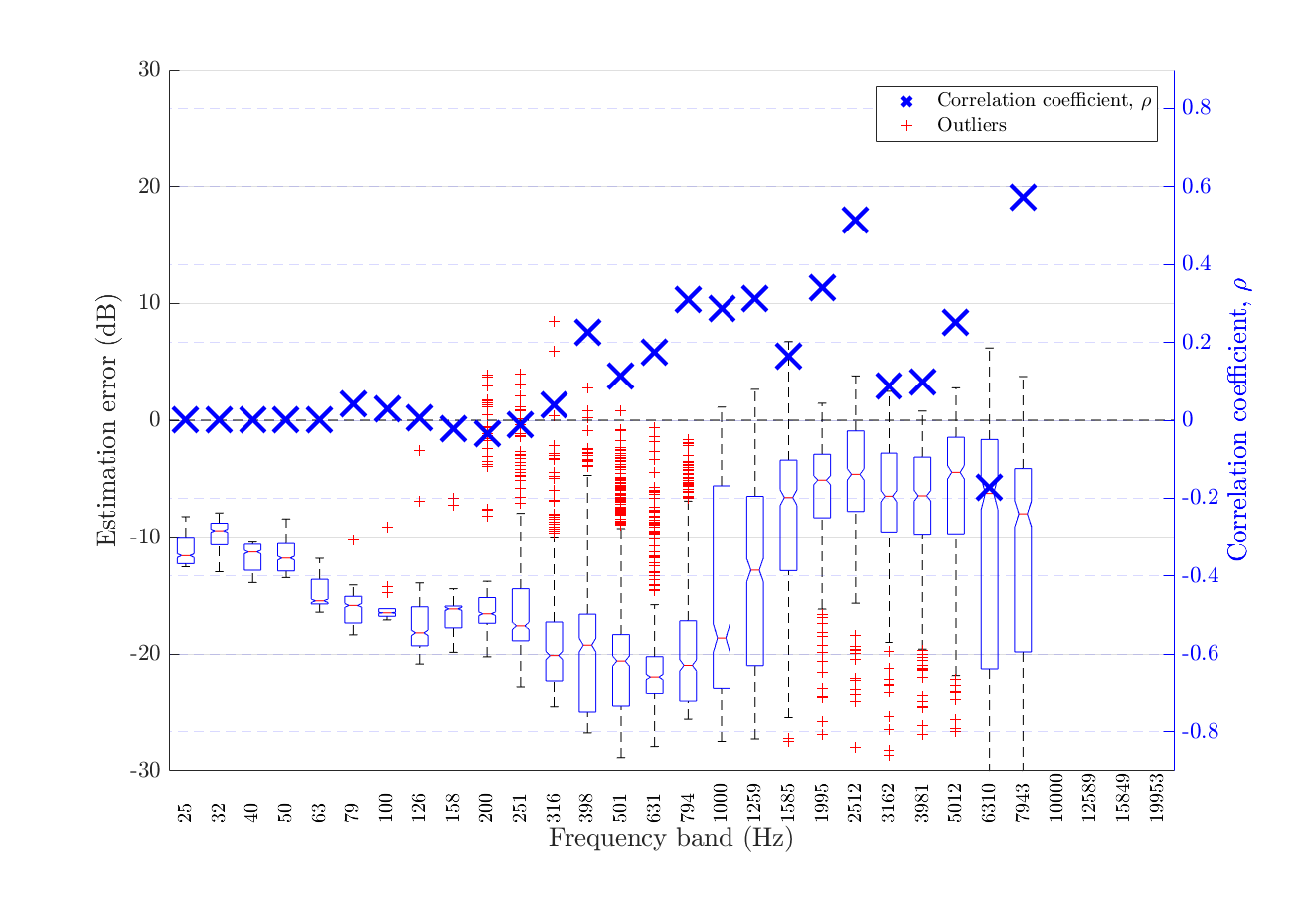,
	width=\figWidthACETR,viewport=45 10 765 530,clip}}%
	\else
	\centerline{\epsfig{figure=FigsACE/ana_eval_gt_partic_results_combined_Phase3_TR_P3S_DRR_dB_12dB_SNR_Babble_sub_ICASSP_2015_DRR_2-ch_Gerkmann_NR_FFT_subband.png,
	width=\figWidthACETR,viewport=45 10 765 530,clip}}%
	\fi
	\caption{{Frequency-dependent \ac{DRR} estimation error in babble noise  at \dBel{12} \ac{SNR} for algorithm \ac{DENBE} with FFT derived subbands~\cite{Eaton2015c}}}%
\label{fig:ACE_DRR_Sub_Babble_12dB_Eaton_FFT}%
\end{figure}%
\begin{table*}[!ht]\small
\caption{
	Frequency-dependent \ac{DRR} estimation error in babble noise  at \dBel{12} \ac{SNR} for algorithm \ac{DENBE} with FFT derived subbands~\cite{Eaton2015c}
}
\vspace{5mm} 
\centering
\begin{tabular}{crrrl}%
\hline%
Freq. band
& Centre Freq. (Hz)
& Bias
& \acs{MSE}
& $\PearsonCC$

\\
\hline
\hline
\ifarXiv
 1 & 25.12 & -11.07 & 124.4 & 0 \\ 
\hline
 2 & 31.62 & -9.93 & 100.8 & 0 \\ 
\hline
 3 & 39.81 & -11.69 & 138.1 & 0 \\ 
\hline
 4 & 50.12 & -11.55 & 135.4 & 0 \\ 
\hline
 5 & 63.10 & -14.82 & 221.5 & 0 \\ 
\hline
 6 & 79.43 & -16.08 & 260.2 & 0.04283 \\ 
\hline
 7 & 100.00 & -16.17 & 262.2 & 0.02861 \\ 
\hline
 8 & 125.89 & -17.69 & 317.2 & 0.005174 \\ 
\hline
 9 & 158.49 & -16.71 & 282.2 & -0.02086 \\ 
\hline
10 & 199.53 & -15.94 & 269.6 & -0.03566 \\ 
\hline
11 & 251.19 & -16.65 & 301.1 & -0.01128 \\ 
\hline
12 & 316.23 & -19.11 & 387.2 & 0.03975 \\ 
\hline
13 & 398.11 & -18.8 & 396.7 & 0.2256 \\ 
\hline
14 & 501.19 & -19.24 & 420.6 & 0.1147 \\ 
\hline
15 & 630.96 & -21.17 & 472.8 & 0.1755 \\ 
\hline
16 & 794.33 & -18.92 & 396.4 & 0.3108 \\ 
\hline
17 & 1000.00 & -14.59 & 291.2 & 0.2861 \\ 
\hline
18 & 1258.93 & -13.35 & 243.5 & 0.3114 \\ 
\hline
19 & 1584.89 & -8.363 & 129.6 & 0.1655 \\ 
\hline
20 & 1995.26 & -6.422 & 68.93 & 0.3393 \\ 
\hline
21 & 2511.89 & -5.361 & 65.83 & 0.5141 \\ 
\hline
22 & 3162.28 & -7.194 & 90.5 & 0.08875 \\ 
\hline
23 & 3981.07 & -7.182 & 78.57 & 0.09841 \\ 
\hline
24 & 5011.87 & -6.73 & 97.75 & 0.2507 \\ 
\hline
25 & 6309.57 & -10.37 & 222.9 & -0.1724 \\ 
\hline
26 & 7943.28 & -10.38 & 177.2 & 0.5722 \\ 
\hline

\else

\fi
\end{tabular}
\end{table*}
%
%
%
\begin{figure}[!ht]
	\ifarXiv
\centerline{\epsfig{figure=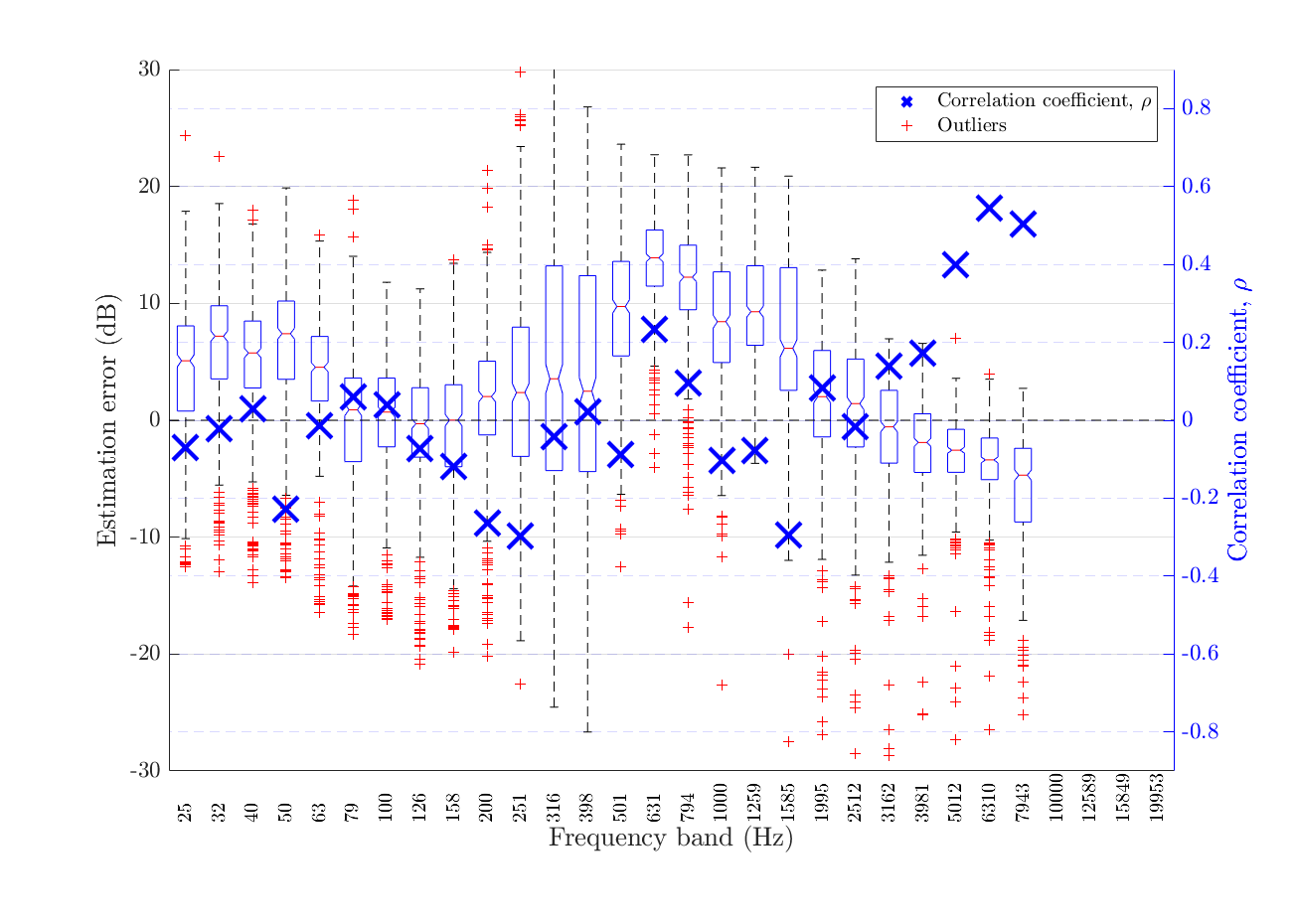,
	width=\figWidthACETR,viewport=45 10 765 530,clip}}%
	\else
	\centerline{\epsfig{figure=FigsACE/ana_eval_gt_partic_results_combined_Phase3_TR_P3S_DRR_dB_12dB_SNR_Babble_sub_ICASSP_2015_DRR_2-ch_Gerkmann_NR_filtered_subband.png,
	width=\figWidthACETR,viewport=45 10 765 530,clip}}%
	\fi
	\caption{{Frequency-dependent \ac{DRR} estimation error in babble noise  at \dBel{12} \ac{SNR} for algorithm \ac{DENBE} with filtered subbands~\cite{Eaton2015c}}}%
\label{fig:ACE_DRR_Sub_Babble_12dB_Eaton_filt}%
\end{figure}%
\begin{table*}[!ht]\small
\caption{
	Frequency-dependent \ac{DRR} estimation error in babble noise  at \dBel{12} \ac{SNR} for algorithm \ac{DENBE} with filtered subbands~\cite{Eaton2015c}
}
\vspace{5mm} 
\centering
\begin{tabular}{crrrl}%
\hline%
Freq. band
& Centre Freq. (Hz)
& Bias
& \acs{MSE}
& $\PearsonCC$

\\
\hline
\hline
\ifarXiv
 1 & 25.12 & 3.524 & 66.65 & -0.06888 \\ 
\hline
 2 & 31.62 & 5.514 & 77.91 & -0.02182 \\ 
\hline
 3 & 39.81 & 4.343 & 61.66 & 0.02824 \\ 
\hline
 4 & 50.12 & 5.904 & 80.68 & -0.2271 \\ 
\hline
 5 & 63.10 & 3.685 & 49.25 & -0.01334 \\ 
\hline
 6 & 79.43 & -0.9334 & 53.69 & 0.06042 \\ 
\hline
 7 & 100.00 & -0.5737 & 42.67 & 0.03834 \\ 
\hline
 8 & 125.89 & -0.8532 & 34.69 & -0.0728 \\ 
\hline
 9 & 158.49 & -1.069 & 42.45 & -0.1184 \\ 
\hline
10 & 199.53 & 1.487 & 43.47 & -0.2627 \\ 
\hline
11 & 251.19 & 2.52 & 83.89 & -0.2978 \\ 
\hline
12 & 316.23 & 4.068 & 154.6 & -0.04307 \\ 
\hline
13 & 398.11 & 2.982 & 117.9 & 0.02072 \\ 
\hline
14 & 501.19 & 9.457 & 129.7 & -0.08717 \\ 
\hline
15 & 630.96 & 13.69 & 202.7 & 0.2336 \\ 
\hline
16 & 794.33 & 11.74 & 165.6 & 0.09492 \\ 
\hline
17 & 1000.00 & 8.25 & 105 & -0.1032 \\ 
\hline
18 & 1258.93 & 9.81 & 119 & -0.07773 \\ 
\hline
19 & 1584.89 & 6.89 & 94.54 & -0.2943 \\ 
\hline
20 & 1995.26 & 1.502 & 44.69 & 0.08286 \\ 
\hline
21 & 2511.89 & 0.6092 & 51.5 & -0.01732 \\ 
\hline
22 & 3162.28 & -1.282 & 28.89 & 0.1388 \\ 
\hline
23 & 3981.07 & -2.301 & 19.14 & 0.1732 \\ 
\hline
24 & 5011.87 & -2.819 & 20.37 & 0.399 \\ 
\hline
25 & 6309.57 & -3.532 & 24.85 & 0.5454 \\ 
\hline
26 & 7943.28 & -6.014 & 59.4 & 0.5041 \\ 
\hline

\else

\fi
\end{tabular}
\end{table*}
%
%
%
%
\clearpage
\subsubsection{Babble noise at \dBel{-1}}
\begin{figure}[!ht]
	\ifarXiv
\centerline{\epsfig{figure=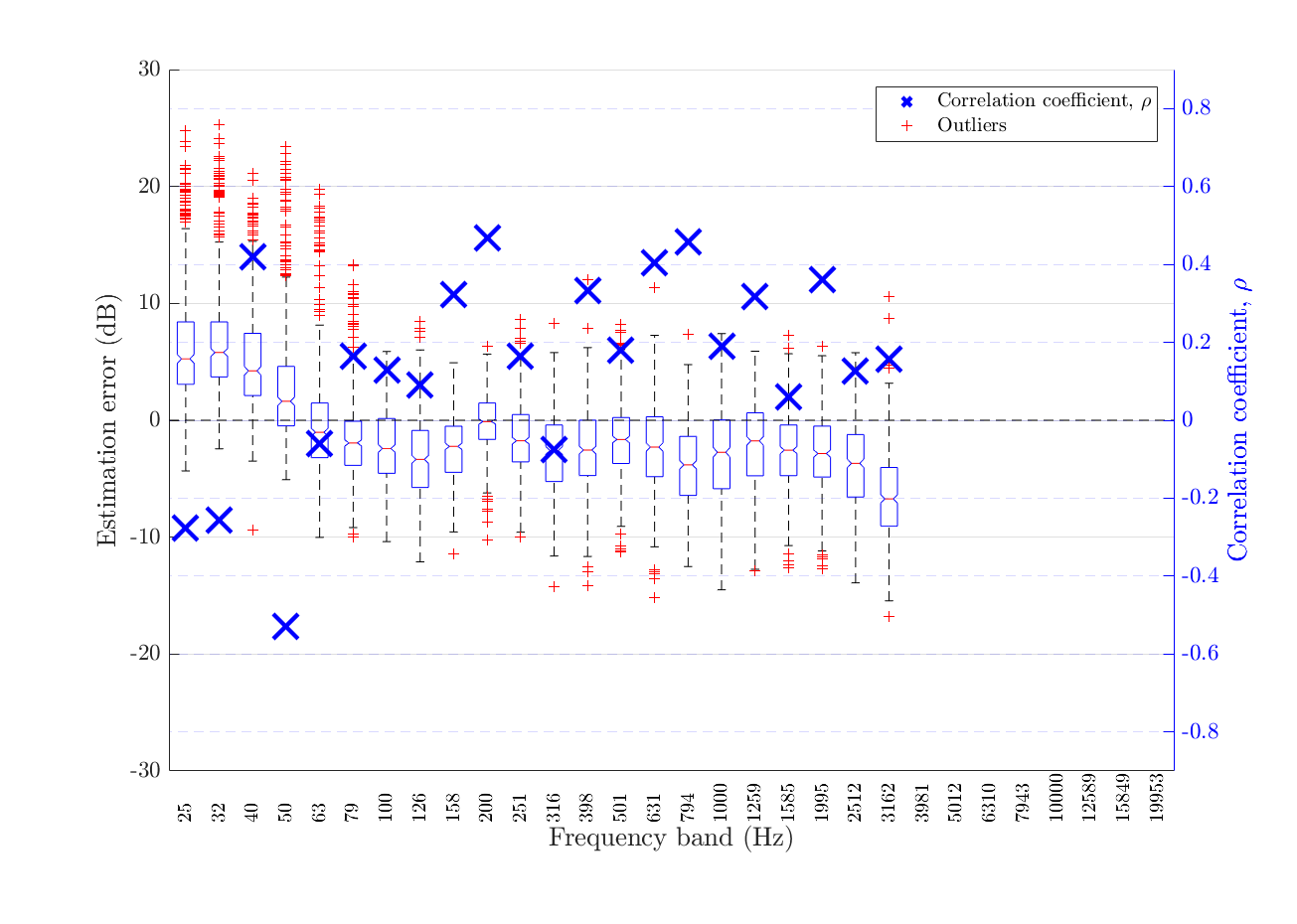,
	width=\figWidthACETR,viewport=45 10 765 530,clip}}%
	\else
	\centerline{\epsfig{figure=FigsACE/ana_eval_gt_partic_results_combined_Phase3_TR_P3S_DRR_dB_-1dB_SNR_Babble_sub_Velocity.png,
	width=\figWidthACETR,viewport=45 10 765 530,clip}}%
	\fi
	\caption{{Frequency-dependent \ac{DRR} estimation error in babble noise at \dBel{-1} \ac{SNR} for algorithm Particle Velocity~\cite{Chen2015}}}%
\label{fig:ACE_DRR_Sub_Babble_-1dB_Velocity}%
\end{figure}%
\begin{table*}[!ht]\small
\caption{
	Frequency-dependent \ac{DRR} estimation error in babble noise at \dBel{-1} \ac{SNR} for algorithm Particle Velocity~\cite{Chen2015}
}
\vspace{5mm} 
\centering
\begin{tabular}{crrrl}%
\hline%
Freq. band
& Centre Freq. (Hz)
& Bias
& \acs{MSE}
& $\PearsonCC$

\\
\hline
\hline
\ifarXiv
 1 & 25.12 & 6.323 & 66.56 & -0.2761 \\ 
\hline
 2 & 31.62 & 6.835 & 72.08 & -0.2555 \\ 
\hline
 3 & 39.81 & 5.145 & 46.98 & 0.4187 \\ 
\hline
 4 & 50.12 & 3.207 & 42.64 & -0.5288 \\ 
\hline
 5 & 63.10 & -0.1255 & 24.6 & -0.06066 \\ 
\hline
 6 & 79.43 & -1.725 & 16.34 & 0.1651 \\ 
\hline
 7 & 100.00 & -2.144 & 15.57 & 0.1279 \\ 
\hline
 8 & 125.89 & -3.258 & 22.74 & 0.09127 \\ 
\hline
 9 & 158.49 & -2.464 & 14.01 & 0.3215 \\ 
\hline
10 & 199.53 & -0.2192 & 5.897 & 0.4681 \\ 
\hline
11 & 251.19 & -1.513 & 11.62 & 0.1635 \\ 
\hline
12 & 316.23 & -2.78 & 19.64 & -0.07476 \\ 
\hline
13 & 398.11 & -2.431 & 18.16 & 0.3329 \\ 
\hline
14 & 501.19 & -1.644 & 13.88 & 0.1786 \\ 
\hline
15 & 630.96 & -2.345 & 18.68 & 0.4048 \\ 
\hline
16 & 794.33 & -3.904 & 27.58 & 0.4582 \\ 
\hline
17 & 1000.00 & -2.931 & 23.96 & 0.1891 \\ 
\hline
18 & 1258.93 & -2.109 & 18.97 & 0.3171 \\ 
\hline
19 & 1584.89 & -2.538 & 17.33 & 0.05895 \\ 
\hline
20 & 1995.26 & -2.865 & 17.91 & 0.3615 \\ 
\hline
21 & 2511.89 & -4.041 & 28.95 & 0.1254 \\ 
\hline
22 & 3162.28 & -6.447 & 56.19 & 0.1557 \\ 
\hline

\else

\fi
\end{tabular}
\end{table*}
%
%
%
\begin{figure}[!ht]
	\ifarXiv
\centerline{\epsfig{figure=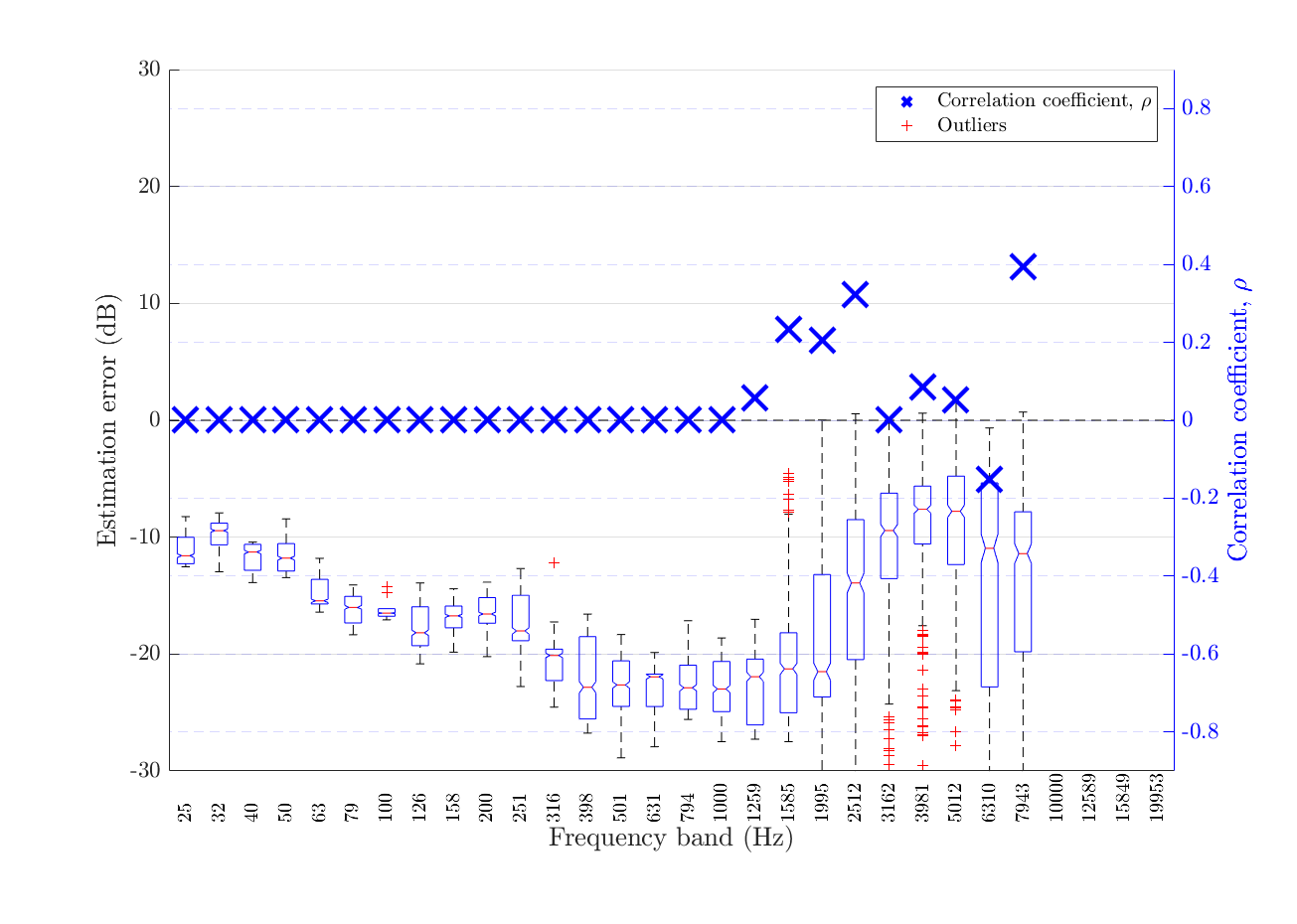,
	width=\figWidthACETR,viewport=45 10 765 530,clip}}%
	\else
	\centerline{\epsfig{figure=FigsACE/ana_eval_gt_partic_results_combined_Phase3_TR_P3S_DRR_dB_-1dB_SNR_Babble_sub_ICASSP_2015_DRR_2-ch_Gerkmann_NR_FFT_subband.png,
	width=\figWidthACETR,viewport=45 10 765 530,clip}}%
	\fi
	\caption{{Frequency-dependent \ac{DRR} estimation error in babble noise  at \dBel{-1} \ac{SNR} for algorithm \ac{DENBE} with FFT derived subbands~\cite{Eaton2015c}}}%
\label{fig:ACE_DRR_Sub_Babble_-1dB_Eaton_FFT}%
\end{figure}%
\begin{table*}[!ht]\small
\caption{
	Frequency-dependent \ac{DRR} estimation error in babble noise  at \dBel{-1} \ac{SNR} for algorithm \ac{DENBE} with FFT derived subbands~\cite{Eaton2015c}
}
\vspace{5mm} 
\centering
\begin{tabular}{crrrl}%
\hline%
Freq. band
& Centre Freq. (Hz)
& Bias
& \acs{MSE}
& $\PearsonCC$

\\
\hline
\hline
\ifarXiv
 1 & 25.12 & -11.07 & 124.4 & 0 \\ 
\hline
 2 & 31.62 & -9.93 & 100.8 & 0 \\ 
\hline
 3 & 39.81 & -11.69 & 138.1 & 0 \\ 
\hline
 4 & 50.12 & -11.55 & 135.4 & 0 \\ 
\hline
 5 & 63.10 & -14.82 & 221.5 & 0 \\ 
\hline
 6 & 79.43 & -16.09 & 260.6 & 0 \\ 
\hline
 7 & 100.00 & -16.18 & 262.6 & 0 \\ 
\hline
 8 & 125.89 & -17.74 & 318.4 & 0 \\ 
\hline
 9 & 158.49 & -16.76 & 283.2 & 0 \\ 
\hline
10 & 199.53 & -16.72 & 282.9 & 0 \\ 
\hline
11 & 251.19 & -17.66 & 321.9 & 0 \\ 
\hline
12 & 316.23 & -19.99 & 410 & 0 \\ 
\hline
13 & 398.11 & -22.2 & 506.2 & 0 \\ 
\hline
14 & 501.19 & -22.78 & 528.4 & 0 \\ 
\hline
15 & 630.96 & -22.89 & 529.6 & 0 \\ 
\hline
16 & 794.33 & -22.2 & 501.5 & 0 \\ 
\hline
17 & 1000.00 & -23.01 & 537.8 & 0 \\ 
\hline
18 & 1258.93 & -22.4 & 512.4 & 0.05738 \\ 
\hline
19 & 1584.89 & -21.07 & 467.4 & 0.2331 \\ 
\hline
20 & 1995.26 & -18.92 & 408.7 & 0.206 \\ 
\hline
21 & 2511.89 & -14.69 & 277.9 & 0.3235 \\ 
\hline
22 & 3162.28 & -11.24 & 173.9 & 0.001717 \\ 
\hline
23 & 3981.07 & -9.009 & 112.8 & 0.08607 \\ 
\hline
24 & 5011.87 & -9.218 & 126.4 & 0.05141 \\ 
\hline
25 & 6309.57 & -13.35 & 263.9 & -0.1515 \\ 
\hline
26 & 7943.28 & -13.38 & 237.1 & 0.3941 \\ 
\hline

\else

\fi
\end{tabular}
\end{table*}
%
%
%
\begin{figure}[!ht]
	\ifarXiv
\centerline{\epsfig{figure=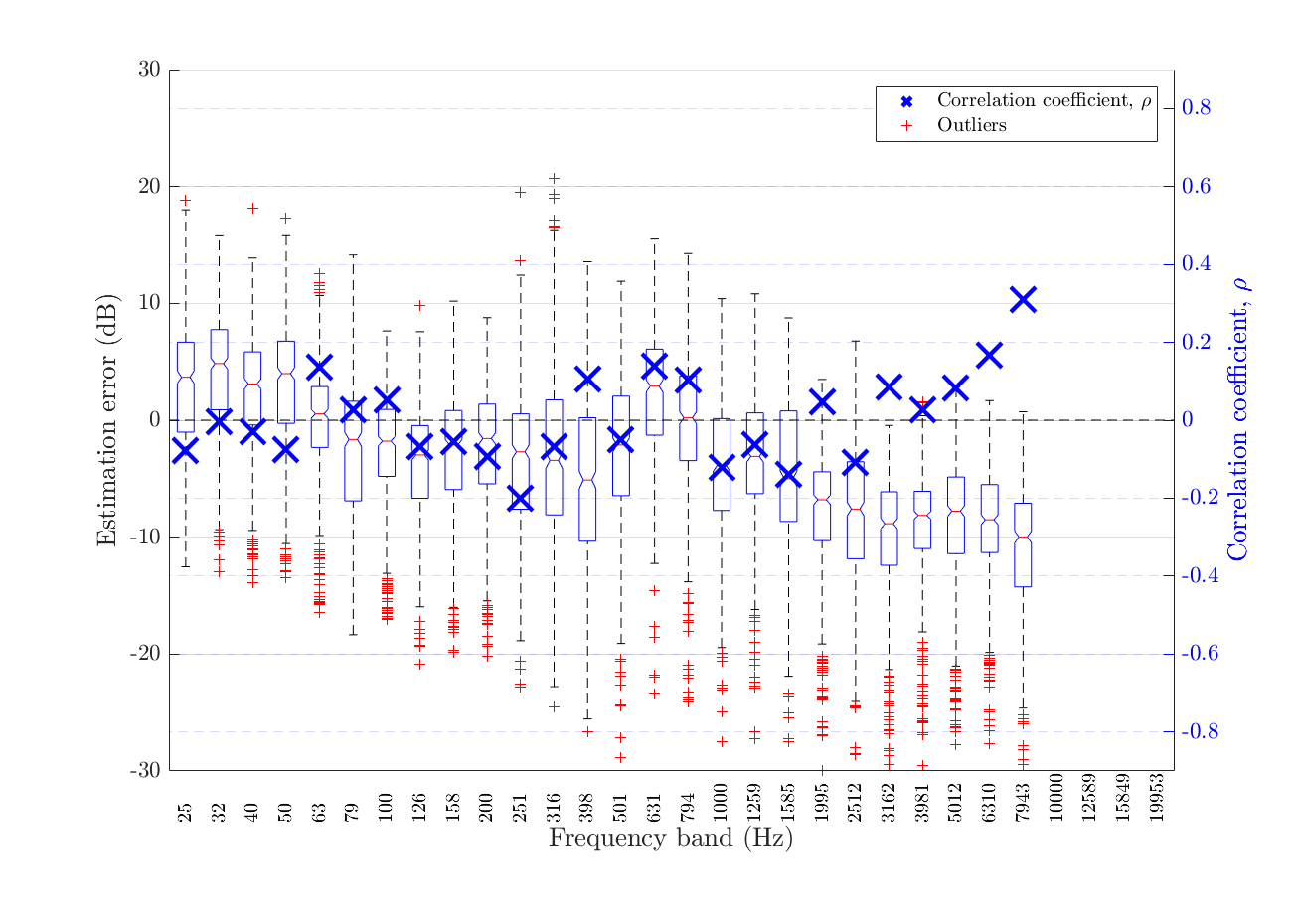,
	width=\figWidthACETR,viewport=45 10 765 530,clip}}%
	\else
	\centerline{\epsfig{figure=FigsACE/ana_eval_gt_partic_results_combined_Phase3_TR_P3S_DRR_dB_-1dB_SNR_Babble_sub_ICASSP_2015_DRR_2-ch_Gerkmann_NR_filtered_subband.png,
	width=\figWidthACETR,viewport=45 10 765 530,clip}}%
	\fi
	\caption{{Frequency-dependent \ac{DRR} estimation error in babble noise  at \dBel{-1} \ac{SNR} for algorithm \ac{DENBE} with filtered subbands~\cite{Eaton2015c}}}%
\label{fig:ACE_DRR_Sub_Babble_-1dB_Eaton_filt}%
\end{figure}%
\begin{table*}[!ht]\small
\caption{
	Frequency-dependent \ac{DRR} estimation error in babble noise  at \dBel{-1} \ac{SNR} for algorithm \ac{DENBE} with filtered subbands~\cite{Eaton2015c}
}
\vspace{5mm} 
\centering
\begin{tabular}{crrrl}%
\hline%
Freq. band
& Centre Freq. (Hz)
& Bias
& \acs{MSE}
& $\PearsonCC$

\\
\hline
\hline
\ifarXiv
 1 & 25.12 & 1.979 & 51.03 & -0.07764 \\ 
\hline
 2 & 31.62 & 3.419 & 50.61 & -0.00371 \\ 
\hline
 3 & 39.81 & 1.894 & 41.34 & -0.02848 \\ 
\hline
 4 & 50.12 & 2.567 & 45.96 & -0.07436 \\ 
\hline
 5 & 63.10 & -0.4269 & 31.73 & 0.1359 \\ 
\hline
 6 & 79.43 & -3.042 & 61.8 & 0.02766 \\ 
\hline
 7 & 100.00 & -3.029 & 42.37 & 0.05269 \\ 
\hline
 8 & 125.89 & -4.269 & 52.43 & -0.06707 \\ 
\hline
 9 & 158.49 & -3.339 & 51.56 & -0.05568 \\ 
\hline
10 & 199.53 & -2.879 & 44.53 & -0.09246 \\ 
\hline
11 & 251.19 & -3.977 & 69.08 & -0.2014 \\ 
\hline
12 & 316.23 & -3.357 & 85.18 & -0.06781 \\ 
\hline
13 & 398.11 & -5.701 & 96.52 & 0.1071 \\ 
\hline
14 & 501.19 & -2.899 & 63.08 & -0.05049 \\ 
\hline
15 & 630.96 & 2.181 & 37.39 & 0.1397 \\ 
\hline
16 & 794.33 & -0.4429 & 41.57 & 0.1041 \\ 
\hline
17 & 1000.00 & -4.534 & 68.66 & -0.1207 \\ 
\hline
18 & 1258.93 & -3.205 & 48.64 & -0.06257 \\ 
\hline
19 & 1584.89 & -4.756 & 70.9 & -0.1387 \\ 
\hline
20 & 1995.26 & -8.265 & 103.4 & 0.04785 \\ 
\hline
21 & 2511.89 & -8.472 & 122.4 & -0.1081 \\ 
\hline
22 & 3162.28 & -10.3 & 146.5 & 0.08453 \\ 
\hline
23 & 3981.07 & -9.382 & 119 & 0.02584 \\ 
\hline
24 & 5011.87 & -8.7 & 106.5 & 0.08226 \\ 
\hline
25 & 6309.57 & -8.815 & 101.9 & 0.1664 \\ 
\hline
26 & 7943.28 & -10.96 & 149.5 & 0.31 \\ 
\hline

\else

\fi
\end{tabular}
\end{table*}
%
%
%
%
%
\clearpage
\subsubsection{Fan noise at \dBel{18}}
\begin{figure}[!ht]
	\ifarXiv
\centerline{\epsfig{figure=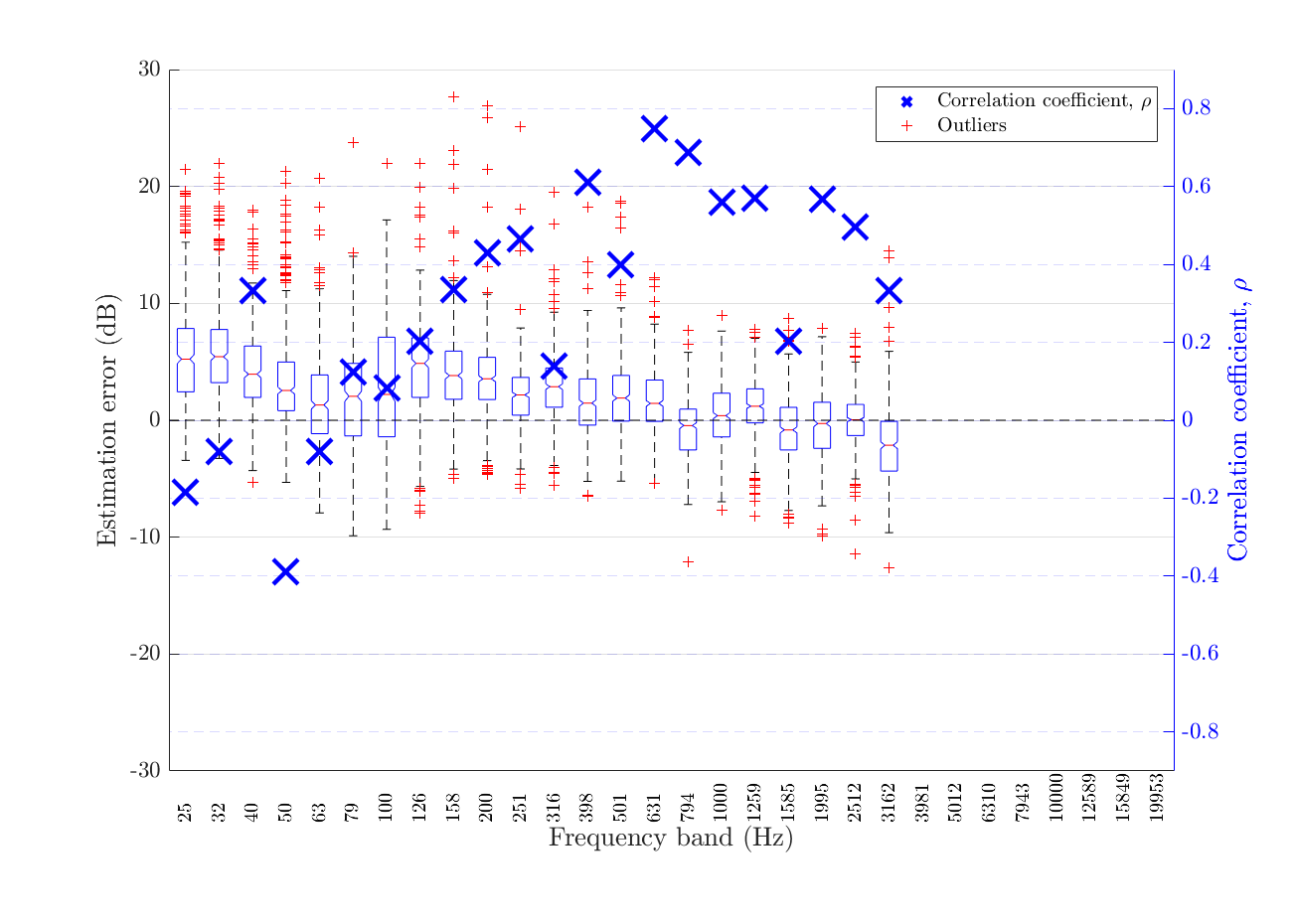,
	width=\figWidthACETR,viewport=45 10 765 530,clip}}%
	\else
	\centerline{\epsfig{figure=FigsACE/ana_eval_gt_partic_results_combined_Phase3_TR_P3S_DRR_dB_18dB_SNR_Fan_sub_Velocity.png,
	width=\figWidthACETR,viewport=45 10 765 530,clip}}%
	\fi
	\caption{{Frequency-dependent \ac{DRR} estimation error in fan noise at \dBel{18} \ac{SNR} for algorithm Particle Velocity~\cite{Chen2015}}}%
\label{fig:ACE_DRR_Sub_Fan_18dB_Velocity}%
\end{figure}%
\begin{table*}[!ht]\small
\caption{
	Frequency-dependent \ac{DRR} estimation error in fan noise at \dBel{18} \ac{SNR} for algorithm Particle Velocity~\cite{Chen2015}
}
\vspace{5mm} 
\centering
\begin{tabular}{crrrl}%
\hline%
Freq. band
& Centre Freq. (Hz)
& Bias
& \acs{MSE}
& $\PearsonCC$

\\
\hline
\hline
\ifarXiv
 1 & 25.12 & 5.568 & 49.61 & -0.186 \\ 
\hline
 2 & 31.62 & 5.965 & 50.92 & -0.08042 \\ 
\hline
 3 & 39.81 & 4.347 & 31.19 & 0.3321 \\ 
\hline
 4 & 50.12 & 3.23 & 26.9 & -0.3896 \\ 
\hline
 5 & 63.10 & 1.687 & 19.34 & -0.0814 \\ 
\hline
 6 & 79.43 & 1.879 & 22.24 & 0.1236 \\ 
\hline
 7 & 100.00 & 2.695 & 34.55 & 0.08174 \\ 
\hline
 8 & 125.89 & 4.451 & 36.98 & 0.2025 \\ 
\hline
 9 & 158.49 & 4.048 & 28.93 & 0.3348 \\ 
\hline
10 & 199.53 & 3.646 & 24.23 & 0.429 \\ 
\hline
11 & 251.19 & 2.164 & 12.43 & 0.4663 \\ 
\hline
12 & 316.23 & 2.854 & 16.33 & 0.1379 \\ 
\hline
13 & 398.11 & 1.763 & 11.81 & 0.6109 \\ 
\hline
14 & 501.19 & 2.169 & 15.36 & 0.3992 \\ 
\hline
15 & 630.96 & 1.783 & 10.47 & 0.7474 \\ 
\hline
16 & 794.33 & -0.5915 & 7.34 & 0.6872 \\ 
\hline
17 & 1000.00 & 0.4346 & 6.893 & 0.5591 \\ 
\hline
18 & 1258.93 & 1.174 & 7.707 & 0.569 \\ 
\hline
19 & 1584.89 & -0.7393 & 8.931 & 0.2038 \\ 
\hline
20 & 1995.26 & -0.3955 & 6.943 & 0.5677 \\ 
\hline
21 & 2511.89 & -0.07322 & 5.205 & 0.496 \\ 
\hline
22 & 3162.28 & -2.109 & 14.92 & 0.3336 \\ 
\hline

\else

\fi
\end{tabular}
\end{table*}
%
%
%
\begin{figure}[!ht]
	\ifarXiv
\centerline{\epsfig{figure=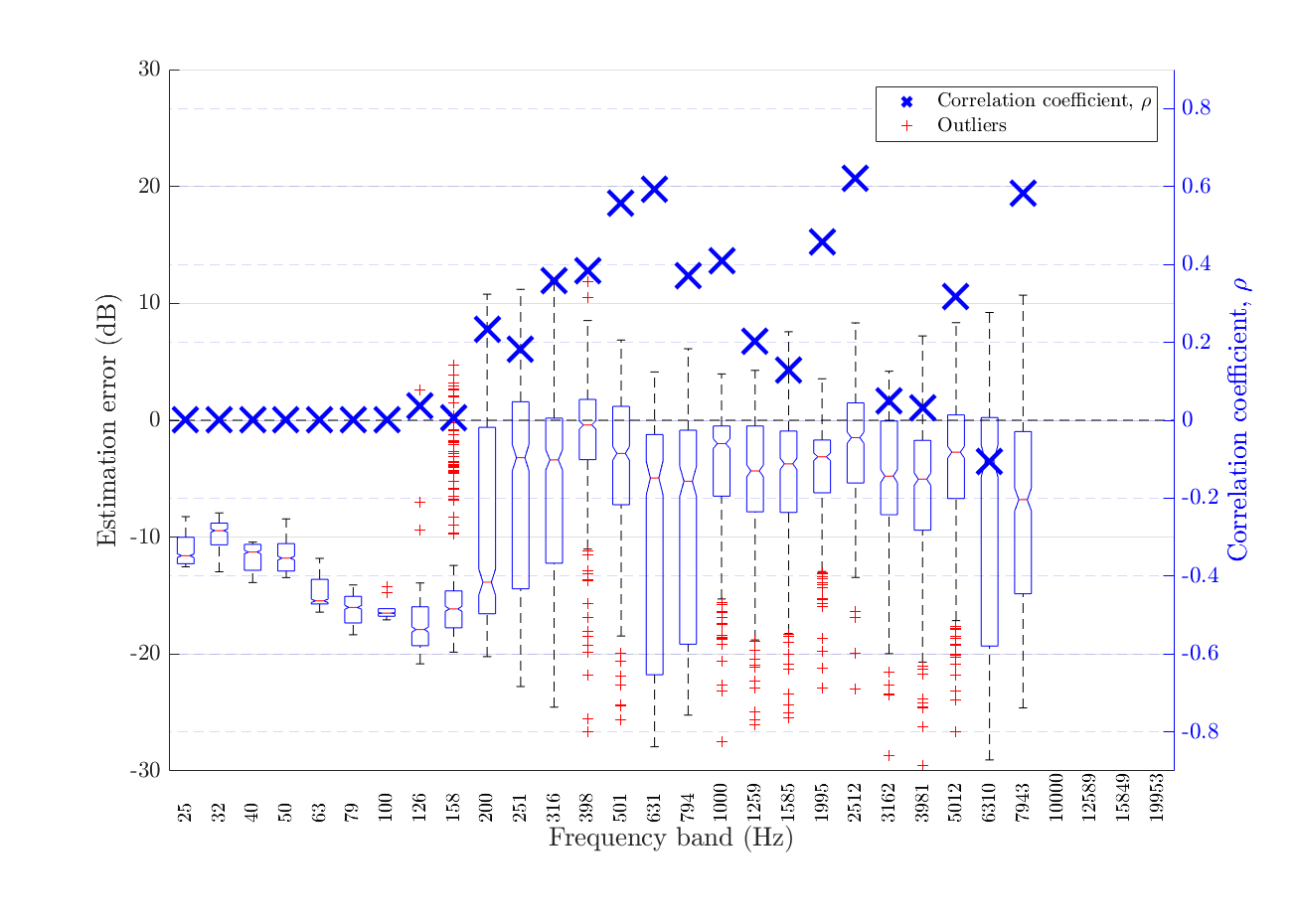,
	width=\figWidthACETR,viewport=45 10 765 530,clip}}%
	\else
	\centerline{\epsfig{figure=FigsACE/ana_eval_gt_partic_results_combined_Phase3_TR_P3S_DRR_dB_18dB_SNR_Fan_sub_ICASSP_2015_DRR_2-ch_Gerkmann_NR_FFT_subband.png,
	width=\figWidthACETR,viewport=45 10 765 530,clip}}%
	\fi
	\caption{{Frequency-dependent \ac{DRR} estimation error in fan noise  at \dBel{18} \ac{SNR} for algorithm \ac{DENBE} with FFT derived subbands~\cite{Eaton2015c}}}%
\label{fig:ACE_DRR_Sub_Fan_18dB_Eaton_FFT}%
\end{figure}%
\begin{table*}[!ht]\small
\caption{
	Frequency-dependent \ac{DRR} estimation error in fan noise  at \dBel{18} \ac{SNR} for algorithm \ac{DENBE} with FFT derived subbands~\cite{Eaton2015c}
}
\vspace{5mm} 
\centering
\begin{tabular}{crrrl}%
\hline%
Freq. band
& Centre Freq. (Hz)
& Bias
& \acs{MSE}
& $\PearsonCC$

\\
\hline
\hline
\ifarXiv
 1 & 25.12 & -11.07 & 124.4 & 0 \\ 
\hline
 2 & 31.62 & -9.93 & 100.8 & 0 \\ 
\hline
 3 & 39.81 & -11.69 & 138.1 & 0 \\ 
\hline
 4 & 50.12 & -11.55 & 135.4 & 0 \\ 
\hline
 5 & 63.10 & -14.82 & 221.5 & 0 \\ 
\hline
 6 & 79.43 & -16.09 & 260.6 & 0 \\ 
\hline
 7 & 100.00 & -16.18 & 262.6 & 0 \\ 
\hline
 8 & 125.89 & -17.65 & 316.3 & 0.03643 \\ 
\hline
 9 & 158.49 & -15.56 & 260.9 & 0.006855 \\ 
\hline
10 & 199.53 & -8.973 & 155.2 & 0.234 \\ 
\hline
11 & 251.19 & -5.995 & 120.8 & 0.1831 \\ 
\hline
12 & 316.23 & -5.986 & 96.83 & 0.3581 \\ 
\hline
13 & 398.11 & -1.476 & 33.87 & 0.3833 \\ 
\hline
14 & 501.19 & -4.451 & 71.67 & 0.5581 \\ 
\hline
15 & 630.96 & -10.04 & 193.9 & 0.5922 \\ 
\hline
16 & 794.33 & -8.864 & 165.3 & 0.3703 \\ 
\hline
17 & 1000.00 & -4.469 & 61.2 & 0.4102 \\ 
\hline
18 & 1258.93 & -5.349 & 70.01 & 0.2037 \\ 
\hline
19 & 1584.89 & -4.49 & 57.63 & 0.1294 \\ 
\hline
20 & 1995.26 & -4.358 & 34.73 & 0.4578 \\ 
\hline
21 & 2511.89 & -2.074 & 22.13 & 0.6219 \\ 
\hline
22 & 3162.28 & -5.059 & 58.43 & 0.05027 \\ 
\hline
23 & 3981.07 & -6.16 & 72.68 & 0.03261 \\ 
\hline
24 & 5011.87 & -5.025 & 90.52 & 0.3185 \\ 
\hline
25 & 6309.57 & -8.276 & 202.8 & -0.1055 \\ 
\hline
26 & 7943.28 & -8.321 & 158.4 & 0.5825 \\ 
\hline

\else

\fi
\end{tabular}
\end{table*}
%
%
%
\begin{figure}[!ht]
	\ifarXiv
\centerline{\epsfig{figure=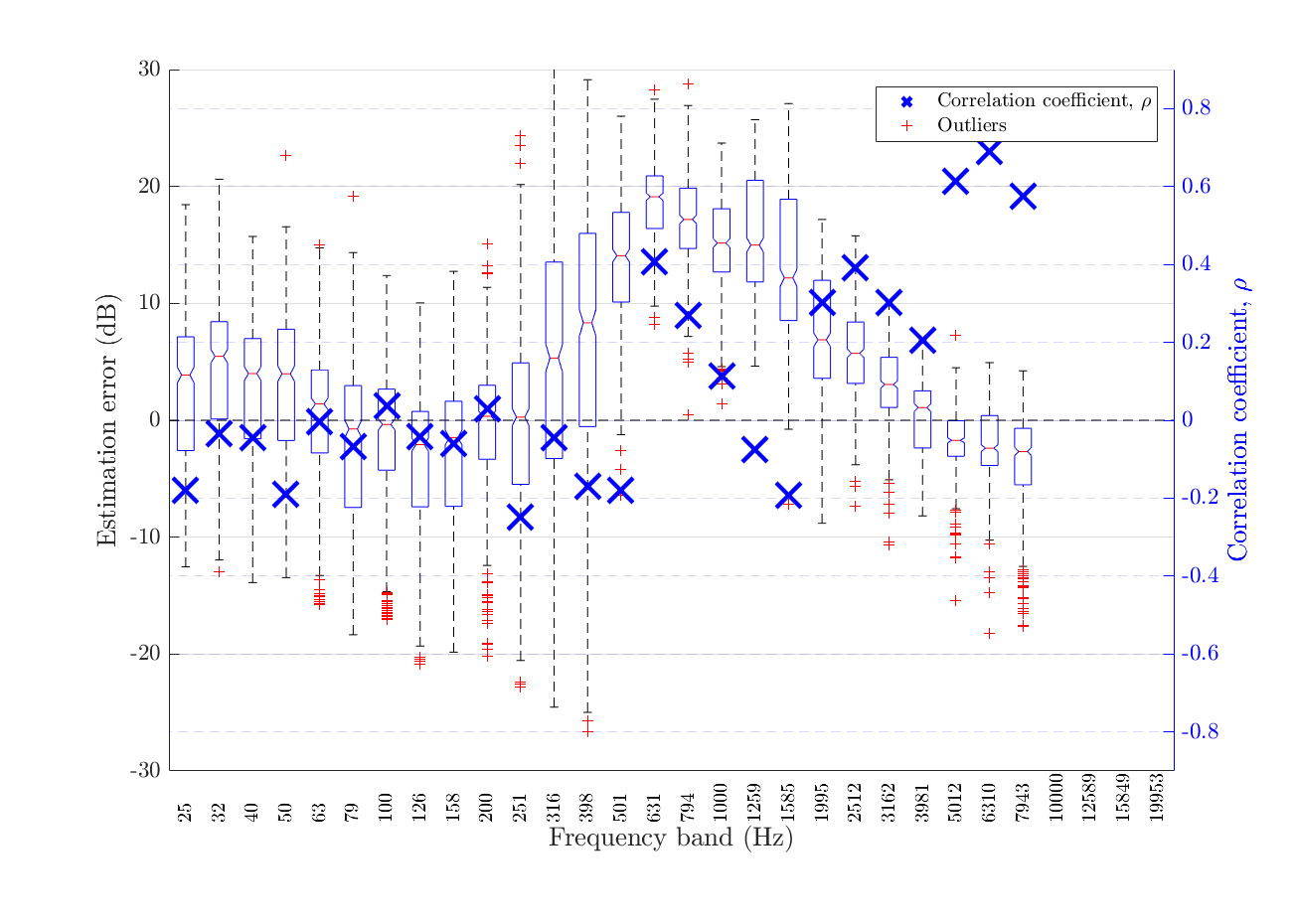,
	width=\figWidthACETR,viewport=45 10 765 530,clip}}%
	\else
	\centerline{\epsfig{figure=FigsACE/ana_eval_gt_partic_results_combined_Phase3_TR_P3S_DRR_dB_18dB_SNR_Fan_sub_ICASSP_2015_DRR_2-ch_Gerkmann_NR_filtered_subband.png,
	width=\figWidthACETR,viewport=45 10 765 530,clip}}%
	\fi
	\caption{{Frequency-dependent \ac{DRR} estimation error in fan noise  at \dBel{18} \ac{SNR} for algorithm \ac{DENBE} with filtered subbands~\cite{Eaton2015c}}}%
\label{fig:ACE_DRR_Sub_Fan_18dB_Eaton_filt}%
\end{figure}%
\begin{table*}[!ht]\small
\caption{
	Frequency-dependent \ac{DRR} estimation error in fan noise  at \dBel{18} \ac{SNR} for algorithm \ac{DENBE} with filtered subbands~\cite{Eaton2015c}
}
\vspace{5mm} 
\centering
\begin{tabular}{crrrl}%
\hline%
Freq. band
& Centre Freq. (Hz)
& Bias
& \acs{MSE}
& $\PearsonCC$

\\
\hline
\hline
\ifarXiv
 1 & 25.12 & 1.877 & 59.57 & -0.18 \\ 
\hline
 2 & 31.62 & 3.281 & 65.71 & -0.03397 \\ 
\hline
 3 & 39.81 & 1.818 & 56.53 & -0.04415 \\ 
\hline
 4 & 50.12 & 1.93 & 66.65 & -0.1904 \\ 
\hline
 5 & 63.10 & -0.3509 & 50.46 & -0.004009 \\ 
\hline
 6 & 79.43 & -2.634 & 69.71 & -0.06639 \\ 
\hline
 7 & 100.00 & -2.127 & 52.01 & 0.03706 \\ 
\hline
 8 & 125.89 & -4.104 & 67.28 & -0.04096 \\ 
\hline
 9 & 158.49 & -3.5 & 64.13 & -0.05915 \\ 
\hline
10 & 199.53 & -1.274 & 46.11 & 0.02888 \\ 
\hline
11 & 251.19 & -0.8414 & 75.93 & -0.249 \\ 
\hline
12 & 316.23 & 5.042 & 166.6 & -0.04575 \\ 
\hline
13 & 398.11 & 7.572 & 164.5 & -0.1703 \\ 
\hline
14 & 501.19 & 13.87 & 223.2 & -0.1801 \\ 
\hline
15 & 630.96 & 18.65 & 358.2 & 0.4069 \\ 
\hline
16 & 794.33 & 17.12 & 307.6 & 0.2691 \\ 
\hline
17 & 1000.00 & 15.18 & 246.9 & 0.1132 \\ 
\hline
18 & 1258.93 & 15.73 & 271.9 & -0.07646 \\ 
\hline
19 & 1584.89 & 13.04 & 207.9 & -0.192 \\ 
\hline
20 & 1995.26 & 7.297 & 77.76 & 0.3027 \\ 
\hline
21 & 2511.89 & 5.92 & 51.58 & 0.3914 \\ 
\hline
22 & 3162.28 & 2.992 & 17.82 & 0.3029 \\ 
\hline
23 & 3981.07 & 0.08042 & 9.421 & 0.2053 \\ 
\hline
24 & 5011.87 & -1.62 & 9.631 & 0.6133 \\ 
\hline
25 & 6309.57 & -2.155 & 13.15 & 0.6909 \\ 
\hline
26 & 7943.28 & -3.579 & 29.57 & 0.5761 \\ 
\hline

\else

\fi
\end{tabular}
\end{table*}
%
%
%
%
\clearpage
\subsubsection{Fan noise at \dBel{12}}
\begin{figure}[!ht]
	\ifarXiv
\centerline{\epsfig{figure=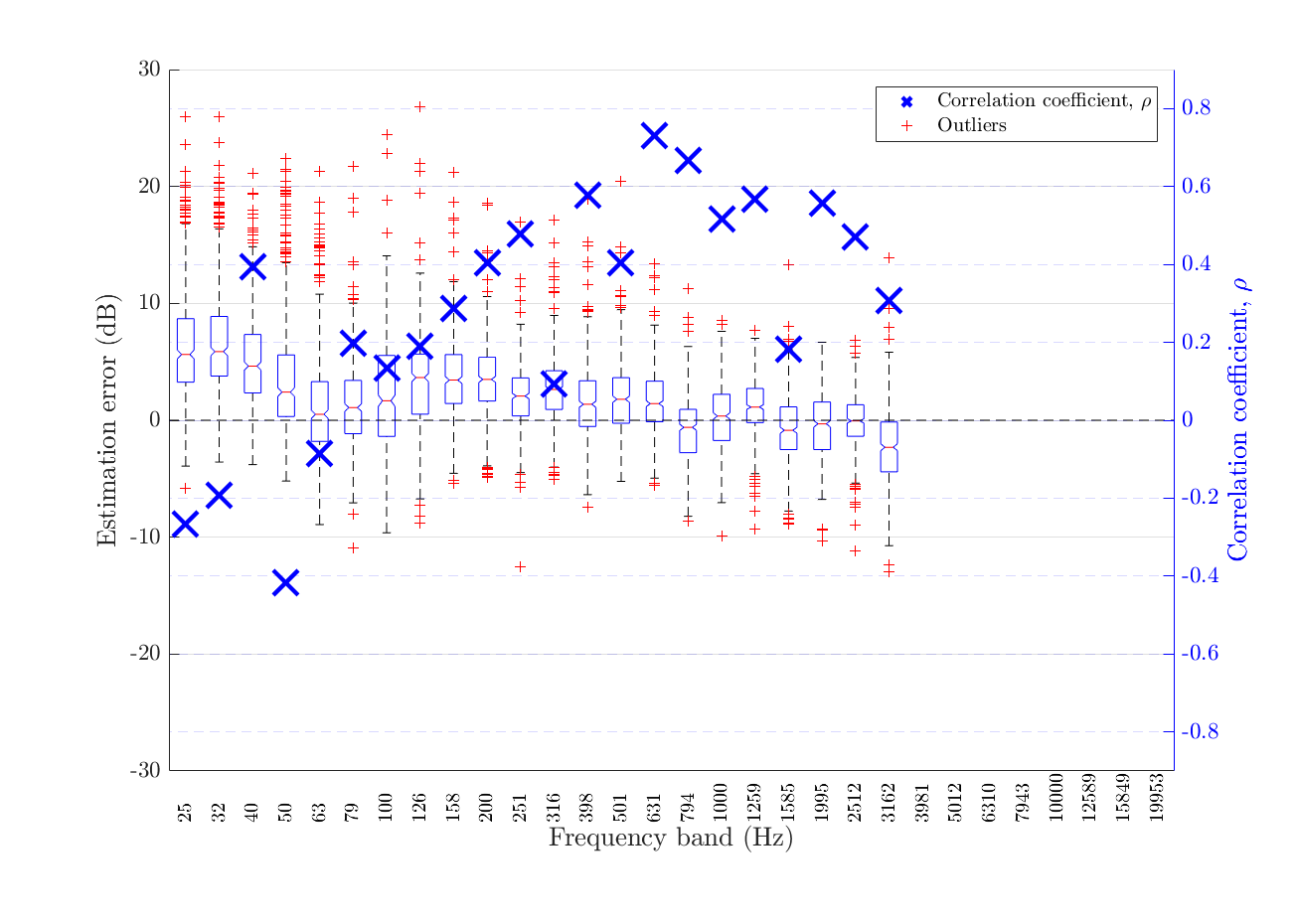,
	width=\figWidthACETR,viewport=45 10 765 530,clip}}%
	\else
	\centerline{\epsfig{figure=FigsACE/ana_eval_gt_partic_results_combined_Phase3_TR_P3S_DRR_dB_12dB_SNR_Fan_sub_Velocity.png,
	width=\figWidthACETR,viewport=45 10 765 530,clip}}%
	\fi
	\caption{{Frequency-dependent \ac{DRR} estimation error in fan noise at \dBel{12} \ac{SNR} for algorithm Particle Velocity~\cite{Chen2015}}}%
\label{fig:ACE_DRR_Sub_Fan_12dB_Velocity}%
\end{figure}%
\begin{table*}[!ht]\small
\caption{
	Frequency-dependent \ac{DRR} estimation error in fan noise at \dBel{12} \ac{SNR} for algorithm Particle Velocity~\cite{Chen2015}
}
\vspace{5mm} 
\centering
\begin{tabular}{crrrl}%
\hline%
Freq. band
& Centre Freq. (Hz)
& Bias
& \acs{MSE}
& $\PearsonCC$

\\
\hline
\hline
\ifarXiv
 1 & 25.12 & 6.373 & 63.43 & -0.2652 \\ 
\hline
 2 & 31.62 & 6.808 & 66.57 & -0.1931 \\ 
\hline
 3 & 39.81 & 5.128 & 41.74 & 0.393 \\ 
\hline
 4 & 50.12 & 3.517 & 35.93 & -0.4176 \\ 
\hline
 5 & 63.10 & 1.242 & 22.04 & -0.08476 \\ 
\hline
 6 & 79.43 & 1.304 & 16.38 & 0.1981 \\ 
\hline
 7 & 100.00 & 2.093 & 26.43 & 0.1344 \\ 
\hline
 8 & 125.89 & 3.115 & 27.96 & 0.1907 \\ 
\hline
 9 & 158.49 & 3.676 & 25.06 & 0.2859 \\ 
\hline
10 & 199.53 & 3.561 & 21.94 & 0.4053 \\ 
\hline
11 & 251.19 & 2.027 & 11.1 & 0.4792 \\ 
\hline
12 & 316.23 & 2.664 & 15.53 & 0.0931 \\ 
\hline
13 & 398.11 & 1.649 & 12.23 & 0.5775 \\ 
\hline
14 & 501.19 & 2.012 & 14.11 & 0.403 \\ 
\hline
15 & 630.96 & 1.736 & 10.56 & 0.7308 \\ 
\hline
16 & 794.33 & -0.7317 & 8.364 & 0.6664 \\ 
\hline
17 & 1000.00 & 0.3256 & 7.451 & 0.5155 \\ 
\hline
18 & 1258.93 & 1.141 & 7.829 & 0.5666 \\ 
\hline
19 & 1584.89 & -0.7265 & 9.257 & 0.1824 \\ 
\hline
20 & 1995.26 & -0.4448 & 7.139 & 0.5573 \\ 
\hline
21 & 2511.89 & -0.1216 & 5.385 & 0.4696 \\ 
\hline
22 & 3162.28 & -2.204 & 15.46 & 0.3065 \\ 
\hline

\else

\fi
\end{tabular}
\end{table*}
%
%
%
\begin{figure}[!ht]
	\ifarXiv
\centerline{\epsfig{figure=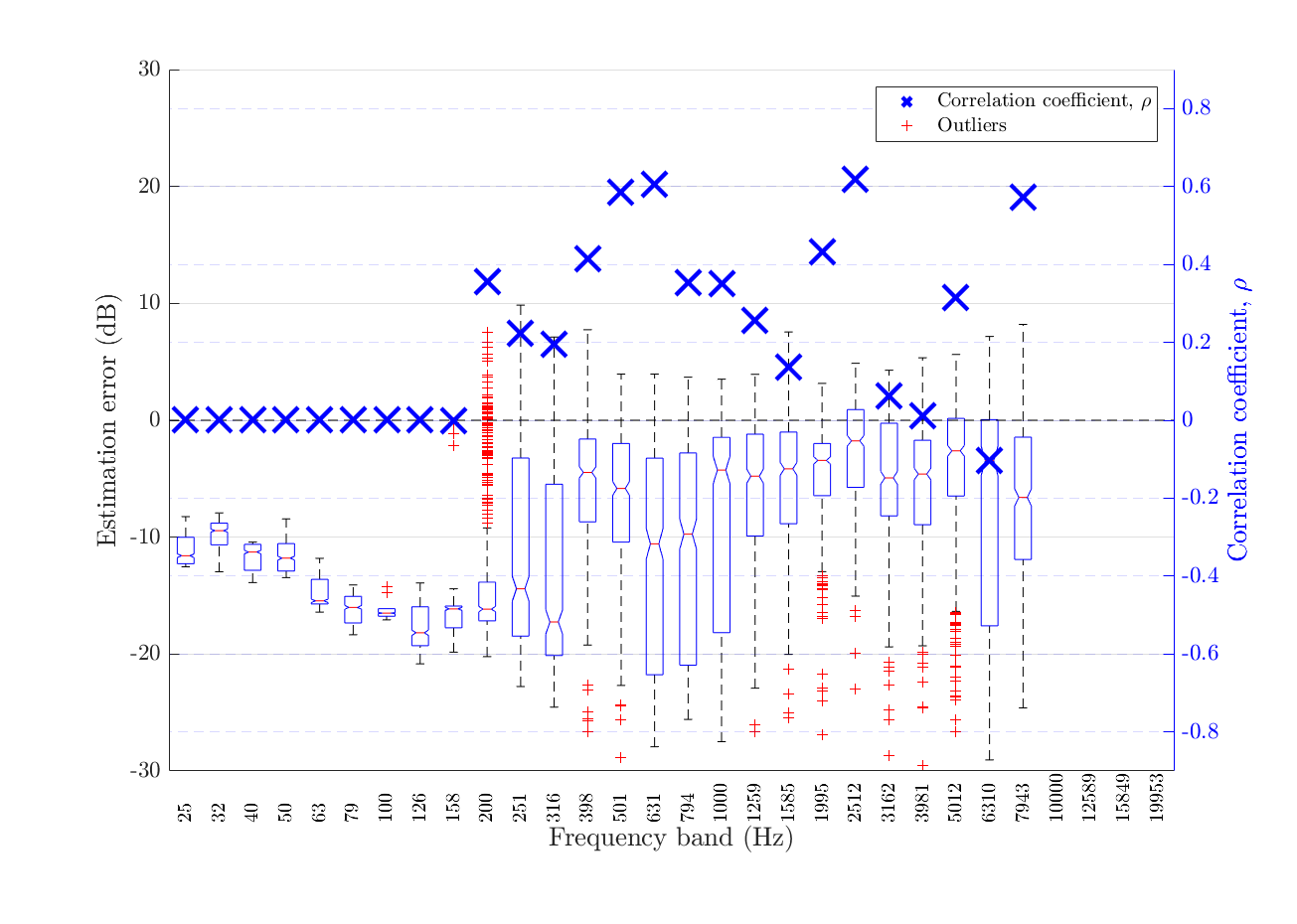,
	width=\figWidthACETR,viewport=45 10 765 530,clip}}%
	\else
	\centerline{\epsfig{figure=FigsACE/ana_eval_gt_partic_results_combined_Phase3_TR_P3S_DRR_dB_12dB_SNR_Fan_sub_ICASSP_2015_DRR_2-ch_Gerkmann_NR_FFT_subband.png,
	width=\figWidthACETR,viewport=45 10 765 530,clip}}%
	\fi
	\caption{{Frequency-dependent \ac{DRR} estimation error in fan noise  at \dBel{12} \ac{SNR} for algorithm \ac{DENBE} with FFT derived subbands~\cite{Eaton2015c}}}%
\label{fig:ACE_DRR_Sub_Fan_12dB_Eaton_FFT}%
\end{figure}%
\begin{table*}[!ht]\small
\caption{
	Frequency-dependent \ac{DRR} estimation error in fan noise  at \dBel{12} \ac{SNR} for algorithm \ac{DENBE} with FFT derived subbands~\cite{Eaton2015c}
}
\vspace{5mm} 
\centering
\begin{tabular}{crrrl}%
\hline%
Freq. band
& Centre Freq. (Hz)
& Bias
& \acs{MSE}
& $\PearsonCC$

\\
\hline
\hline
\ifarXiv
 1 & 25.12 & -11.07 & 124.4 & 0 \\ 
\hline
 2 & 31.62 & -9.93 & 100.8 & 0 \\ 
\hline
 3 & 39.81 & -11.69 & 138.1 & 0 \\ 
\hline
 4 & 50.12 & -11.55 & 135.4 & 0 \\ 
\hline
 5 & 63.10 & -14.82 & 221.5 & 0 \\ 
\hline
 6 & 79.43 & -16.09 & 260.6 & 0 \\ 
\hline
 7 & 100.00 & -16.18 & 262.6 & 0 \\ 
\hline
 8 & 125.89 & -17.74 & 318.4 & 0 \\ 
\hline
 9 & 158.49 & -16.66 & 281.5 & -0.001611 \\ 
\hline
10 & 199.53 & -13.78 & 225.6 & 0.3551 \\ 
\hline
11 & 251.19 & -11.38 & 199.2 & 0.2235 \\ 
\hline
12 & 316.23 & -13.58 & 253.6 & 0.1957 \\ 
\hline
13 & 398.11 & -5.869 & 78.75 & 0.414 \\ 
\hline
14 & 501.19 & -7.438 & 106.1 & 0.5862 \\ 
\hline
15 & 630.96 & -12.22 & 230.3 & 0.6053 \\ 
\hline
16 & 794.33 & -11.73 & 223.1 & 0.3537 \\ 
\hline
17 & 1000.00 & -8.183 & 144 & 0.3512 \\ 
\hline
18 & 1258.93 & -6.818 & 99.92 & 0.2574 \\ 
\hline
19 & 1584.89 & -4.966 & 65.99 & 0.1365 \\ 
\hline
20 & 1995.26 & -4.839 & 43.62 & 0.4311 \\ 
\hline
21 & 2511.89 & -2.527 & 25.36 & 0.6183 \\ 
\hline
22 & 3162.28 & -5.205 & 60.28 & 0.0624 \\ 
\hline
23 & 3981.07 & -6.064 & 71.88 & 0.0111 \\ 
\hline
24 & 5011.87 & -5.174 & 90.29 & 0.3143 \\ 
\hline
25 & 6309.57 & -8.275 & 195.9 & -0.1023 \\ 
\hline
26 & 7943.28 & -8.247 & 151.4 & 0.5723 \\ 
\hline

\else

\fi
\end{tabular}
\end{table*}
%
%
%
\begin{figure}[!ht]
	\ifarXiv
\centerline{\epsfig{figure=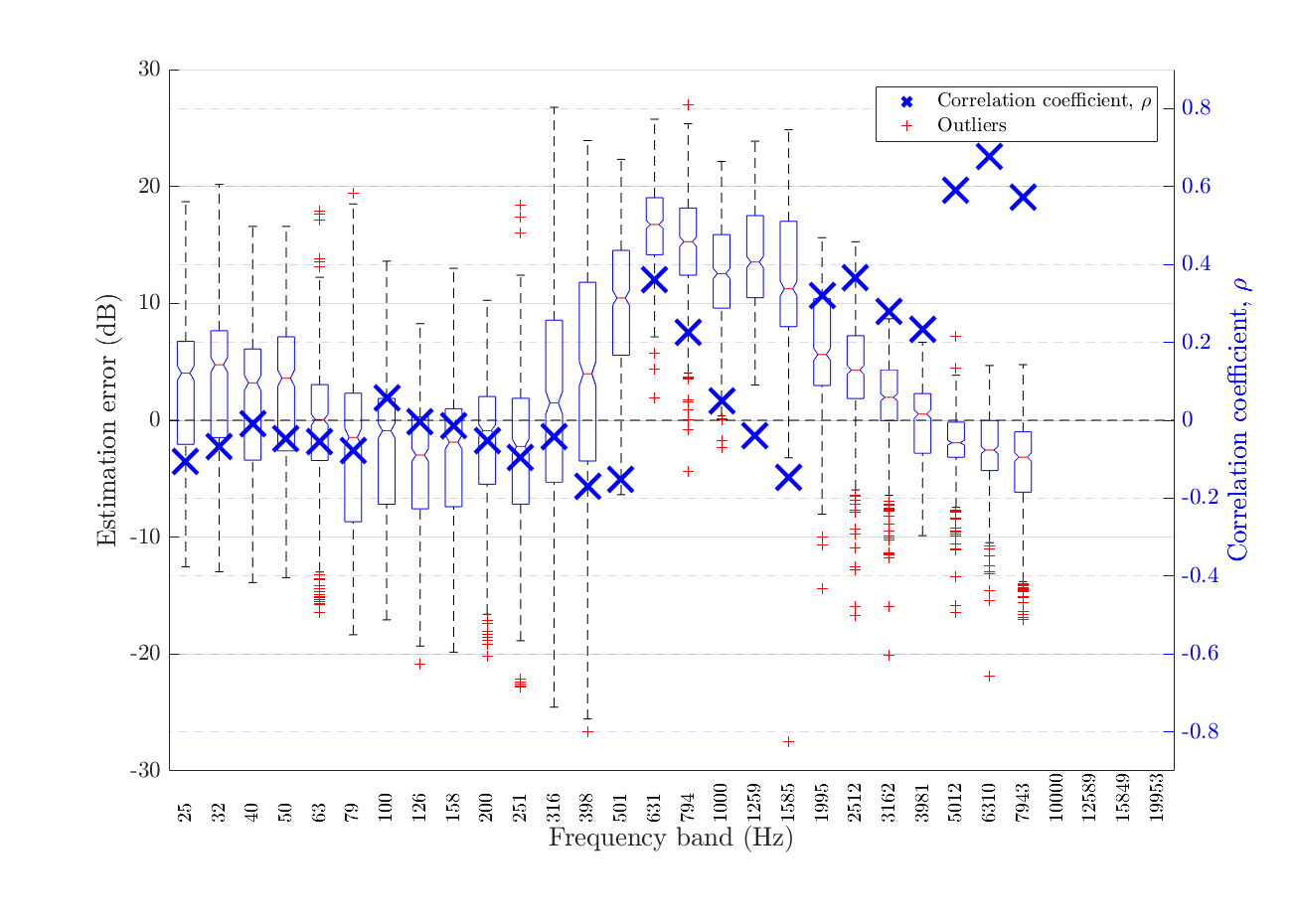,
	width=\figWidthACETR,viewport=45 10 765 530,clip}}%
	\else
	\centerline{\epsfig{figure=FigsACE/ana_eval_gt_partic_results_combined_Phase3_TR_P3S_DRR_dB_12dB_SNR_Fan_sub_ICASSP_2015_DRR_2-ch_Gerkmann_NR_filtered_subband.png,
	width=\figWidthACETR,viewport=45 10 765 530,clip}}%
	\fi
	\caption{{Frequency-dependent \ac{DRR} estimation error in fan noise  at \dBel{12} \ac{SNR} for algorithm \ac{DENBE} with filtered subbands~\cite{Eaton2015c}}}%
\label{fig:ACE_DRR_Sub_Fan_12dB_Eaton_filt}%
\end{figure}%
\begin{table*}[!ht]\small
\caption{
	Frequency-dependent \ac{DRR} estimation error in fan noise  at \dBel{12} \ac{SNR} for algorithm \ac{DENBE} with filtered subbands~\cite{Eaton2015c}
}
\vspace{5mm} 
\centering
\begin{tabular}{crrrl}%
\hline%
Freq. band
& Centre Freq. (Hz)
& Bias
& \acs{MSE}
& $\PearsonCC$

\\
\hline
\hline
\ifarXiv
 1 & 25.12 & 1.945 & 57.33 & -0.1064 \\ 
\hline
 2 & 31.62 & 2.648 & 61.43 & -0.06851 \\ 
\hline
 3 & 39.81 & 0.732 & 50.87 & -0.009295 \\ 
\hline
 4 & 50.12 & 1.479 & 55.93 & -0.04758 \\ 
\hline
 5 & 63.10 & -1.065 & 47.53 & -0.05469 \\ 
\hline
 6 & 79.43 & -3.138 & 76.39 & -0.07837 \\ 
\hline
 7 & 100.00 & -3.221 & 60.88 & 0.05739 \\ 
\hline
 8 & 125.89 & -4.621 & 67.87 & -0.003219 \\ 
\hline
 9 & 158.49 & -3.731 & 64.3 & -0.01393 \\ 
\hline
10 & 199.53 & -2.962 & 56.44 & -0.05119 \\ 
\hline
11 & 251.19 & -3.522 & 73.98 & -0.09599 \\ 
\hline
12 & 316.23 & 1.612 & 118.3 & -0.0424 \\ 
\hline
13 & 398.11 & 3.888 & 110.4 & -0.1705 \\ 
\hline
14 & 501.19 & 9.92 & 134.4 & -0.1507 \\ 
\hline
15 & 630.96 & 16.39 & 281.5 & 0.3608 \\ 
\hline
16 & 794.33 & 14.98 & 244 & 0.2259 \\ 
\hline
17 & 1000.00 & 12.57 & 180.4 & 0.05034 \\ 
\hline
18 & 1258.93 & 13.99 & 215.4 & -0.03996 \\ 
\hline
19 & 1584.89 & 11.76 & 175.6 & -0.1455 \\ 
\hline
20 & 1995.26 & 6.072 & 61.61 & 0.3206 \\ 
\hline
21 & 2511.89 & 4.316 & 41.11 & 0.3652 \\ 
\hline
22 & 3162.28 & 1.643 & 17.93 & 0.2787 \\ 
\hline
23 & 3981.07 & -0.2957 & 10.14 & 0.2337 \\ 
\hline
24 & 5011.87 & -1.847 & 11.5 & 0.5915 \\ 
\hline
25 & 6309.57 & -2.485 & 16.07 & 0.6766 \\ 
\hline
26 & 7943.28 & -3.96 & 32.92 & 0.5712 \\ 
\hline

\else

\fi
\end{tabular}
\end{table*}
%
%
%
%
\clearpage
\subsubsection{Fan noise at \dBel{-1}}
\begin{figure}[!ht]
	\ifarXiv
\centerline{\epsfig{figure=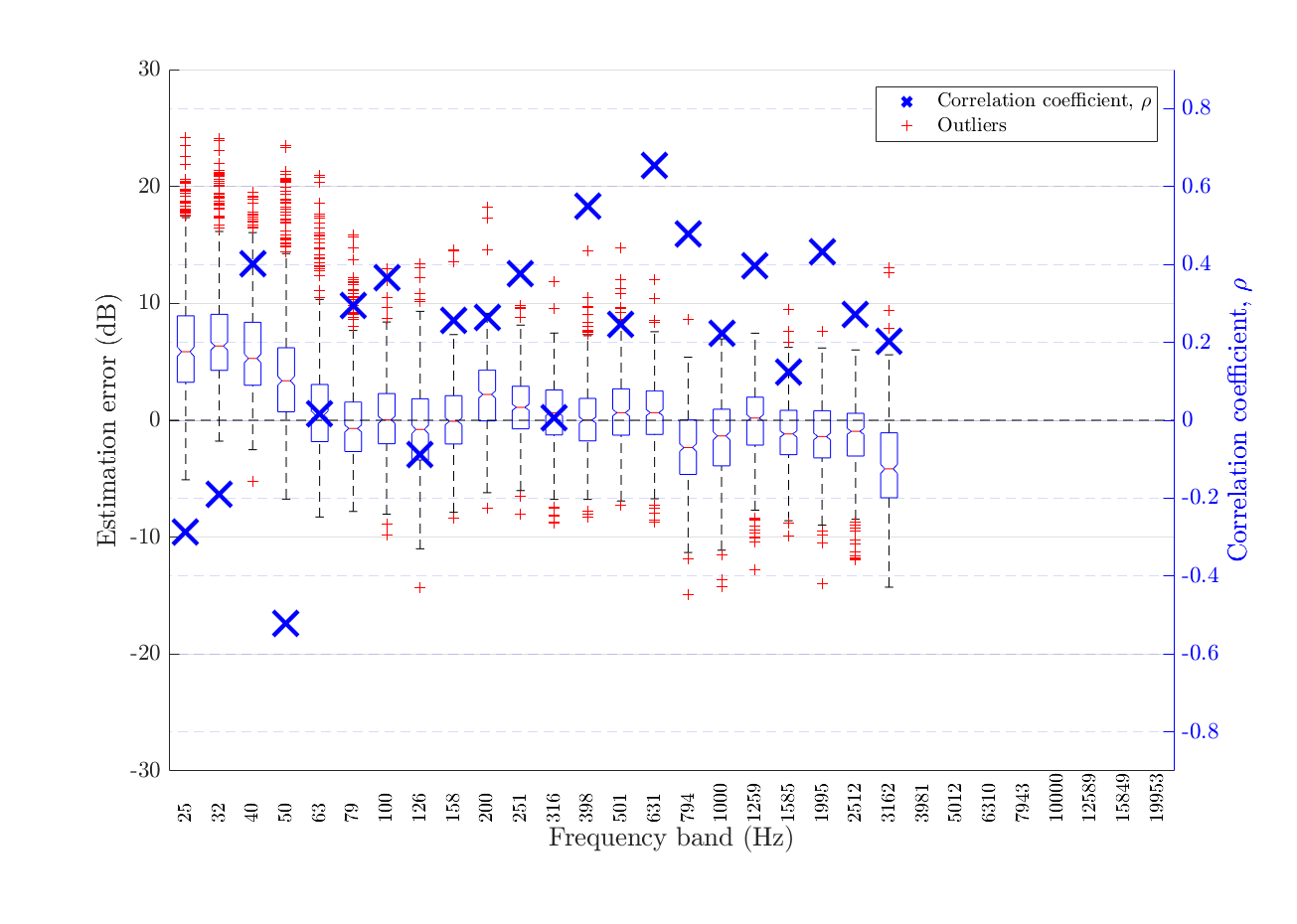,
	width=\figWidthACETR,viewport=45 10 765 530,clip}}%
	\else
	\centerline{\epsfig{figure=FigsACE/ana_eval_gt_partic_results_combined_Phase3_TR_P3S_DRR_dB_-1dB_SNR_Fan_sub_Velocity.png,
	width=\figWidthACETR,viewport=45 10 765 530,clip}}%
	\fi
	\caption{{Frequency-dependent \ac{DRR} estimation error in fan noise at \dBel{-1} \ac{SNR} for algorithm Particle Velocity~\cite{Chen2015}}}%
\label{fig:ACE_DRR_Sub_Fan_-1dB_Velocity}%
\end{figure}%
\begin{table*}[!ht]\small
\caption{
	Frequency-dependent \ac{DRR} estimation error in fan noise at \dBel{-1} \ac{SNR} for algorithm Particle Velocity~\cite{Chen2015}
}
\vspace{5mm} 
\centering
\begin{tabular}{crrrl}%
\hline%
Freq. band
& Centre Freq. (Hz)
& Bias
& \acs{MSE}
& $\PearsonCC$

\\
\hline
\hline
\ifarXiv
 1 & 25.12 & 6.8 & 72.14 & -0.287 \\ 
\hline
 2 & 31.62 & 7.455 & 78.9 & -0.19 \\ 
\hline
 3 & 39.81 & 6.075 & 55.28 & 0.4025 \\ 
\hline
 4 & 50.12 & 4.507 & 49.85 & -0.5202 \\ 
\hline
 5 & 63.10 & 1.368 & 25.83 & 0.01604 \\ 
\hline
 6 & 79.43 & -0.09664 & 15.99 & 0.2945 \\ 
\hline
 7 & 100.00 & 0.228 & 11.85 & 0.3664 \\ 
\hline
 8 & 125.89 & -0.6912 & 16.52 & -0.08779 \\ 
\hline
 9 & 158.49 & 0.03129 & 11.16 & 0.2558 \\ 
\hline
10 & 199.53 & 2.087 & 14.03 & 0.2636 \\ 
\hline
11 & 251.19 & 1.118 & 9.479 & 0.377 \\ 
\hline
12 & 316.23 & 0.4814 & 10.35 & 0.00513 \\ 
\hline
13 & 398.11 & 0.2001 & 9.138 & 0.5485 \\ 
\hline
14 & 501.19 & 0.9111 & 11.69 & 0.2467 \\ 
\hline
15 & 630.96 & 0.6737 & 9.642 & 0.653 \\ 
\hline
16 & 794.33 & -2.361 & 17.37 & 0.4788 \\ 
\hline
17 & 1000.00 & -1.593 & 15.72 & 0.2228 \\ 
\hline
18 & 1258.93 & -0.1792 & 10.81 & 0.3957 \\ 
\hline
19 & 1584.89 & -1.115 & 10.13 & 0.1225 \\ 
\hline
20 & 1995.26 & -1.244 & 10.04 & 0.4323 \\ 
\hline
21 & 2511.89 & -1.36 & 10.79 & 0.2724 \\ 
\hline
22 & 3162.28 & -3.863 & 30 & 0.2019 \\ 
\hline

\else

\fi
\end{tabular}
\end{table*}
%
%
%
\begin{figure}[!ht]
	\ifarXiv
\centerline{\epsfig{figure=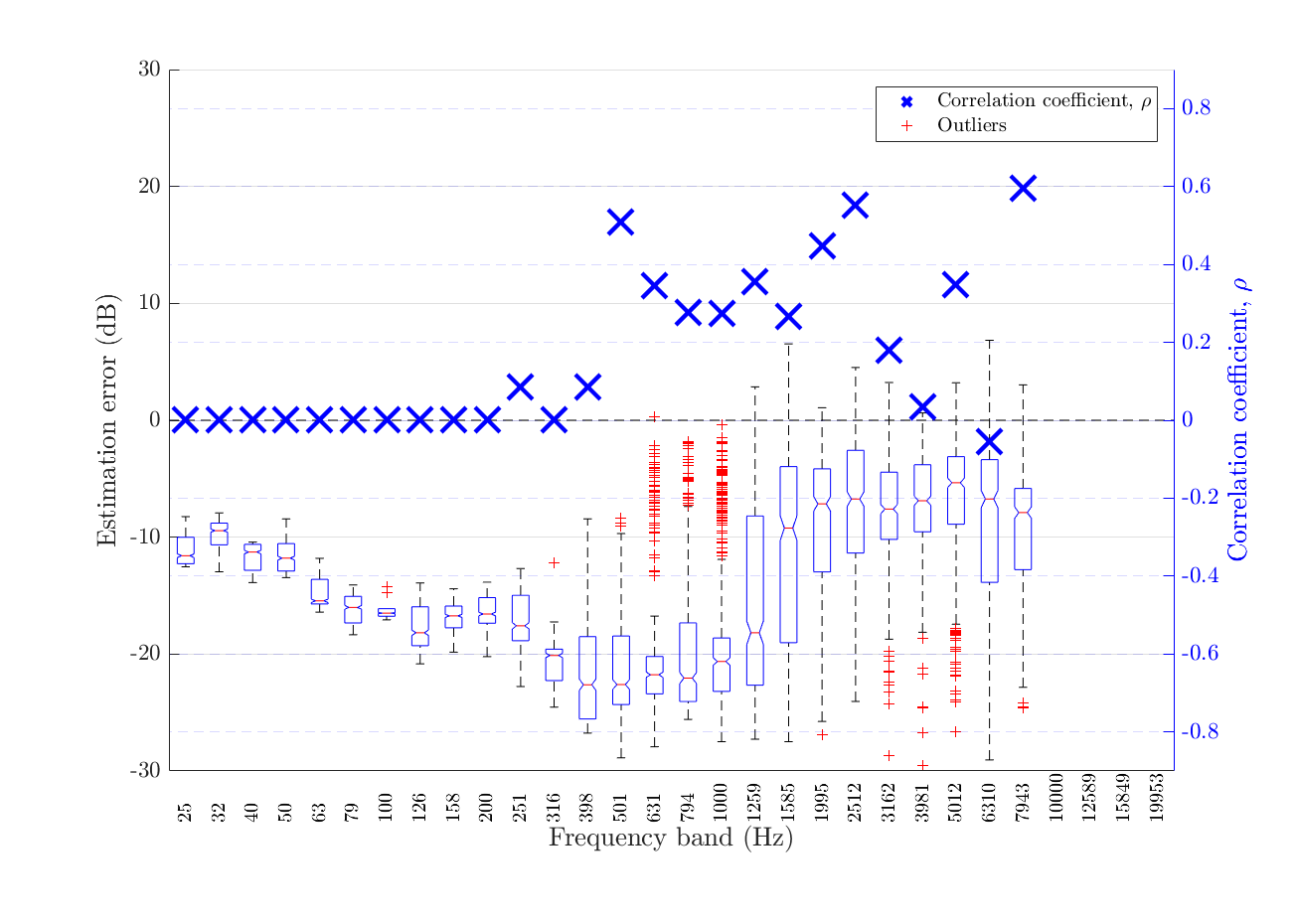,
	width=\figWidthACETR,viewport=45 10 765 530,clip}}%
	\else
	\centerline{\epsfig{figure=FigsACE/ana_eval_gt_partic_results_combined_Phase3_TR_P3S_DRR_dB_-1dB_SNR_Fan_sub_ICASSP_2015_DRR_2-ch_Gerkmann_NR_FFT_subband.png,
	width=\figWidthACETR,viewport=45 10 765 530,clip}}%
	\fi
	\caption{{Frequency-dependent \ac{DRR} estimation error in fan noise  at \dBel{-1} \ac{SNR} for algorithm \ac{DENBE} with FFT derived subbands~\cite{Eaton2015c}}}%
\label{fig:ACE_DRR_Sub_Fan_-1dB_Eaton_FFT}%
\end{figure}%
\begin{table*}[!ht]\small
\caption{
	Frequency-dependent \ac{DRR} estimation error in fan noise  at \dBel{-1} \ac{SNR} for algorithm \ac{DENBE} with FFT derived subbands~\cite{Eaton2015c}
}
\vspace{5mm} 
\centering
\begin{tabular}{crrrl}%
\hline%
Freq. band
& Centre Freq. (Hz)
& Bias
& \acs{MSE}
& $\PearsonCC$

\\
\hline
\hline
\ifarXiv
 1 & 25.12 & -11.07 & 124.4 & 0 \\ 
\hline
 2 & 31.62 & -9.93 & 100.8 & 0 \\ 
\hline
 3 & 39.81 & -11.69 & 138.1 & 0 \\ 
\hline
 4 & 50.12 & -11.55 & 135.4 & 0 \\ 
\hline
 5 & 63.10 & -14.82 & 221.5 & 0 \\ 
\hline
 6 & 79.43 & -16.09 & 260.6 & 0 \\ 
\hline
 7 & 100.00 & -16.18 & 262.6 & 0 \\ 
\hline
 8 & 125.89 & -17.74 & 318.4 & 0 \\ 
\hline
 9 & 158.49 & -16.76 & 283.2 & 0 \\ 
\hline
10 & 199.53 & -16.72 & 282.9 & 0 \\ 
\hline
11 & 251.19 & -17.65 & 321.3 & 0.0866 \\ 
\hline
12 & 316.23 & -19.99 & 410 & 0 \\ 
\hline
13 & 398.11 & -22.1 & 502.3 & 0.08582 \\ 
\hline
14 & 501.19 & -21.71 & 484.2 & 0.509 \\ 
\hline
15 & 630.96 & -21.11 & 472.3 & 0.346 \\ 
\hline
16 & 794.33 & -20.42 & 446.5 & 0.2772 \\ 
\hline
17 & 1000.00 & -19.35 & 421.9 & 0.2746 \\ 
\hline
18 & 1258.93 & -15.99 & 322.6 & 0.3556 \\ 
\hline
19 & 1584.89 & -11.44 & 211.3 & 0.2663 \\ 
\hline
20 & 1995.26 & -9.818 & 149.7 & 0.4475 \\ 
\hline
21 & 2511.89 & -7.989 & 113.3 & 0.5531 \\ 
\hline
22 & 3162.28 & -7.629 & 86.07 & 0.1787 \\ 
\hline
23 & 3981.07 & -7.42 & 80.61 & 0.03488 \\ 
\hline
24 & 5011.87 & -7.202 & 96.03 & 0.3475 \\ 
\hline
25 & 6309.57 & -10.12 & 195.3 & -0.05608 \\ 
\hline
26 & 7943.28 & -10.14 & 155.3 & 0.5944 \\ 
\hline

\else

\fi
\end{tabular}
\end{table*}
%
%
%
\begin{figure}[!ht]
	\ifarXiv
\centerline{\epsfig{figure=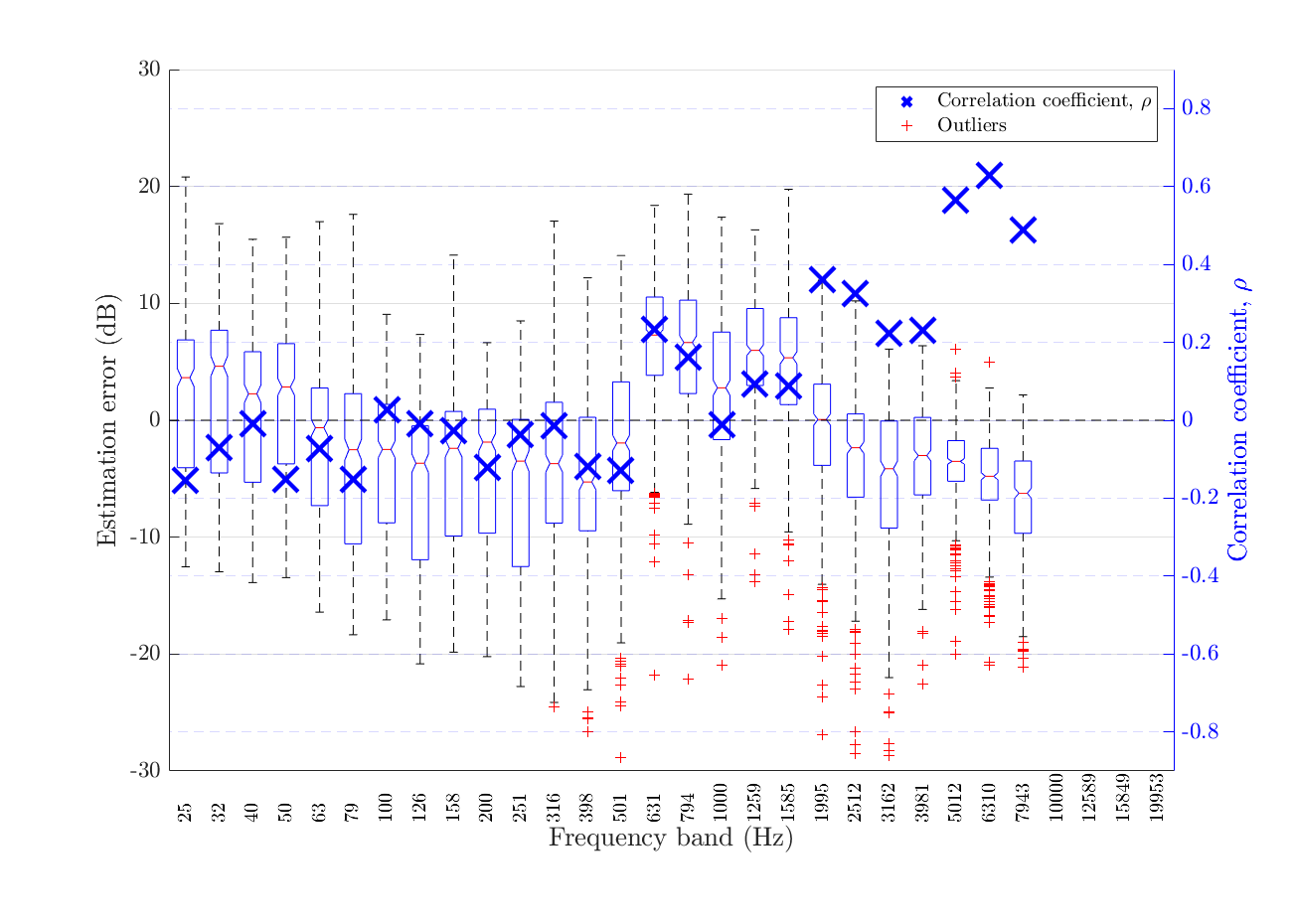,
	width=\figWidthACETR,viewport=45 10 765 530,clip}}%
	\else
	\centerline{\epsfig{figure=FigsACE/ana_eval_gt_partic_results_combined_Phase3_TR_P3S_DRR_dB_-1dB_SNR_Fan_sub_ICASSP_2015_DRR_2-ch_Gerkmann_NR_filtered_subband.png,
	width=\figWidthACETR,viewport=45 10 765 530,clip}}%
	\fi
	\caption{{Frequency-dependent \ac{DRR} estimation error in fan noise  at \dBel{-1} \ac{SNR} for algorithm \ac{DENBE} with filtered subbands~\cite{Eaton2015c}}}%
\label{fig:ACE_DRR_Sub_Fan_-1dB_Eaton_filt}%
\end{figure}%
\begin{table*}[!ht]\small
\caption{
	Frequency-dependent \ac{DRR} estimation error in fan noise  at \dBel{-1} \ac{SNR} for algorithm \ac{DENBE} with filtered subbands~\cite{Eaton2015c}
}
\vspace{5mm} 
\centering
\begin{tabular}{crrrl}%
\hline%
Freq. band
& Centre Freq. (Hz)
& Bias
& \acs{MSE}
& $\PearsonCC$

\\
\hline
\hline
\ifarXiv
 1 & 25.12 & 1.193 & 59.71 & -0.1535 \\ 
\hline
 2 & 31.62 & 2.184 & 62.21 & -0.07125 \\ 
\hline
 3 & 39.81 & 0.1626 & 53.79 & -0.009991 \\ 
\hline
 4 & 50.12 & 0.8293 & 54.66 & -0.1506 \\ 
\hline
 5 & 63.10 & -2.534 & 63.23 & -0.07202 \\ 
\hline
 6 & 79.43 & -3.812 & 87.12 & -0.1506 \\ 
\hline
 7 & 100.00 & -4.249 & 67.75 & 0.02707 \\ 
\hline
 8 & 125.89 & -5.983 & 89.82 & -0.009966 \\ 
\hline
 9 & 158.49 & -4.554 & 78.42 & -0.02563 \\ 
\hline
10 & 199.53 & -4.572 & 75.23 & -0.1223 \\ 
\hline
11 & 251.19 & -5.887 & 95.82 & -0.03686 \\ 
\hline
12 & 316.23 & -4.265 & 85.56 & -0.01496 \\ 
\hline
13 & 398.11 & -5.381 & 95.37 & -0.1178 \\ 
\hline
14 & 501.19 & -1.976 & 58.09 & -0.1292 \\ 
\hline
15 & 630.96 & 6.768 & 71.35 & 0.2326 \\ 
\hline
16 & 794.33 & 6.021 & 72.37 & 0.162 \\ 
\hline
17 & 1000.00 & 2.513 & 48.16 & -0.01271 \\ 
\hline
18 & 1258.93 & 5.994 & 58.64 & 0.09297 \\ 
\hline
19 & 1584.89 & 4.96 & 62.21 & 0.08827 \\ 
\hline
20 & 1995.26 & -0.658 & 35.28 & 0.3601 \\ 
\hline
21 & 2511.89 & -3.467 & 51.26 & 0.3254 \\ 
\hline
22 & 3162.28 & -5.099 & 63.05 & 0.2232 \\ 
\hline
23 & 3981.07 & -3.56 & 31.75 & 0.2298 \\ 
\hline
24 & 5011.87 & -3.682 & 24.87 & 0.5644 \\ 
\hline
25 & 6309.57 & -5.057 & 40.37 & 0.6292 \\ 
\hline
26 & 7943.28 & -6.982 & 71.56 & 0.4892 \\ 
\hline

\else

\fi
\end{tabular}
\end{table*}
%
%
%
%
\clearpage
\balance
\bibliographystyle{IEEEtran}
\bibliography{../SapBibTex/sapref}

\begin{thebibliography}{10}
\providecommand{\url}[1]{#1}
\def\UrlFont{\rmfamily}
\providecommand{\newblock}{\relax}
\providecommand{\bibinfo}[2]{#2}
\providecommand\BIBentrySTDinterwordspacing{\spaceskip=0pt\relax}
\providecommand\BIBentryALTinterwordstretchfactor{4}
\providecommand\BIBentryALTinterwordspacing{\spaceskip=\fontdimen2\font plus
\BIBentryALTinterwordstretchfactor\fontdimen3\font minus
  \fontdimen4\font\relax}
\providecommand\BIBforeignlanguage[2]{{%
\expandafter\ifx\csname l@#1\endcsname\relax
\typeout{** WARNING: IEEEtran.bst: No hyphenation pattern has been}%
\typeout{** loaded for the language `#1'. Using the pattern for}%
\typeout{** the default language instead.}%
\else
\language=\csname l@#1\endcsname
\fi
#2}}

\bibitem{Eaton2015a}
J.~Eaton, N.~D. Gaubitch, A.~H. Moore, and P.~A. Naylor, ``The {ACE}
  {C}hallenge - corpus description and performance evaluation,'' in \emph{Proc.
  {IEEE} Workshop on Applications of Signal Processing to Audio and Acoustics
  ({WASPAA})}, New Paltz, NY, USA, 2015.

\bibitem{Eaton2015d}
\BIBentryALTinterwordspacing
------, ``Proc. {ACE} challenge workshop, a satellite of {IEEE}-{WASPAA},'' New
  Paltz, NY, USA, 2015. [Online]. Available:
  \url{http://arxiv.org/abs/1510.00383}
\BIBentrySTDinterwordspacing

\bibitem{Lollmann2010}
H.~W. L{\"{o}}llmann, E.~Yilmaz, M.~Jeub, and P.~Vary, ``An improved algorithm
  for blind reverberation time estimation,'' in \emph{Proc. Intl. on Workshop
  Acoust. Echo and Noise Control ({IWAENC})}, Tel-Aviv, Israel, Aug. 2010, pp.
  1--4.

\bibitem{Parada2015}
P.~P. Parada, D.~Sharma, T.~{van Waterschoot}, and P.~A. Naylor, ``Evaluating
  the non-intrusive room acoustics algorithm with the {ACE} challenge,'' in
  \emph{Proc. {ACE} Challenge Workshop, a satellite of {IEEE}-{WASPAA}}, New
  Paltz, NY, USA, Oct. 2015.

\bibitem{Gaubitch2012}
N.~D. Gaubitch, H.~W. L{\"{o}}llmann, M.~Jeub, T.~H. Falk, P.~A. Naylor,
  P.~Vary, and M.~Brookes, ``Performance comparison of algorithms for blind
  reverberation time estimation from speech,'' in \emph{Proc. Intl. Workshop on
  Acoustic Signal Enhancement ({IWAENC})}, Aachen, Germany, Sept. 2012.

\bibitem{Jeub2011}
M.~Jeub, C.~M. Nelke, C.~Beaugeant, and P.~Vary, ``Blind estimation of the
  coherent-to-diffuse energy ratio from noisy speech signals,'' in \emph{Proc.
  European Signal Processing Conf. (EUSIPCO)}, Barcelona, Spain, 2011, pp.
  1347--1351.

\bibitem{Eaton2015c}
J.~Eaton and P.~A. Naylor, ``Direct-to-reverberant ratio estimation on the
  {ACE} corpus using a two-channel beamformer,'' in \emph{Proc. {ACE} Challenge
  Workshop, a satellite of {IEEE}-{WASPAA}}, New Paltz, NY, USA, Oct. 2015.

\bibitem{Chen2015}
H.~Chen, P.~N. Samarasinghe, T.~D. Abhayapala, and W.~Zhang, ``Estimation of
  the direct-to-reverberant energy ratio using a spherical microphone array,''
  in \emph{Proc. {ACE} Challenge Workshop, a satellite of {IEEE}-{WASPAA}}, New
  Paltz, NY, USA, Oct. 2015.

\bibitem{Lollmann2015}
H.~W. L\"{o}llmann, A.~Brendel, P.~Vary, and W.~Kellermann, ``Single-channel
  maximum-likelihood {$T_{60}$} estimation exploiting subband information,'' in
  \emph{Proc. {ACE} Challenge Workshop, a satellite of {IEEE}-{WASPAA}}, New
  Paltz, NY, USA, Oct. 2015.

\bibitem{ISO_3382}
\emph{Acoustics - Measurement of the Reverberation Time of Rooms with Reference
  to Other Acoustical Parameters}, Intl. Org. for Standardization ({ISO})
  Recommendation {ISO}-3382, May 2009.

\bibitem{Prego2015}
T.~de~M.~Prego, A.~A. de~Lima, R.~Zambrano-L\'{o}pez, and S.~L. Netto, ``Blind
  estimators for reverberation time and direct-to-reverberant energy ratio
  using subband speech decomposition,'' in \emph{Proc. {IEEE} Workshop on
  Applications of Signal Processing to Audio and Acoustics ({WASPAA})}, New
  Paltz, NY, USA, 2015.

\bibitem{Eaton2015b}
J.~Eaton and P.~A. Naylor, ``Reverberation time estimation on the {ACE} corpus
  using the {SDD} method,'' in \emph{Proc. {ACE} Challenge Workshop, a
  satellite of {IEEE}-{WASPAA}}, New Paltz, NY, USA, Oct. 2015.

\bibitem{Eaton2013}
J.~Eaton, N.~D. Gaubitch, and P.~A. Naylor, ``Noise-robust reverberation time
  estimation using spectral decay distributions with reduced computational
  cost,'' in \emph{Proc. {IEEE} Intl. Conf. on Acoustics, Speech and Signal
  Processing ({ICASSP})}, Vancouver, Canada, May 2013, pp. 161--165.

\bibitem{Xiong2015}
F.~Xiong, S.~Goetze, and B.~T. Meyer, ``Joint estimation of reverberation time
  and direct-to-reverberation ratio from speech using auditory inspired
  features,'' in \emph{Proc. {ACE} Challenge Workshop, a satellite of
  {IEEE}-{WASPAA}}, New Paltz, NY, USA, Oct. 2015.

\bibitem{Senoussaoui2015}
M.~Senoussaoui, J.~F. Santos, and T.~H. Falk, ``{SRMR} variants for improved
  blind room acoustics characterization,'' in \emph{Proc. {ACE} Challenge
  Workshop, a satellite of {IEEE}-{WASPAA}}, New Paltz, NY, USA, Oct. 2015.

\bibitem{Santos2014}
J.~F. Santos, M.~Senoussaoui, and T.~H. Falk, ``An improved non-intrusive
  intelligibility metric for noisy and reverberant speech,'' in \emph{Proc.
  Intl. Workshop on Acoustic Signal Enhancement ({IWAENC})}, Antibes, France,
  Sept. 2014, pp. 55--59.

\bibitem{Lim2015}
F.~Lim, M.~R.~P. Thomas, and I.~J. Tashev, ``Blur kernel estimation approach to
  blind reverberation time estimation,'' in \emph{Proc. {IEEE} Intl. Conf. on
  Acoustics, Speech and Signal Processing ({ICASSP})}, no.~40, Brisbane,
  Australia, Apr. 2015, pp. 41--45.

\bibitem{Lim2015a}
F.~Lim, M.~R.~P. Thomas, P.~A. Naylor, and I.~J. Tashev, ``Acoustic blur kernel
  with sliding window for blind estimation of reverberation time,'' in
  \emph{Proc. {IEEE} Workshop on Applications of Signal Processing to Audio and
  Acoustics ({WASPAA})}, New Paltz, NY, USA, Oct. 2015.

\bibitem{Falk2010a}
T.~H. Falk, C.~Zheng, and W.-Y. Chan, ``A non-intrusive quality and
  intelligibility measure of reverberant and dereverberated speech,''
  \emph{{IEEE} Trans. Audio, Speech, Lang. Process.}, vol.~18, no.~7, pp.
  1766--1774, Sept. 2010.

\bibitem{Wen2008}
J.~Y.~C. Wen, E.~A.~P. Habets, and P.~A. Naylor, ``Blind estimation of
  reverberation time based on the distribution of signal decay rates,'' in
  \emph{Proc. {IEEE} Intl. Conf. on Acoustics, Speech and Signal Processing
  ({ICASSP})}, Las Vegas, USA, Apr. 2008, pp. 329--332.

\bibitem{Hioka2015}
Y.~Hioka and K.~Niwa, ``{PSD} estimation in beamspace for estimating
  direct-to-reverberant ratio from a reverberant speech signal,'' in
  \emph{Proc. {ACE} Challenge Workshop, a satellite of {IEEE}-{WASPAA}}, New
  Paltz, NY, USA, Oct. 2015.

\bibitem{Hioka2012}
Y.~Hioka, K.~Furuya, K.~Niwa, and Y.~Haneda, ``Estimation of
  direct-to-reverberation energy ratio based on isotropic and homogeneous
  propagation model,'' in \emph{Proc. Intl. Workshop on Acoustic Signal
  Enhancement ({IWAENC})}, Sept. 2012.

\bibitem{Hioka2011}
Y.~Hioka, K.~Niwa, S.~Sakauchi, K.~Furuya, and Y.~Haneda, ``Estimating
  direct-to-reverberant energy ratio using {D/R} spatial correlation matrix
  model,'' \emph{{IEEE} Trans. Audio, Speech, Lang. Process.}, vol.~19, no.~8,
  pp. 2374--2384, Nov. 2011.

\bibitem{Eaton2015}
J.~Eaton, A.~H. Moore, P.~A. Naylor, and J.~Skoglund, ``Direct-to-reverberant
  ratio estimation using a null-steered beamformer,'' in \emph{Proc. {IEEE}
  Intl. Conf. on Acoustics, Speech and Signal Processing ({ICASSP})}, Brisbane,
  Australia, Apr. 2015, pp. 46--50.

\bibitem{Falk2009}
T.~H. Falk and W.-Y. Chan, ``Temporal dynamics for blind measurement of room
  acoustical parameters,'' \emph{{IEEE} Trans. Instrum. Meas.}, vol.~59, no.~4,
  pp. 978--989, 2010.

\end{thebibliography}
\label{sec:bibliography}
\end{sloppy}
\end{document}